\documentclass[preprint,11pt]{aastex}

\usepackage{graphics}
\usepackage{natbib}
\usepackage{multirow}
\usepackage{rotating}
\usepackage{amsfonts}
\usepackage{amsmath}
\usepackage{enumerate}
\usepackage{tabularx}
\usepackage{natbib,amsmath,amssymb,graphicx,lscape,longtable,multirow,float,url,footnote}
\usepackage{url}

\usepackage{array}
\newcolumntype{+}{>{\global\let\currentrowstyle\relax}}
\newcolumntype{^}{>{\currentrowstyle}}

\graphicspath{{figs/}}

\begin{document}

\title{Retrieving Neptune's aerosol properties from Keck OSIRIS observations. I. Dark regions}

\shortauthors{S. H. Luszcz-Cook, I. de Pater, K. deKleer, M. Adamkovics, \$ H.B. Hammel}
\slugcomment{{\sc Published in Icarus: 15 September 2016. \\ https://doi.org/10.1016/j.icarus.2016.04.032}} 

\author{
  S. H. Luszcz-Cook\altaffilmark{1},
  K. de Kleer\altaffilmark{2},
  I. de Pater\altaffilmark{2,}\altaffilmark{3},
  M. Adamkovics\altaffilmark{2},
  H.B. Hammel\altaffilmark{4}
}

\altaffiltext{1}{Astrophysics Department, American Museum of Natural History, Central Park West at 79th Street, New York, NY 10024, USA}
\altaffiltext{2}{Astronomy Department, University of California, Berkeley, CA 94720, USA}
\altaffiltext{3}{SRON Netherlands Institute for Space Research, 3584 CA Utrecht, and Faculty of Aerospace Engineering, Delft University of Technology, 2629 HS Delft, The Netherlands}
 \altaffiltext{4}{AURA, 1212 New York Ave. NW, Suite 450, Washington, DC 20005}


\begin{abstract}

We present and analyze three-dimensional data cubes of Neptune from the OSIRIS integral-field spectrograph on the 10-meter W.M. Keck II telescope, from 26 July 2009. These data have a spatial resolution of 0.035''/pixel and spectral resolution of R$\sim$3800 in the H (1.47--1.80 $\mu$m) and K (1.97--2.38 $\mu$m) broad bands. We focus our analysis on regions of Neptune's atmosphere that are near-infrared dark -- that is, free of discrete bright cloud features. We use a forward model coupled to a Markov chain Monte Carlo algorithm to retrieve properties of Neptune's aerosol structure and methane profile above $\sim4$ bar in these near-infrared dark regions.

We construct a set of high signal-to-noise spectra spanning a range of viewing geometries to constrain the vertical structure of Neptune's aerosols in a cloud-free latitude band from 2--12$^\circ$N. We find that Neptune's cloud opacity at these wavelengths is dominated by a compact, optically thick cloud layer with a base near 3 bar.  Using the pyDISORT algorithm for the radiative transfer and assuming a Henyey-Greenstein phase function, we observe this cloud to be composed of low albedo (single scattering albedo $ = 0.45^{+0.01}_{-0.01}$), forward scattering (asymmetry parameter $g = 0.50^{+0.02}_{-0.02}$) particles, with an assumed characteristic size of $\sim 1 \mu$m. Above this cloud, we require an aerosol layer of smaller ($\sim 0.1 \mu$m) particles forming a vertically extended haze, which reaches from the upper troposphere ($0.59^{+0.04}_{-0.03}$ bar) into the stratosphere. The particles in this haze are brighter (single scattering albedo $=0.91^{+0.06}_{-0.05}$) and more isotropically scattering (asymmetry parameter $g= 0.24^{+0.02}_{-0.03}$) than those in the deep cloud.  When we extend our analysis to 18 cloud-free locations from 20$^\circ$N to 87$^\circ$S, we observe that the optical depth in aerosols above 0.5 bar decreases by a factor of 2--3 or more at mid- and high-southern latitudes relative to low latitudes. 

We also consider Neptune's methane (CH$_4$) profile, and find that our retrievals indicate a strong preference for a low methane relative humidity at pressures where methane is expected to condense.  When we include in our fits a parameter for methane depletion below the CH$_4$ condensation pressure, our preferred solution at most locations is for a methane relative humidity below 10\% near the tropopause in addition to methane depletion down to 2.0--2.5 bar. We tentatively identify a trend of lower CH$_4$ columns above 2.5 bar at mid- and high-southern latitudes over low latitudes, qualitatively consistent with what is found by \cite{kark11a}, and similar to, but weaker than, the trend observed for Uranus.


\vspace{\baselineskip}
\small{NOTICE: this is the authors' version of a work that was accepted for publication in Icarus. Changes resulting from the publishing process may not be reflected in this document. Figures have been compressed; for higher resolution figures, please contact the corresponding author at shcook@amnh.org.}
\end{abstract}
\clearpage

\section{Introduction}

Nearly three decades after the flyby of {\it Voyager 2}, Neptune's upper troposphere and stratosphere remain enigmatic in terms of composition, circulation, and spatial and temporal variability. Aerosols represent a key probe of the atmospheric physics at these altitudes: aerosols trace dynamical motions associated with large-scale circulation patterns and localized storms, and influence the structure of the atmosphere by contributing opacity and by modifying the composition and thermal profile. 

\begin{figure}[htb!]
\begin{center}$
\begin{array}{l r}
\includegraphics[width=0.49\textwidth]{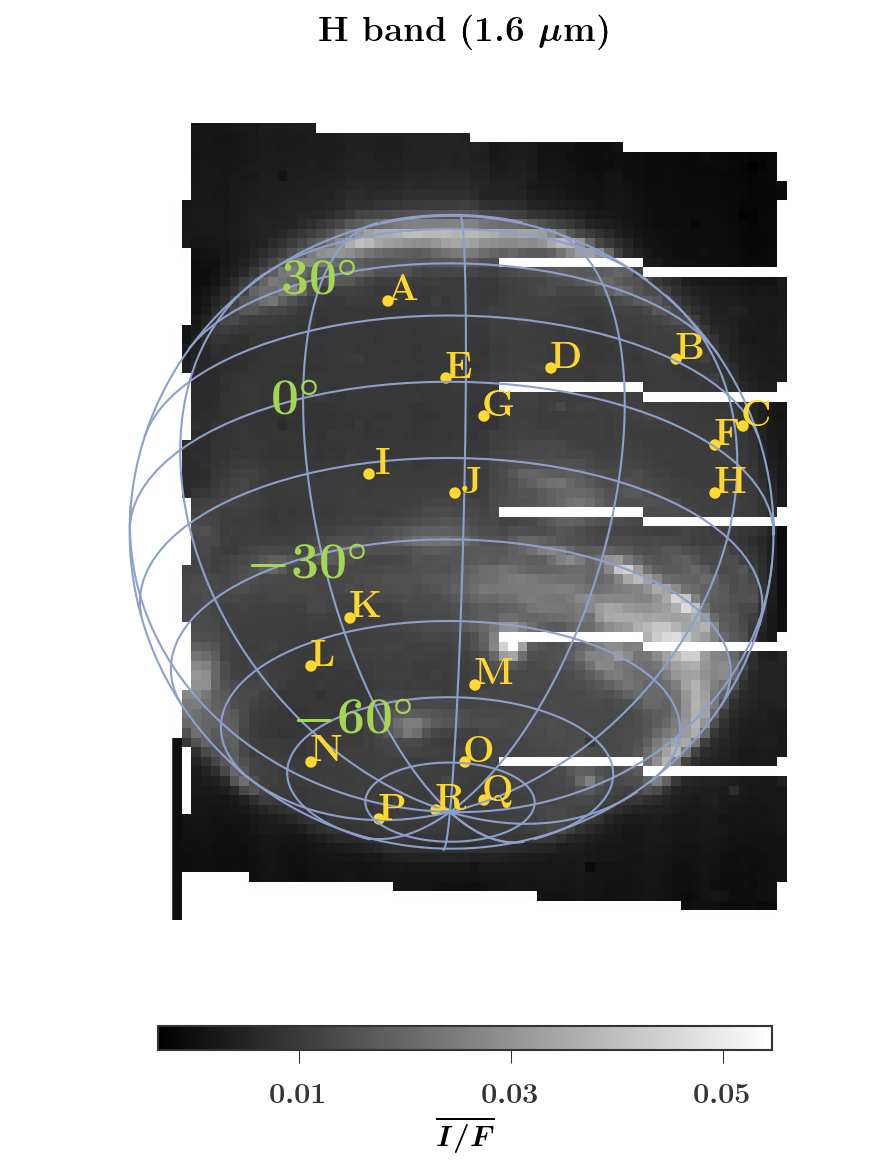}
\includegraphics[width=0.49\textwidth]{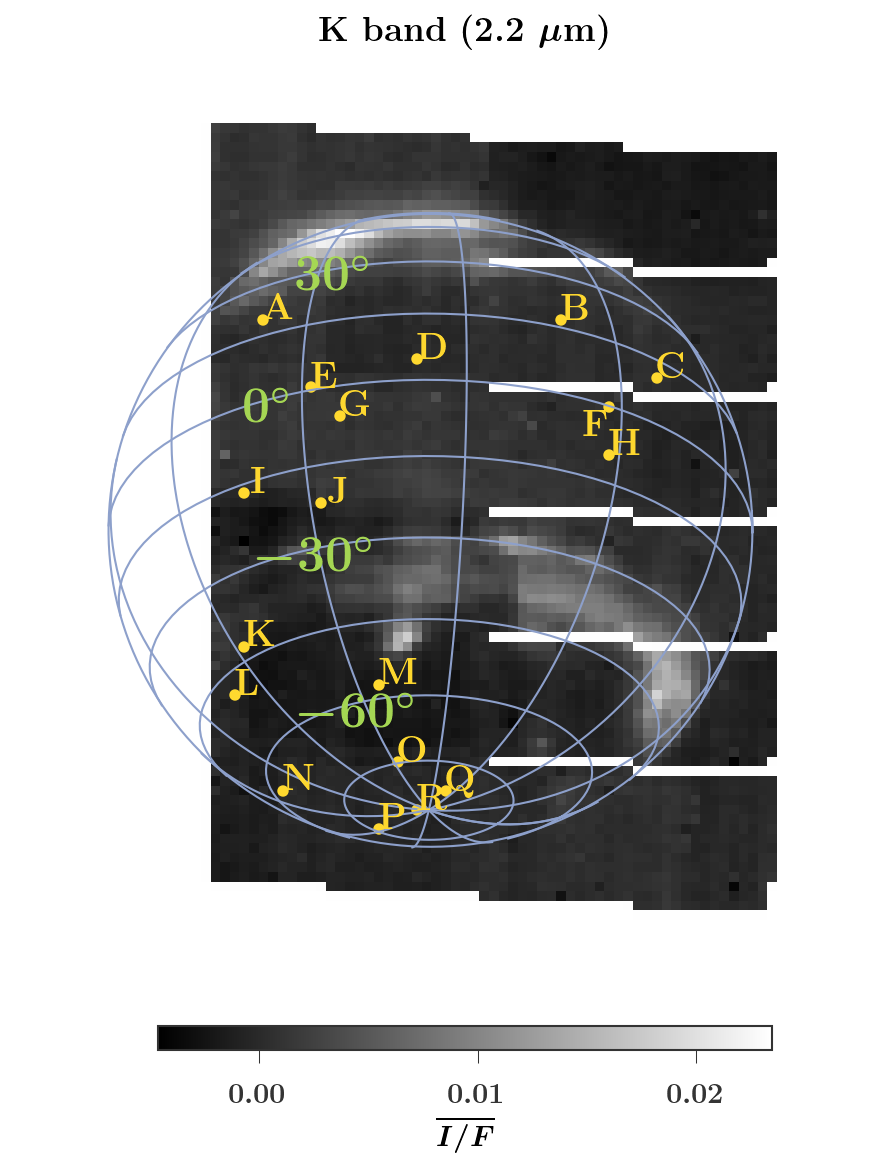}
\end{array}
$
\end{center}
\caption[18 dark regions]{\label{fig:DRcubes}\small  Mosaicked and calibrated 26 July 2009 data cubes, averaged over wavelength. Color bars indicate average reflectance, in units of $I/F$. Pixels which are flagged at all wavelengths (see Section \ref{sec:obs_dr}) are shown in white. The labels  indicate the 18 locations analyzed in Section \ref{sec:DR}. The locations are labeled A--R in order of decreasing latitude. To view the wavelength-averaged cubes without feature labels, see Supplementary Fig. \ref{fig:cubes}. }
\end{figure}

Numerous photometric and spectroscopic studies have targeted the distribution and properties of Neptune's aerosols; the majority of these efforts have focused on the discrete, bright clouds, which dominate Neptune's near-infrared (NIR) appearance  \citep[e.g.,][]{roe01a, srom01c, gibbard03,max03, luszcz10,irwin11,irwin14,depater14}.  These clouds tend to be concentrated in latitude bands \citep[][Fig. \ref{fig:DRcubes}]{smith89}, and are typically found to be at altitudes near or below the tropopause, which may suggest a composition of methane ice \citep{roe01a,srom01c,max03,gibbard03, luszcz10,kark11a}. Methane has an equilibrium condensation temperature of  $\sim$ 80K, which corresponds roughly to the 1.5-bar level in Neptune's atmosphere. Methane ice clouds should therefore form at or above this level. In some cases, a fraction of the observed clouds appear to be well (more than a scale height) above the tropopause  \citep{gibbard03,irwin11,depater14}. It is possible that these clouds originate from the settling of hydrocarbons produced by photochemistry \citep{gibbard03}; from localized upwelling of methane within an anticyclone \citep{depater14}; or from a mix of methane ice and settling hydrocarbon haze particles \citep{irwin11}. Latitudinal trends in the vertical location of clouds have been observed: \cite{gibbard03} note an increase in the altitudes of clouds from the south pole to northern midlatitudes. Higher resolution studies  indicate more complex patterns in cloud altitudes \citep{irwin11,fitzpatrick14,depater14}. 

In this paper we focus on Neptune's three-dimensional aerosol structure within NIR ``dark regions'' -- regions free of discrete bright clouds, rather than on the bright clouds themselves. In particular, we seek to characterize the other layers of aerosols that are both expected and observed in Neptune's atmosphere, including global cloud decks, tropospheric haze, and stratospheric haze. The properties and distribution of aerosols in dark regions have implications for Neptune's global circulation pattern and the mechanisms of cloud and haze production, and serve as a baseline atmospheric structure that can be used when interpreting the spectra of discrete bright clouds.

Thermochemical equilibrium models indicate that Neptune's condensible gases should form a series of global cloud layers in the troposphere \citep{depater91}. Based on these models and radio observations, we expect the presence of an H$_2$S ice cloud with a base pressure of 6--7 bar (equilibrium condensation temperature of $\sim140$K).  Initially, photometric and spectroscopic data suggested the presence of an optically thick cloud with a top around 3--5 bar  \citep{hammel89, baines90, baines95b, baines95a}, which could be the putative H$_2$S ice cloud. Since then, several analyses have presupposed a cloud with these properties  \citep[e.g.][]{srom01c, gibbard02, luszcz10}. However, other results indicate this cloud may be optically thin, or may not exist at all, with the tropospheric opacity instead arising from vertically extended haze \citep{srom01c,kark11a,kark11b}. If an optically thick cloud deck is indeed present at these depths,  it must have a very low single scattering albedo in the NIR (less than 0.1--0.2 at 1.6 microns -- \cite{srom01c, roe01b}). 

Above the H$_2$S cloud deck, thermochemical models predict CH$_4$ condensation \citep{depater91}. Some atmosphere models posit a transparent, horizontally homogenous layer of condensed methane in addition to any discrete, NIR-bright methane ice clouds \citep{baines90,baines94,baines95b,srom01b}. The base of this methane haze is presumed to be near the methane condensation level. \cite{srom01b} determined that the global average optical depth of this layer is 0.11 at 0.75 microns, higher than the values of 0.03 and 0.05 inferred by \cite{baines90} and \cite{baines94}, respectively for the global average, but less than the values found for latitudes of 22--30$^\circ$S  by \cite{pryor92,baines94}, assuming isotropic scattering. Assuming the methane haze is composed of $\gtrapprox 1 \mu$m Mie scatterers \citep{conrath91,baines94, burgdorf03},  the NIR optical depth of this haze is predicted to be comparable to that at 0.75 $\mu$m. Other atmospheric models have a very different aerosol structure in the upper troposphere:  \cite{gibbard02} found that a haze at an altitude of 0.3 bar provides a good fit to their NIR imaging data of cloud-free regions, although more complex, multi-layer aerosol distributions are also permitted. \cite{irwin11} treat NIR-dark regions in the same ways as  bright regions-- with a two-layer aerosol structure consisting of a cloud deck at 2 bar and a horizontally varying upper cloud deck at 0.02--0.2 bar. In their 2014 reanalysis of these data, they model a dark belt near 10$^\circ$S and placed the upper cloud at 0.3 bar. The extended haze model of \cite{kark11a} is free of aerosols above the 1.4 bar pressure level, aside from discrete clouds, although \cite{kark11b} consider the possibility of an extended upper tropospheric haze between the tropopause and 1.4 bar, which is 10 times thinner than their main haze layer and is latitudinally variable. 

In the stratosphere, methane is photodissociated by solar ultraviolet photons, which results in the production of a number of hydrocarbons-- primarily ethane and acetylene \citep{romani93}. As these hydrocarbons settle, they condense into optically thin haze layers. The observations of \cite{pryor92} and \cite{baines94} indicate that the global average optical depth for the stratospheric aerosols is of order 0.1 at 0.75 $\mu$m, with a preferred mean particle radius of 0.2 $\mu$m.  Using Mie theory to extrapolate to NIR wavelengths, we find the predicted opacity of such particles is $<0.02$ at 1.6 $\mu$m. \cite{kark11a} find the stratospheric haze optical depth to be an order of magnitude lower than \cite{pryor92}; they note that the discrepancy is primarily due to different assumptions for the single scattering albedo by the two authors.  Meridional variability -- specifically a deficit in stratospheric haze near the equator -- was observed by \cite{baines94}.
 
\begin{figure}
\begin{center}$
\begin{array}{ccc}
\includegraphics[width=0.43\textwidth]{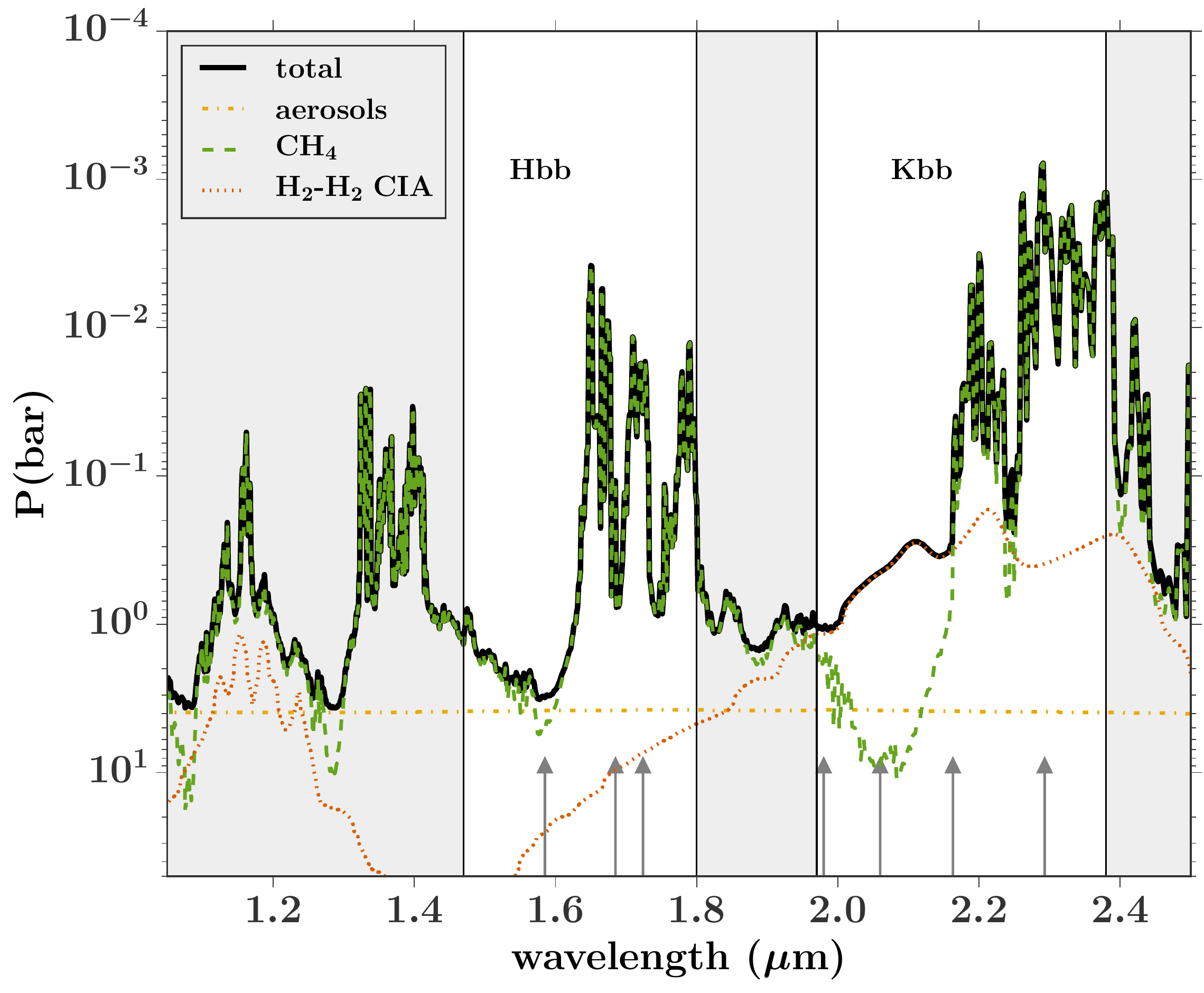} \includegraphics[width=0.28\textwidth]{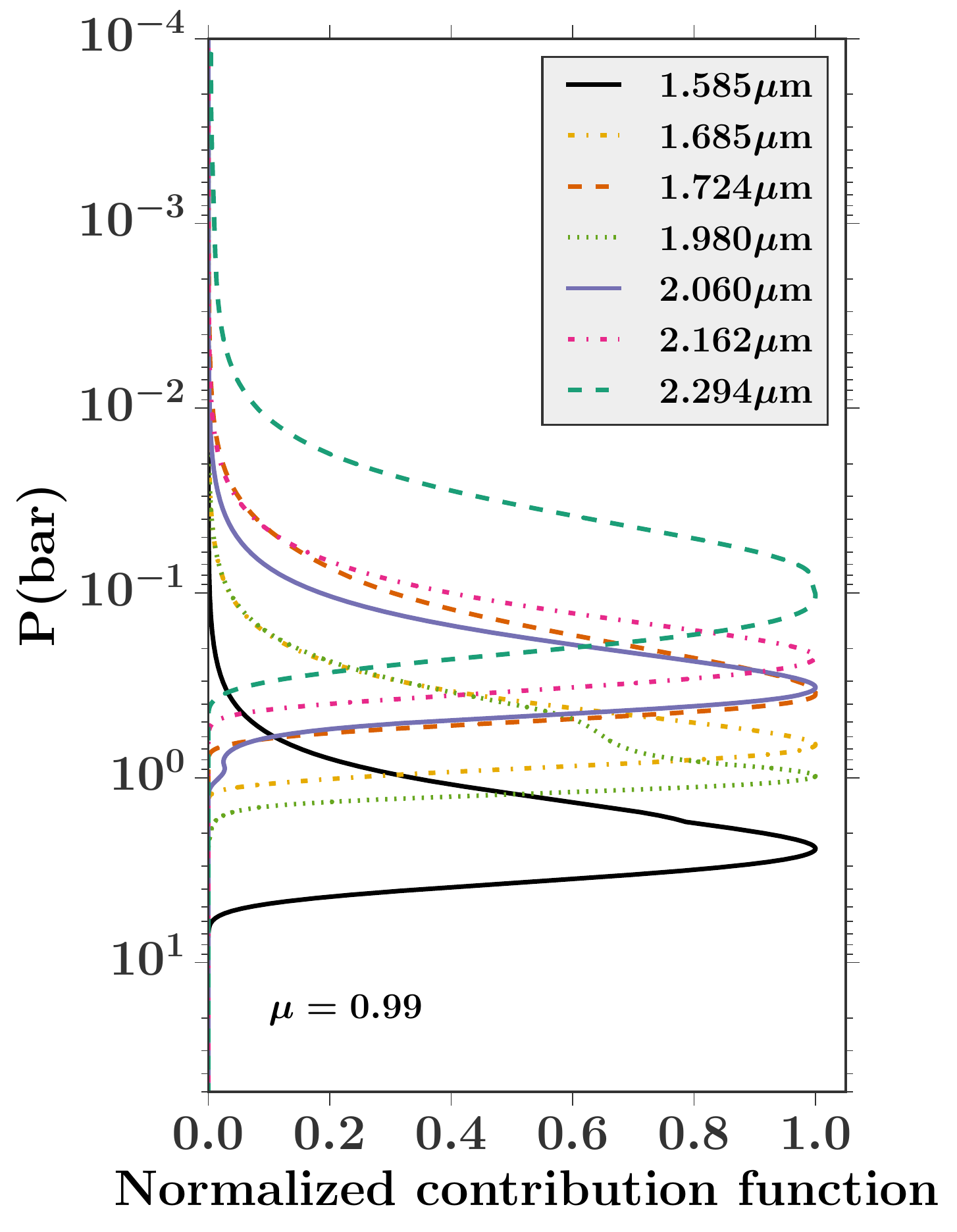}\includegraphics[width=0.28\textwidth]{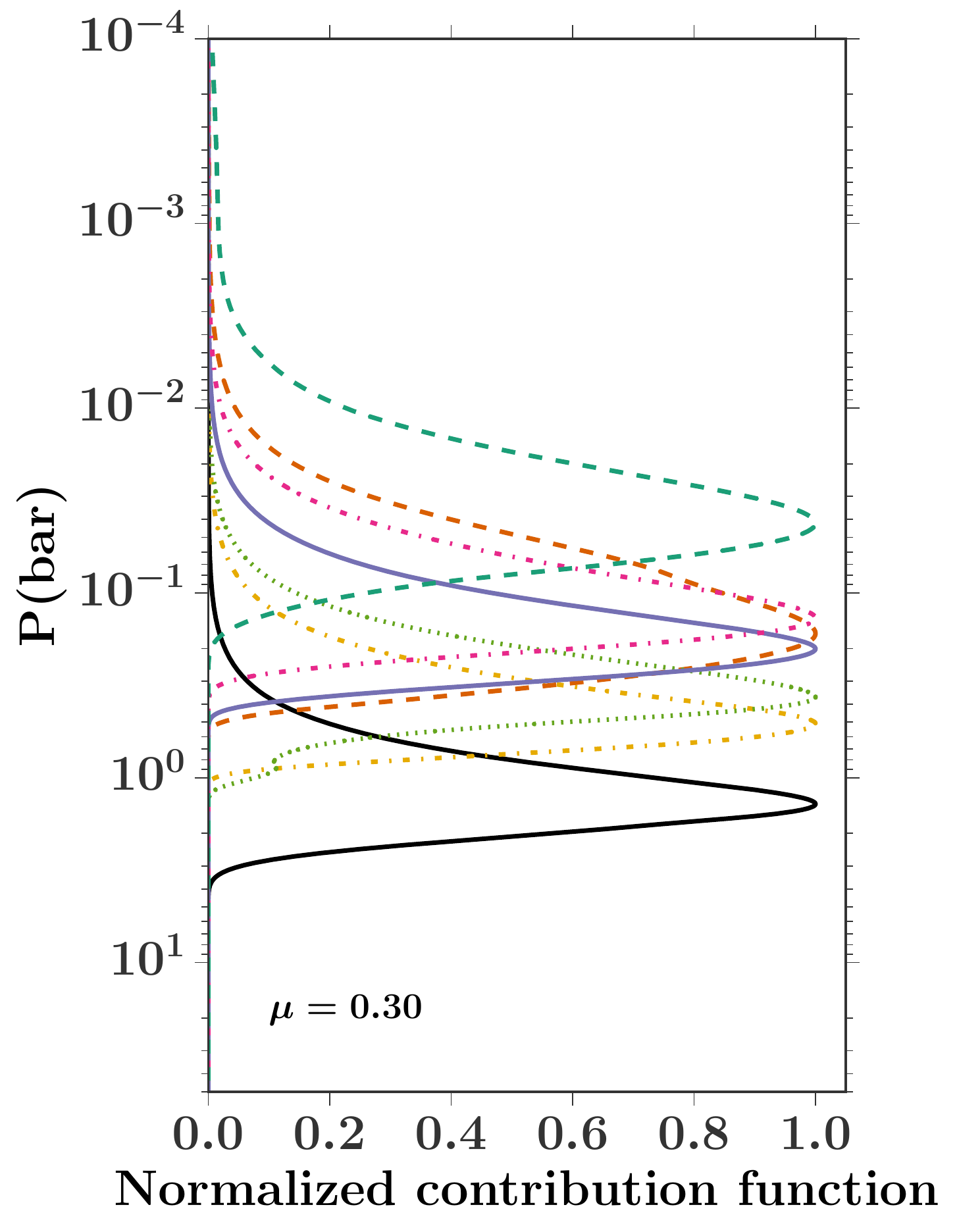}\\

\end{array}$
\end{center}
\caption[Transmission and contribution functions - 1L]{\label{fig:1L_contrib} \small Transmission and contribution functions for Neptune. Left: transmission depth as a function of wavelength, for a sample model atmosphere. In this model, our nominal thermal profile and initial methane profile, as described in Appendix \ref{app:model}, and a single, optically thick tropospheric aerosol layer (the best-fit {\it1L\_nom} model -- Section \ref{sec:1L}) are used.   The curves indicate the pressure at which the one-way optical depth is 1.0; the black solid curve includes all absorbers, while the other curves are for single absorbers. Note that while CH$_4$ absorption dominates at a majority of wavelengths, other wavelengths, particularly in  K band, are dominated by H$_2$ CIA or aerosols. Middle and right: normalized contribution functions for the wavelengths indicated by grey arrows in the transmission plot, for two different viewing geometries (disk center and $\mu=0.3$, respectively). The wavelengths are indicated in the legend. These functions are based on the same atmospheric model as the transmission plot. Note how the spectra are sensitive to the atmosphere from several bar up to $\sim$10 mbar, with sensitivity to the lowest pressure levels arising from wavelengths in K band. Note also how the sensitivity changes as a function of viewing geometry.   }
\end{figure}

With the objective of improving the characterization of Neptune's aerosols, we have obtained H- and K- broadband (1.47 -- 2.38 $\mu$m)  spectroscopic observations of Neptune with the OSIRIS AO-assisted integral field spectrograph on the Keck II telescope. The NIR is well suited to studying the distribution of aerosols in the upper troposphere and lower stratosphere, since wavelength variations in methane opacity result in sensitivity to a broad range of atmospheric depths (Fig. \ref{fig:1L_contrib}). Using these data, we retrieve Neptune's aerosol properties in dark regions at latitudes from 20$^\circ$N to 87$^\circ$S, considering the effects of signal to noise (S/N) and viewing geometry changes (center-to-limb variations) on our results. Our observations offer several advantages over previous studies for characterizing Neptune's dark regions: whereas traditional slit spectroscopy does not facilitate the clean separation of the quiescent background from the bright features that typically dominate the observed NIR intensity, OSIRIS produces three-dimensional data cubes ($x$, $y$, and wavelength); these data offer moderate (R$\sim3800$) spectral resolution at very good spatial resolution of 0.06--0.08'' across more than 90\% of Neptune's visible hemisphere, and are therefore well-suited for extracting spectra of dark regions that are relatively free of contamination from nearby bright features. Furthermore, in contrast to slit spectroscopy, the OSIRIS cubes are ideal for selecting spectra at desired latitudes and viewing geometries.  These data represent more than a factor of two increase in spatial resolution over the NIFS integral field spectrograph observations previously reported by \cite{irwin11,irwin14} and the first published integral field spectrograph observations of Neptune in K band. As demonstrated in Fig. \ref{fig:1L_contrib}, the addition of the K-band data increases our sensitivity to higher altitudes (see also Supplementary Fig. \ref{fig:2L_contrib}). 

The structure of this paper is as follows: in Section \ref{sec:obs_dr} we describe the observations and data reduction. Section \ref{sec:modeling} presents an overview of the radiative transfer code and atmospheric retrieval procedure. In Section \ref{sec:limb}, the latitude band from 2--12$^\circ$N is analyzed using binned spectra spanning a range of viewing geometries. In Section \ref{sec:mu8}, we evaluate the effects of the vertical temperature and CH$_4$ profiles on our derived aerosol structure for a high signal-to-noise (S/N), binned spectrum. Section \ref{sec:DR} broadens our investigation to additional locations on Neptune. A summary and discussion are presented in Sections \ref{sec:summ} and \ref{sec:disc}.

\section{Observations and data reduction} \label{sec:obs_dr}
\subsection{Observations}
We observed Neptune in the H (1.47 -- 1.80 $\mu$m) and K (1.97--2.38 $\mu$m) broadband filters (referred to as Hbb and Kbb, respectively) on 25 and 26 July 2009 with the near-infrared imaging spectrometer OSIRIS on the 10-meter W. M. Keck II telescope. The intention was to obtain data from both hemispheres of Neptune, which are observable on consecutive nights due to Neptune's rotation period of $\sim16$ hours \citep{kark11c}. However, the seeing conditions on 25 July 2009 were unstable, resulting in poor quality data on that night, particularly in H band. Therefore we present only the data from 26 July 2009. 

In broad band mode, $\sim16\times64$ lenslets in the OSIRIS integral field spectrograph image sampler are illuminated. We selected an instrumental plate scale of 0.035''; at this resolution, a 0.56''$\times$2.24'' patch of the sky is sampled with each exposure. The light passing through each individual lenslet is diffracted to produce a moderate-resolution (R$\sim3800$) spectrum at each of 1019 locations. 

Neptune's angular diameter on 26 July 2009 was 2.35'' and its distance from Earth was 29.1 AU. At this distance, each 0.035'' pixel corresponds to 740 km at disk center. We oriented the exposures so that the long direction of each frame pointed east-west (parallel to the equator) along the planet. Since Neptune's diameter was slightly larger than the length of the field of view, the western limb of the planet was not observed. A set of eight 300-second exposures were taken in each of the Hbb and Kbb filters. The first and last frames of each set were `sky' frames, for subtracting the infrared background. The remaining exposures were positioned to obtain a series of contiguous horizontal slices starting at Neptune's south pole and stepping across the disk until the northern limb was reached.  During the night, several A0-type stars with known J, H, and K magnitudes were also observed, to serve as telluric and photometric standards. Two exposures were taken of each star, dithering between the two halves of the field of view so that the exposures could serve as sky frames for one another. The airmass of the Hbb observations of Neptune was 1.3--1.4; the airmass of the Kbb observations was 1.2.

\subsection{Initial processing}
The initial processing of the scientific and calibration data is accomplished using version 2.3 of the OSIRIS data reduction pipeline\footnote{http://irlab.astro.ucla.edu/osiris/pipeline.html}, which performs sky subtraction, cleans the data of cosmic rays and common instrumental issues, extracts spectra from the raw frames, and combines the individual science exposures into mosaicked Hbb and Kbb spectral cubes that cover more than 90\% of Neptune's visible hemisphere. This last step is performed using the telescope right ascension and declination, which has an estimated astrometric accuracy of 40 mas\footnote{https://www2.keck.hawaii.edu/inst/osiris/OSIRIS\_Manual\_v2.3.pdf}. The pixel values of locations covered by multiple pointings are computed by averaging the pixel values of the individual exposures.  The data reduction pipeline also produces mask cubes that flag bad pixels identified by the pipeline. We flag 30 additional pixels near the edge of each of the unmosaicked frames, which appear anomalously bright due to lenslet masking errors; and 5 edge columns of the mosaicked Kbb cube, which are also anomalously bright. The initial Kbb mosaic exhibited brightness fluctuations between the individual exposures used to create the mosaic, indicative of an imperfect sky subtraction due to short-timescale fluctuations in the sky brightness. To improve the sky subtraction, at each wavelength we take the median of all unflagged background (off-disk) pixels in each individual exposure, to create a median background spectrum for that exposure. These background spectra are subtracted from the appropriate pixels in the mosaicked Kbb cube. A similar correction in Hbb is not possible, because Neptune itself is much brighter at these wavelengths, so that pixels near to the disk are dominated by flux from the nearby planet rather than by the sky. 

\subsection{Flux calibration}\label{sec:calib}
In order to calibrate our target data we use the calibration star observed under the most similar atmospheric conditions (in time and airmass) to those of the Neptune observations (SAO164840, airmass=1.2).  After initial processing of the pair of star exposures in each filter, stellar spectra are extracted using aperture photometry, then divided by a model of an A0 star spectrum that is calibrated to the 2MASS J, H and K magnitudes \citep{kurucz} and adjusted for reddening \citep{fitzpatrick99}. The two calibration spectra are averaged and corrected for the difference in airmass between the science target and calibrator observations. Then, each pixel in the science data cubes is divided by this stellar spectrum to correct for telluric lines and convert from units of counts into units of flux density.  Finally, we convert the data from observed flux density into units of $I/F$, defined as  \citep{hammel89}:
\[
\frac{I}{F} =\frac{r^2}{\Omega}\frac{F_N}{F_\odot}
\]
where $r$ is Neptune's heliocentric distance in AU, $\pi F_\odot$ is the sun's flux density at Earth's orbit \citep{colina96}, $F_N$ is Neptune's observed flux density, and $\Omega$ is the solid angle subtended by a pixel on the detector. By this definition,  $I/F=1$ for uniformly diffuse scattering from a Lambert surface when viewed at normal incidence. The reduced and calibrated data, averaged over wavelength, are shown in Fig. \ref{fig:DRcubes}.

\subsection{Spatial and spectral resolution}\label{sec:resolution}
The spatial resolution of the data is estimated from the stellar spectra. We find that the full width at half maximum (FWHM) of the point spread function (PSF) core is 2.0--2.2 pixels in Hbb, and 1.7--1.8 pixels in Kbb. We note that the PSF is non-Gaussian, having in addition to the narrow core a broad component (halo, FWHM $\sim$6--10 pixels) which complicates the interpretation of discrete features and the nearby background- this will be discussed further in Paper II. In this work, we avoid whenever possible locations that are within 10 pixels of identified bright features, to minimize their influence on our dark-region spectra. 

Prior to analysis, the data are smoothed to a spectral resolution of R$\sim$600 to increase the signal to noise (S/N) and match the resolution of our methane absorption coefficients (Appendix \ref{app:opacity}). Additional binning of the Kbb data, to increase S/N, is performed for the analysis in Section \ref{sec:DR}. 
\subsection{Data uncertainties}\label{sec:quality}

\begin{figure}[h!]
\begin{center}
\includegraphics[width=0.7\textwidth]{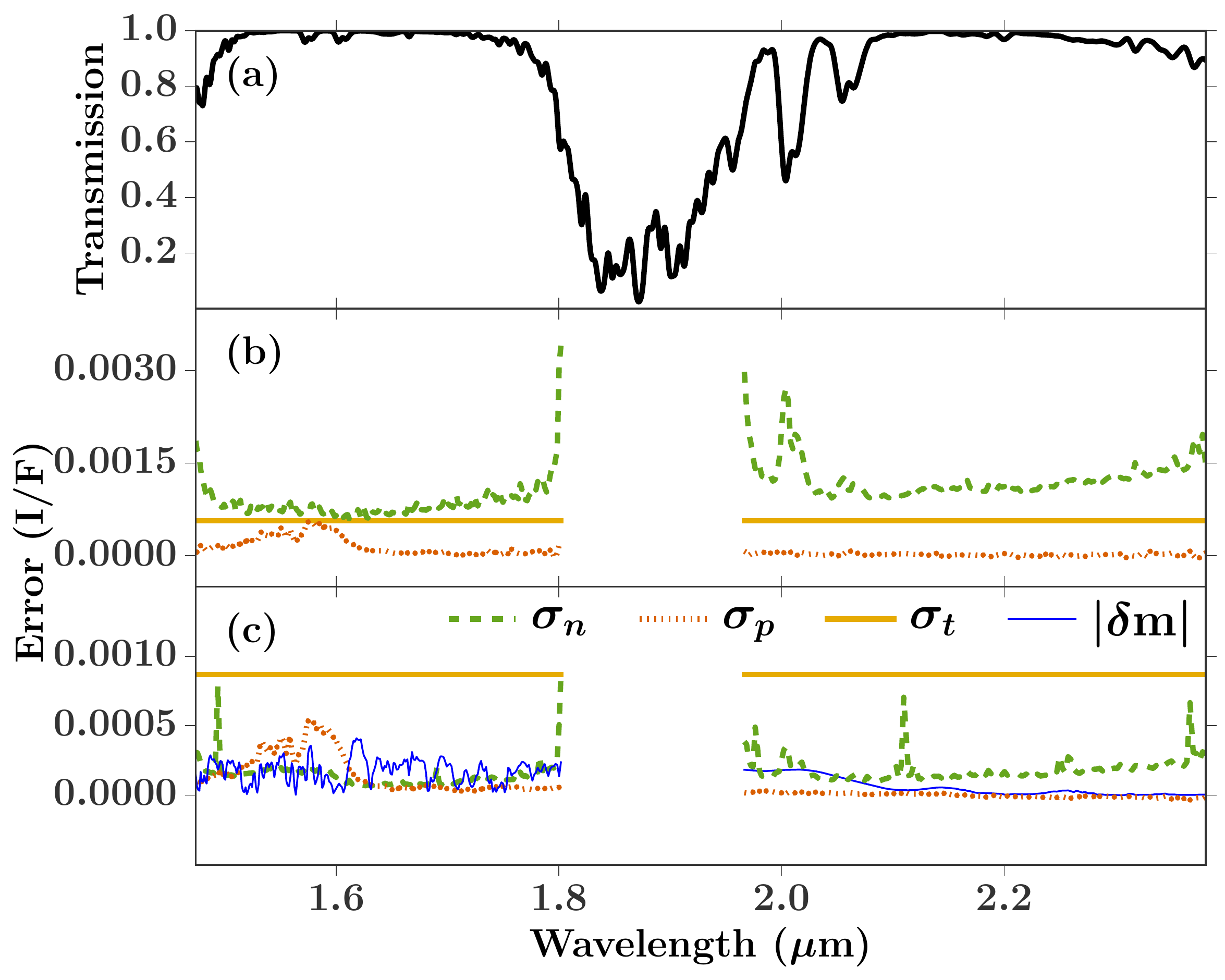} \\
\end{center}
\caption[Uncertainties]{\label{fig:error} \small Contributions to the uncertainties. Panel (a): a transmission spectrum of the Earth \citep{lord92}. Panel (b): uncertainty contributions for a single pixel from a dark region. $\sigma_n$ is the wavelength-dependent noise, as described in Section \ref{sec:quality}; $\sigma_p$ is the relative photometry error, which makes the smallest contribution to the total error, and $\sigma_t$ is the model tolerance, which represents any unknown uncertainties, as described in Section \ref{sec:retrieval}. For a single pixel, the dominant error term is $\sigma_n$ at all wavelengths. Panel (c): uncertainty contributions for the average $\mu=0.8$ spectrum used in the limb-darkening analysis (Section \ref{sec:limb}) and high S/N dataset analysis (Section \ref{sec:mu8}). In this case,  $\sigma_n$ is determined from the data used to produce the average spectrum. $\sigma_p$ is greater than $\sigma_n$ in the Hbb reflectivity peak, with $\sigma_t$ dominating at all wavelengths. In panel (c), the absolute difference between the two stream model and the more accurate pyDISORT model, $|\delta m|$, is shown by a thin blue line, for the case discussed in Appendix \ref{app:disort}.  }
\end{figure}

To measure the wavelength-dependent noise in the data ($\sigma_n$), we use the unflagged pixels in each cube (Hbb and Kbb) that lie outside of Neptune's disk as determined by image navigation (Section \ref{sec:nav}). An uncertainty spectrum is calculated as the standard deviation of all background pixels at each wavelength, after rejecting outlier pixels (pixels with a value more than 30 and 15 standard deviations away from the mean background in the Hbb and Kbb filters, respectively). This rejection typically discards 0--3 pixels per wavelength, or $\leq0.03$\% of the background pixels at a given wavelength. The mean uncertainty in I/F in a single pixel due to random noise is 0.001. The uncertainties are highest at wavelengths where the atmospheric transmission is low: near the edges of the bands and in the CO$_2$ features at 2.0 and 2.06 microns (Fig. \ref{fig:error}).

A second contribution to the uncertainty is systematic, introduced by the photometric calibration and telluric correction. We consider two components to the uncertainty in the calibrator spectrum, which we call ``overall photometry'' and ``relative photometry'' uncertainties. The overall photometry uncertainty refers to gross errors in the photometry due to, e.g., uncertainties in the H and K magnitudes of the calibrator star or loss of stellar flux off the edges of the detector. Based on comparisons with other calibrator stars observed on the same night and previous experience our conservative estimate for these uncertainties is 15--20\%. These errors would systematically offset all wavelengths in a given band, and we choose not to incorporate this term into the data uncertainty estimate, but do test the effects of such errors on our model fits. This is done by selecting a spectrum from the data and generating a grid of test spectra with offsets of $\pm20\%$ in the Hbb and/or Kbb photometry from the nominal calibration. We then perform a sample analysis, akin to those discussed in Section \ref{sec:DR}, for each of the test spectra and compare the results. We find that variations in the model results between test spectra are generally less than the parameter uncertainties from the model retrievals (Section \ref{sec:retrieval}). The one exception is that the single scattering albedos of the aerosols increase/decrease by 15--20\% when the Hbb photometry is adjusted by $\pm 20\%$, respectively. 

The relative photometry error ($\sigma_p$) refers to the random noise and small slopes (in wavelength) introduced by the calibrator spectrum, which we infer and characterize based on observed differences between the pair of spectra extracted for the calibrator star. We estimate $\sigma_p$ by determining the difference between our two stellar calibration spectra at each wavelength. We find that the differences between the two calibration spectra are correlated in wavelength -- that is, data from neighboring wavelengths are more likely to show similar relative offsets than data from widely separated wavelengths. However, there is not enough information in our calibration spectra to remove these systematic effects. Fortunately, differences between our calibration spectra are small, averaging 1.7\% of the flux in Hbb and 1.8\% in Kbb. We therefore include in our uncertainty estimate the term  $\sigma_{p,\lambda}=0.017\times(I/F)_\lambda$ in H band and $\sigma_{p,\lambda}=0.018\times(I/F)_\lambda$ in K band.  In the spectrum of a single pixel, the relative photometry error is small compared to the random error; however, when spectra are averaged, the relative photometry errors can dominate, especially at the brightest wavelengths (Fig. \ref{fig:error}). The calculation of the overall uncertainty term from these contributions is discussed in Section \ref{sec:retrieval}. 

\subsection{Image navigation}\label{sec:nav}
The brightness and variability of Neptune's clouds present a challenge for automated methods of image navigation. Therefore, to determine planetocentric latitude, longitude and emission angle ($\theta$, defined as the angle between the line of sight and the local normal to the planetary surface, specified throughout by $\mu \equiv \cos \theta$ ) for each pixel in our data, we fit a circle to the limb of planet by eye using the wavelength-collapsed Hbb and Kbb cubes. By testing small perturbations to the disk center positions, we estimate the accuracy of our centering to be  $\lesssim 1$ pixel -- roughly half the FWHM of the PSF core. The JPL Horizons\footnote{\url{http://ssd.jpl.nasa.gov/?horizons}} ephemerides provide the  sub-observer latitude (-28.7$^\circ$) and longitude (152$^\circ$ and 126$^\circ$ at the midpoint of the Hbb and Kbb observations, respectively).  The phase angle of the observations was 0.7$^\circ$. 

\section{Modeling}\label{sec:modeling}

To analyze our OSIRIS spectra, we use a Python-based atmospheric retrieval code which has two main components: a forward model, which performs the radiative transfer, and a retrieval algorithm to determine the posterior probability distribution of the model parameters.  In this section we present a brief overview of our retrieval code and a summary of the parameters most relevant to the analysis described in Sections \ref{sec:limb} -- \ref{sec:DR}. A more detailed description of the model may be found in Appendix \ref{app:model}.

\subsection{Forward model}\label{sec:forward}
\begin{figure}
\begin{center}$
\begin{array}{lcr}
\includegraphics[width=0.32\textwidth]{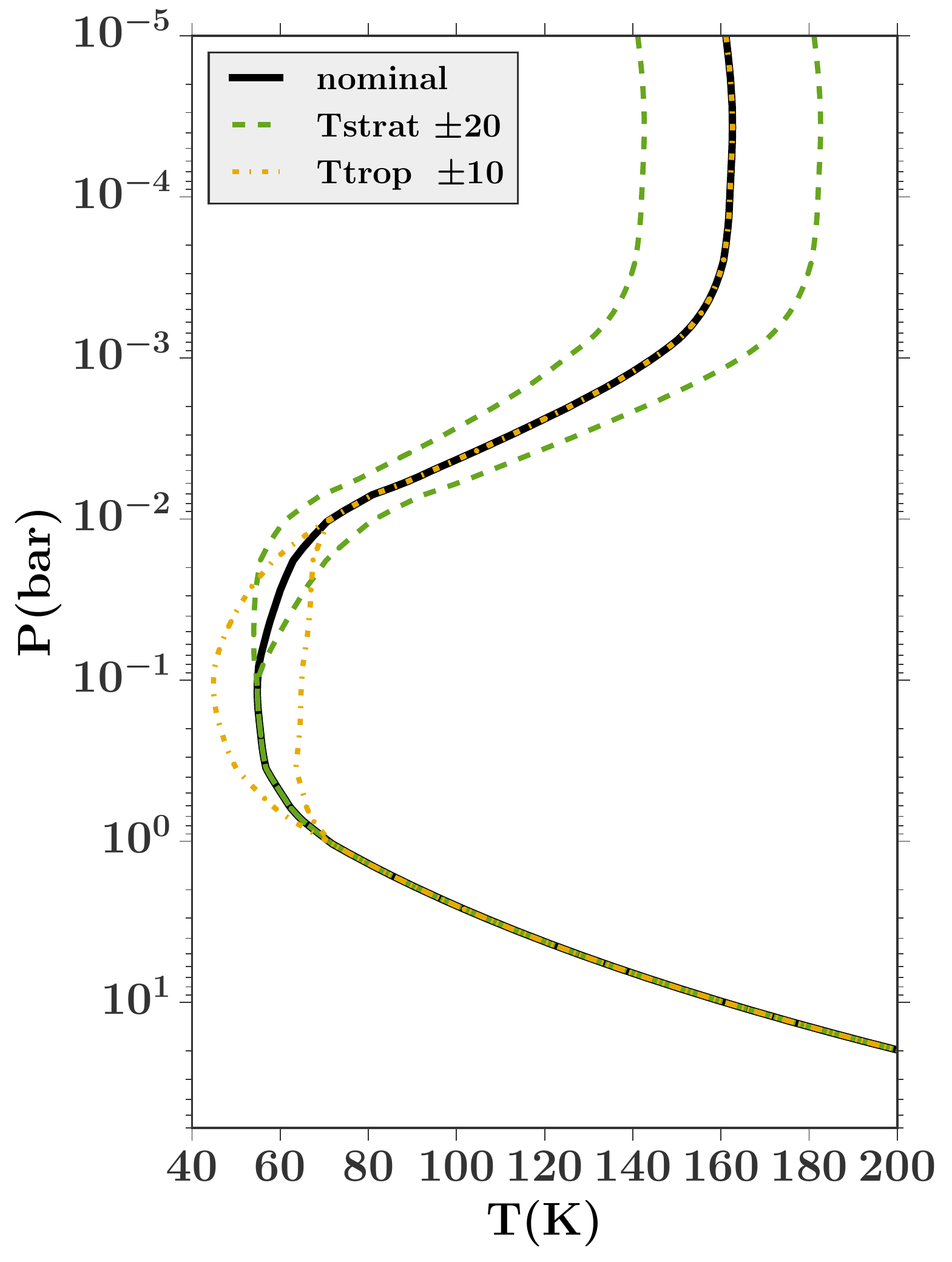} \includegraphics[width=0.32\textwidth]{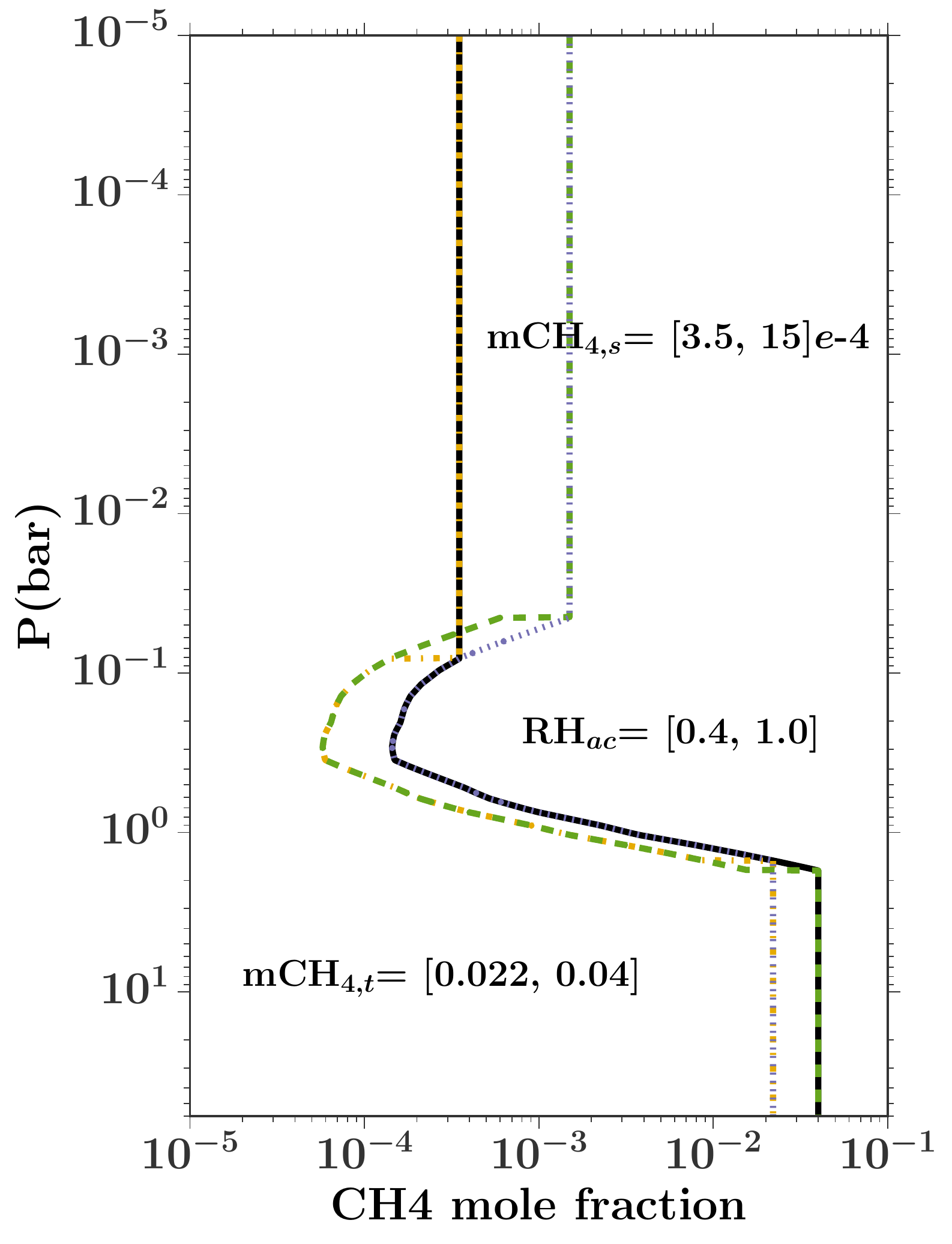} \includegraphics[width=0.32\textwidth]{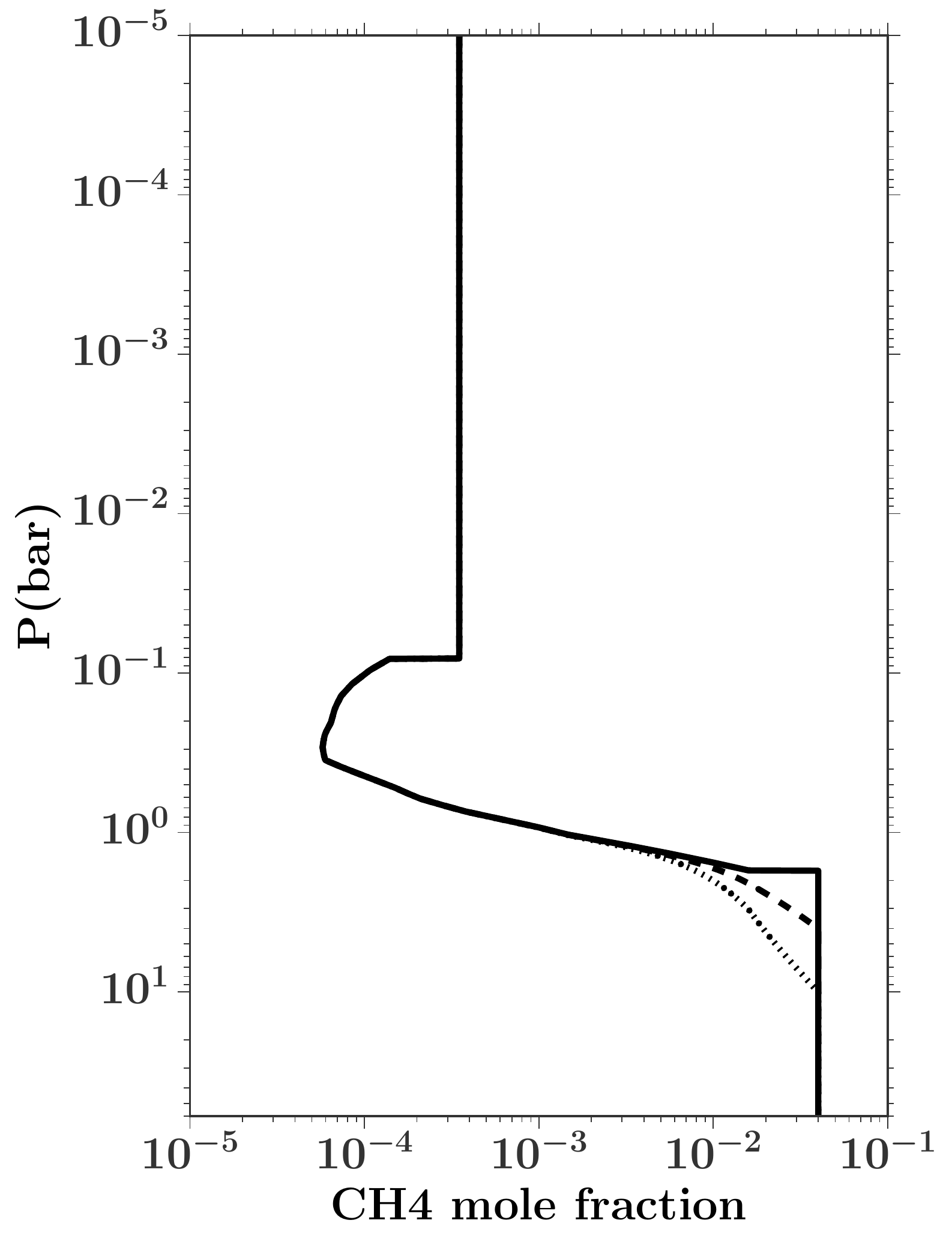} 
\end{array}$
\end{center}
\caption[Thermal and CH$_4$ profiles]{\label{fig:TP_CH4} \small Illustration of the thermal and CH$_4$ profiles used in this study. Left: nominal thermal profile (black, solid) and perturbations to the tropospheric and stratospheric temperatures, as indicated in the legend. Center: four examples of the parameterized CH$_4$ profile. Our initial profile of  {\it mCH$_{4,t}=0.04$}, {\it mCH$_{4,s}=3.5 \times 10^{-4}$} and  $RH_{ac}=100\%$ is shown by a black solid line. The profile selected after testing a range of profiles, referred to as {\it SN\_CH$_4$grid$_E$}, which has {\it mCH$_{4,t}=0.04$}, {\it mCH$_{4,s}=3.5 \times 10^{-4}$}, and  $RH_{ac}=40\%$, is shown by an orange dot-dashed line. Right: illustration of methane depletion. Depletion depths of 4.0 (dashed line) and 10.0 bar (dotted line) are shown with the un-depleted profile (solid line). }
\end{figure}

We construct a $\sim$100-layer model atmosphere that extends from pressures of 20 bar up to $10^{-5}$ bar from an input thermal structure (temperature-pressure profile); atmospheric composition as a function of depth; gas opacities as a function of temperature and pressure; and a description of the aerosols.  The nominal thermal and composition profiles for this study (Appendix \ref{app:model}) are illustrated in Fig. \ref{fig:TP_CH4}. Our model accepts any number of aerosol layers, which can be placed at any depth within the model atmosphere and can overlap one another -- a detailed description of the aerosol layer parameterization is also available in Appendix \ref{app:model}. The key model parameters for this study, and their default values, are summarized in Table \ref{table:model}; these parameters involve perturbations to the nominal thermal and methane profiles and variations on the aerosol properties and distribution. 

Given a model atmosphere, we solve the radiative transfer equation using either a two-stream approximation or a Python implementation of the discrete ordinate method for radiative transfer (pyDISORT).  The two stream radiative transfer code is a two-point quadrature method, following \cite{toon89}; this code has been previously used in NIR studies of Jupiter, Uranus, and Neptune \citep{depater10, depater10b, dekleer15,luszcz10}. The pyDISORT software was introduced by \cite{adamkovics15} for use with Titan, and is an implementation of CDISORT\footnote{www.libradtran.org}. The NIR scattering phase functions of Neptune's aerosols are poorly known; for simplicity, we assume single Henyey-Greenstein phase functions in our pyDISORT calculations, which are expanded into infinite series of Legendre polynomials. The radiative transfer is performed using four moments (and streams), which is the number we find adequately represents this simple phase function. 

The two-stream approximation sacrifices computational accuracy for the sake of speed, an essential compromise for this analysis: using two stream, a typical retrieval for a single model and single spectrum takes of order one day to complete on one of our 12-core machines. An equivalent pyDISORT run with four streams is a factor of $\sim3$ slower, with computation time increasing roughly as the third power of the number of streams. In the interest of time, our initial retrievals (Sections \ref{sec:1L} -- \ref{sec:3L}) utilize the two stream algorithm. The effects and limitations of using this approximation are examined in Appendix \ref{app:disort} and briefly reviewed in Section \ref{sec:disort}; in each subsequent retrieval, the adopted radiative transfer algorithm is clearly indicated.

\begin{minipage}{1.1\textwidth}
\setlength\LTleft{-0.5in}
\setlength\LTright{-1in plus 1 fill}
\renewcommand{\footnoterule}{}
\renewcommand{\thempfootnote}{\it \alph{mpfootnote}}
 \renewcommand{\thefootnote}{\it \alph{footnote}}
\begin{center}
\small
\setlength{\tabcolsep}{0.05in}
\begin{longtable}{+l  ^l ^l ^l ^l ^l }
                                                            
\caption[Model parameters table]{\small Overview of model parameters \label{table:model}} \tabularnewline

 & Parameter	& Definition	& Default value	& Sections where investigated \\
Thermal profile &	& 	& 	&	 \\
			& $\delta$\it{TP\_trop}		& tropopause temperature offset (K)	 &  no offset & \ref{sec:TPtests}; \ref{sec:DR_TP}\\
			& $\delta$\it{TP\_strat}		& stratosphere temperature offset (K) &  no offset &\ref{sec:TPtests} \\
CH$_4$ profile	&	& 	& 	&	 \\
			& {\it mCH$_{4,t}$}		& deep troposphere CH$_4$ mole fraction & 0.04 &  \ref{sec:CH4tests} \\
			& {\it mCH$_{4,s}$}		& upper stratosphere CH$_4$ mole fraction& 0.00035 & \ref{sec:CH4tests}\\
			 &$RH_{ac}$ 		& CH$_4$ relative humidity (near tropopause) &1.0\footnote{After Section \ref{sec:CH4tests}, we adopt a new nominal of  $RH_{ac} =0.4$} & \ref{sec:CH4tests}; \ref{sec:depletion}\\
			 &  $P_{dep}$		& CH$_4$ depletion pressure depth (bar) & no depletion &  \ref{sec:depletion}\\
Aerosols\footnote{parameters may be varied in one or all aerosol layers}		&		& 	& 	&	 \\
			&{\it P$_{max}$} & maximum (bottom) pressure	 (bar) & free & \ref{sec:limb}; \ref{sec:mu8}; \ref{sec:DR}\\
			&{\it P$_{min}$}	 & minimum (top) pressure (bar)& $10^{-5}$ & \ref{sec:1L}\\
			&{\it h$_{frac}$}	& fractional scale height& free & \ref{sec:limb}; \ref{sec:mu8}; \ref{sec:defaulthaze}; \ref{sec:DR_TP}\\
			&$\tau$		& optical depth at 1.6 $\mu$m &free & \ref{sec:limb}; \ref{sec:mu8}; \ref{sec:DR}\\
			&$r_p$		& particle radius ($\mu$m)\footnote{radius of particles at the peak of the particle size distribution, which is described in Appendix \ref{app:model}} & 0.1, 0.5 or 1.0 & \ref{sec:2L}\\
			& $g$		&phase function asymmetry parameter&  free & \ref{sec:limb}; \ref{sec:mu8nom}\\
			& $\omega_{0}$ & single scattering albedo	& free &  \ref{sec:limb}; \ref{sec:mu8nom}\\
\hline

\end{longtable}
 \end{center}
\end{minipage}

\subsection{Retrieval}\label{sec:retrieval}
As in \cite{dekleer15}, we pair our forward model with a Python implementation of the \cite{goodman10} affine-invariant Markov chain Monte Carlo (MCMC) ensemble sampler called {\it emcee} \citep{foreman13}, as a means of estimating model parameter values and uncertainties. Like any MCMC algorithm, {\it emcee} generates a sampling approximation to the posterior probability function by constructing one -- or in this case, an ensemble of -- chain(s) sampled from the desired probability distribution by random walk. In contrast to the simpler and more common Metropolis-Hastings MCMC sampling, this algorithm is more efficient, insensitive to the aspect ratio of the distribution, requires much less hand-tuning, and is easily parallelized. The details of our retrieval method are described in Appendix \ref{app:model}. One modification of the retrievals in \cite{dekleer15} is that our uncertainty estimate now includes a free parameter, $\sigma_t$, which is a model ``tolerance", representing any unknown uncertainties, particularly deficiencies in the model (for example, in the model opacities or in assumptions for the composition or thermal structure) that prevent the model from matching the data. The full model uncertainty is thus defined as: 

\begin{eqnarray}
\sigma^2=\sigma_n^2+\sigma_p^2+\sigma_t^2
\end{eqnarray}
where $\sigma_n$ is the random noise component and $\sigma_p$ is the relative photometry uncertainty, which is a scale factor multiplied by the model reflectivities (Section \ref{sec:quality}). In practice, we find that including $\sigma_t$ does not appreciably change the estimated values of other model parameters, but improves convergence and causes the parameter uncertainties to be more realistic. 

\begin{figure*}[t]
\begin{center}$
\includegraphics[width=0.6\textwidth]{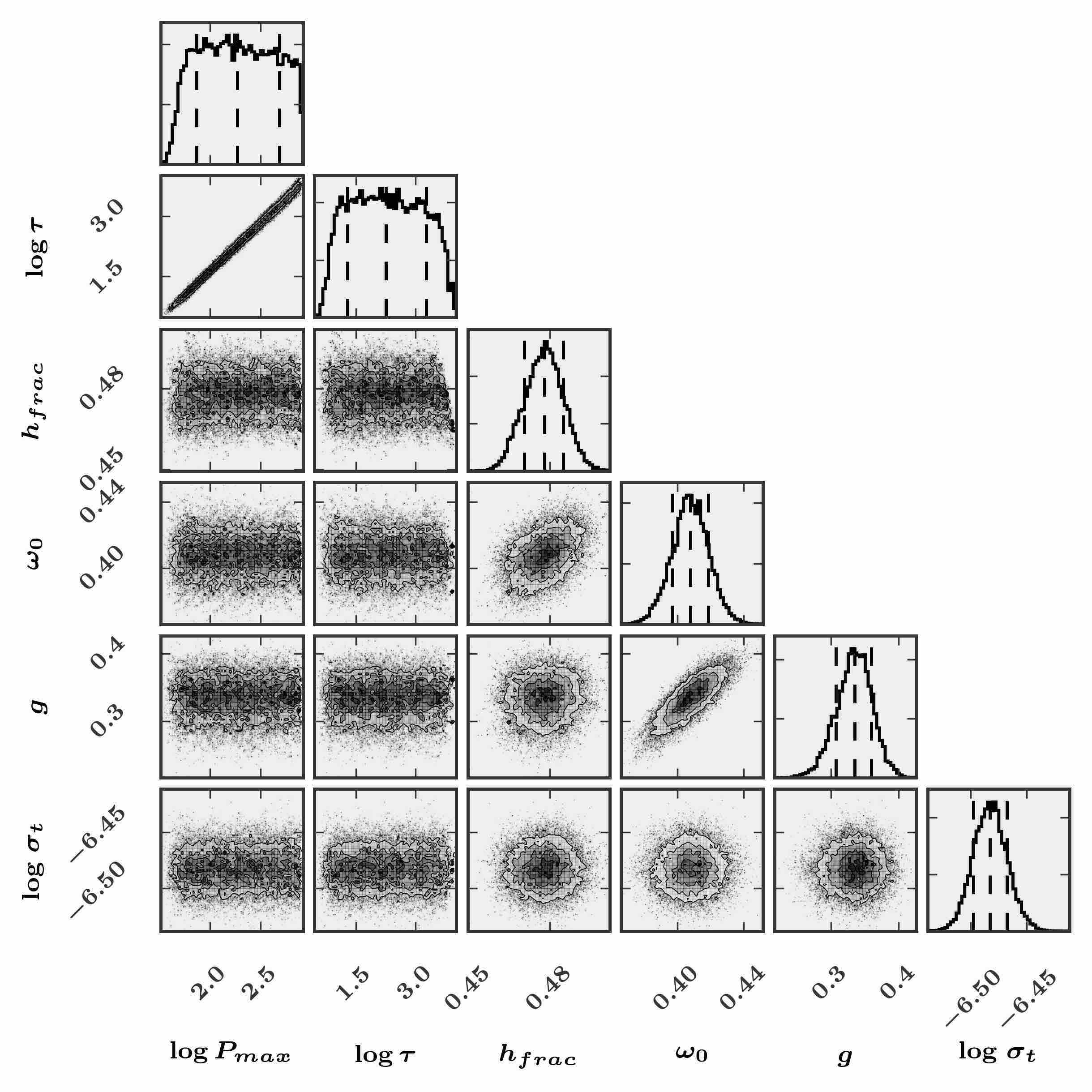}
$
\end{center}
\caption[Triangle plot (1L)]{\label{fig:1L_triangle} \small Demonstration of the MCMC retrieval method for parameter estimation, for the {\it 1L\_nom} model case. Scatterplots show the values taken on by each pair of parameters for each walker and step after burn-in. Histograms show the 1D marginalized probability distributions for each parameter, with lines indicating the 16th, 50th, and 84th percentiles of the distribution. Note the large degree of correlation between some parameters, and the deviation of the distribution from multivariate normal. }
\end{figure*}

Once we have used {\it emcee} to approximate the posterior probability distribution for a given model and dataset, we report the 16th, 50th, and 84th percentiles from the marginalized parameter distributions (e.g., Table \ref{table:1L}). In some cases, we also provide plots of the one- and two-dimensional projections of the posterior probability distributions (e.g., Fig. \ref{fig:1L_triangle}). We frequently refer to the 50th percentile of the marginalized distributions as the model ``best-fit'' for convenience, and refer to the 16th and 84th percentiles of the marginalized distributions as the ``1$\sigma$ parameter uncertainties.'' We note that these best-fit and uncertainty values quantify the spread of parameter values {\it for a given model}, and do not necessarily encompass the range of values a parameter might take for different atmospheric models (different model assumptions and/or free parameters). To compare different models, we calculate the Deviance Information Criterion \citep[$DIC$,][]{spiegelhalter02}. The $DIC$ includes a term for goodness-of-fit with a penalty term for model complexity. The $DIC$ for a single model is not meaningful, and may be a positive or negative number. However, differences in the $DIC$ provide a metric for evaluating the relative success of different models for the same dataset, with a difference in the $DIC$ ($\Delta DIC$) of 10 or more indicating a preference for the model with the lower $DIC$.  See Appendix \ref{app:model} for more information on the $DIC$. We note that a large value of $\sigma_t$ may be another indicator that a model is not a good match to the data, as it implies that the differences between the data and model are not well captured by the known data uncertainties ($\sigma_n$ and $\sigma_p$). 

\section{Aerosol properties and structure: 2--12$^\circ$N}\label{sec:limb}
\begin{figure}[htb!]
\begin{center}$
\includegraphics[width=0.5\textwidth]{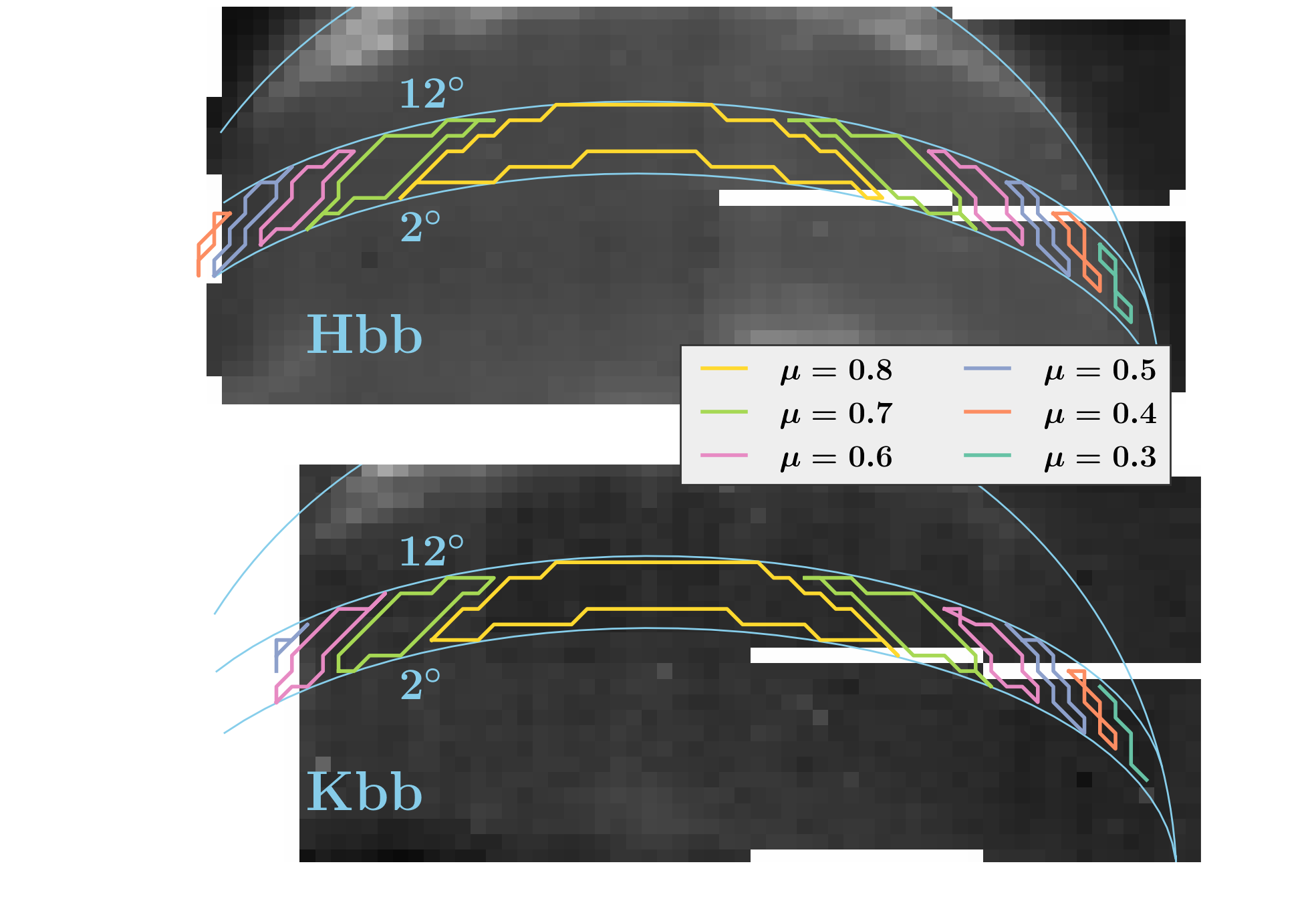}
$
\end{center}
\caption[Limb regions]{\label{fig:limb}\small Regions used in Section \ref{sec:limb}. Each outlined region represents an interval of 0.1 in $\mu$, centered at the values specified in the figure. Within each bin, a mean spectrum and uncertainty spectrum is calculated. These six spectra are shown in, e.g., Fig. \ref{fig:1L_hgfree_modelfits}.   }
\end{figure}

In this section, we focus our analysis on constraining the properties and structure of Neptune's aerosols. We take advantage of the spatial resolution of our data to select out only spectra in the latitude range of 2--12$^\circ$N. Across this range, discrete clouds are generally not observed n Neptune (and none are observed in our data). By selecting out a relatively thin latitude band, we hope to minimize inherent variations in the properties of the atmosphere (temperature, composition, and aerosol structure).  We separate the data into bins at intervals of 0.1 in the cosine of the emission angle, centered at $\mu=$ 0.3, 0.4, 0.5, 0.6, 0.7, and 0.8. The bins contain seven ($\mu=0.3$) to 106 ($\mu=0.8$) individual spectra (Fig. \ref{fig:limb}). Within each bin, we calculate the mean spectrum (considering only unflagged pixels) and the standard deviation as a function of wavelength, which is divided by the square root of the number of input spectra to estimate the noise in the averaged spectrum. The uncertainties estimated in this way are somewhat larger than ($1.1$--$2.2$ times)  those found from the background pixels (Section \ref{sec:quality}); this may be due to real  variations in the atmospheric properties within the latitude and $\mu$ ranges included in each bin.  We define $\sigma_n$ for the six mean spectra using these higher uncertainty estimates. 

We then perform a series of retrievals, considering the six spectra simultaneously and assuming they all reflect the same atmosphere in terms of aerosol structure, temperature, and composition. This dataset is referred to as the `limb-darkening' dataset, since it contains information on the aerosol structure that is revealed due to changes in viewing geometry. We use the default thermal structure  (Appendix \ref{app:TP}); set {\it mCH$_{4,t}=0.04$}, {\it mCH$_{4,s}=3.5\times 10^{-4}$},  and$RH_{ac}=100\%$ (refer to Table \ref{table:model} for definitions); and assume no shallow tropospheric methane depletion (Appendix \ref{app:composition}). In Sections \ref{sec:1L} -- \ref{sec:3L} we use the two-stream algorithm to solve the radiative transfer equations; in Section \ref{sec:disort} we evaluate the limitations of this approximation and present a retrieval using pyDISORT.

\subsection{Single aerosol layer ({\it 1L}) models} \label{sec:1L}

\begin{minipage}{1.1\textwidth}
\setlength\LTleft{-0.25in}
\setlength\LTright{-1in plus 1 fill}
\renewcommand{\footnoterule}{}
\renewcommand{\thempfootnote}{\it \alph{mpfootnote}}
 \renewcommand{\thefootnote}{\it \alph{footnote}}
\begin{center}
\small
\setlength{\tabcolsep}{0.05in}
\begin{longtable}{+l  | ^l ^l ^l ^l ^l  ^l ^l ^l}
\caption[Fit results- 1L models]{Fit results, single aerosol layer ({\it 1L}) models \label{table:1L}} \tabularnewline

\hline 
run name		& $P_{max}$ (bar)	 &$P_{min}$ (bar)						&	$\tau$ 		&$h_{frac}$				&$\omega_0$			    	&	$g$					&  $\log{\sigma_t}$			& 	$DIC$\\

{\it1L\_nom} 	&$10^{+5}_{-3}$ &  $0.001$  	&	$9^{+20}_{-6}$                 &   $0.478^{+0.007}_{-0.007}$ &   $0.41^{+0.01}_{-0.01}$ &   $0.34^{+0.02}_{-0.03}$ &   $-6.48^{+0.02}_{-0.02}$ &$-25641$ \\
{\it1L\_pmin1.4} 	&$10^{+5}_{-3}$ & 1.4\footnote{fixed at the \cite {kark11a} value.}  &	 $11^{+20}_{-7}$ &   $0.471^{+0.009}_{-0.009}$ &   $0.40^{+0.01}_{-0.01}$ &   $0.33^{+0.03}_{-0.03}$ &   $-6.44^{+0.02}_{-0.01}$ &$-25430$ \\
{\it1L\_pminfree} 	&$9^{+5}_{-3}$ &   $0.007^{+0.01}_{-0.005}$ &   $8^{+20}_{-5}$ &   $0.478^{+0.007}_{-0.007}$ &   $0.41^{+0.01}_{-0.01}$ &   $0.34^{+0.02}_{-0.03}$ &   $-6.48^{+0.02}_{-0.01}$ &$-25642$ \\

\end{longtable}
 \end{center}
\end{minipage}

Considering the uncertainty in Neptune's aerosol structure, our approach is to determine the simplest structure that can reproduce our data. Therefore, we begin by testing models with only a single layer of aerosols, defined by its maximum pressure ($P_{max}$), total optical depth ($\tau$), single scattering albedo ($\omega_0$), phase function asymmetry factor ($g$), and fractional scale height ($h_{frac}$). Previous models of Neptune's cloud-free regions or disk-averaged aerosol structure \citep[e.g.][]{baines94,kark11a,kark11b,irwin11,irwin14} find that Neptune's aerosol opacity is greatest in the troposphere, at pressures $>1$ bar, where particle sizes $ \gtrapprox 1\mu$m are indicated \citep{conrath91}; therefore we assume that the extinction cross section is given by a particle distribution with $r_p=1\mu$m (Appendix \ref{app:aerosols}).

\subsubsection{ {\it1L\_nom} model}
\begin{figure}[h!]
\begin{center}$
\begin{array}{cccccc}
\includegraphics[width=0.45\textwidth]{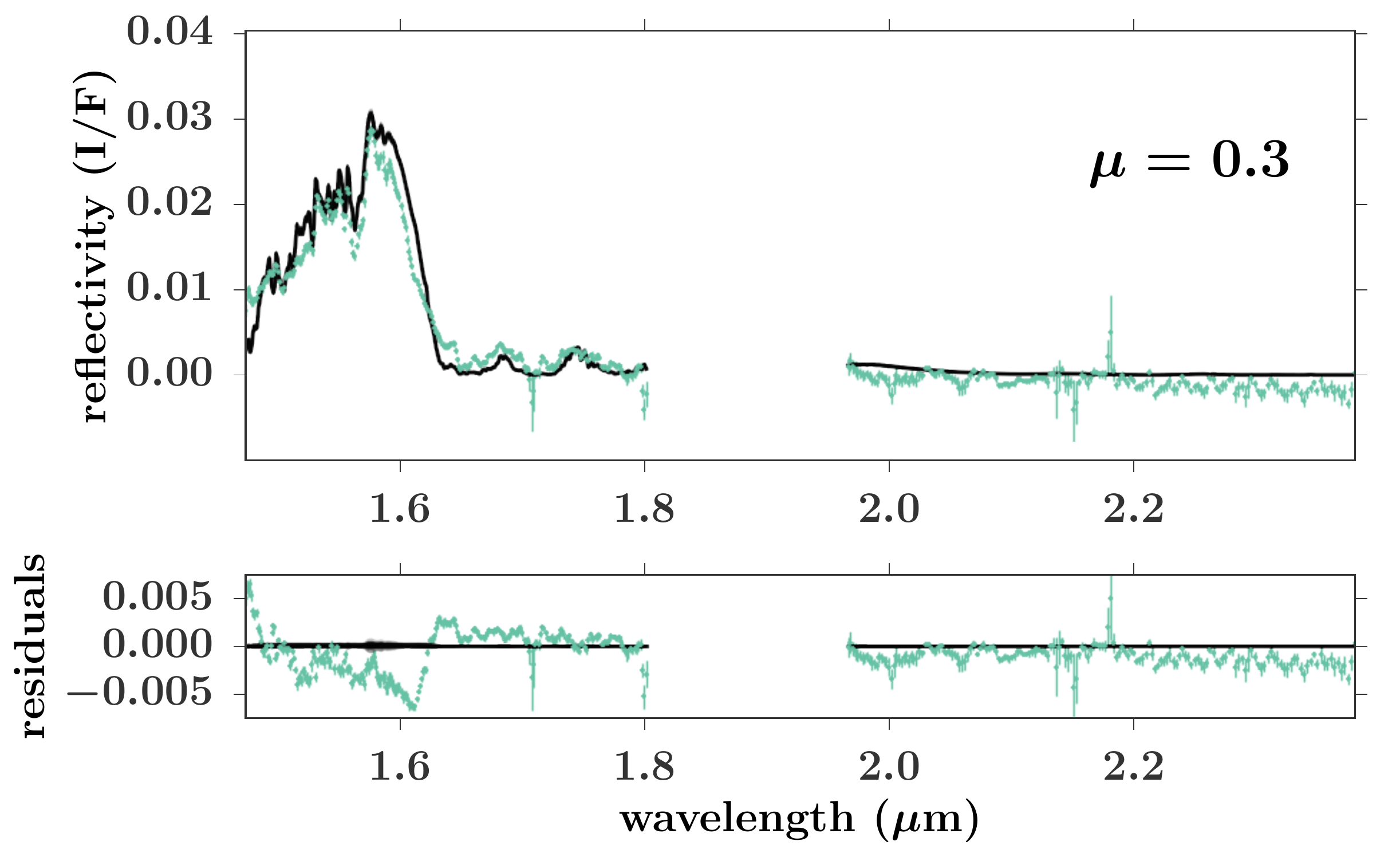} \includegraphics[width=0.45\textwidth]{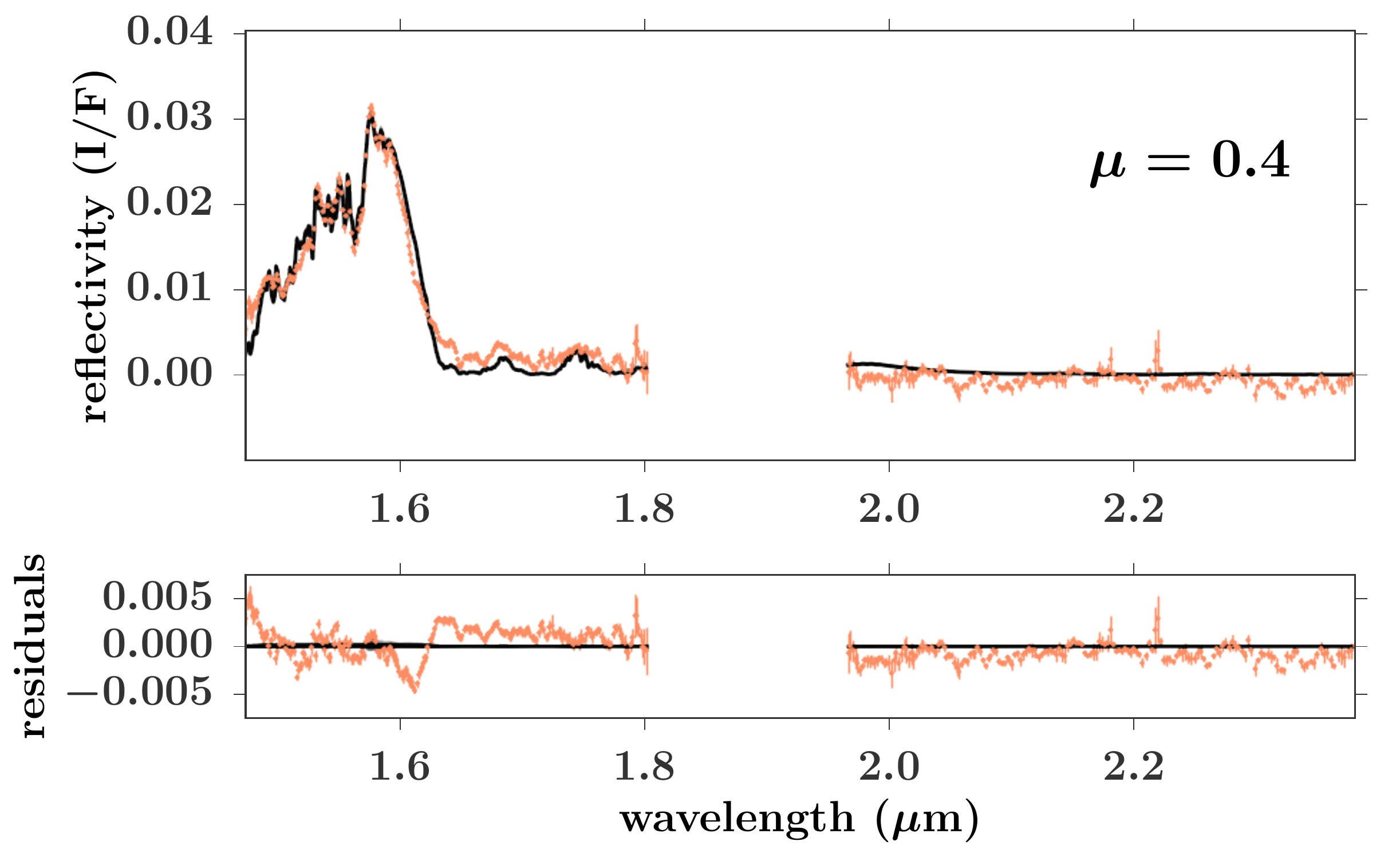}\\
\includegraphics[width=0.45\textwidth]{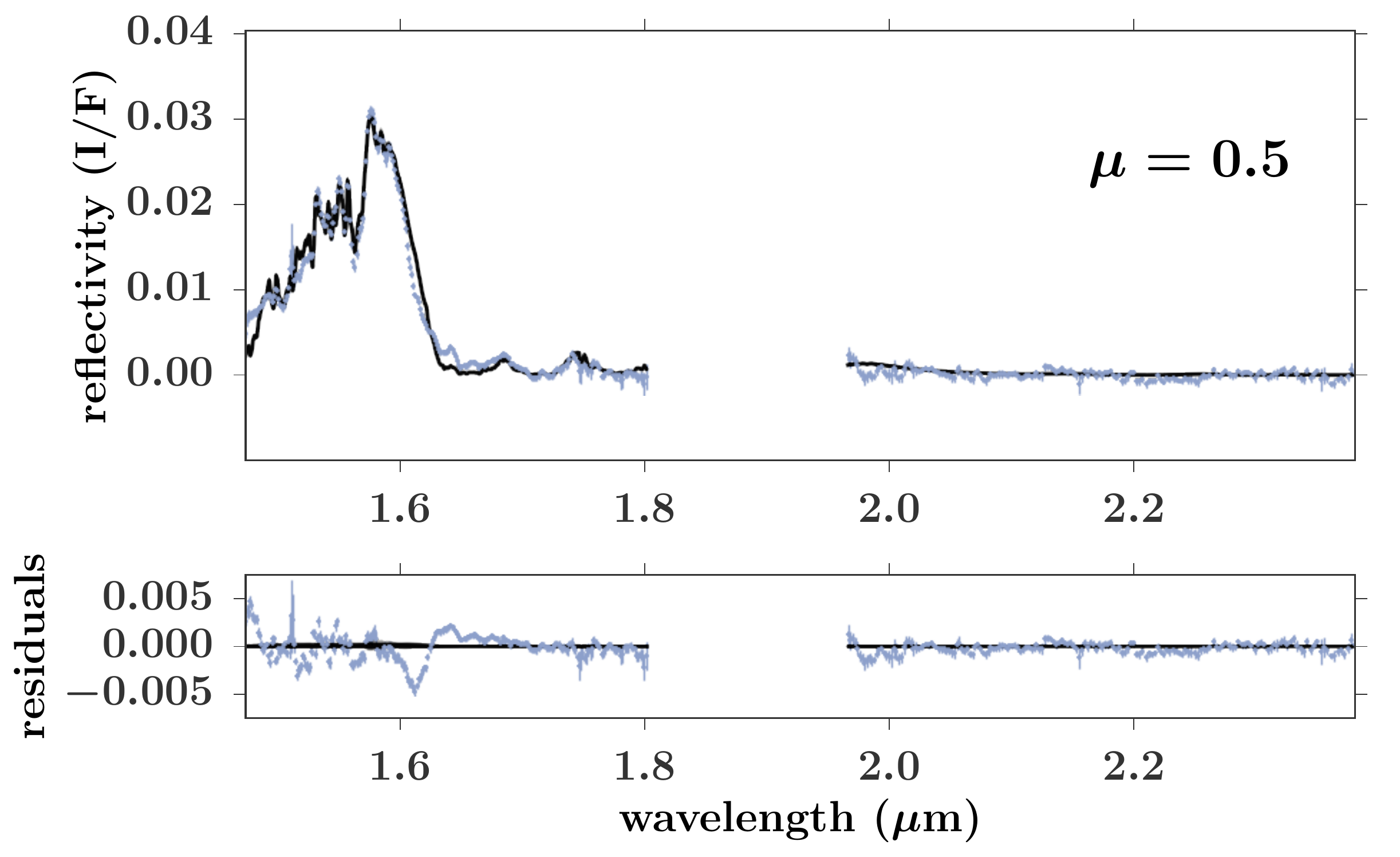}\includegraphics[width=0.45\textwidth]{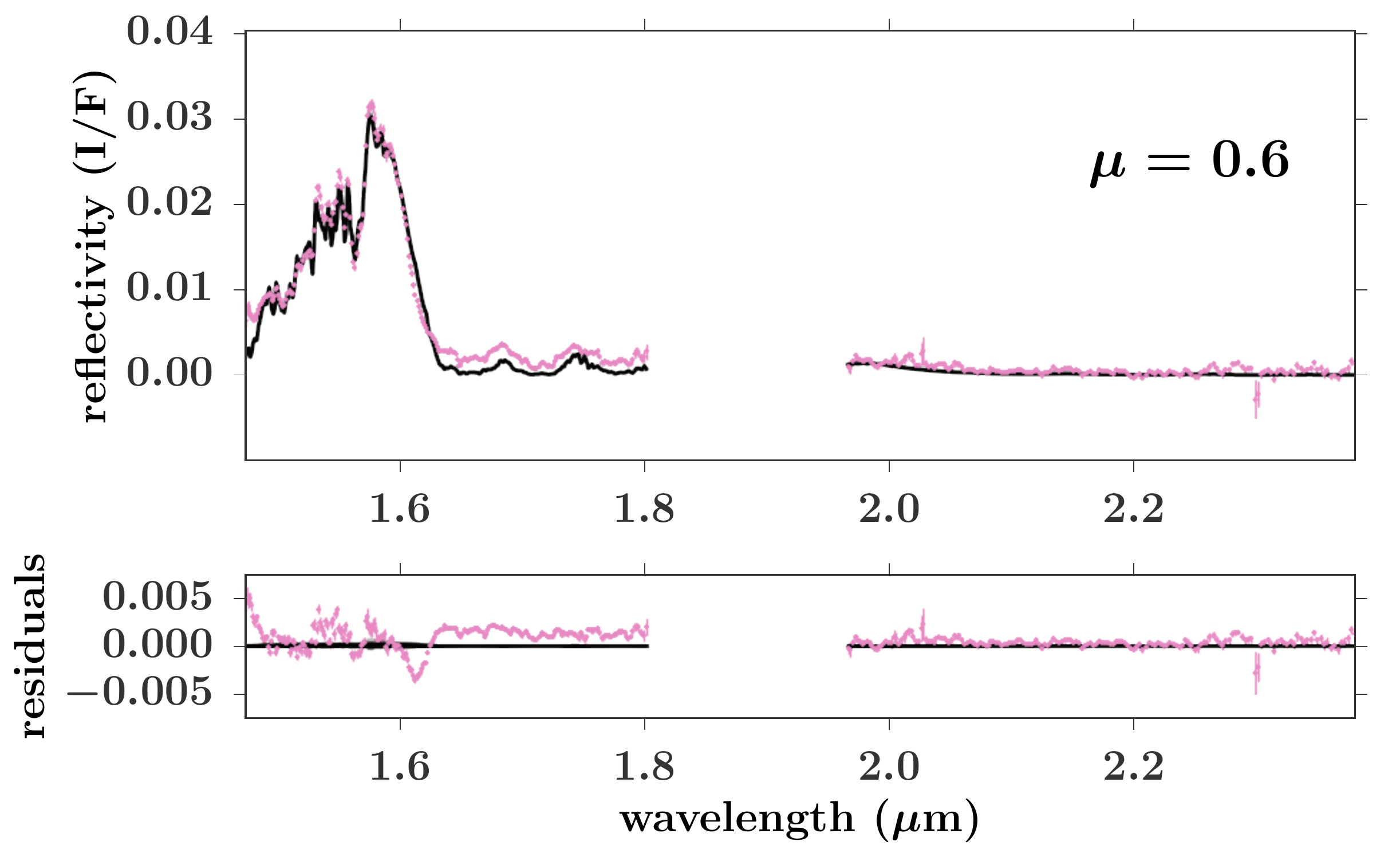}\\
\includegraphics[width=0.45\textwidth]{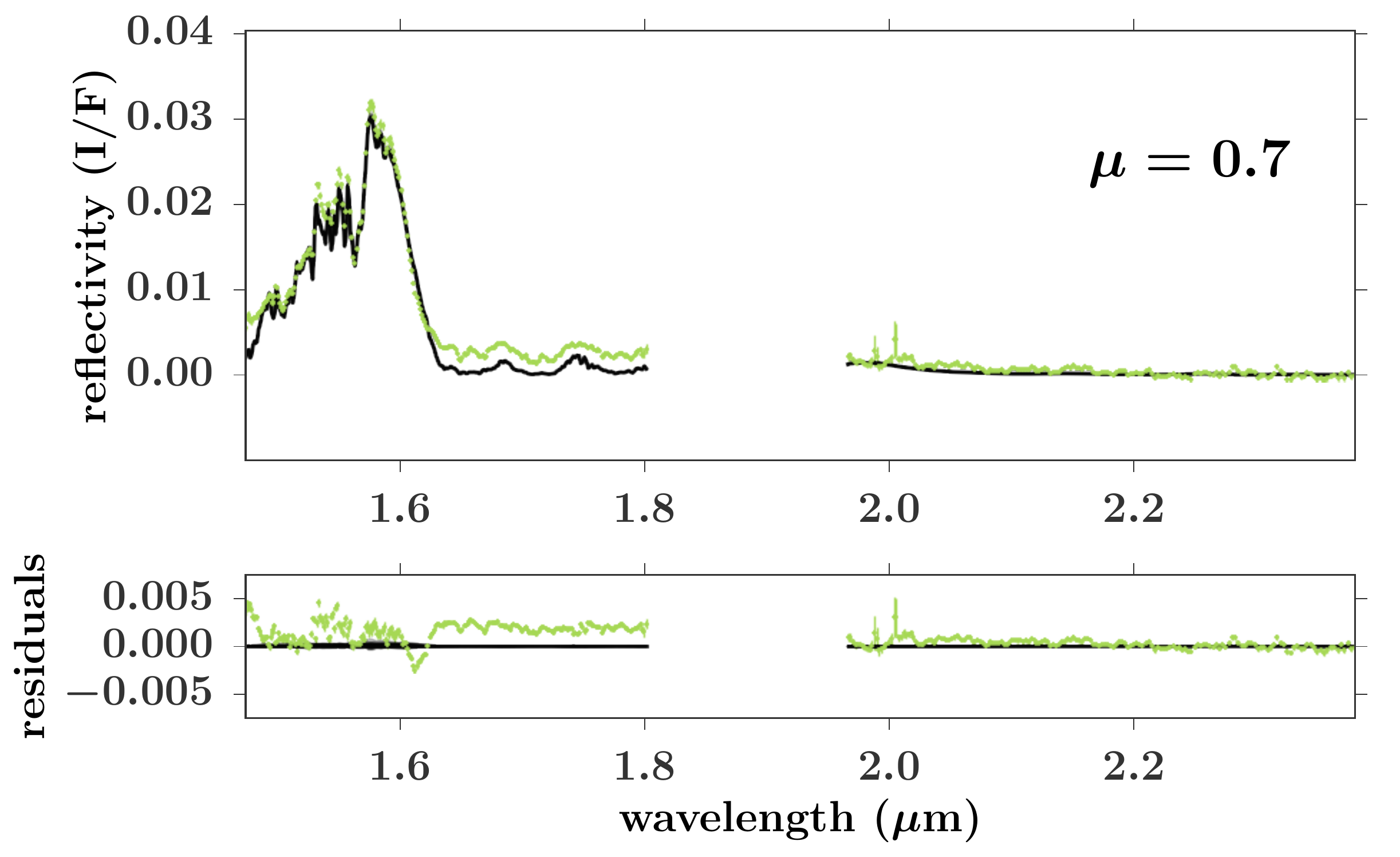}\includegraphics[width=0.45\textwidth]{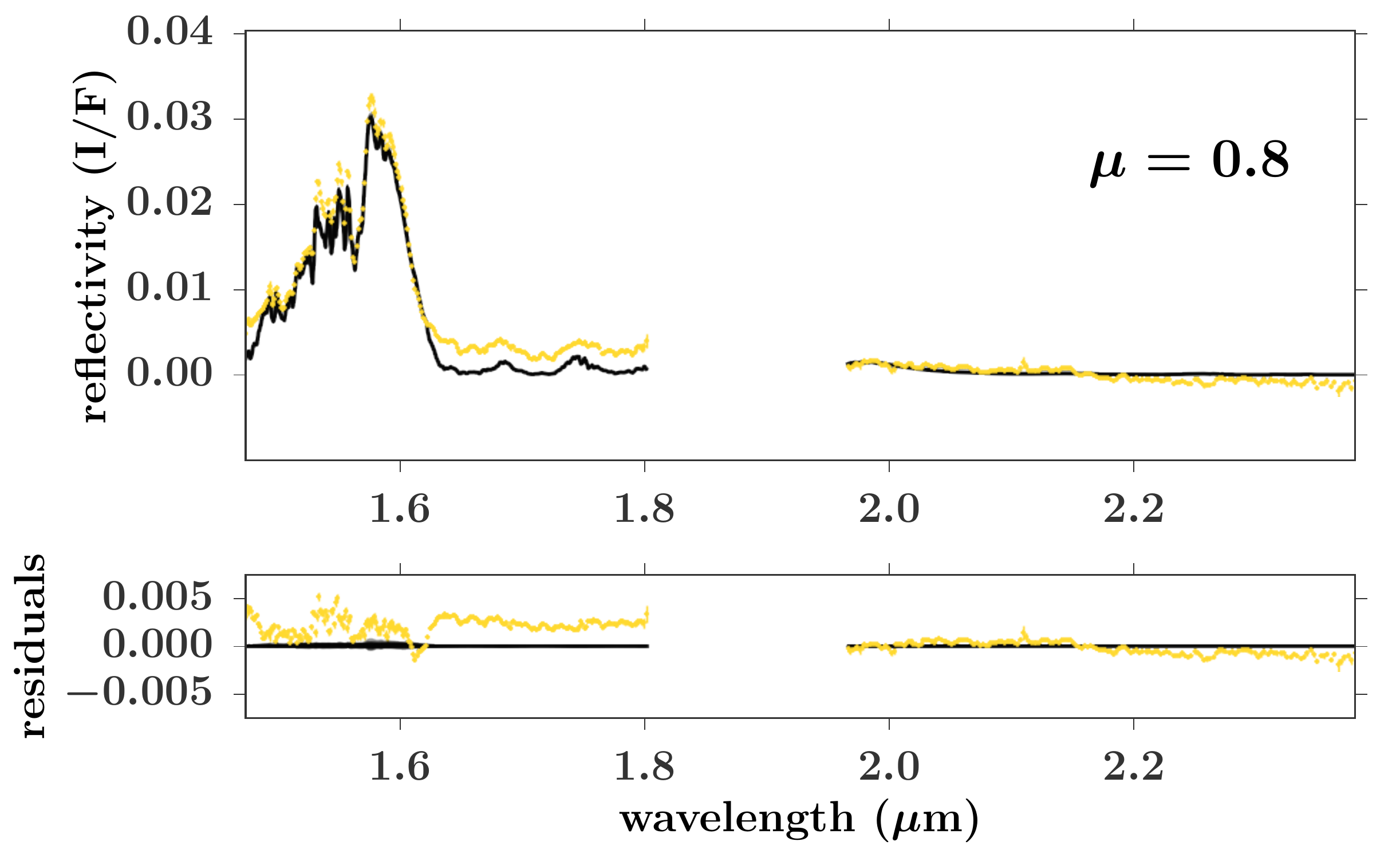}\\
\end{array}$
\end{center}
\caption[{\it1L\_nom}  model fits to limb darkening data]{\label{fig:1L_hgfree_modelfits} \small The six mean spectra used to investigate limb darkening behavior and a projection of the {\it1L\_nom} model results onto these data. The average value of $\mu$ for each spectrum is indicated: the top left plot is near the limb, and the bottom right plot is towards disk center. The colors of the data points match those in Fig. \ref{fig:limb}, and the error bars reflect the $\sigma_n$ and $\sigma_p$ contributions to the uncertainty only. The model representing the 50th percentile in the posterior probability distribution (``best fit'') is shown as a thick black line; 100 other samples from the posterior probability distribution have been randomly selected and are shown as thinner black lines to illustrate the scatter in the models.  Beneath each spectrum is a plot of the fit residuals, in which the best-fit model has been subtracted from the data and from the random models.}
\end{figure}

We first perform a retrieval in which the aerosol layer top pressure $P_{min}$ is set to 1 mbar.  This parameterization permits either a single, compact cloud deck (for $h_{frac} < 1$) or a haze that extends from the base pressure to the top of the observable atmosphere (for $h_{frac} \sim 1$).  The results from this retrieval are summarized in Table \ref{table:1L} and Fig. \ref{fig:1L_hgfree_modelfits}, and the aerosol distribution corresponding to these results is illustrated in Fig. \ref{fig:taubar}. We retrieve an optically thick, tropospheric aerosol layer with a deep base at 4--30 bar and a moderate scale height of $\sim 0.5$ times the gas scale height. As illustrated by Fig. \ref{fig:1L_triangle}, $P_{max}$ and $\tau$ are highly correlated, with a deeper cloud base pressure corresponding to a higher total optical depth, such that the optical depth per bar at pressures that the observations are sensitive to (Fig. \ref{fig:1L_contrib}) does not vary appreciably between models (Fig. \ref{fig:taubar}). Figure \ref{fig:1L_triangle} also shows that the posterior probability distributions of $P_{max}$ and $\tau$ are roughly flat across a broad range of values. This makes sense, since adding additional cloud opacity below an already optically thick cloud should not influence the resulting model spectrum. 

The retrieved value of the asymmetry factor $g$ is $0.34^{+0.02}_{-0.03}$, in contrast to a 1.6 $\mu$m value of 0.76 predicted by Mie theory for 1-$\mu$m particles (see Fig. \ref{fig:particles}).  The retrieved cloud is also fairly dark, with $\omega_0 \approx 0.4$ for the best fit. However, as we will show in Section \ref{sec:disort}, $\omega_0$ and $g$ are strongly influenced by the use of two stream for the radiative transfer, and the results for these parameters should be interpreted with great caution for all models which rely on the two-stream algorithm.

\subsubsection{  {\it1L\_pmin1.4} and  {\it1L\_pminfree} models}

\begin{figure}
\begin{center}$
\begin{array}{ccc}
\includegraphics[width=0.3\textwidth]{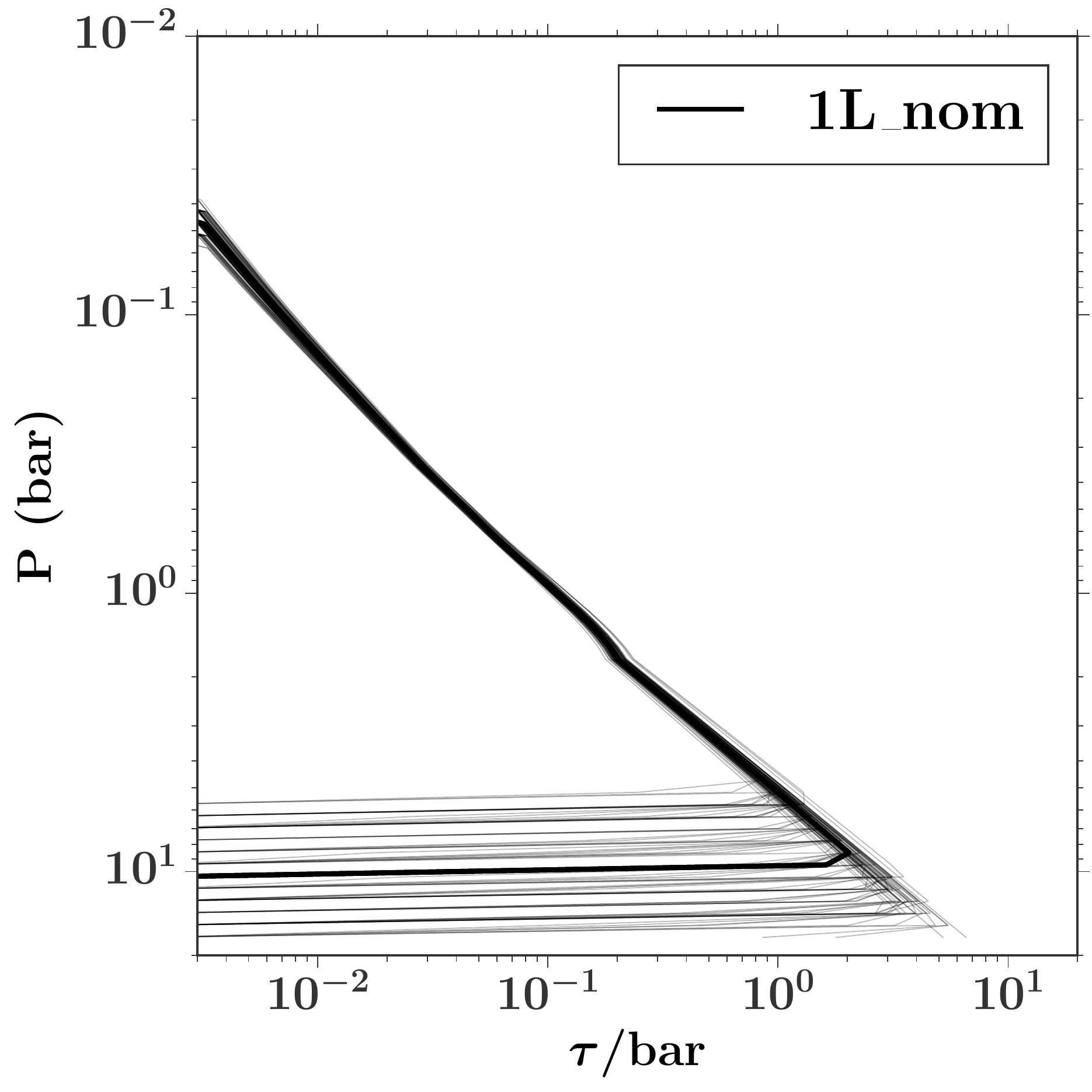} \includegraphics[width=0.3\textwidth]{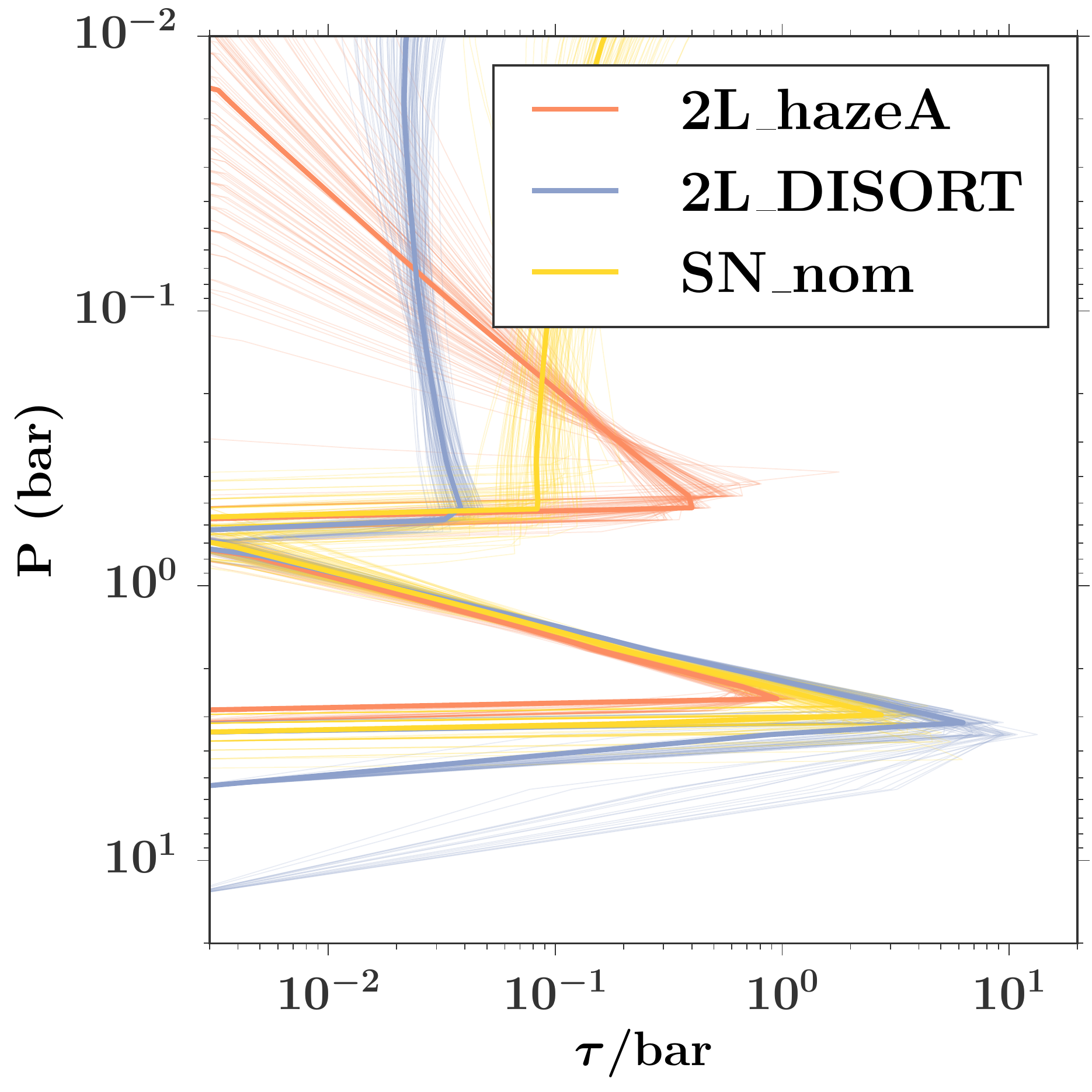} \includegraphics[width=0.3\textwidth]{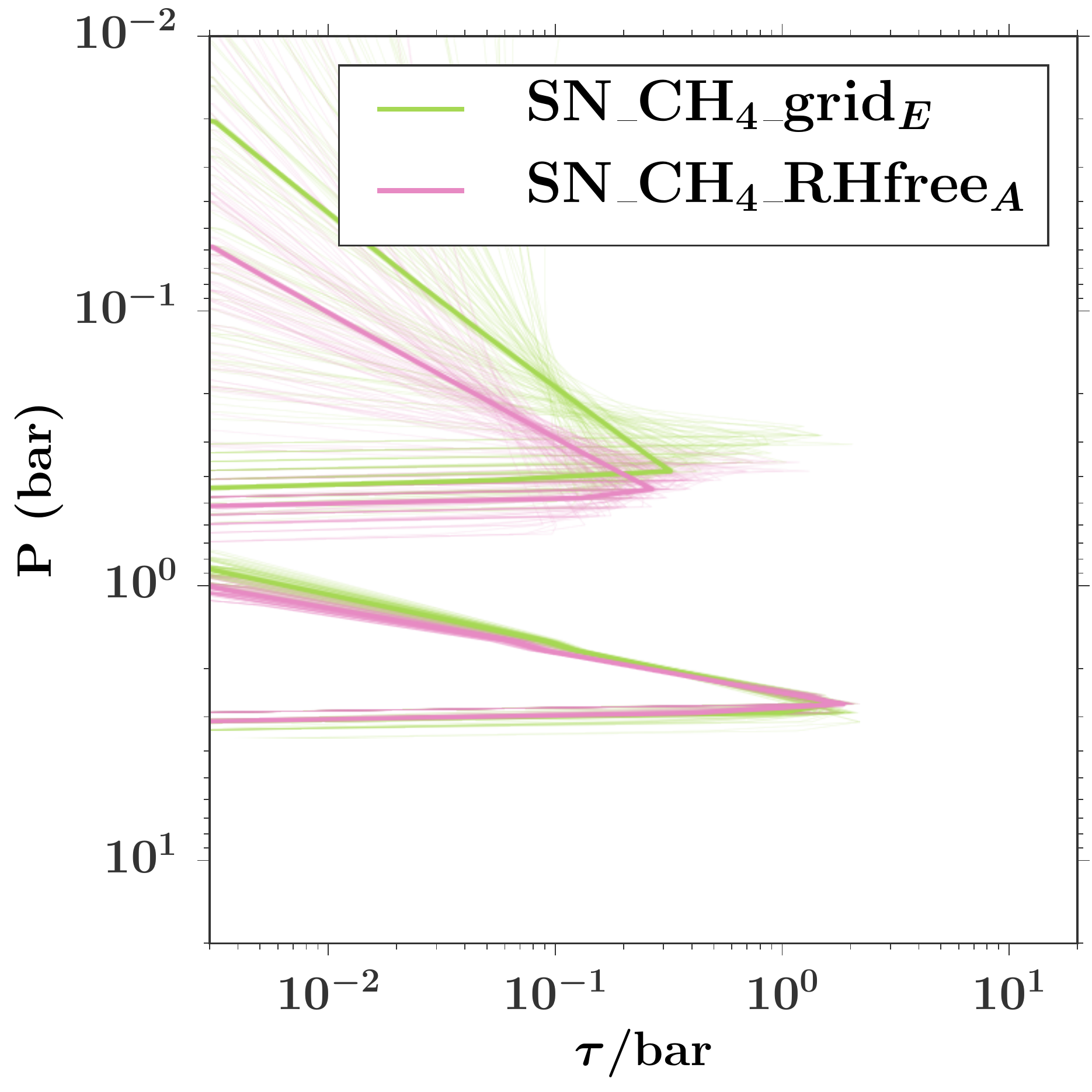}\\
\end{array}$
\end{center}
\caption[$\tau$/bar for 1L and 2L models]{\label{fig:taubar} \small Optical depth per bar ($\tau$/bar) as a function of pressure for a selection of models to the mean spectra from 2--12$^\circ$N. In each case, the best-fit model is shown as a thick line, and 100 random samples from the posterior probability distribution are shown as thinner lines of the same color. Left: the {\it1L\_nom} model as described in Section \ref{sec:1L}. Center: three two-layer retrievals. The {\it 2L\_hazeA} retrieval, and the equivalent pyDISORT retrieval ({\it 2L\_DISORT}) are shown in orange and blue, respectively -- these models are described in Section \ref{sec:2L}. The third model ({\it SN\_nom}, yellow) is a similar retrieval to the {\it 2L\_hazeA} case, except using the $\mu=0.8$ data only (see Section \ref{sec:mu8}). Right: two variations of the {\it SN\_nom} case, using alternative CH$_4$ profiles, as described in Section \ref{sec:CH4tests}. }

\end{figure}

As shown in Fig. \ref{fig:taubar}, the {\it1L\_nom} model is not sensitive to the precise value of the parameter $P_{min}$, since the aerosol opacity drops off below the altitude of the tropopause. However, this model setup does not capture a scenario in which the aerosols are vertically uniform below some cutoff depth in the troposphere, as inferred by \cite{kark11a, kark11b}. These authors found that Neptune's aerosols are uniformly distributed between 1.4 bar and a pressure of at least 10 bar. We now consider whether this type of aerosol structure is compatible with our observations. In the {\it1L\_pmin1.4} retrieval, we  perform a run with the same free parameters as in the  {\it1L\_nom} retrieval, but set $P_{min}$ to the \cite{kark11a} haze upper bound of 1.4 bar. As shown in Table \ref{table:1L}, the retrieved aerosol properties for the  {\it1L\_pmin1.4} retrieval are consistent with the {\it1L\_nom} case; however, the fit quality as measured by the $DIC$ is significantly worse when the aerosols are restricted to $P > 1.4$ bar. Finally, we consider the case where $P_{min}$ is a free parameter in the retrieval, to determine whether our data indicate a preference for a different value of $P_{min}$, and whether this parameter influences the probability distributions of other model parameters. This run is referred to as the {\it1L\_pminfree} retrieval. We find that, as in the {\it1L\_pmin1.4} case, the retrieved aerosol properties are consistent with the {\it1L\_nom} retrieval. The fit quality does not improve when $P_{min}$ is allowed to be a free parameter in the retrieval.

\subsection{Two aerosol layer (2L) models}\label{sec:2L}

We next consider models containing two layers of aerosols. As in the $1L$ fits, we allow $P_{max}$, $\tau$,  $h_{frac}$, $\omega_0$, and $g$ to be free parameters with flat (for $h_{frac}$, $\omega_0$,  $g$) or log-flat (for $P_{max}$, $\tau$) priors. For simplicity, we restrict $\log{P_{max,\alpha}} (bar)=[-6.8,0.9]$ ($P_{max,\alpha}$ between 1 mbar and 2.5 bar) and  $\log{P_{max,\beta}} (bar)=[-2.3,2.3]$ ($P_{max,\beta}$ between 0.1 and 10 bar).  For the bottom layer (layer $\beta$) we assume the extinction cross section  is given by the particle distribution with $r_p=1 \mu$m. For the top layer (layer $\alpha$), we consider three particle distributions, with characteristic particle sizes of  $r_p = 0.1, 0.5,$ and 1.0 $\mu$m (Appendix \ref{app:aerosols}). 

This functional form for the aerosol distribution encompasses the parameterizations of several previous efforts: for example, \cite{srom01b} find that the background aerosol structure consists of a homogeneous reflecting layer at 3.8 bar and a transparent reflecting layer near 1.3 bar.  \cite{gibbard02} assume an optically thick cloud layer at $P \geq 3.8$ bar and find that the addition of a thin haze layer at 0.3 bar matches their cloud-free region data well, although more complex aerosol structures are also deemed possible. \cite{irwin11} find that a dark region near the equator (their Pixel 2, Fig. 7) in their Gemini data from September 2009 are well-matched by a moderate optical thickness  cloud at $\sim2$ bar, and a thinner cloud between 50 and 100 mbar. In their 2014 reanalysis, these authors revise the depth of the upper cloud to near 300 mbar.

\begin{figure}[h]
\begin{center}$
\begin{array}{cccccc}
\includegraphics[width=0.5\textwidth]{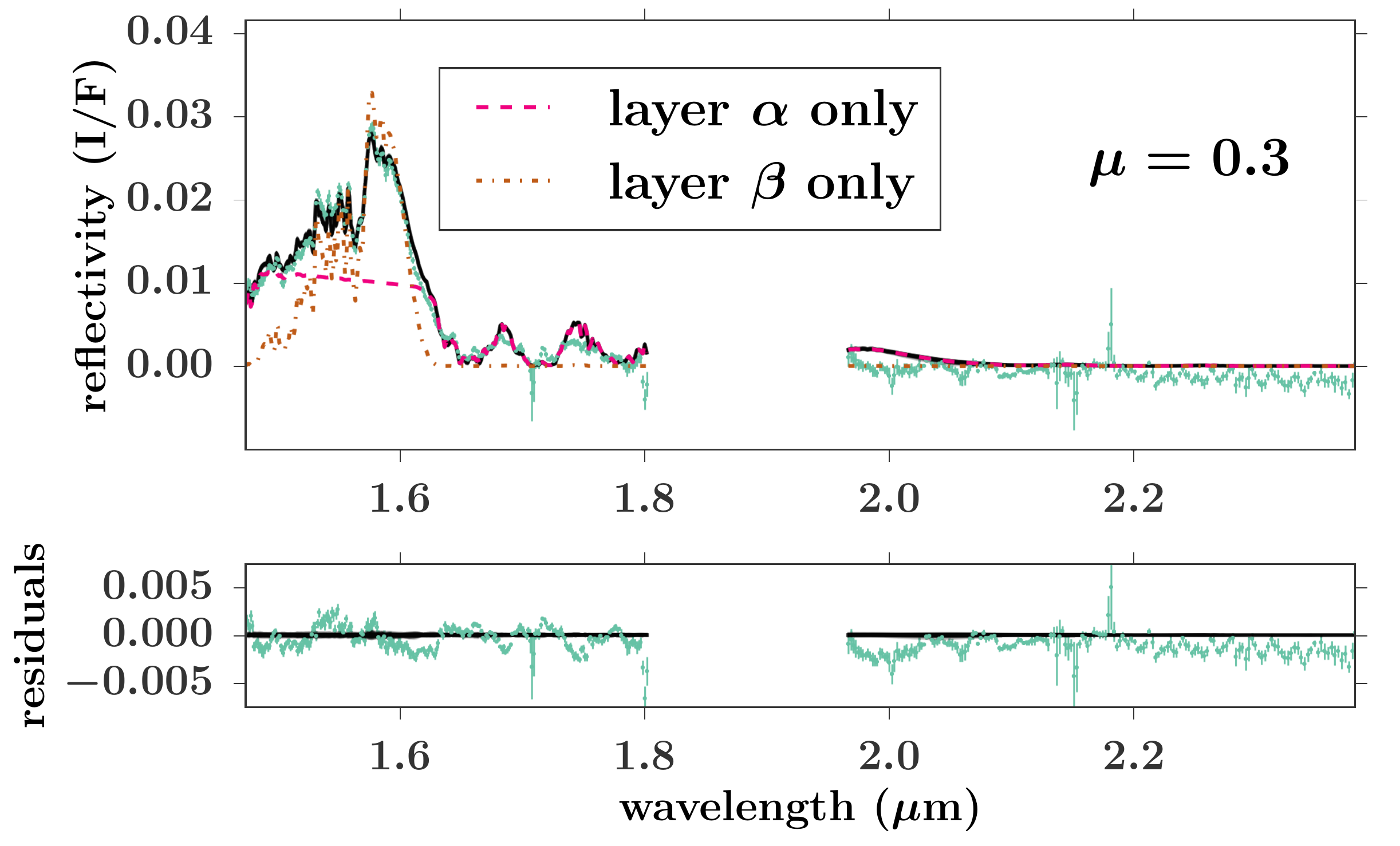} \includegraphics[width=0.5\textwidth]{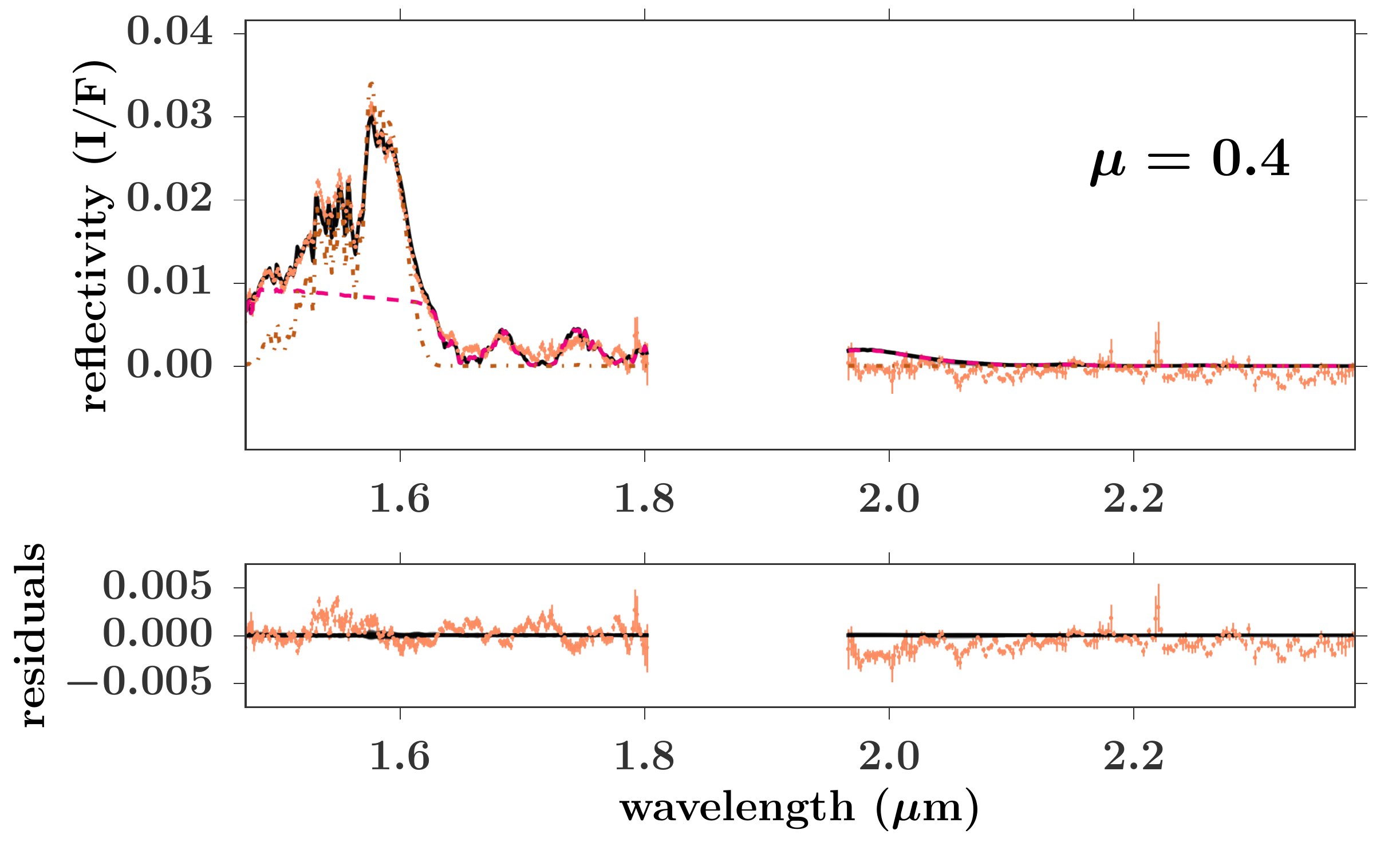}\\
\includegraphics[width=0.5\textwidth]{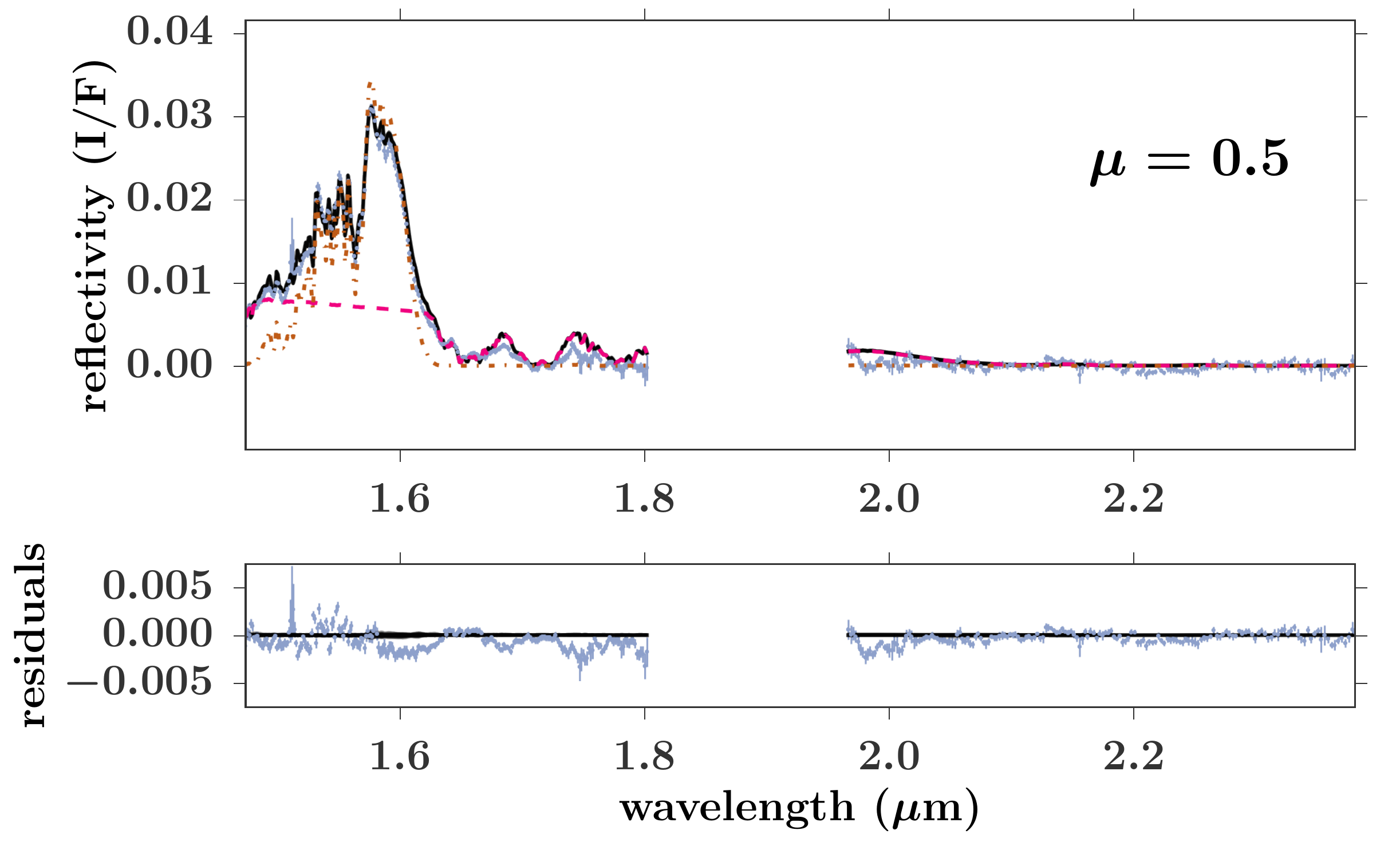}\includegraphics[width=0.5\textwidth]{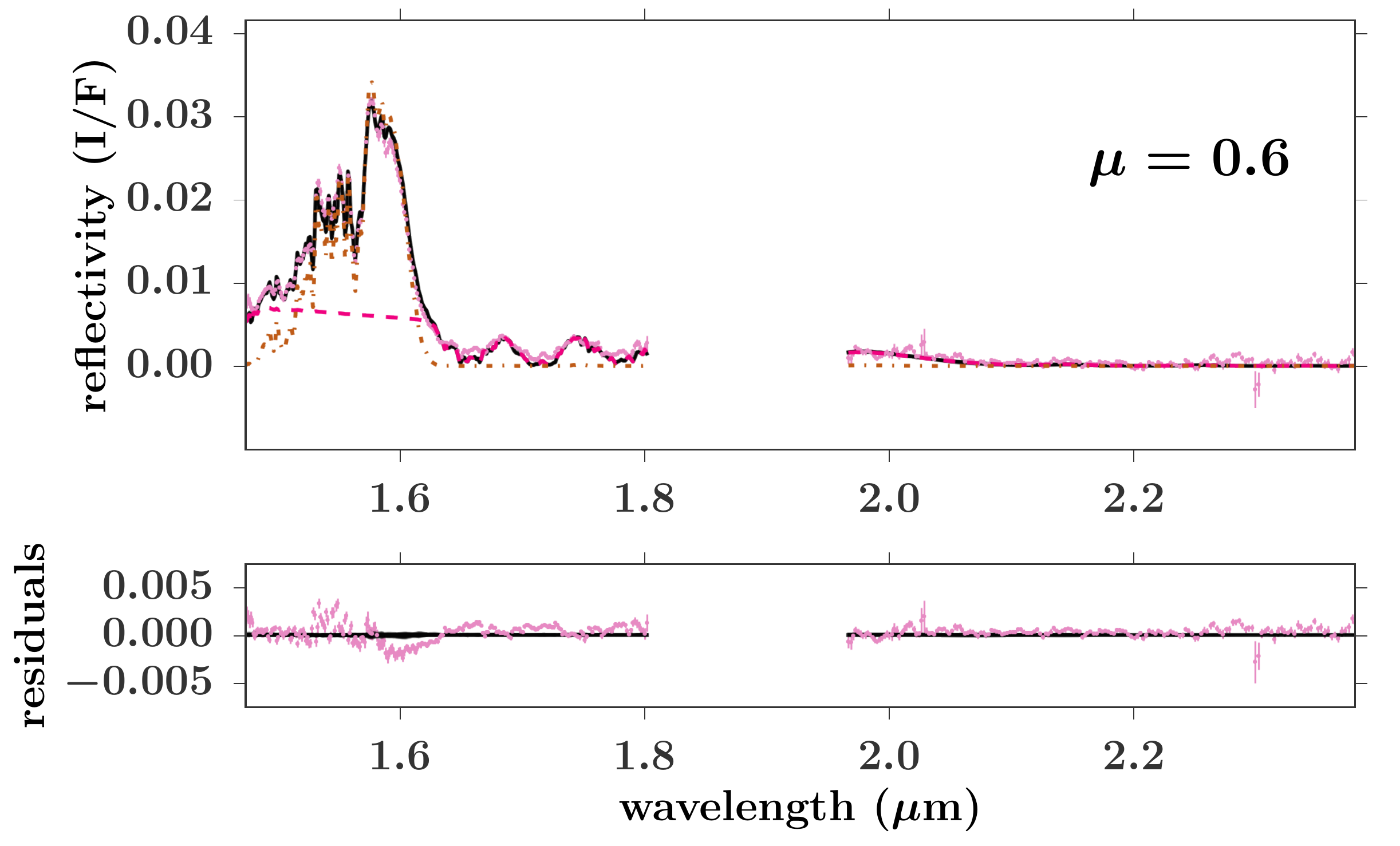}\\
\includegraphics[width=0.5\textwidth]{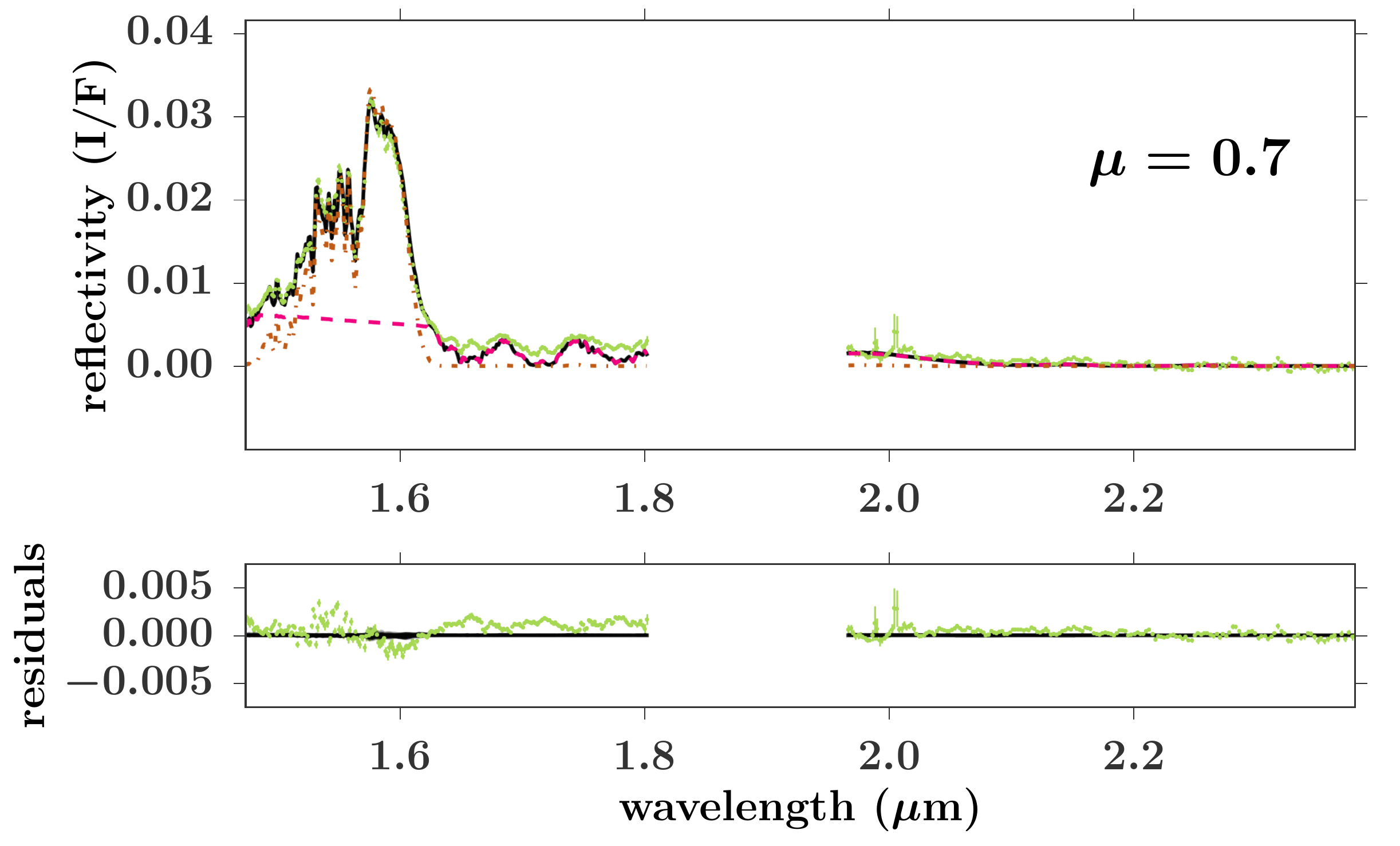}\includegraphics[width=0.5\textwidth]{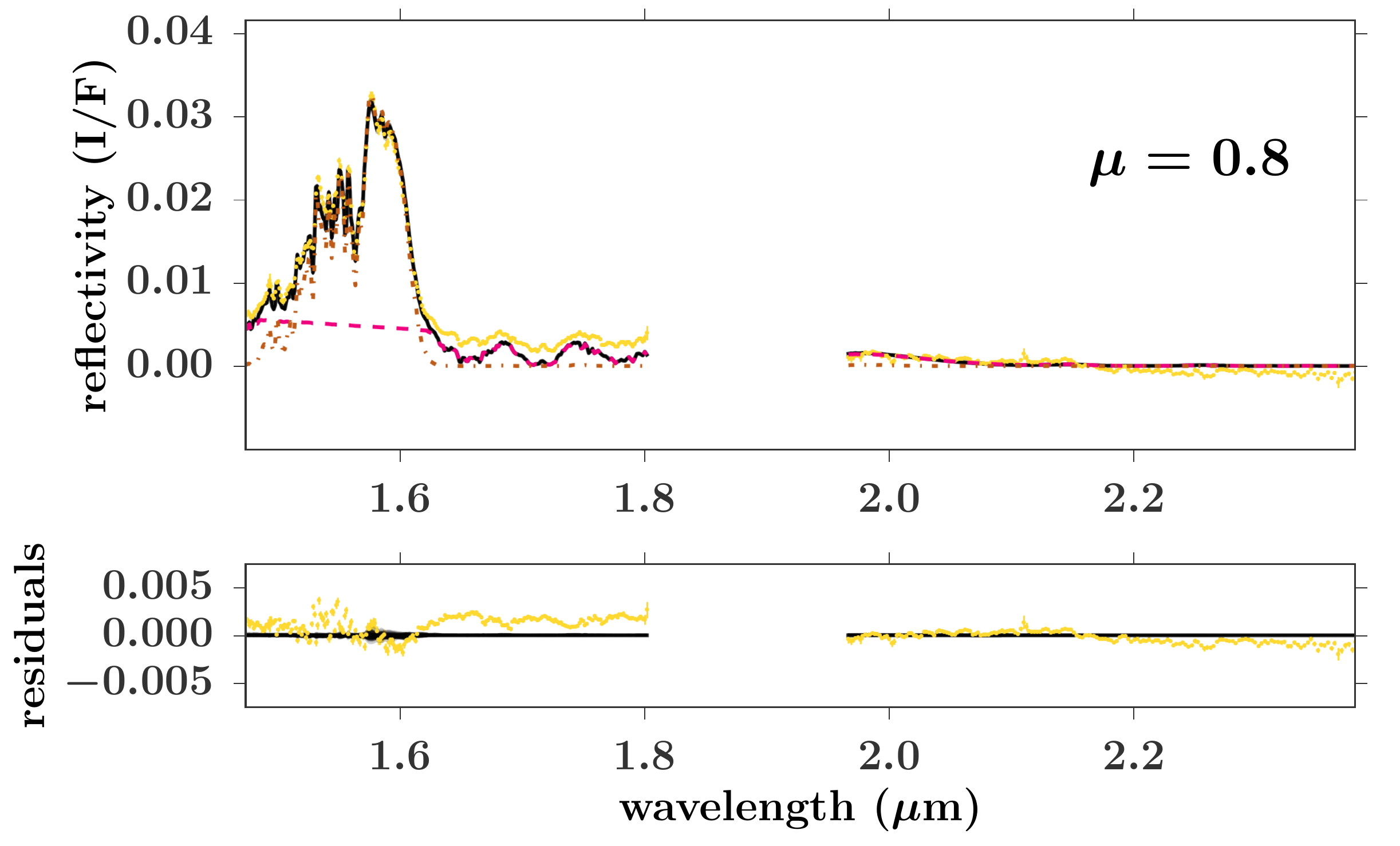}\\
\end{array}$
\end{center}
\caption[{\it 2L\_hazeA}  model fits to limb darkening data]{\label{fig:2L_nom_modelfits} \small Limb-darkening data vs. models, as in Fig. \ref{fig:1L_hgfree_modelfits}. Models represent the {\it 2L\_hazeA} case. In addition to the model representing the 50th percentile of the posterior probability distribution and 100 randomly selected fits, models including just the top aerosol layer (layer $\alpha$) and just the bottom aerosol layer (layer $\beta$) from the best-fit solution are shown (magenta dashed line and brown dot-dashed line, respectively), to illustrate the separate contributions of the two aerosol layers.  } 
\end{figure}

\begin{figure*}[h]
\begin{center}$
\includegraphics[width=0.9\textwidth]{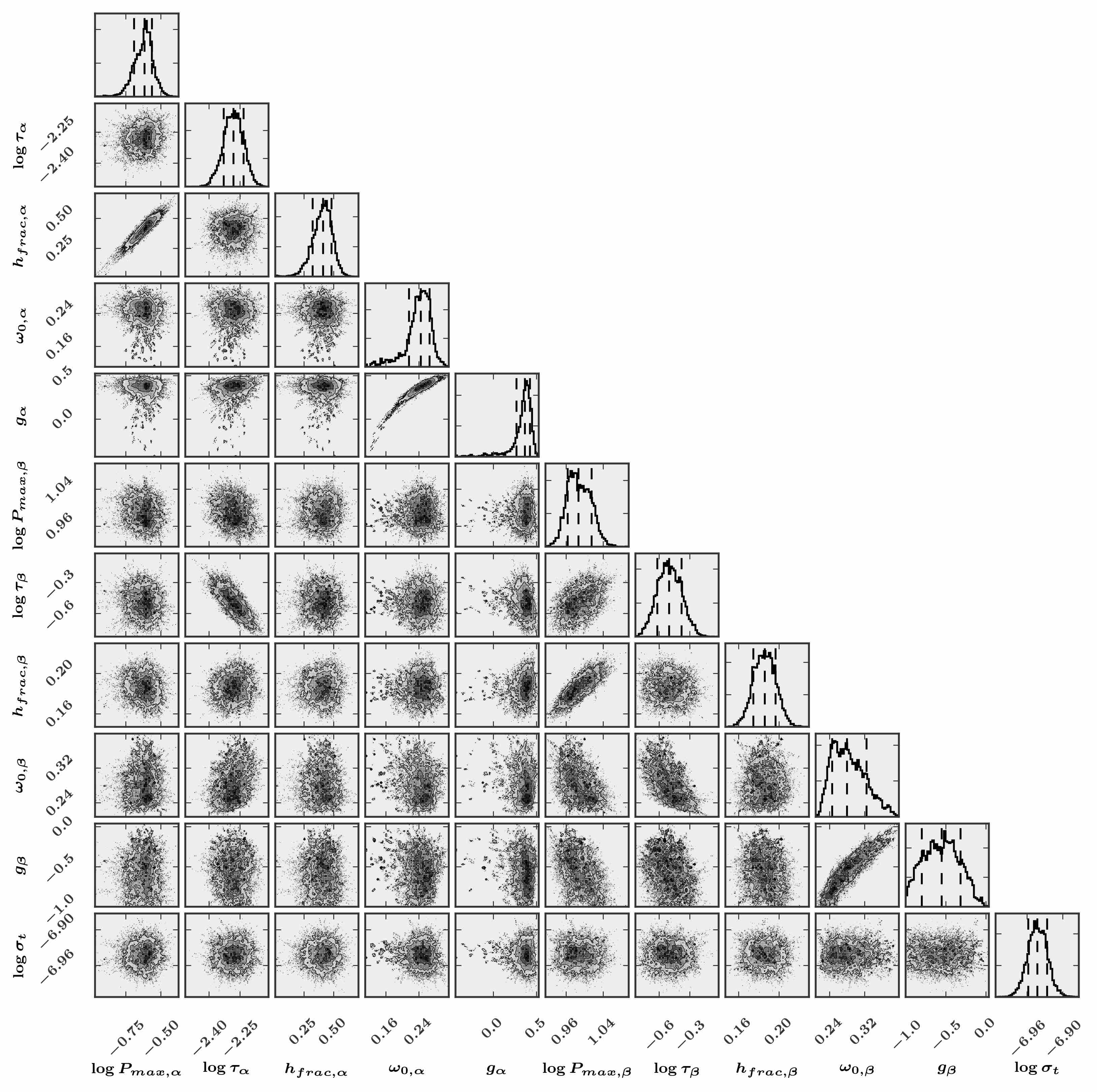}
$
\end{center}
\caption[Triangle plot -- 2L]{\label{fig:2L_triangle} \small 1- and 2-D posterior probability distributions for {\it 2L\_hazeA} model case. }
\end{figure*}

The retrieved parameters for our initial {\it 2L} fits are summarized in the top of Table \ref{table:2L}:  the $r_p = 0.1, 0.5,$ and 1.0$\mu$m cases for the top aerosol layer are labeled {\it 2L\_hazeA, B} and {\it C}, respectively. We find a preference for small particles in the top aerosol layer ({\it 2L\_hazeA}). All three cases result in a similar aerosol structure, consisting of an optically thin ($\tau_\alpha \approx 0.1$) upper aerosol layer with a base around 0.5 bar, and a more optically thick ($\tau_\beta \sim$ 0.3--0.6) layer near 3 bar. In all cases the bottom layer is found to be compact. The  {\it 2L\_hazeA} models are illustrated relative to the data in Fig. \ref{fig:2L_nom_modelfits}, and the posterior probability distributions for this retrieval are shown in Fig. \ref{fig:2L_triangle}. The model fits show a clear improvement over the single aerosol layer ({\it 1L}) models, which is reflected in the substantial decrease in the $DIC$ ($\Delta DIC < -2000$). The improvements are concentrated at the shortest wavelengths and the $\sim$1.65--1.8 $\mu$m spectral region. 

Figure \ref{fig:taubar} permits a visual comparison of the aerosol distributions for the best one-layer retrieval ({\it1L\_nom}) and the {\it 2L\_hazeA} retrieval. We observe that aerosol layer $\beta$ in the {\it 2L\_hazeA} model is qualitatively similar to the single cloud layer in the {\it1L\_nom} retrieval.  Fig. \ref{fig:2L_nom_modelfits} shows, in addition to spectra corresponding to the best fit and randomly selected  {\it 2L\_hazeA}  models, two models that correspond to only layer $\alpha$ or layer $\beta$ of the best-fit {\it 2L\_hazeA} model. These plots highlight that layer $\alpha$ is primarily responsible for the improved fit at $\lambda \lessapprox1.5$ $\mu$m and $\lambda \sim$1.65--1.8 $\mu$m. 

\begin{landscape}
\begin{minipage}{1.1\textwidth}
\setlength\LTleft{-1in}
\setlength\LTright{-1in plus 1 fill}
\renewcommand{\footnoterule}{}
\renewcommand{\thempfootnote}{\it \alph{mpfootnote}}
 \renewcommand{\thefootnote}{\it \alph{footnote}}
\begin{center}
\small
\setlength{\tabcolsep}{0.05in}
\begin{longtable}{+l  ^l | ^l ^l ^l ^l ^l | ^l ^l ^l ^l ^l |^l ^l}
\caption[2L table]{\small Results of two aerosol layer retrievals from limb darkening dataset. \label{table:2L}} \tabularnewline

   run name				&$r_{p,\alpha}$ ($\mu$m) &\multicolumn{3}{l }{layer $\alpha$ (top)}	 	&			    & &\multicolumn{3}{l }{ layer $\beta$ (bottom)	} 	    &	& & $\log{\sigma_t}$	& 	$DIC$\\
			& & $P_{max,\alpha}$ 	 &$\tau_\alpha$ &$h_{frac,\alpha}$		&$\omega_{0,\alpha}$			    &	$g_\alpha$& $P_{max,\beta}$  &$\tau_\beta$ &$h_{frac,\beta}$		&$\omega_{0,\beta}$			    &	$g_\beta$&  & 	\\
			& & (bar)	 & &	&			    &	&  (bar) & &		&			    &	&  & 	\\
{\it 2L\_hazeA}& 0.1	&	$0.54^{+0.03}_{-0.04}$ &   $0.102^{+0.005}_{-0.005}$ &   $0.41^{+0.07}_{-0.09}$ &   $0.24^{+0.02}_{-0.03}$ &   $0.37^{+0.06}_{-0.1}$ &   $2.68^{+0.08}_{-0.06}$ &   $0.61^{+0.08}_{-0.07}$ &   $0.19^{+0.01}_{-0.01}$ &   $0.28^{+0.05}_{-0.03}$ &   $-0.6^{+0.2}_{-0.2}$ &   $-6.94^{+0.02}_{-0.02}$ &$-27684$ \\ 
{\it 2L\_hazeB}& 0.5	& $0.51^{+0.02}_{-0.01}$ &   $0.138^{+0.005}_{-0.006}$ &   $0.04^{+0.04}_{-0.02}$ &   $0.16^{+0.02}_{-0.02}$ &   $0.24^{+0.09}_{-0.09}$ &   $2.94^{+0.1}_{-0.07}$ &   $0.40^{+0.05}_{-0.04}$ &   $0.26^{+0.02}_{-0.02}$ &   $0.37^{+0.03}_{-0.03}$ &   $-0.8^{+0.2}_{-0.1}$ &   $-6.81^{+0.02}_{-0.02}$ &$-27142$ \\
{\it 2L\_hazeC}&1.0	& $0.51^{+0.03}_{-0.01}$ &   $0.144^{+0.006}_{-0.006}$ &   $0.04^{+0.04}_{-0.02}$ &   $0.14^{+0.02}_{-0.02}$ &   $0.1^{+0.1}_{-0.1}$ &   $2.91^{+0.09}_{-0.07}$ &   $0.33^{+0.04}_{-0.03}$ &   $0.27^{+0.02}_{-0.02}$ &   $0.45^{+0.04}_{-0.04}$ &   $-0.6^{+0.1}_{-0.2}$ &   $-6.82^{+0.02}_{-0.02}$ &$-27150$ \\
 \hline
{\it 2L\_DISORT}& 0.1 &$0.59^{+0.04}_{-0.03}$ &   $0.019^{+0.002}_{-0.001}$ &   $0.85^{+0.07}_{-0.06}$ &   $0.91^{+0.06}_{-0.05}$ &   $0.24^{+0.02}_{-0.03}$ &   $3.3^{+0.4}_{-0.3}$ &   $5^{+4}_{-2}$ &   $0.170^{+0.01}_{-0.009}$ &   $0.45^{+0.01}_{-0.01}$ &   $0.50^{+0.02}_{-0.02}$ &   $-7.03^{+0.02}_{-0.02}$ &$-28016$ \\ \\\hline
\hline

\end{longtable}
 \end{center}
\end{minipage}
\end{landscape}

\subsection{Atmospheres with more than two aerosol layers}\label{sec:3L}
Residuals in our two-layer model fits are greater than our estimated data uncertainties; one potential contribution to this discrepancy is that a two-layer aerosol model is insufficient to describe the true distribution of aerosols. We attempted to model the atmosphere with three aerosol layers using a number of prior distributions and initial values for the parameters. We found that the inclusion of a third aerosol layer caused the retrieval to become unstable: parameters had multimodal posterior probability distributions that depended substantially on the parameter initialization. Aerosol layers frequently overlapped substantially and appeared less physically compelling than our two-layer solutions. Therefore, we restrict ourselves to two aerosol layers for the remainder of the analysis. 

\subsection{Limb-darkening analysis with pyDISORT}\label{sec:disort}

In Appendix \ref{app:disort}, we show that two stream is accurate only when modeling a single spectrum at a single viewing geometry, and when the two stream asymmetry factor $g$ is allowed to be a free parameter in the retrieval. The tests described in Appendix \ref{app:disort} strongly suggest that while two stream is sufficient for estimating the vertical profile of aerosols, it is  not well suited to fully take advantage of the information content of our binned, limb-darkening dataset -- that is, to place the best possible constraints on the optical properties of the aerosols from the changes in the spectrum due to variations in viewing geometry. Despite the added computational cost, we require a more accurate radiative transfer solver to achieve this goal. 

We therefore perform a new retrieval of the aerosol properties for the limb-darkening dataset using the pyDISORT algorithm with four streams (section \ref{sec:forward}, Appendix \ref{app:model}). This run, referred to as the  {\it 2L\_DISORT}  model, corresponds exactly to the {\it 2L\_hazeA} run, except for the use of the more accurate radiative transfer solver.  The main purpose of the {\it 2L\_DISORT} retrieval is to accurately constrain Neptune's aerosol properties (including $\omega_0$ and $g$)  from 2--12$^\circ$N. Secondary goals are to determine 1) whether the use of the two stream approximation contributes to the residuals in our earlier model fits, and 2) if and how the use of the two stream approximation influences the retrieved values of model parameters.

 \begin{figure*}[t]
\begin{center}$
\includegraphics[width=0.9\textwidth]{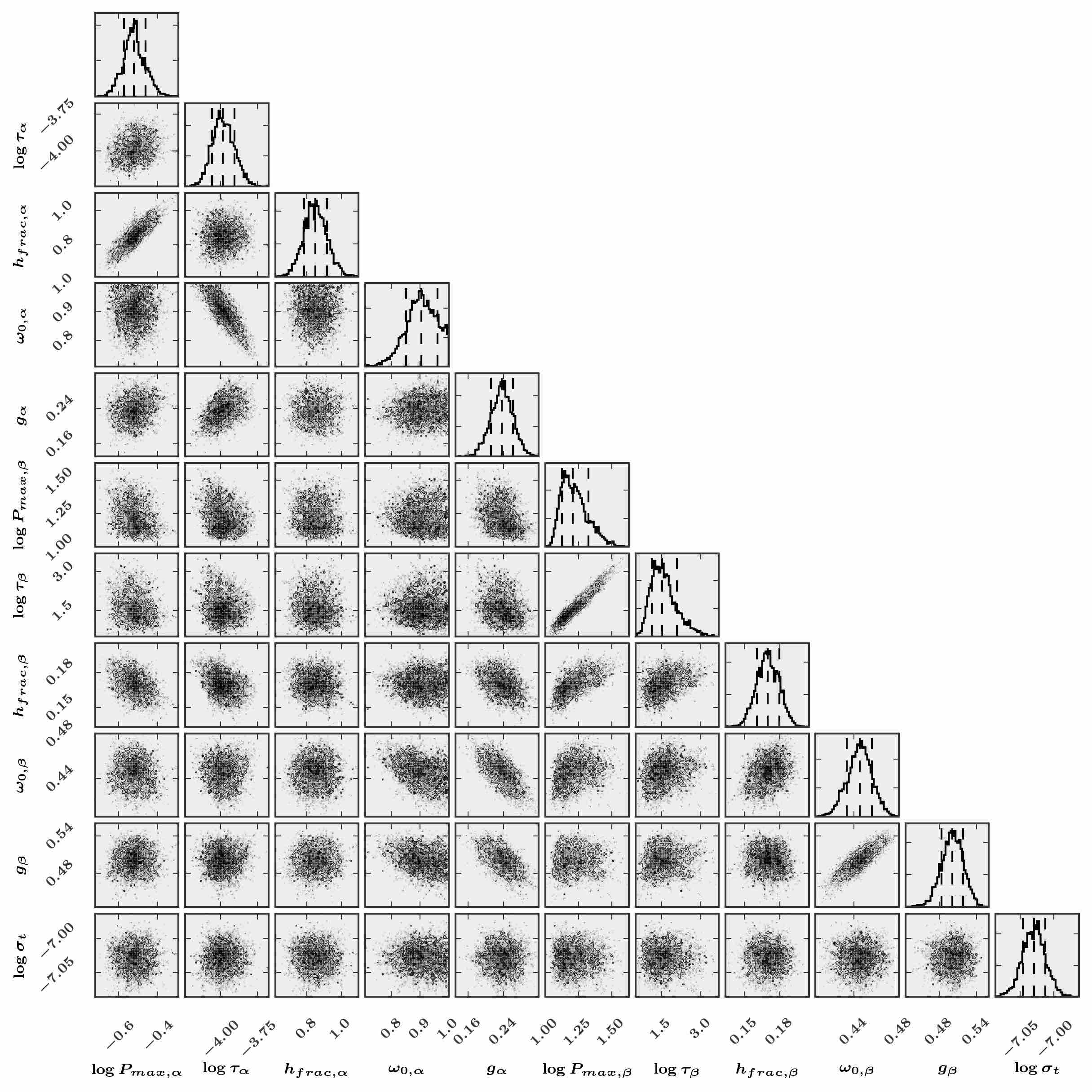}
$
\end{center}
\caption[Triangle plot -- 2L]{\label{fig:disort_triangle} \small 1- and 2-D posterior probability distributions for {\it 2L\_DISORT} model case. }
\end{figure*}

\begin{figure}
\begin{center}$
\begin{array}{cccccc}
\includegraphics[width=0.5\textwidth]{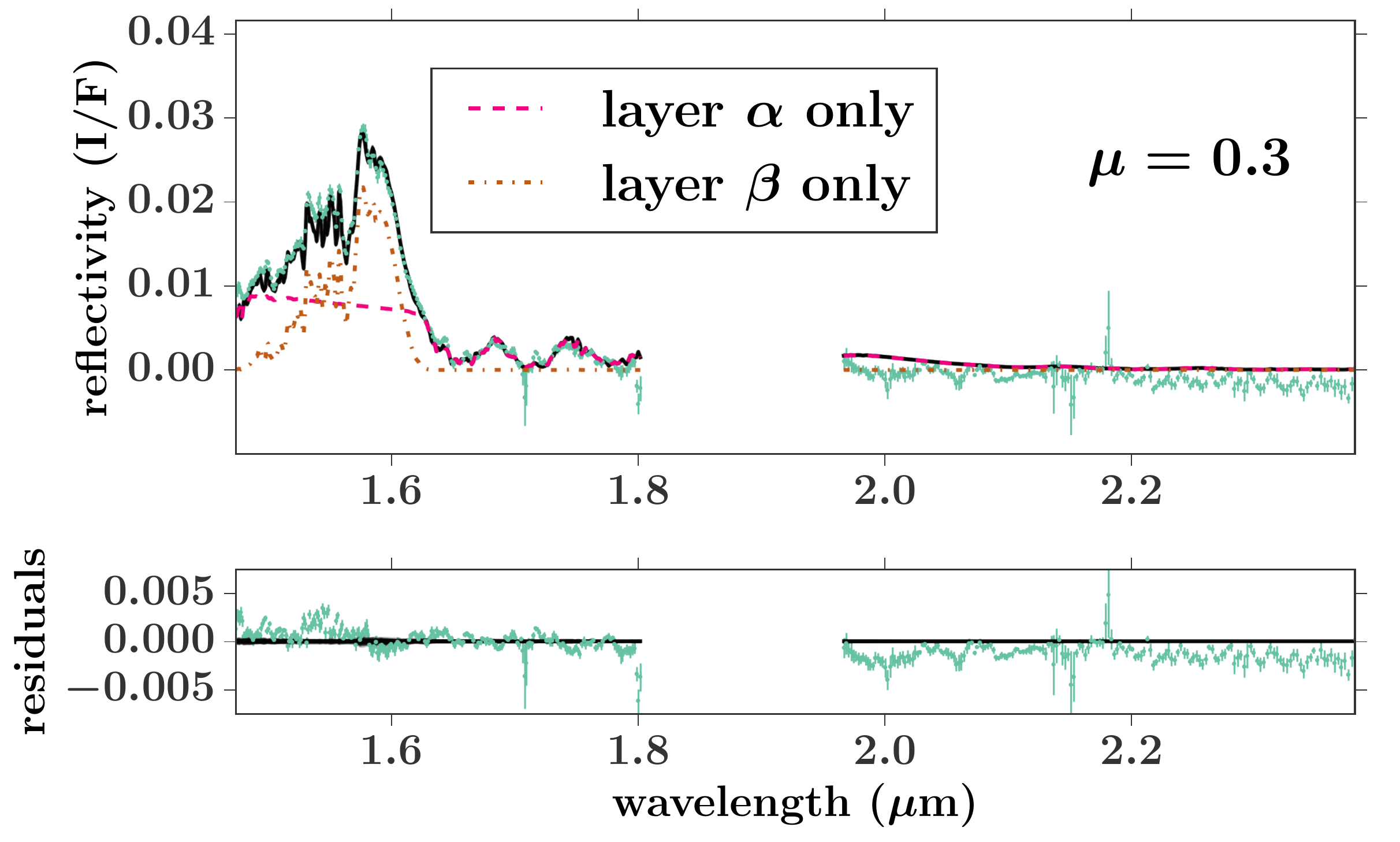} \includegraphics[width=0.5\textwidth]{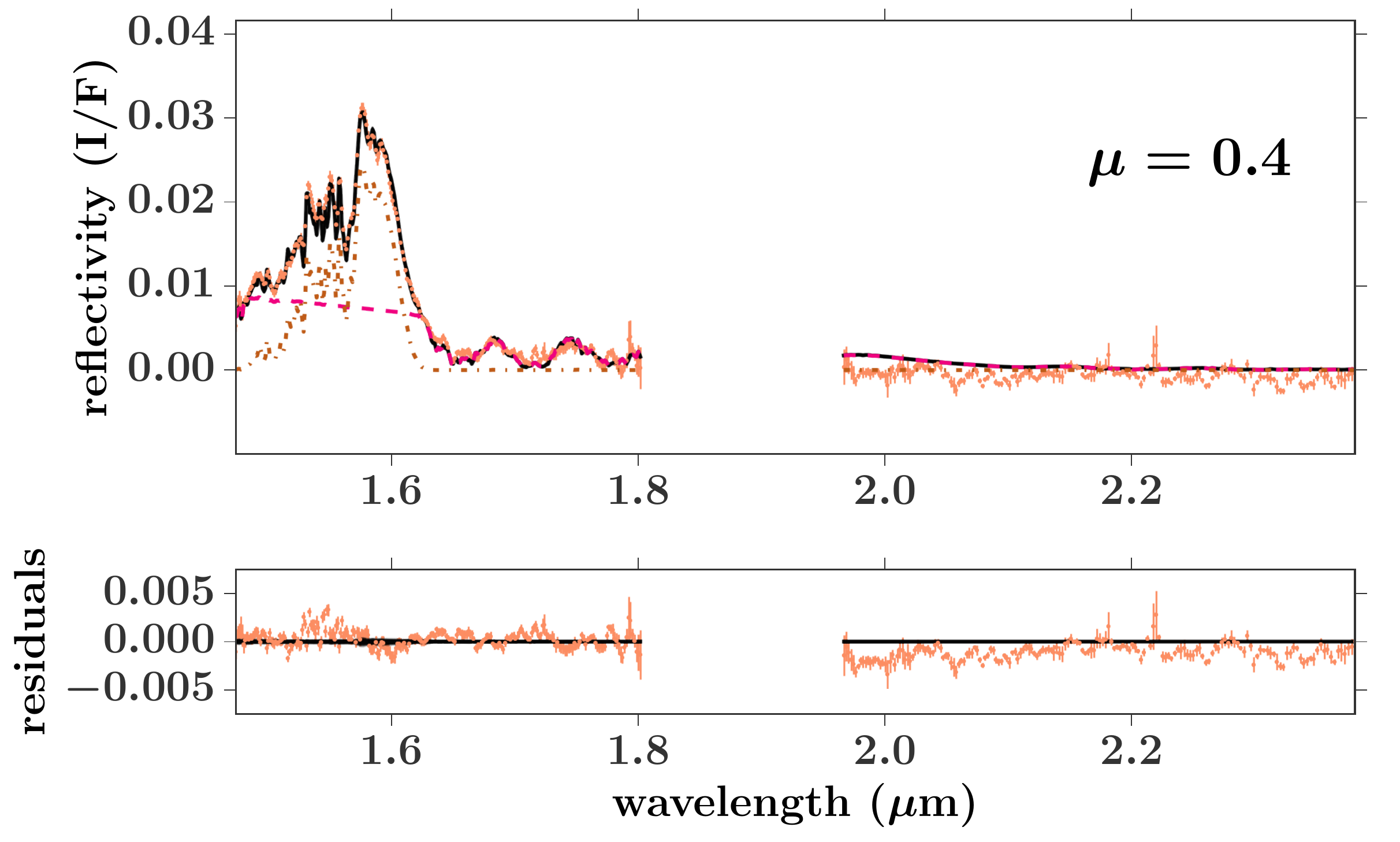}\\
\includegraphics[width=0.5\textwidth]{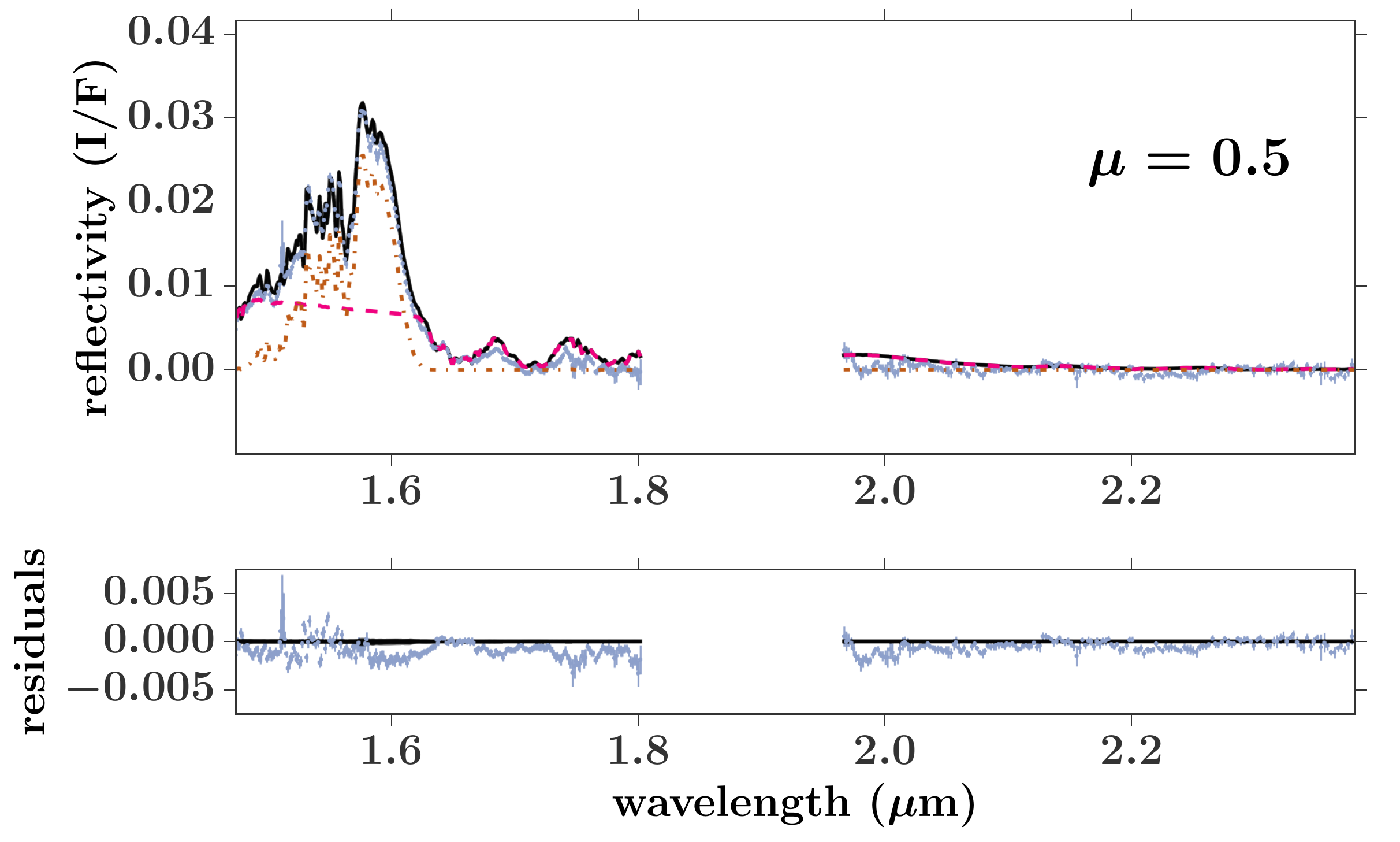}\includegraphics[width=0.5\textwidth]{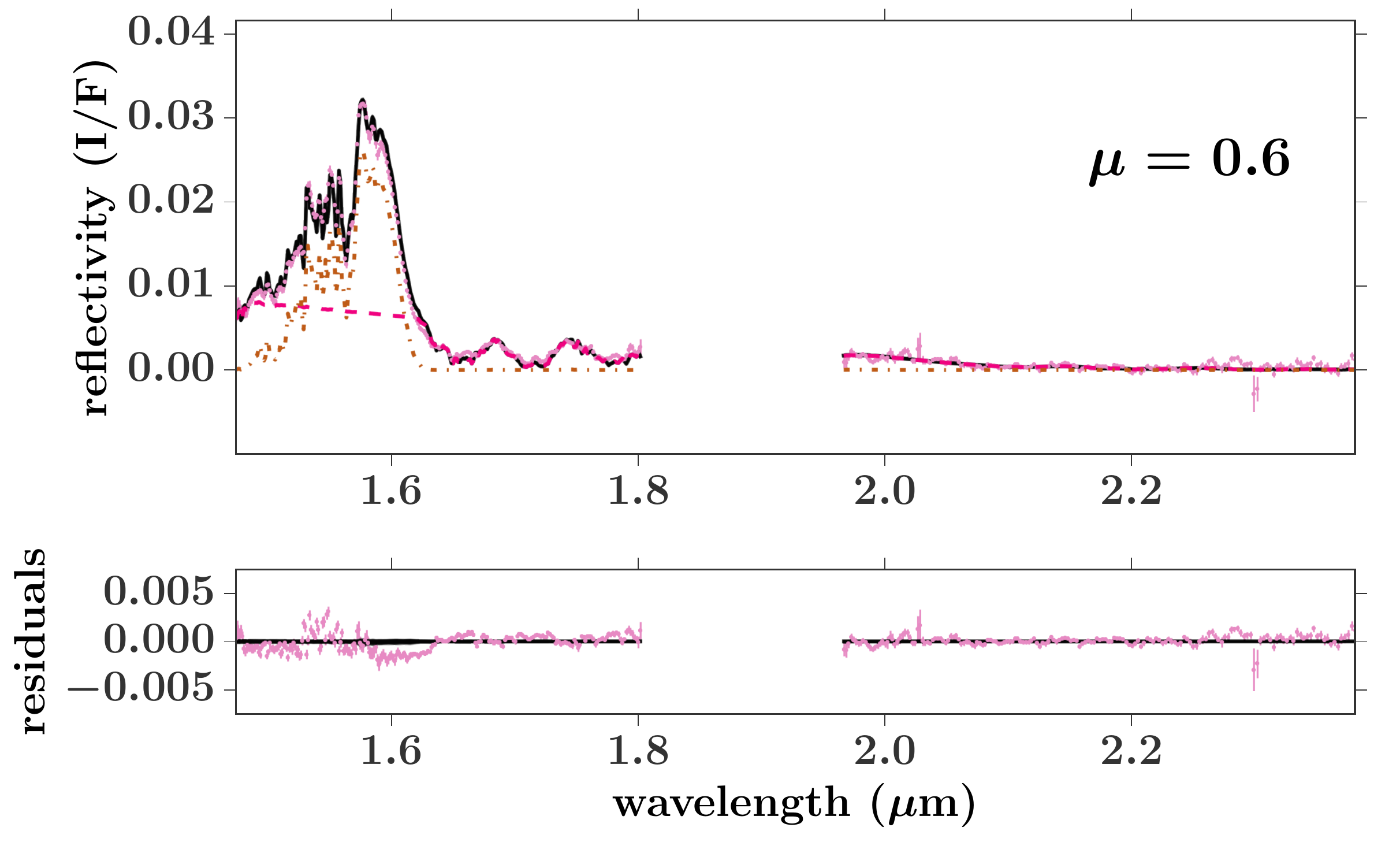}\\
\includegraphics[width=0.5\textwidth]{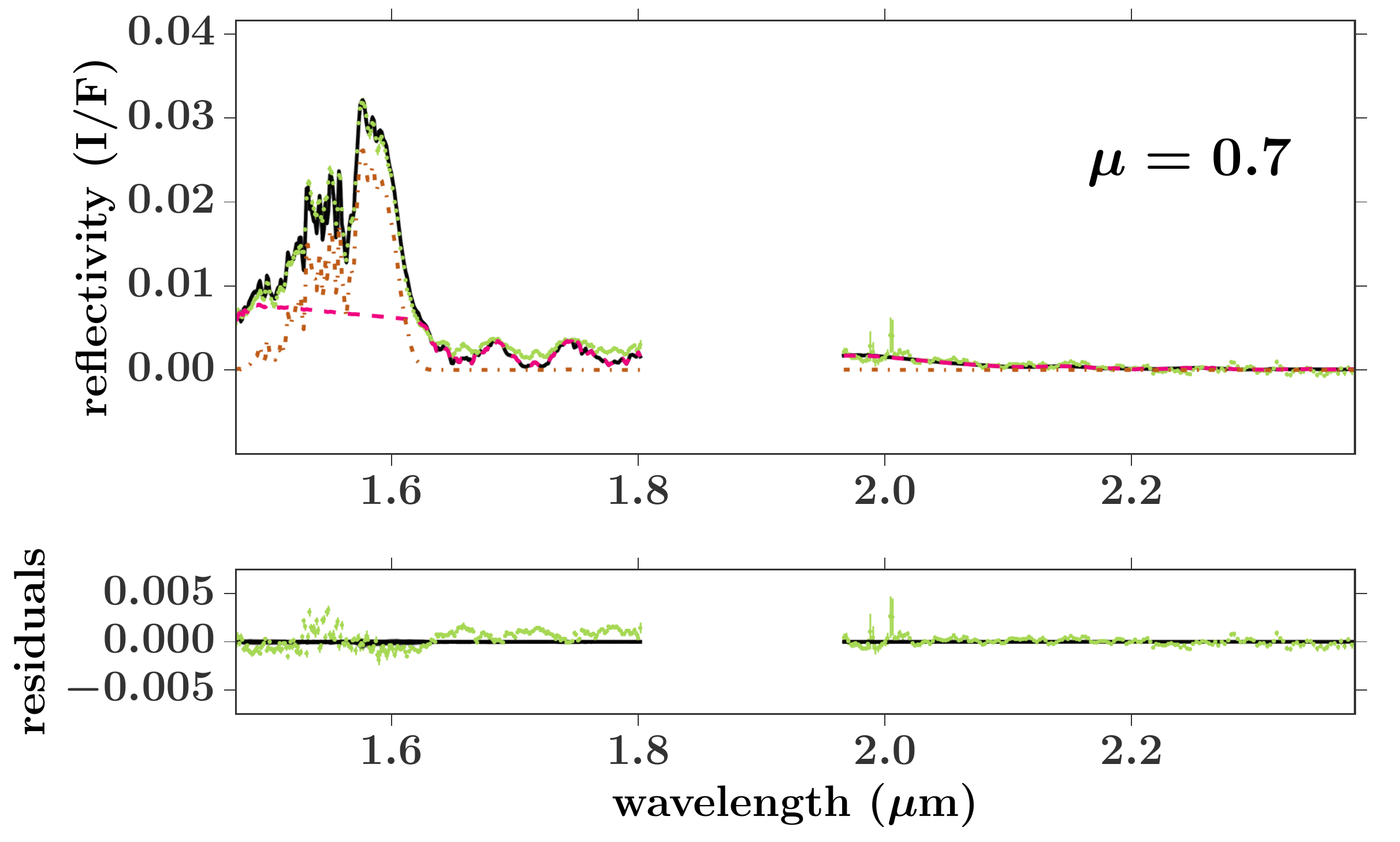}\includegraphics[width=0.5\textwidth]{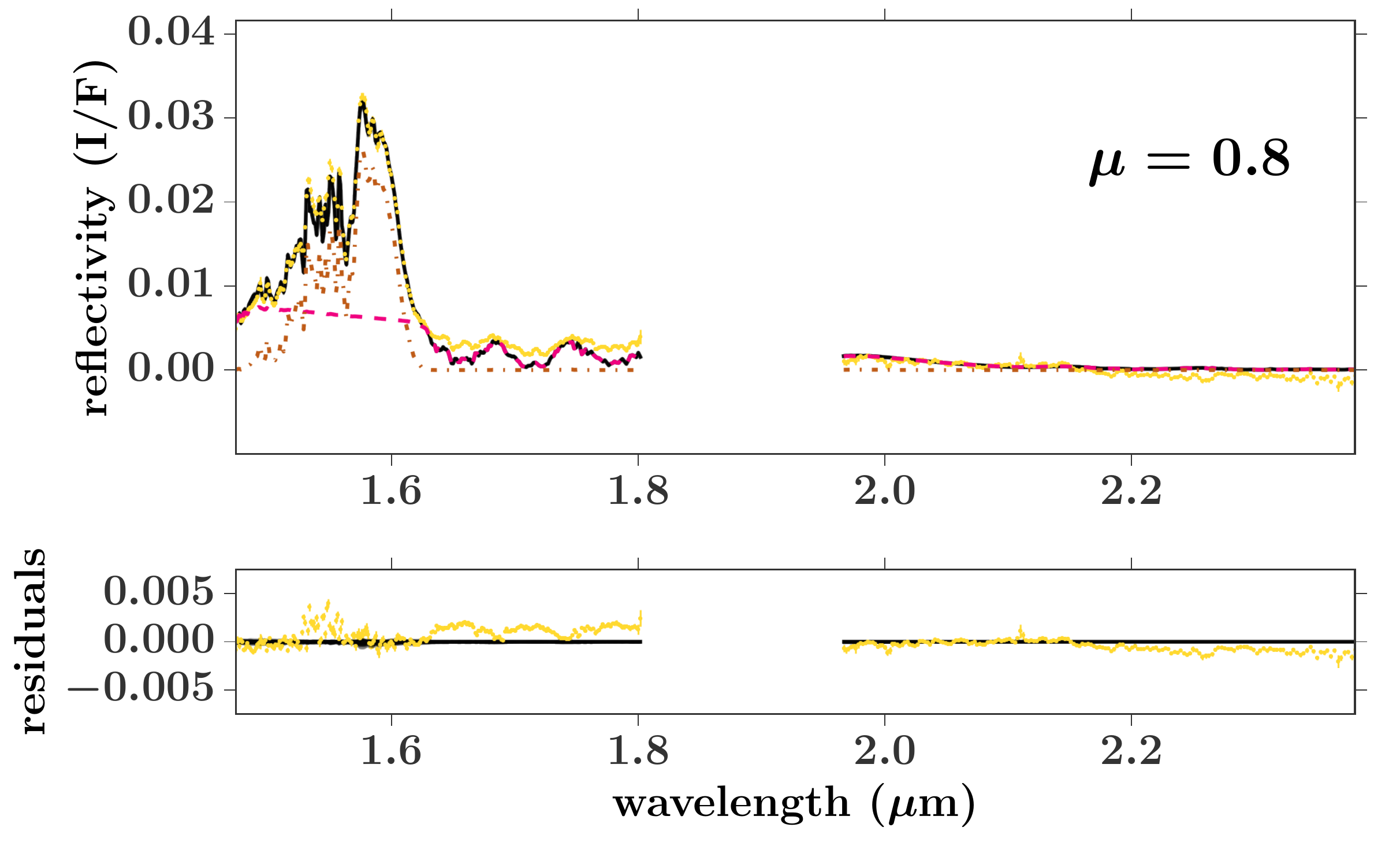}\\
\end{array}$
\end{center}
\caption[{\it 2L\_DISORT}  model fits to limb darkening data]{\label{fig:DISORTfits} \small Limb-darkening data vs. the {\it 2L\_DISORT} models (best fit and 100 randomly selected fits, as in previous cases). As in Fig. \ref{fig:2L_nom_modelfits},  models that include just the top aerosol layer (layer $\alpha$) or just the bottom aerosol layer (layer $\beta$) from the best-fit solution are shown (magenta dashed line and brown dot-dashed line, respectively), to illustrate the separate contributions of the two aerosol layers.  } 
\end{figure}

The results of the  {\it 2L\_DISORT} run are presented in Table \ref{table:SN} and Fig. \ref{fig:disort_triangle}; the models are shown relative to the data in Fig. \ref{fig:DISORTfits}. We find that the more accurate radiative transfer solver produces a better fit to the data than two stream with $\Delta DIC = -332$ for the {\it 2L\_DISORT} retrieval relative to the  {\it 2L\_hazeA} case. While the parameter probability distributions retrieved by the two algorithms are generally different with statistical significance, we find that the aerosol structure derived by the two algorithms is qualitatively similar, consisting of an optically thin ($\tau_\alpha \lessapprox 0.1$) and vertically extended upper layer with a base at 0.5--0.6 bar, and a higher optical depth, more vertically compact bottom aerosol layer near 3 bar. The  {\it 2L\_DISORT} retrieval favors a totally optically thick ($\tau_\beta > 1$) bottom aerosol layer. As observed in the {\it 1L} retrievals, the base of an optically thick cloud is not precisely constrained and is strongly correlated with layer optical depth. The upper aerosol layer is observed to be more vertically extended in the  {\it 2L\_DISORT} solution, with a factor of two larger aerosol scale height than in the best-fit two stream model. A visual comparison of the retrieved aerosol distributions for the two models can be found in Fig. \ref{fig:taubar}. 

Consistent with our expectations from Appendix \ref{app:disort}, we find that the retrieved optical properties of the aerosols are most affected by the choice of radiative transfer solver. In the  {\it 2L\_DISORT} case, a moderately forward scattering phase function ($g_\beta = 0.50^{+0.02}_{-0.02}$) is preferred for the deeper aerosol layer, as compared to the backscattering phase function suggested by two stream ($g_\beta = -0.6^{+0.2}_{-0.2}$).  Using pyDISORT, we retrieve a  more reflective, but still relatively dark, bottom cloud ($\omega_{0,\beta}=0.45^{+0.01}_{-0.01}$). For the upper aerosol layer, the {\it 2L\_DISORT} retrieval indicates nearly perfectly reflecting aerosols, with $\omega_{0,\alpha}=0.91^{+0.06}_{-0.05}$ -- much brighter than what was found by the  {\it 2L\_hazeA}  retrieval.  

\section{Atmospheric composition and temperature, 2--12$^\circ$N }\label{sec:mu8}

We now investigate the effects of varying our assumed methane and thermal profiles. The full limb-darkening dataset described in the previous section, while ideally suited for placing center-to-limb constraints on the scattering properties of Neptune's aerosols, is expensive to fit in terms of CPU hours: each retrieval to the limb-darkening dataset involves simultaneously fitting six spectra, and the variation in viewing geometry necessitates the use of the slower pyDISORT radiative transfer algorithm. Therefore, for this next set of retrievals, we use only the $\mu=0.8$  spectrum from the limb-darkening dataset, which has the highest S/N of the six binned spectra from the previous section. We refer to this set of retrievals as the $SN$ retrievals, to reflect that they are performed on the `high S/N' dataset.

\subsection{Nominal model, high S/N dataset}\label{sec:mu8nom}

\begin{figure}[h!]
\begin{center}$
\begin{array}{cccccc}
\includegraphics[width=0.5\textwidth]{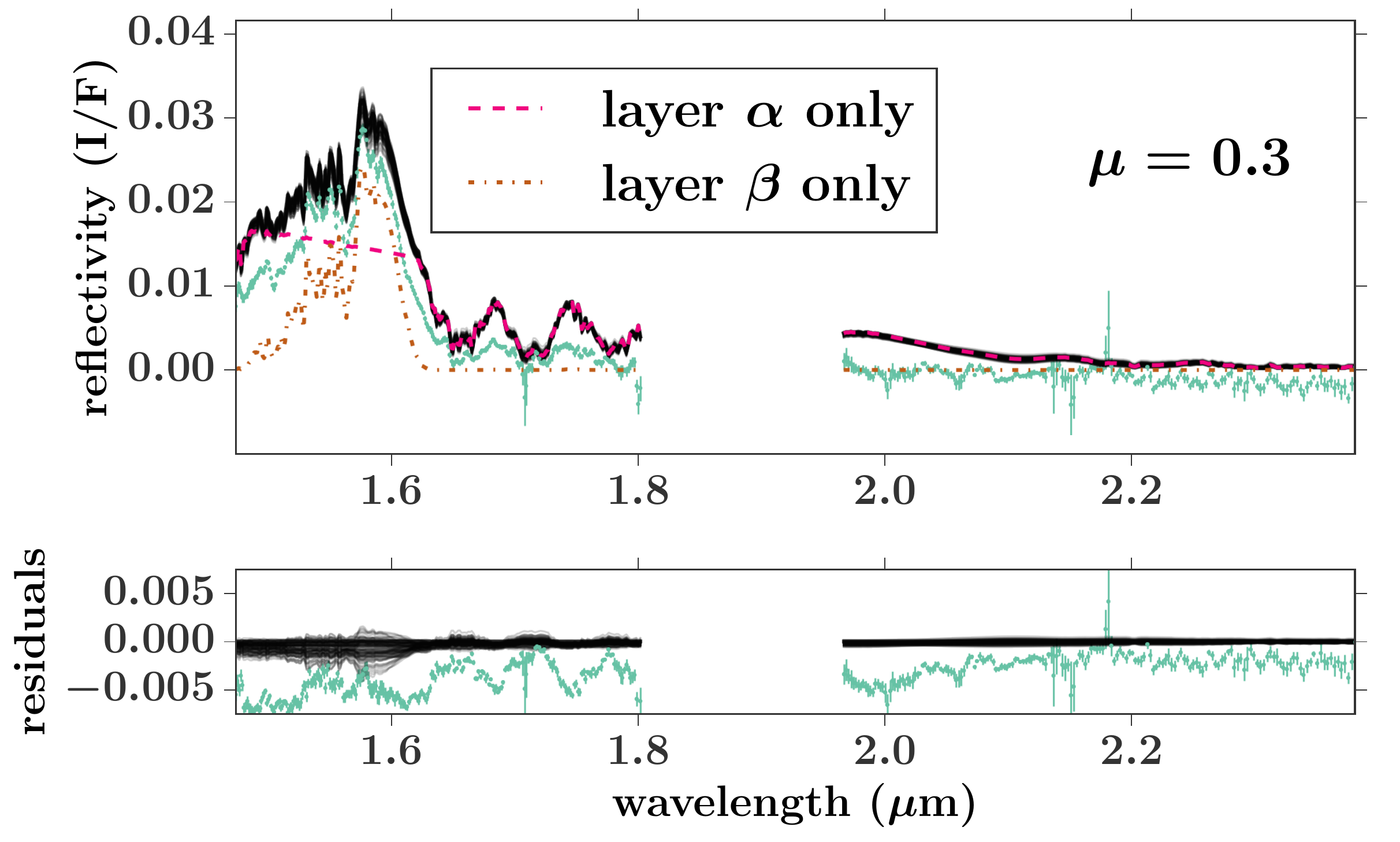} \includegraphics[width=0.5\textwidth]{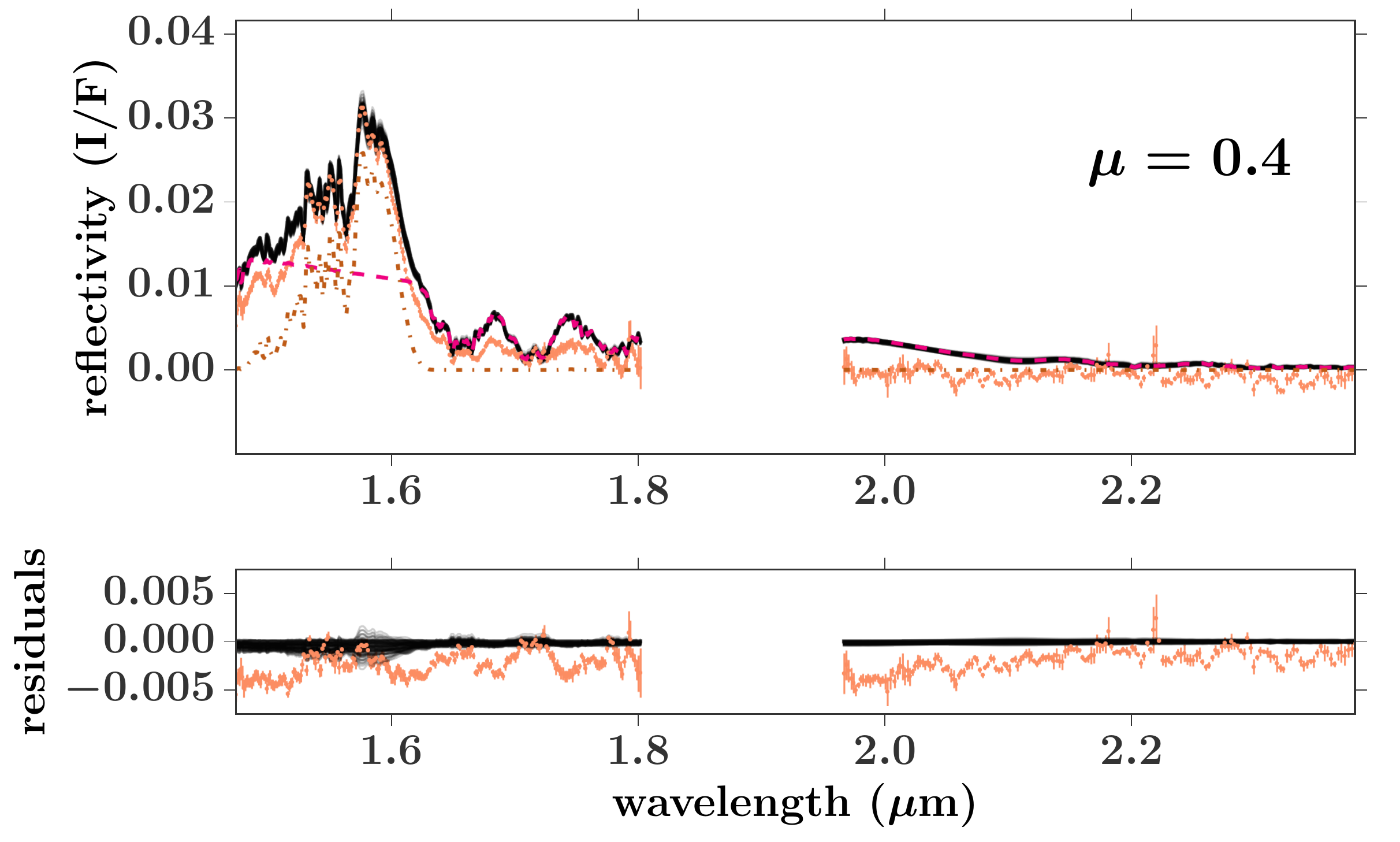}\\
\includegraphics[width=0.5\textwidth]{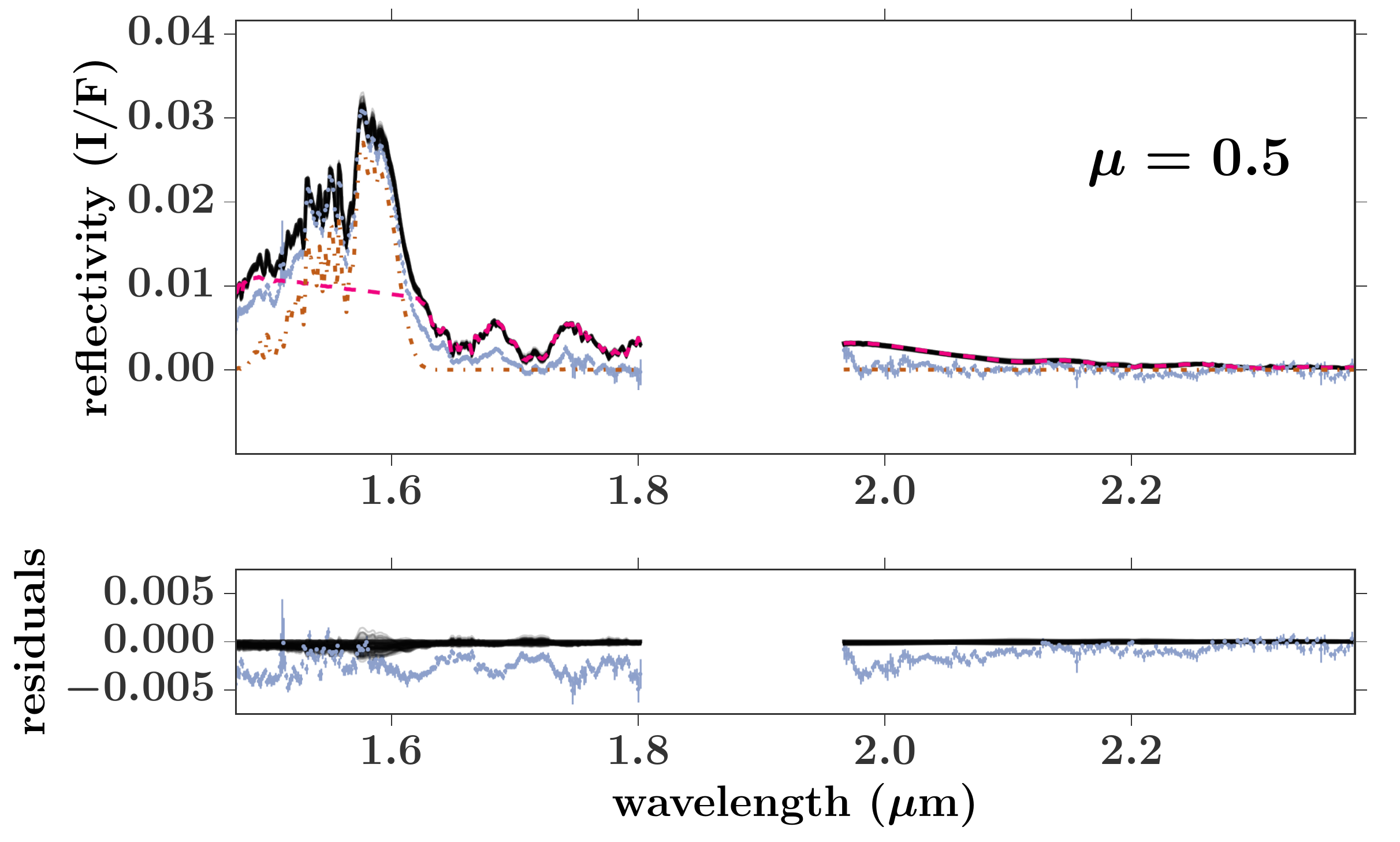}\includegraphics[width=0.5\textwidth]{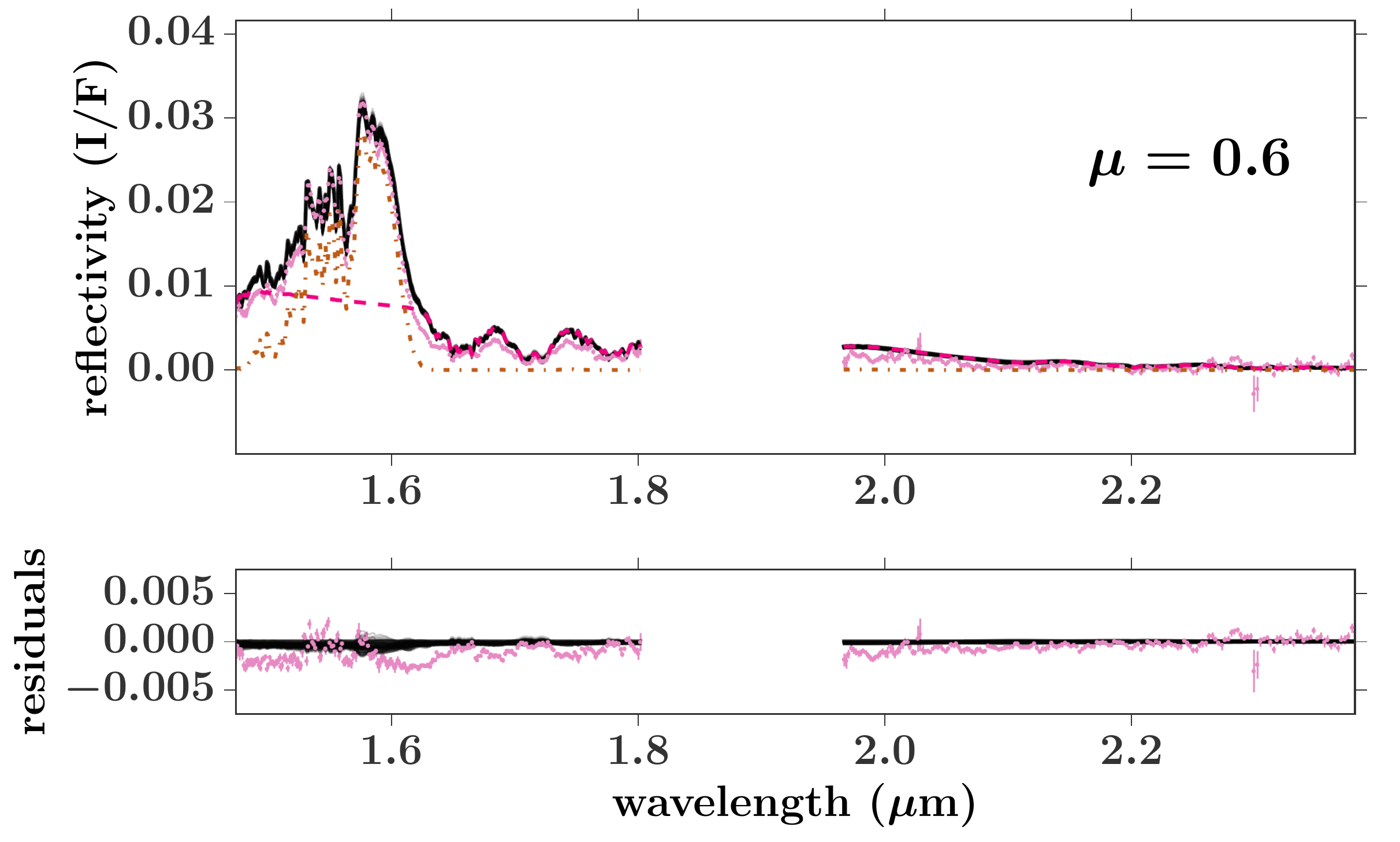}\\
\includegraphics[width=0.5\textwidth]{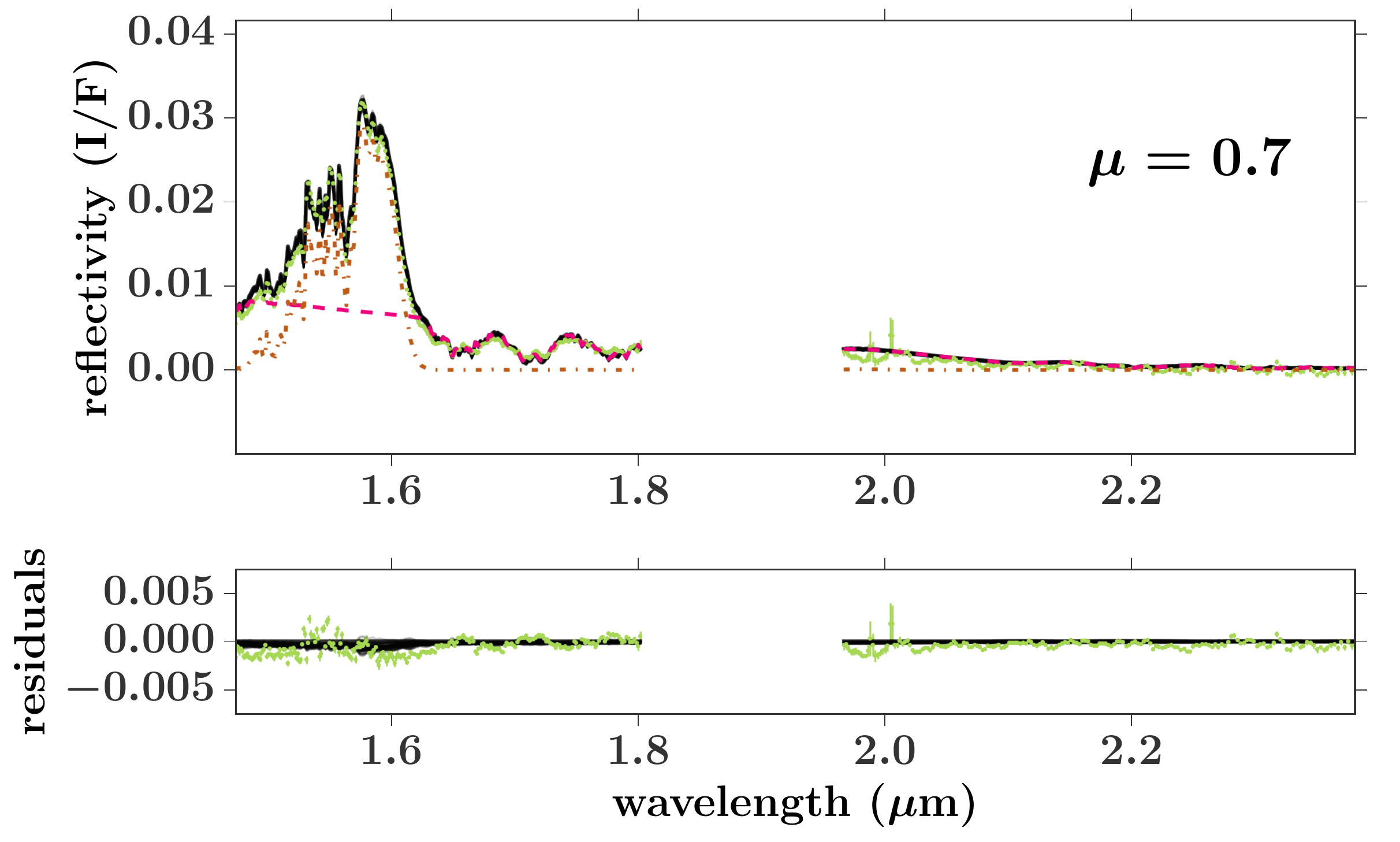}\includegraphics[width=0.5\textwidth]{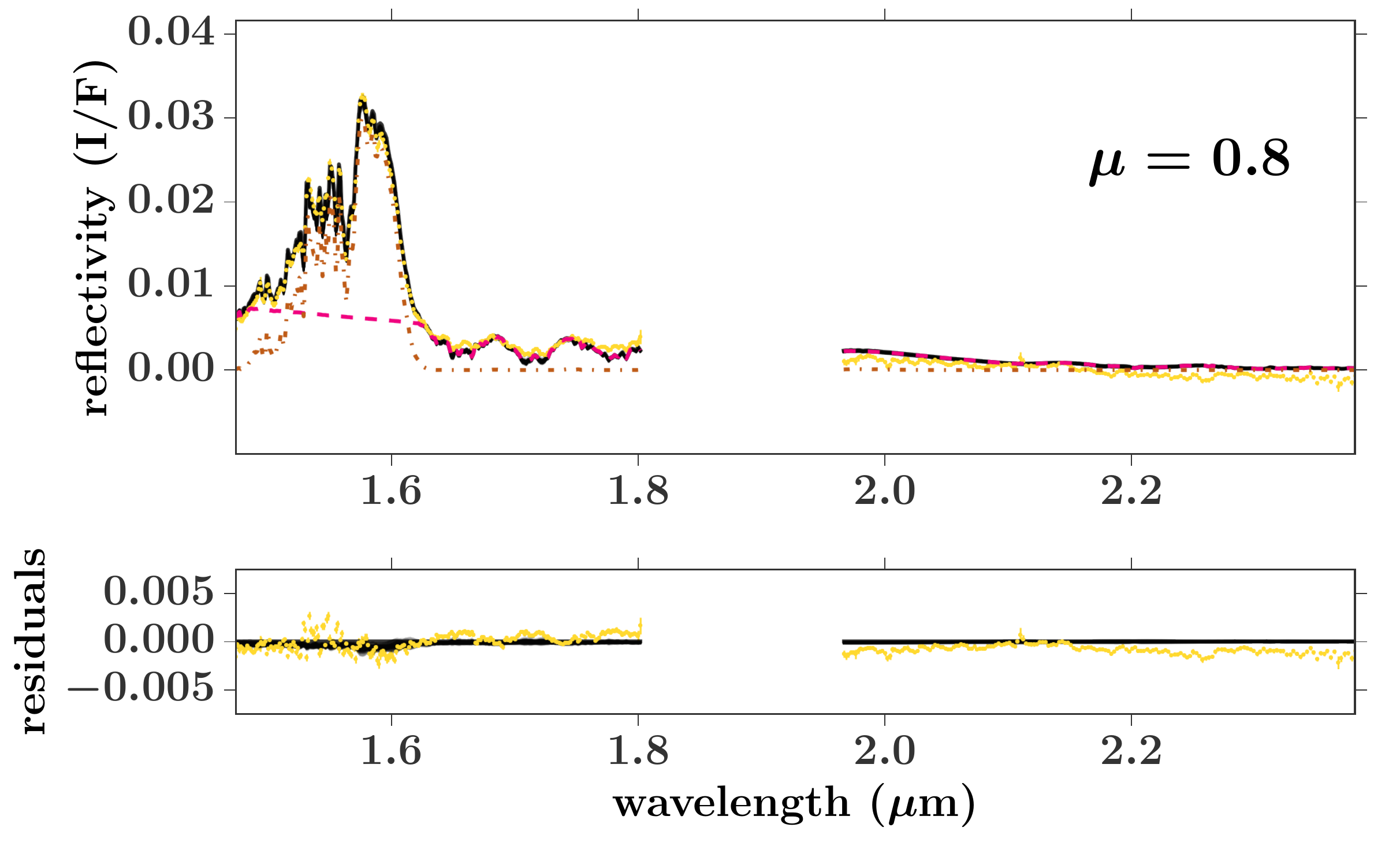}\\
\end{array}$
\end{center}
\caption[  {\it SN\_nom}  model fits relative to the limb-darkening dataset]{\label{fig:SN_nom_modelfits} \small The {\it SN\_nom} models. In this case, retrieval of the aerosol properties is performed using the $\mu=0.8$ data only (bottom right), as described in  Section \ref{sec:mu8}. Model spectra for all six emission angles are shown relative to the six spectra from the 2--12$^\circ$N latitude band. } 
\end{figure}

Since the $SN$ dataset consists of only a single spectrum at a single viewing geometry, we expect (from Appendix \ref{app:disort}) that we should be able to use the more computationally-efficient two stream algorithm to perform the radiative transfer, as long as we are interested primarily in the aerosol structure (as opposed to the aerosol scattering properties). Before exploring the role of composition and temperatures in our model fits, we first establish a nominal model to the $SN$ dataset. This {\it SN\_nom} retrieval has the same model parameters as the {\it 2L\_hazeA} and {\it 2L\_DISORT} runs, but is performed using two stream on the single spectrum of the $SN$ dataset only. The results of this retrieval are presented in Table \ref{table:SN} and illustrated in Fig. \ref{fig:SN_nom_modelfits}. Encouragingly, we find that, with the exception of the optical properties, the parameter values retrieved using the  {\it SN\_nom} model are in good agreement with those from the more rigorous {\it 2L\_DISORT} retrieval described in the previous section. In particular, the retrieved base pressures of both aerosol layers agree to within 1$\sigma$ for the two runs.  The  {\it SN\_nom} retrieval finds aerosol layer $\alpha$ to be slightly more optically thick ($\tau_\alpha=0.050^{+0.01}_{-0.009}$ vs. $\tau_\alpha=0.019^{+0.002}_{-0.001}$  for {\it 2L\_DISORT}) and vertically extended ($h_{frac,\alpha}=1.2^{+0.2}_{-0.2}$ vs. $h_{frac,\alpha}=0.85^{+0.07}_{-0.06}$  for {\it 2L\_DISORT}). Both models find that aerosol layer $\beta$ is optically thick and vertically compact. Figure \ref{fig:taubar} illustrates the aerosol distributions for the {\it SN\_nom} model relative to the  {\it 2L\_DISORT} and {\it 2L\_hazeA} solutions.

Consistent with the results of the tests described in Appendix \ref{app:disort},  the values of the optical parameters ($\omega_0, g$) retrieved by the {\it SN\_nom} model differ from both the  {\it 2L\_DISORT} and {\it 2L\_hazeA} values. The {\it SN\_nom} optical parameters should be interpreted as the two-stream equivalent to the pyDISORT optical parameters {\it for $\mu=0.8$ only}. As illustrated by Fig. \ref{fig:SN_nom_modelfits}, the  {\it SN\_nom} fit quality is excellent for  the $\mu=0.8$ data for which the retrieval was performed, but degrades towards higher emission angles. While the optical parameters from the {\it SN\_nom} model should not be extended to more accurate radiative transfer formulations (or other viewing geometries), the parameters describing the aerosol structure should be robust. Furthermore, the {\it SN\_nom} retrieval is expected to be more accurate than the  {\it 2L\_hazeA} retrieval,  since the {\it 2L\_hazeA}  solution represents a forced compromise in which two stream attempts to accommodate the range of viewing geometries, for which a single set of two-stream $\omega_0, g$ values is likely not appropriate. 

\begin{landscape}
\begin{minipage}{1.1\textwidth}
\setlength\LTleft{-1in}
\setlength\LTright{-1in plus 1 fill}
\renewcommand{\footnoterule}{}
\renewcommand{\thempfootnote}{\it \alph{mpfootnote}}
 \renewcommand{\thefootnote}{\it \alph{footnote}}
\begin{center}
\tiny
\setlength{\tabcolsep}{0.05in}
\begin{longtable}{+l ^l ^l ^l | ^l ^l ^l ^l ^l | ^l ^l ^l ^l ^l |^l ^l ^r}
\caption[High S/N results table]{\small Results of two aerosol layer retrievals from the high S/N $\mu=0.8$ dataset.  \label{table:SN}} \tabularnewline

   run name				&{\it mCH$_{4,t}$}	& {\it mCH$_{4,s}$}	& $RH_{ac}$	&\multicolumn{3}{l }{layer $\alpha$ (top)}	 	&			    & &\multicolumn{3}{l }{ layer $\beta$ (bottom)	} 	    &	& & $\log{\sigma_t}$	& 	$DIC$& $\Delta DIC$\footnote{change in $DIC$ relative to the {\it SN\_nom} model fit; other models of the {\it SN} dataset may also be compared directly using the $DIC$.  A negative change of 10 or more is interpreted as an improvement in the fit quality. A positive change of $>10$ implies a worsening of the fit, whereas changes of less than $\pm10$ are not conclusively better of worse.}\\
   		&	& 	& & $P_{max,\alpha}$ (bar)	 &$\tau_\alpha$ &$h_{frac,\alpha}$		&$\omega_{0,\alpha}$\footnote{$\omega_0$ and $g$ are retrieved in the {\it SN\_nom} case and fixed near the best-fit {\it SN\_nom} values in the remaining model cases. Two stream was used for all fits summarized in this table, and $\omega_0$ and $g$ are only appropriate for two stream at this particular viewing geometry.}			    &	$g_\alpha$& $P_{max,\beta}$ (bar)	 &$\tau_\beta$ &$h_{frac,\beta}$		&$\omega_{0,\beta}$			    &	$g_\beta$&  & & 	\\

{\it SN\_nom}&$0.040$& $3.5e-4$& 			$1.0$ 		&	$0.55^{+0.07}_{-0.07}$ &   $0.050^{+0.01}_{-0.009}$ &   $1.2^{+0.2}_{-0.2}$ &   $0.23^{+0.04}_{-0.04}$ &   $-0.7^{+0.2}_{-0.2}$ &   $3.1^{+0.3}_{-0.2}$ &   $1.8^{+0.9}_{-0.5}$ &   $0.18^{+0.02}_{-0.02}$ &   $0.29^{+0.03}_{-0.03}$ &   $0.0^{+0.1}_{-0.2}$ &   $-7.11^{+0.04}_{-0.04}$ &$-4800$ &  \\
\hline
{\it SN\_CH$_4$grid$_A$}&$0.022$& $3.5e-4$& $0.4$             &$0.41^{+0.07}_{-0.07}$ &   $0.053^{+0.002}_{-0.002}$ &   $0.4^{+0.2}_{-0.2}$ & $0.2$&$-0.7$&  $3.6^{+0.2}_{-0.2}$ &   $1.07^{+0.1}_{-0.09}$ &   $0.23^{+0.02}_{-0.02}$ & $0.3$&$0.0$&  $-7.28^{+0.04}_{-0.04}$ &$-4923$ & $-123$ \\
{\it SN\_CH$_4$grid$_B$}&$0.022$& $3.5e-4$& 			$1.0$ 		&$0.53^{+0.07}_{-0.06}$ &   $0.052^{+0.003}_{-0.003}$ &   $1.2^{+0.2}_{-0.2}$ & $0.2$&$-0.7$&  $3.8^{+0.6}_{-0.3}$ &   $1.2^{+0.5}_{-0.2}$ &   $0.27^{+0.03}_{-0.03}$ & $0.3$&$0.0$&  $-7.14^{+0.04}_{-0.04}$ &$-4821$ & $-21$\\
{\it SN\_CH$_4$grid$_C$}&$0.022$& $1.5e-3$& 			$0.4$ 		&$0.45^{+0.08}_{-0.07}$ &   $0.057^{+0.002}_{-0.002}$ &   $0.5^{+0.3}_{-0.2}$ &  $0.2$&$-0.7$&  $3.5^{+0.2}_{-0.2}$ &   $1.04^{+0.1}_{-0.09}$ &   $0.20^{+0.02}_{-0.02}$ &$0.3$&$0.0$&   $-7.20^{+0.04}_{-0.04}$ &$-4868$ & $-68$\\
{\it SN\_CH$_4$grid$_D$}&$0.022$& $1.5e-3$& 			$1.0$ 		&$0.68^{+0.1}_{-0.08}$ &   $0.053^{+0.004}_{-0.003}$ &   $1.8^{+0.3}_{-0.2}$ &   $0.2$&$-0.7$& $4.0^{+1.0}_{-0.4}$ &   $1.3^{+2.0}_{-0.3}$ &   $0.28^{+0.03}_{-0.04}$ &$0.3$&$0.0$&     $-7.08^{+0.04}_{-0.04}$ &$-4771$ &$29$ \\
{\it SN\_CH$_4$grid$_E$}&$0.040$& $3.5e-4$& 			$0.4$ 		&$0.41^{+0.06}_{-0.07}$ &   $0.054^{+0.001}_{-0.001}$ &   $0.4^{+0.2}_{-0.2}$ &  $0.2$&$-0.7$&   $2.94^{+0.1}_{-0.09}$ &   $1.08^{+0.1}_{-0.09}$ &   $0.16^{+0.01}_{-0.01}$ & $0.3$&$0.0$&   $-7.27^{+0.04}_{-0.04}$ &$-4921$& $-121$ \\
{\it SN\_CH$_4$grid$_F$}&$0.040$& $1.5e-3$& 			$0.4$ 		&$0.45^{+0.08}_{-0.08}$ &   $0.058^{+0.002}_{-0.002}$ &   $0.5^{+0.3}_{-0.2}$ &    $0.2$&$-0.7$& $2.9^{+0.1}_{-0.1}$ &   $1.04^{+0.1}_{-0.09}$ &   $0.14^{+0.01}_{-0.01}$ & $0.3$&$0.0$&  $-7.20^{+0.04}_{-0.04}$ &$-4867$ & $-66$\\
{\it SN\_CH$_4$grid$_G$}&$0.040$& $1.5e-3$& 			$1.0$ 		&$0.67^{+0.1}_{-0.09}$ &   $0.055^{+0.003}_{-0.003}$ &   $1.7^{+0.2}_{-0.2}$ &   $0.2$&$-0.7$& $3.1^{+0.8}_{-0.2}$ &   $1.3^{+2.0}_{-0.2}$ &   $0.20^{+0.02}_{-0.03}$ &  $0.3$&$0.0$&   $-7.06^{+0.03}_{-0.04}$ &$-4753$ & $47$\\
\hline
{\it SN\_CH$_4$RHfree$_A$}&$0.040$& $3.5e-4$& $0.016^{+0.01}_{-0.008}$ &   $0.49^{+0.07}_{-0.07}$ &   $0.046^{+0.002}_{-0.004}$ &   $0.3^{+0.1}_{-0.2}$ &  $0.2$&$-0.7$&  $2.88^{+0.06}_{-0.07}$ &   $0.96^{+0.06}_{-0.06}$ &   $0.135^{+0.01}_{-0.009}$ &  $0.3$&$0.0$&   $-7.47^{+0.04}_{-0.04}$ &$-5052$  & $-252$ \\
{\it SN\_CH$_4$RHfree$_B$}&$0.040$& $1.5e-3$&$8e-4^{+4e-4}_{-3e-4}$ &   $0.35^{+0.08}_{-0.05}$ &   $0.033^{+0.003}_{-0.003}$ &   $0.2^{+0.2}_{-0.1}$ &  $0.2$&$-0.7$&  $2.92^{+0.05}_{-0.05}$ &   $0.97^{+0.05}_{-0.05}$ &   $0.159^{+0.008}_{-0.008}$ &  $0.3$&$0.0$&   $-7.47^{+0.04}_{-0.04}$ &$-5058$&$-258$  \\
{\it SN\_CH$_4$stratfree}&$0.040$&$2.2e-4^{+4e-5}_{-2e-5}$ &$0.4$&   $0.42^{+0.06}_{-0.07}$ &   $0.053^{+0.001}_{-0.002}$ &   $0.4^{+0.2}_{-0.2}$ &   $0.2$&$-0.7$& $3.0^{+0.1}_{-0.1}$ &   $1.1^{+0.1}_{-0.1}$ &   $0.16^{+0.01}_{-0.01}$ &    $0.3$&$0.0$&$-7.29^{+0.04}_{-0.04}$ &$-4928$&$-128$\\
\hline
{\it SN\_Ttrop}           &$0.040$& $3.5e-4$& 			$0.4$ 		&$0.51^{+0.06}_{-0.05}$ &   $0.054^{+0.002}_{-0.002}$ &   $1.0^{+0.2}_{-0.1}$ &  $0.2$&$-0.7$ &   $2.9^{+0.1}_{-0.1}$ &   $1.1^{+0.1}_{-0.1}$ &   $0.16^{+0.01}_{-0.01}$ &  $0.3$&$0.0$&   $-7.21^{+0.04}_{-0.04}$ &$-4872$ &$-72$ \\
{\it SN\_Tstratp20} 	&$0.040$& $3.5e-4$& 			$0.4$ 		&$0.36^{+0.05}_{-0.04}$ &   $0.052^{+0.001}_{-0.001}$ &   $0.2^{+0.1}_{-0.1}$ &  $0.2$&$-0.7$ &   $2.97^{+0.1}_{-0.09}$ &   $1.09^{+0.1}_{-0.09}$ &   $0.16^{+0.01}_{-0.01}$ &  $0.3$&$0.0$&   $-7.31^{+0.04}_{-0.04}$ &$-4941$&$-141$ \\
{\it SN\_Tstratm20}	&$0.040$& $3.5e-4$& 			$0.4$ 		&$0.44^{+0.06}_{-0.07}$ &   $0.054^{+0.001}_{-0.001}$ &   $0.5^{+0.2}_{-0.2}$ &  $0.2$&$-0.7$ &   $2.9^{+0.1}_{-0.1}$ &   $1.1^{+0.1}_{-0.1}$ &   $0.15^{+0.01}_{-0.01}$ &  $0.3$&$0.0$&   $-7.27^{+0.04}_{-0.04}$ &$-4914$&$-114$ \\

\hline
\hline

\end{longtable}
 \end{center}
\end{minipage}
\end{landscape}

\subsection{Varying the CH$_4$ profile}\label{sec:CH4tests}

We now use the {\it SN} dataset  to explore the relationship between our model fits and our choice of vertical methane  profile. Specifically, we wish to determine whether our data indicate a preference for particular values of  {\it mCH$_{4,t}$}, {\it mCH$_{4,s}$},  and $RH_{ac}$, as defined in Appendix \ref{app:composition}; and how the values of these parameters influence the retrieved aerosol structure. For these retrievals, we utilize two stream for the radiative transfer. We assume a two-layer aerosol structure, allowing $P_{max}$, $\tau$, and $h_{frac}$ for each layer to vary, while fixing $\omega_0$ and $g$ to the best-fit values from the {\it SN\_nom} retrieval.  We perform a set of retrievals equivalent to the {\it SN\_nom} fit, but fix the methane parameters to different combinations of values from the following set:  {\it mCH$_{4,t}=0.22, 0.04$}; {\it mCH$_{4,s}=3.5 \times 10^{-4},15 \times 10^{-4}$};  and $RH_{ac}=40\%,100\%$. Since the \cite{kark11a} results do not indicate any methane depletion at these latitudes, we do not consider methane depletion here.  The case where  {\it mCH$_{4,t}=0.04$}; {\it mCH$_{4,s}=3.5 \times 10^{-4}$};  $RH_{ac}=100\%$ corresponds to the {\it SN\_nom} case and is not rerun. The seven remaining cases are labeled {\it SN\_CH$_4$grid$_{A-F}$} and the results are reported in Table \ref{table:SN}. 

 \begin{figure}[h]
\begin{center}
\includegraphics[width=0.5\textwidth]{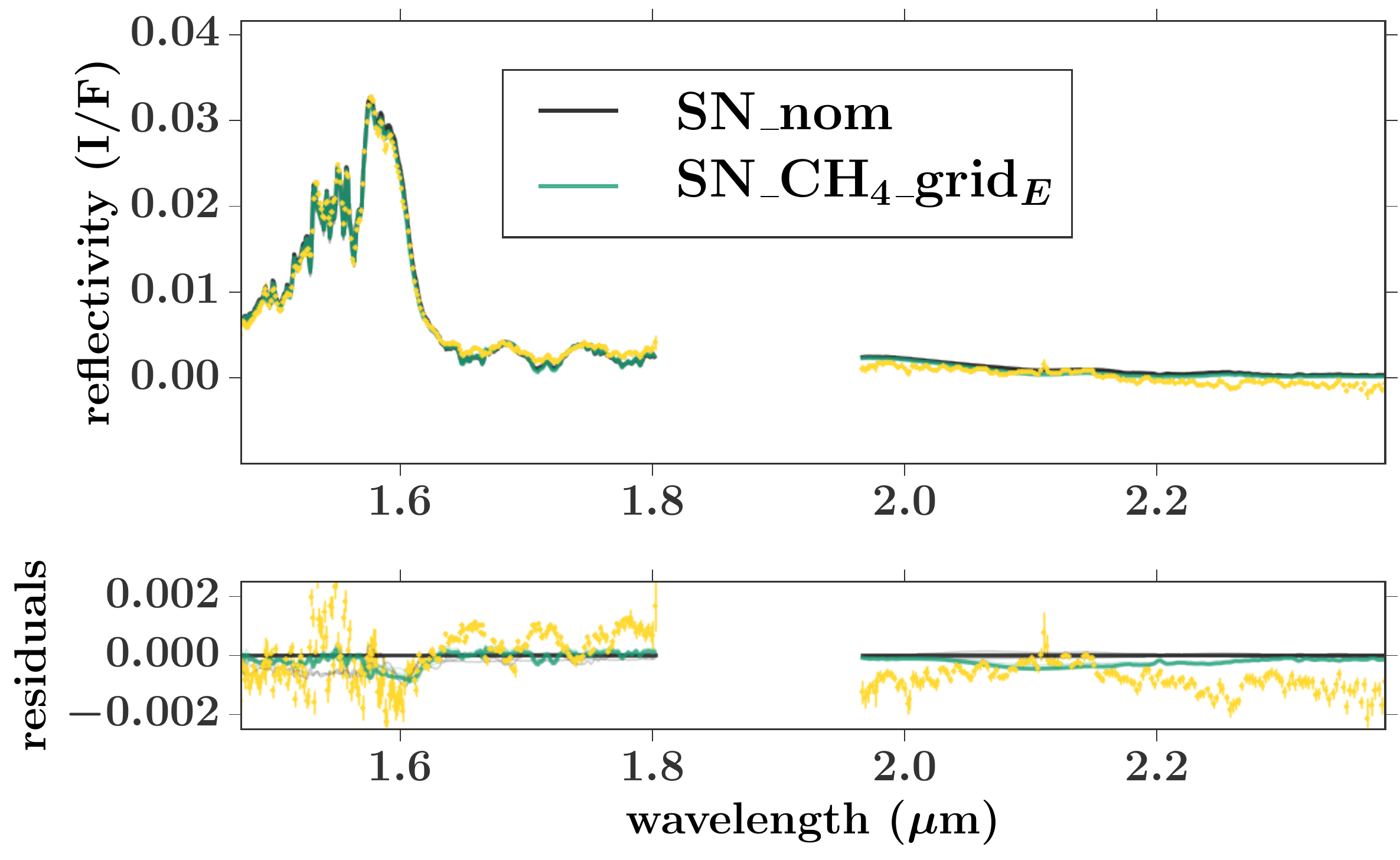} \\
\end{center}
\caption[{\it SN\_CH$_4$grid$_E$}  model fits vs {\it SN\_nom}  fits]{\label{fig:CH4_modelfits} \small Comparison of the {\it SN\_CH$_4$grid$_E$} model fits (green) to the data (yellow) and the {\it SN\_nom}  fits (black), for the $\mu=0.8$ high S/N spectrum. The residuals are shown relative to the best-fit  {\it SN\_nom} model; best-fit models and 100 random samples from the posterior probability are shown. }
\end{figure}

We find that, for all eight of the CH$_4$ profiles considered, the derived aerosol structure is qualitatively consistent: in all cases, we find aerosol layer $\alpha$ to have a base pressure of 0.4--0.7 bar and a 1.6-$\mu$m optical depth of 0.05--0.06. Aerosol layer $\beta$ is vertically compact ($h_{frac, \beta} < 0.3$), has an optical depth near unity, and has a base pressure of 3--4 bar.  

We also observe that the choice of CH$_4$ profile systematically affects the retrieved probability distributions of the properties describing the distribution of aerosols, with all three of the parameters describing the CH$_4$ profile having some effect on our retrievals. The parameters that show the strongest dependences on the adopted CH$_4$ profile are the base pressures of both aerosol layers and the scale height of the upper aerosol layer. The retrieved base pressure of the deeper aerosol layer ($\beta$) depends on the assumed methane abundance in the troposphere, with lower values of {\it mCH$_{4,t}$} resulting in higher retrieved base pressures of layer $\beta$. This result is unsurprising, since a lower methane mixing ratio implies that one must traverse to greater depths to encounter the same total methane column. A secondary trend of our retrievals is that, for a given choice of  {\it mCH$_{4,t}$} and {\it mCH$_{4,s}$}, the retrieval using the lower methane relative humidity near the tropopause favors a deeper, but slightly more extended and optically thick aerosol layer $\beta$. 

The retrieved properties of the upper aerosol layer are independent of the deep tropospheric methane mole fraction {\it mCH$_{4,t}$} but do depend on the other two methane parameters: higher values of $RH_{ac}$ result in retrievals with a more vertically extended upper haze layer ($h_{frac, \alpha}>1$ when $RH_{ac}=100\%$ vs. $h_{frac, \alpha}\sim$ 0.4--0.5 when $RH_{ac}=40\%$).  For a given value of $RH_{ac}$ and  {\it mCH$_{4,t}$}, the best-fit model of layer $\alpha$ with the higher value of {\it mCH$_{4,s}$} has a deeper base, a higher total optical depth, and a higher fractional scale height. However, the differences are small ($<1.5\sigma$).

Using the $DIC$, we next evaluate whether our analysis favors a particular choice of methane profile. We find that we are not able to constrain the deep methane abundance, as parameterized by  {\it mCH$_{4,t}$}, at all: pairs of models which differ only in this methane parameter -- for example,  {\it SN\_CH$_4$grid$_A$} and {\it SN\_CH$_4$grid$_E$} -- have mutually consistent values of the {\it DIC}. This is unsurprising, considering that the transmission and contribution functions for the wavelengths of our data (Fig. \ref{fig:1L_contrib}) indicate that the pressures influenced by  {\it mCH$_{4,t}$} contribute little at the wavelengths of our observations. Our poor constraint on {\it mCH$_{4,t}$} translates into an uncertainty in $P_{max,\beta}$ beyond what is captured by the statistical uncertainty reported for an individual retrieval. 

The methane parameter that we most strongly constrain is $RH_{ac}$: we find that all four models with $RH_{ac}=40\%$ have significantly lower (better) values of the $DIC$ than any of the four models with $RH_{ac}=100\%$.  As shown in Fig. \ref{fig:CH4_modelfits}, the fit improvement is most evident in  the reflectivity peak near 1.6 $\mu$m. To further investigate Neptune's relative humidity, we perform two additional {\it SN} retrievals in which we allow $RH_{ac}$ to be a free parameter (referred to as  {\it SN\_CH$_4$RHfree$_{A,B}$}). For these retrievals,  {\it mCH$_{4,t}$} remains fixed at 0.04, and we assume {\it mCH$_{4,s}$} is either $3.5 \times 10^{-4}$ (case A) or {\it $15 \times 10^{-4}$} (case B). The results are summarized in Table \ref{table:SN}.  We find that both of the {\it RHfree} retrievals show an improvement over the {\it SN\_CH$_4$grid} models, with $\Delta DIC=-252$ and $-258$ from the {\it SN\_nom} retrieval, and  $\Delta DIC <-100$ from the best {\it SN\_CH$_4$grid} retrieval. The retrieved relative humidities are near zero ($\lessapprox$2--3\%). Such a low methane relative humidity is inconsistent with, e.g, \cite{kark11a}; however, we note that our observations are most sensitive to the CH$_4$ abundance at altitudes above the CH$_4$ condensation level ($\sim1.5$ bar), whereas \cite{kark11a} are most sensitive to the methane abundance at deeper pressures -- this is discussed further in Section \ref{sec:disc}. By comparing the {\it SN\_CH$_4$grid$_{E,F}$} and {\it SN\_CH$_4$RHfree$_{A,B}$} retrievals (Table \ref{table:SN}, Fig. \ref{fig:taubar}), we note that a decrease of $RH_{ac}$ from 40\% to $\lessapprox$2--3\% has little influence on the retrieval of the aerosol distribution:  with the exception of $\tau_\alpha$, the aerosol parameter distributions for these four retrievals agree at the $\sim1.5\sigma$ level. 

Our {\it SN\_CH$_4$grid}  retrievals appear to indicate a dependence of the {\it DIC} on the value of {\it mCH$_{4,s}$}, with models having a lower stratospheric methane abundance resulting in better fits for a given tropospheric methane abundance and relative humidity. However, the two {\it SN\_CH$_4$RHfree} runs, differing only in the value of  {\it mCH$_{4,s}$}, have equivalent {\it DIC} values  ($|\Delta DIC| <10$). This leads us to suspect that the preference for a smaller value of {\it mCH$_{4,s}$} in the {\it CH$_4$grid} models may not actually be an indication of a lower value of the methane mixing ratio throughout the stratosphere, but could instead reflect the preference for a lower methane abundance near the tropopause. We try one additional retrieval ({\it SN\_CH$_4$stratfree}) in which {\it mCH$_{4,t}$} and  $RH_{ac}$ are fixed (at 0.04 and 0.4, respectively) but  {\it mCH$_{4,s}$} is added as a free parameter. The retrieved value of  {\it mCH$_{4,s}$} is only $2.2e-4$, lower than our nominal choice of  $3.5e-4$ which itself is at the low end of the values determined by previous authors.  However, the {\it DIC} indicates no improvement over the {\it SN\_CH$_4$grid$_E$} retrieval, and the aerosol structure is effectively unchanged by decreasing  {\it mCH$_{4,s}$} from $3.5e-4$  to $2.2e-4$. 

In the retrievals that follow, we adopt the  {\it SN\_CH$_4$grid$_E$} methane profile as our new nominal model. This profile has a CH$_4$ relative humidity of 0.4 near the tropopause; as discussed above, our {\it RHfree} retrievals indicate a humidity that is lower by more than an order of magnitude. While we have no reason to reject this finding, we adopt the more moderate value of  {\it RHfree} for consistency with previous studies \citep{kark11a, irwin11}, to simplify comparison with these works. As noted previously, this choice has little effect on the retrieval of the aerosol distribution, with the exception that models with the higher value of $RH_{ac}$ have higher -- but qualitatively similar -- values of $\tau_\alpha$. We revisit Neptune's methane profile in Section \ref{sec:depletion}.

\subsection{Varying the thermal profile}\label{sec:TPtests}

We perform a final set of model runs on the {\it SN} data set, to investigate the influence of possible errors in the adopted thermal profile. In our model, the thermal profile influences the gas opacity coefficients as well as the CH$_4$ profile, through its effect on the saturation vapor pressure curve of methane. Using the {\it SN\_CH$_4$grid$_E$} methane profile, we consider three perturbed thermal profiles. In the {\it SN\_Ttrop} case, we adjust the tropopause temperature from its nominal value of 54.9 K to the value of 56.4 K, as found by \cite{orton07} (`case B'). In  {\it SN\_Tstratp20} and  {\it SN\_Tstratm20} we increased/decreased, respectively, the temperature in the stratosphere by 20 K at 1 mbar. The details of these adjustments are described in Appendix \ref{app:TP}. The results of these model runs are shown in Table \ref{table:SN}. We find that the posterior probability distributions for all parameters are consistent with (within $1\sigma$ of) those found for the unperturbed thermal profile, with the exception that increasing the temperature of the tropopause results in a higher value of $h_{frac,\alpha}$. The  {\it 2L\_Tstratp20} retrieval has a $\Delta DIC=-20$ relative to the  {\it SN\_CH$_4$grid$_E$} retrieval; the other two modified thermal profiles result in a decreased fit quality relative to the  {\it SN\_CH$_4$grid$_E$} retrieval. 

\section{Individual cloud-free regions}\label{sec:DR}
We follow our intensive analysis of the limb darkening data set with a broader study of many cloud-free regions across Neptune, spanning latitudes from 20$^\circ$N to 87$^\circ$S and viewing geometries from $\mu=0.26$ to $\mu=0.99$. This analysis permits the consideration of spatial variations in the distribution of aerosols, temperature, and methane abundance. It also allows us to evaluate the relative information content of a single spectrum at a single viewing geometry relative to the extensive limb-darkening dataset. 

Our approach for selecting individual locations for study is as follows: for every 10$^\circ$ in latitude, we seek two locations that are 1) dark (cloud-free),  2) similar in latitude but with different viewing geometries, 3) at longitudes for which we have both Hbb and Kbb data,  and 4) at least 10 pixels from any bright clouds, to avoid contamination due to the broad PSF halo (Section \ref{sec:resolution}). At some latitudes (e.g., 20$^\circ$N and 60$^\circ$S), nearby bright bands lead us to relax the final criterion; at other latitudes (e.g., 30$^\circ$S) these bright bands prevent the identification of any suitably dark locations. Further, at high southern latitudes, the prevalence of bright clouds has resulted in a more uneven latitude spacing of selected suitable dark locations. Our final sample consists of 18 specific locations, shown and labeled A--R in  Fig. \ref{fig:DRcubes}. Two of the locations (C and D) are  within the original limb-darkening latitude band from 2--12$^\circ$N. 

To extract a spectrum for each of these locations, we first extract the Hbb spectrum from a single pixel and record the latitude and cosine of the emission angle, $\mu_H$, for that pixel. We then extract a Kbb spectrum corresponding to a similar physical location on Neptune in the following way: we determine the time interval between the Hbb and Kbb observations of the relevant latitude. Next, we estimate the expected rotation for a patch of atmosphere at the given latitude using the planetary rotation rate and the mean zonal wind profile from \cite{srom93}. Finally, we estimate the Kbb {\it X,Y} location using the Hbb {\it X,Y} location and latitude, the centers of the Hbb and Kbb cubes, and the expected amount of rotation. The data from the two bands are considered a single `location' for the purposes of atmospheric retrieval, and are modeled simultaneously, with the caveat that we use the  $\mu_H$ and $\mu_K$ value appropriate for each band.  We note that the feature coordinates are imprecise due to the $\lesssim 1$ pixel uncertainty in the image centering. A one-pixel error in position would correspond to roughly 3$^\circ$ error in latitude/longitude for a feature near disk center, with larger errors possible at other viewing geometries \citep[see Fig. 4 of][]{martin12}. Large-scale meridional trends retrieved by our analysis should be robust to possible navigation errors.  Our retrieval procedure also ignores the finite width of the PSF for our observations: models are produced for the singular $\mu_H$ and $\mu_K$ corresponding to the expected viewing geometry of the extracted data pixels, neglecting any changes in the measured intensity due to contributions from the wings of the PSF. To estimate the errors introduced by this approximation, we compared a synthetic, unsmoothed model cube to one convolved with the PSF, as estimated from stellar spectra. We find that for the three locations at the highest Hbb viewing angles (B,C,P), the mean error introduced by neglecting the finite width of the PSF is comparable to the mean uncertainty due to random noise ($\bar{\sigma_n} \approx 0.001$). Further, at locations with moderate Hbb viewing angles (F,H,N,Q,R), the {\it maximum} error introduced by this approximation is greater than $\bar{\sigma_n}$. We do not incorporate an estimate of this error -- which is wavelength, location, and model-dependent --  into our analysis: doing so would decrease the unknown component of the uncertainty ($\sigma_T$) for high viewing angle models, and would in some cases alter the relative weighting of the data points in the retrievals: errors from ignoring the width of the PSF are greatest in the 1.6-$\mu$m I/F peak.

\begin{figure}[htb!]
\begin{center}$
\includegraphics[width=0.8\textwidth]{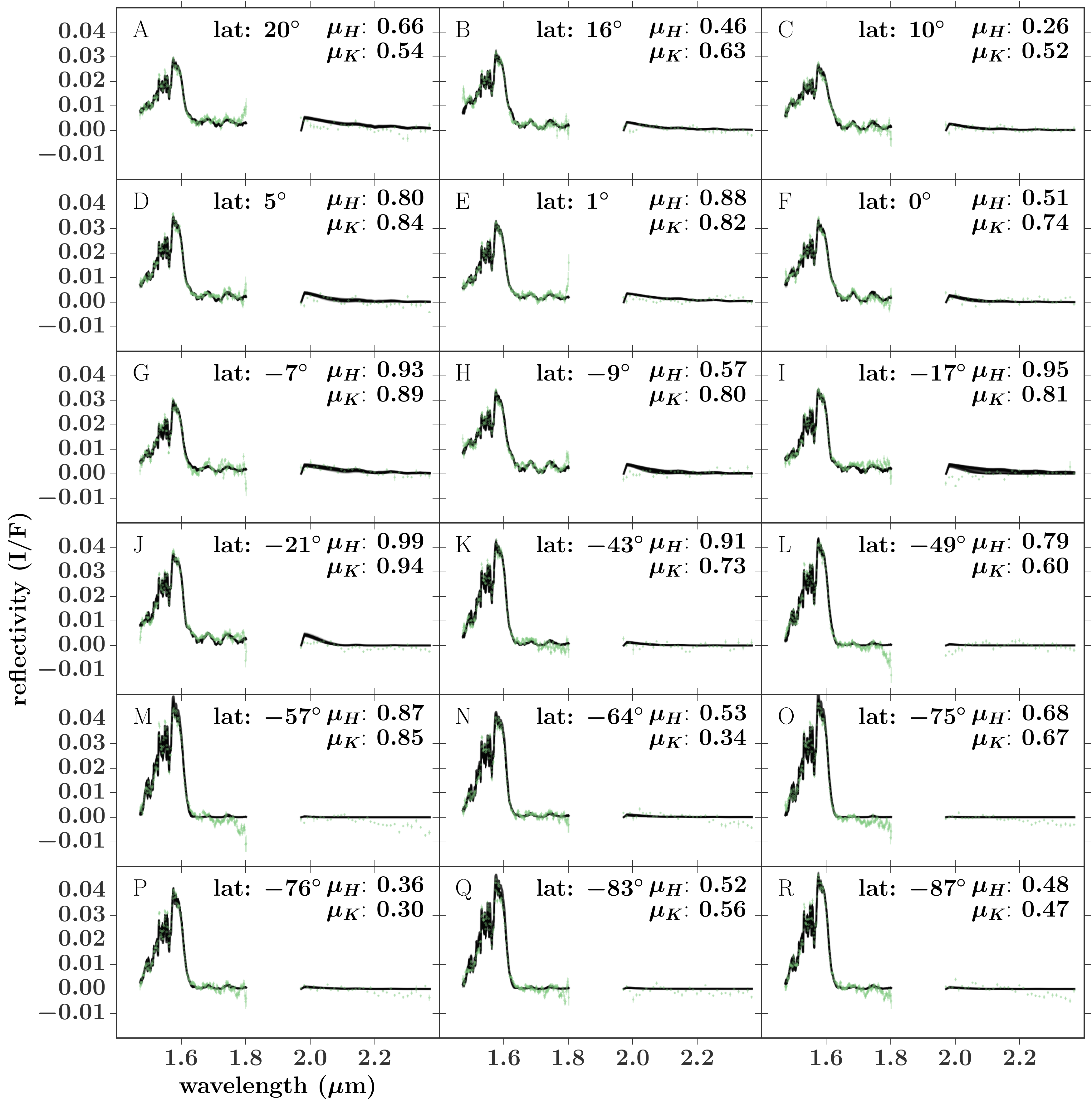}
$
\end{center}
\caption[Initial fits -- cloud-free regions]{\label{fig:DRfits}  \small Spectra for the 18 cloud-free locations identified in Section \ref{sec:DR} (green), and initial fits to the aerosol distributions as described in Section \ref{sec:defaulthaze} and Table \ref{table:DR}. The best fit and 100 random solutions drawn from the posterior probability distribution are shown in black. }
\end{figure}

The spectra corresponding to the 18 selected locations are shown in Fig. \ref{fig:DRfits}. Visual inspection of these plots indicates a north-south trend in the Hbb spectral shape and peak intensity: spectra at northern latitudes tend to have higher 1.5 $\mu$m intensities and lower peak intensities than spectra from southern latitudes.

\subsection{Local aerosol distribution}\label{sec:defaulthaze}
\begin{figure}[htb!]
\begin{center}$
\includegraphics[width=0.9\textwidth]{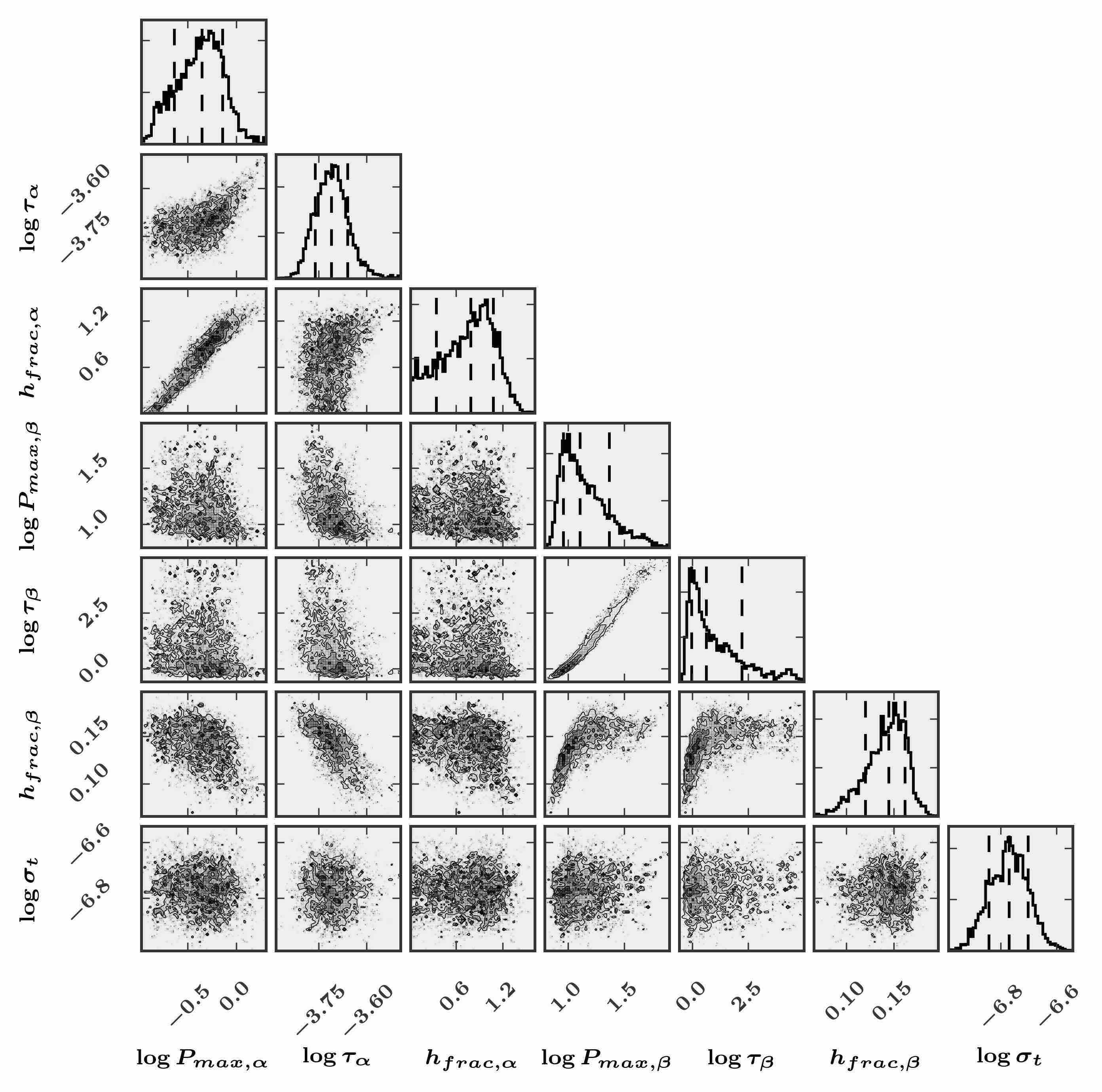}
$
\end{center}
\caption[Triangle plot -- location H initial fit]{\label{fig:triangle_H} \small  1D and 2D posterior probability distributions for the initial retrieval of the atmospheric properties for location H, a typical, well-behaved retrieval (Section \ref{sec:defaulthaze}).   }
\end{figure}

\begin{figure}[htb!]
\begin{center}$
\includegraphics[width=0.9\textwidth]{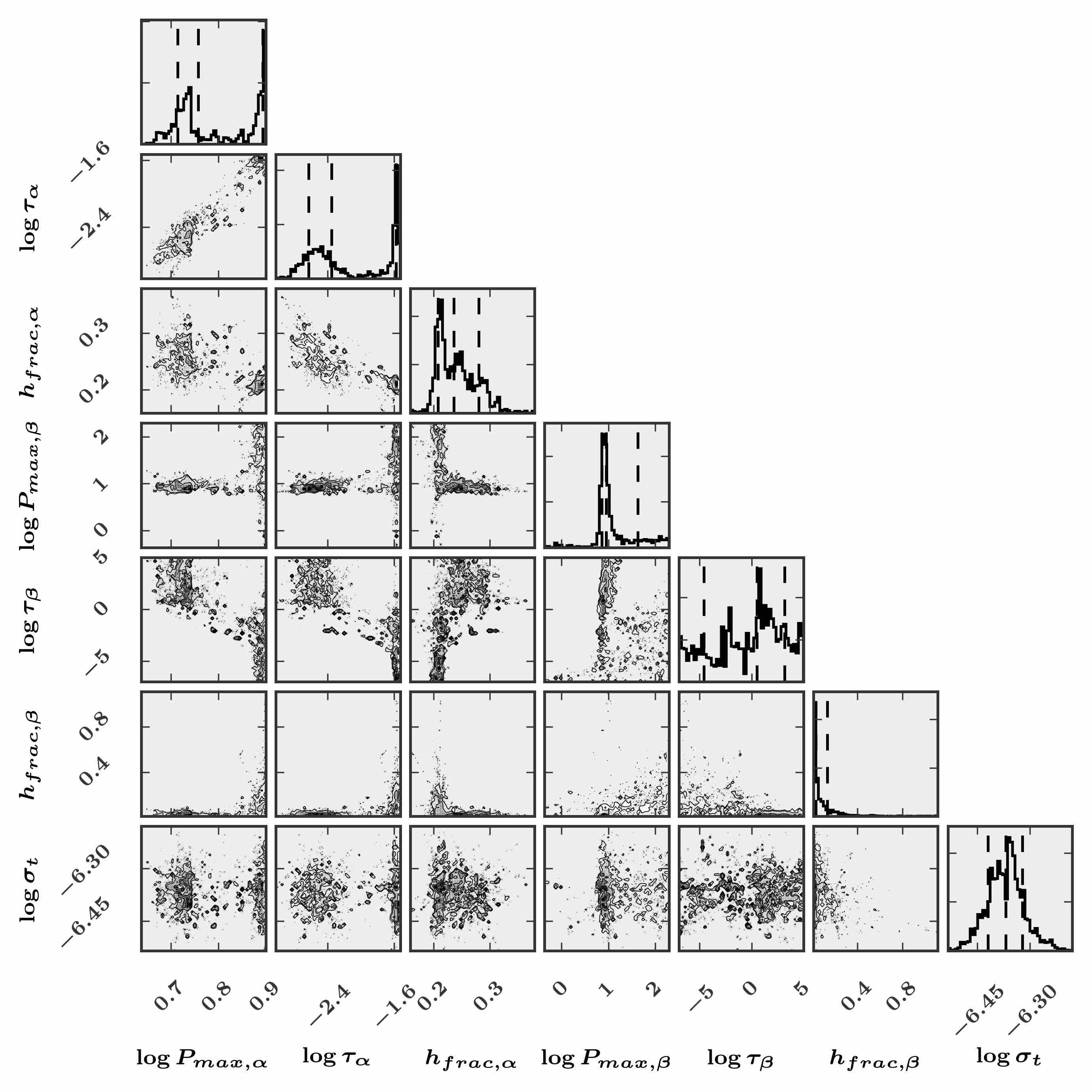}
$
\end{center}
\caption[Triangle plot -- location O initial fit]{\label{fig:triangle_O} \small  1D and 2D posterior probability distributions for the initial retrieval of the atmospheric properties for location O (Section \ref{sec:defaulthaze}). The retrieved posterior probability distribution is double-peaked in $P_{max,\alpha}$ and $\tau_\alpha$; $\tau_\beta$ is poorly constrained and correlated with the solution for aerosol layer $\alpha$.   }
\end{figure}

\begin{figure}[h!]
\begin{center}$
\includegraphics[width=0.8\textwidth]{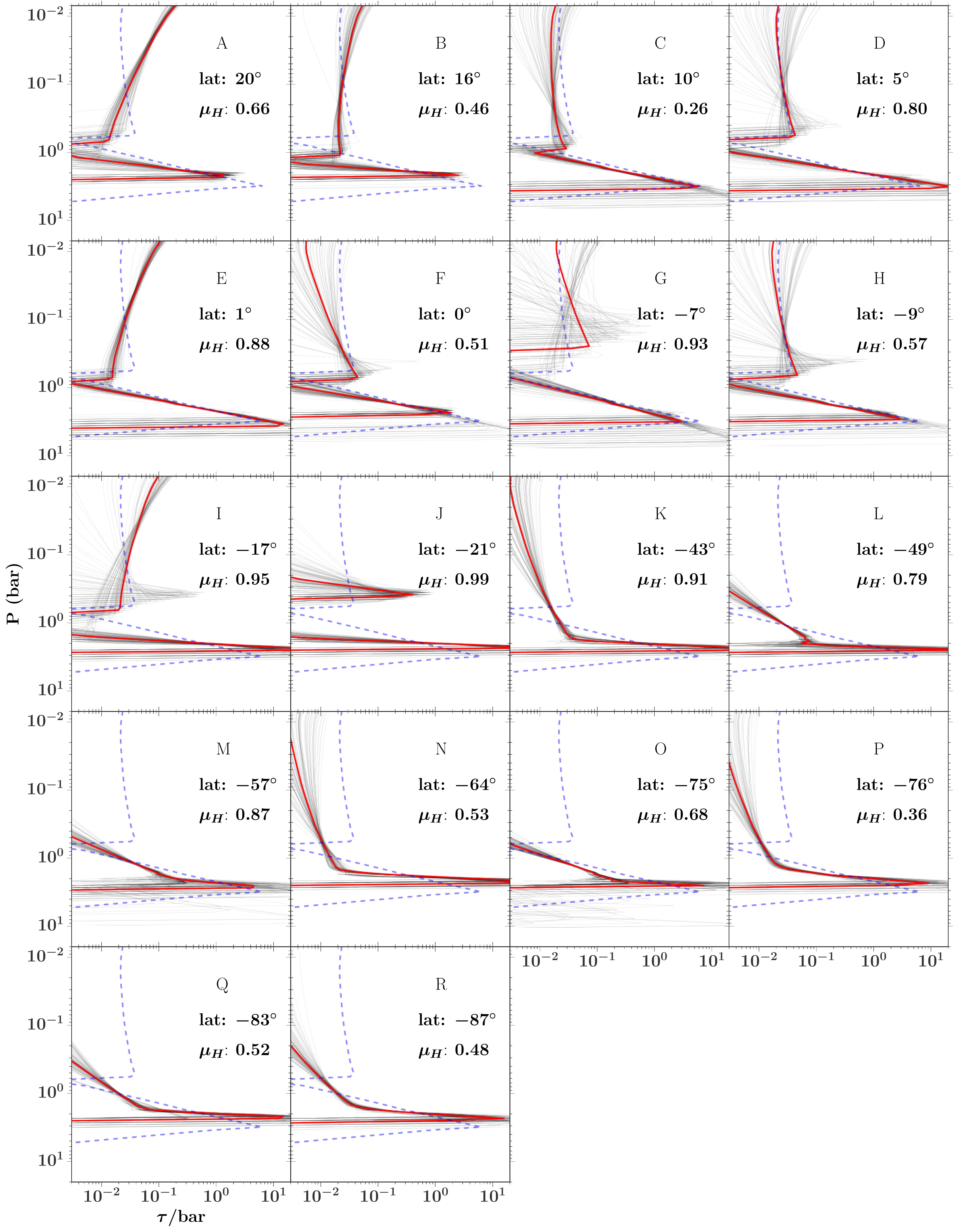}
$
\end{center}
\caption[$\tau$/bar, initial fits to cloud-free regions]{\label{fig:DRtaubar} \small  Optical depth per bar for the best fit and 100 random solutions, for the initial set of  aerosol-only retrievals for the 18 dark locations. For clarity, the best-fit solution is shown as a thick red line, and the 100 random solutions are shown as thin black lines. The lavendar dashed line indicates the best-fit  {\it 2L\_DISORT} solution, shown for reference. }
\end{figure}

\begin{figure}[h!]
\begin{center}$
\includegraphics[width=0.85\textwidth]{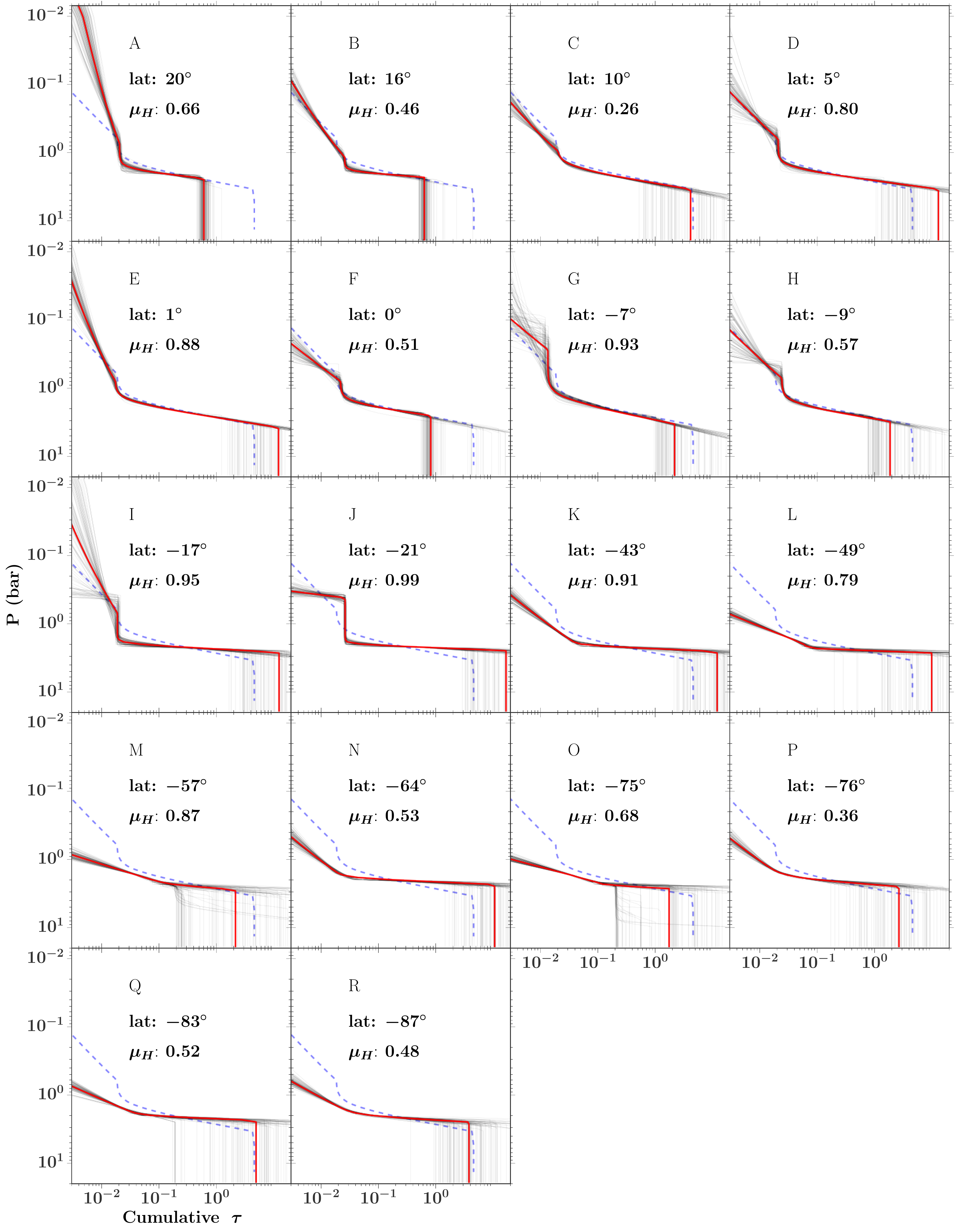}
$
\end{center}
\caption[Cumulative optical depth, initial fits to cloud-free regions]{\label{fig:DRtauC} \small  Cumulative optical depth as a function of pressure, for the initial aerosol-only fits to the 18 dark locations. Colors and line styles are defined as in Fig. \ref{fig:DRtaubar}. }
\end{figure}

\begin{figure}[h!]
\begin{center}$
\begin{array}{cccc}
\includegraphics[width=0.5\textwidth]{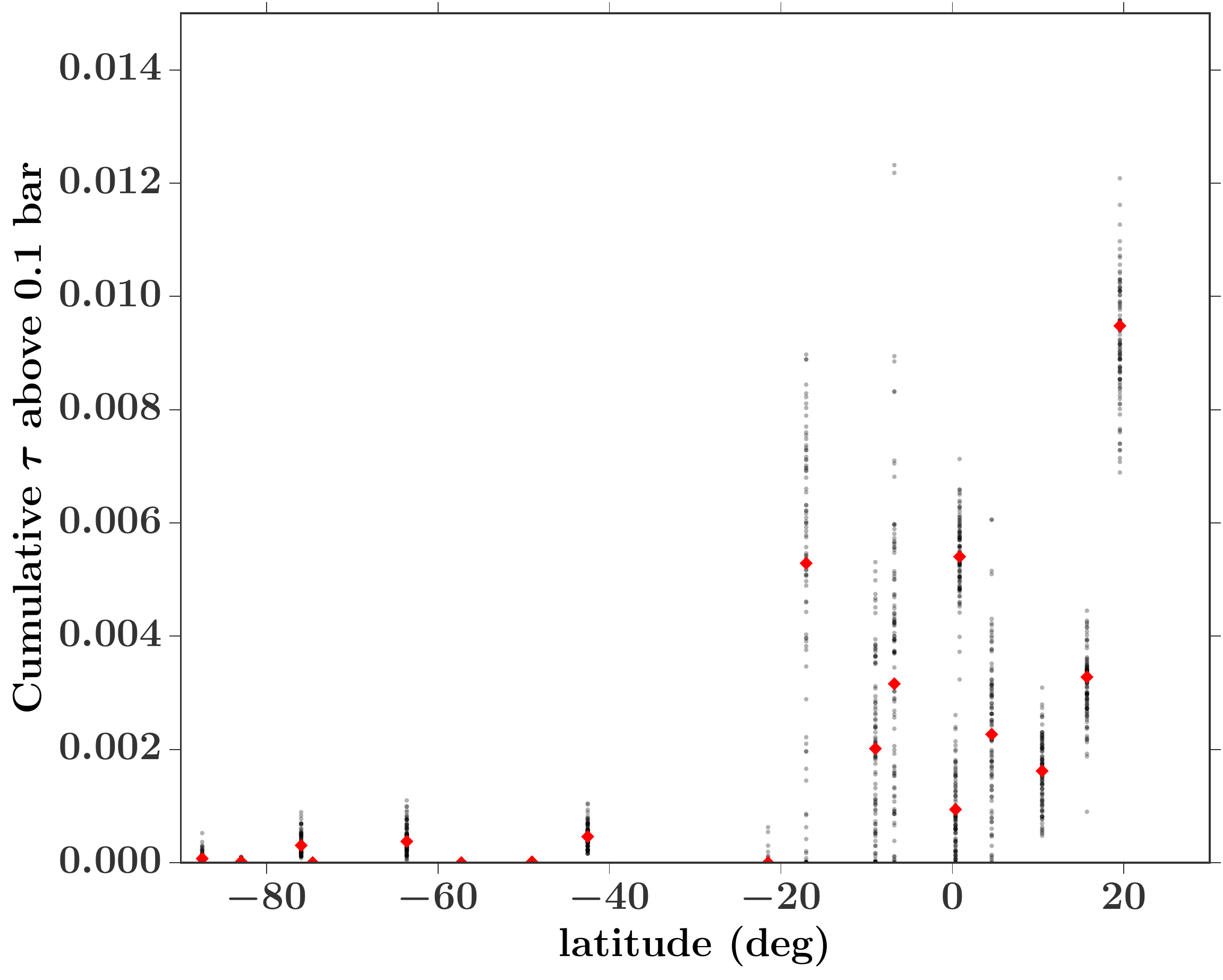} \includegraphics[width=0.5\textwidth]{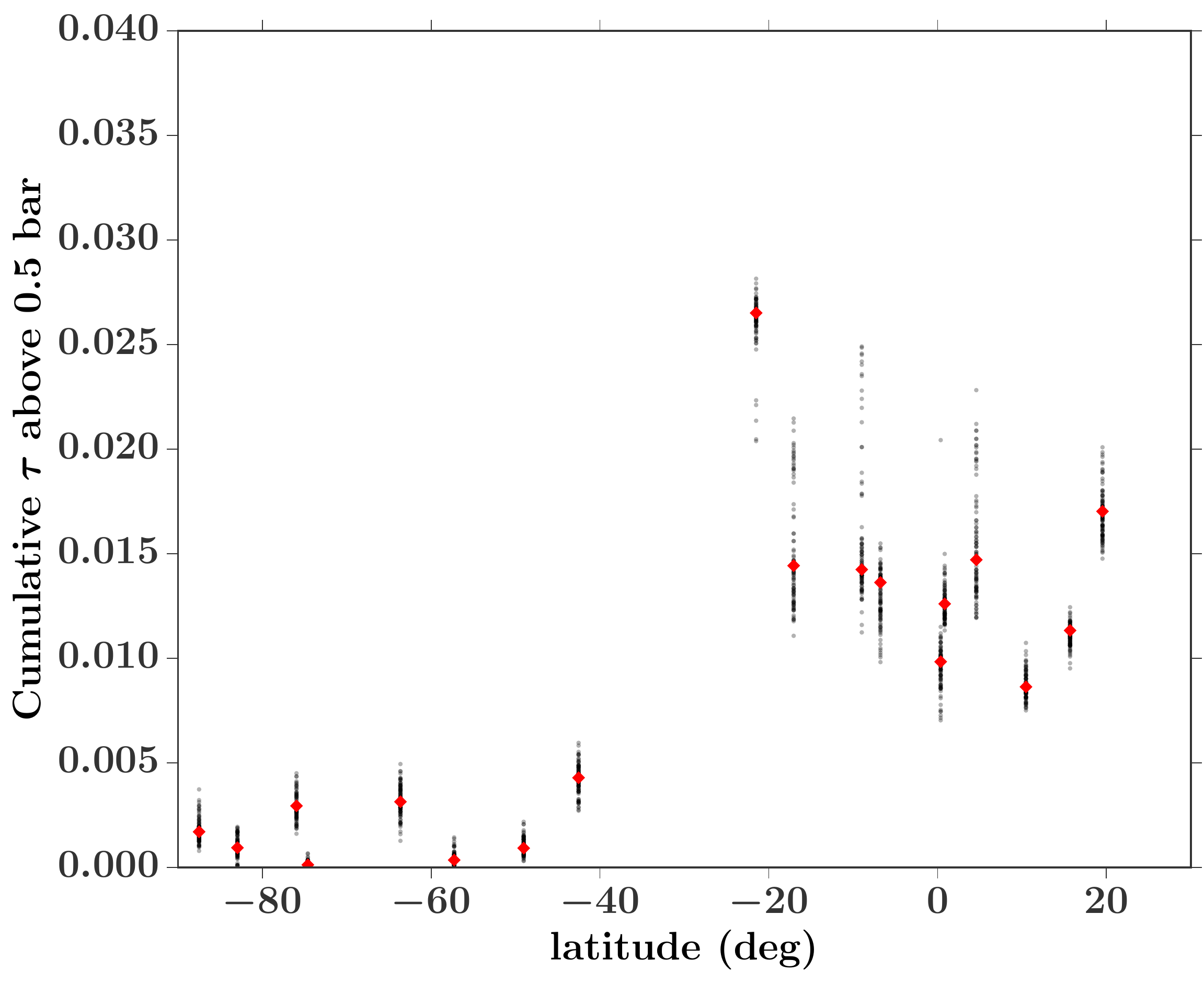}\\
\includegraphics[width=0.5\textwidth]{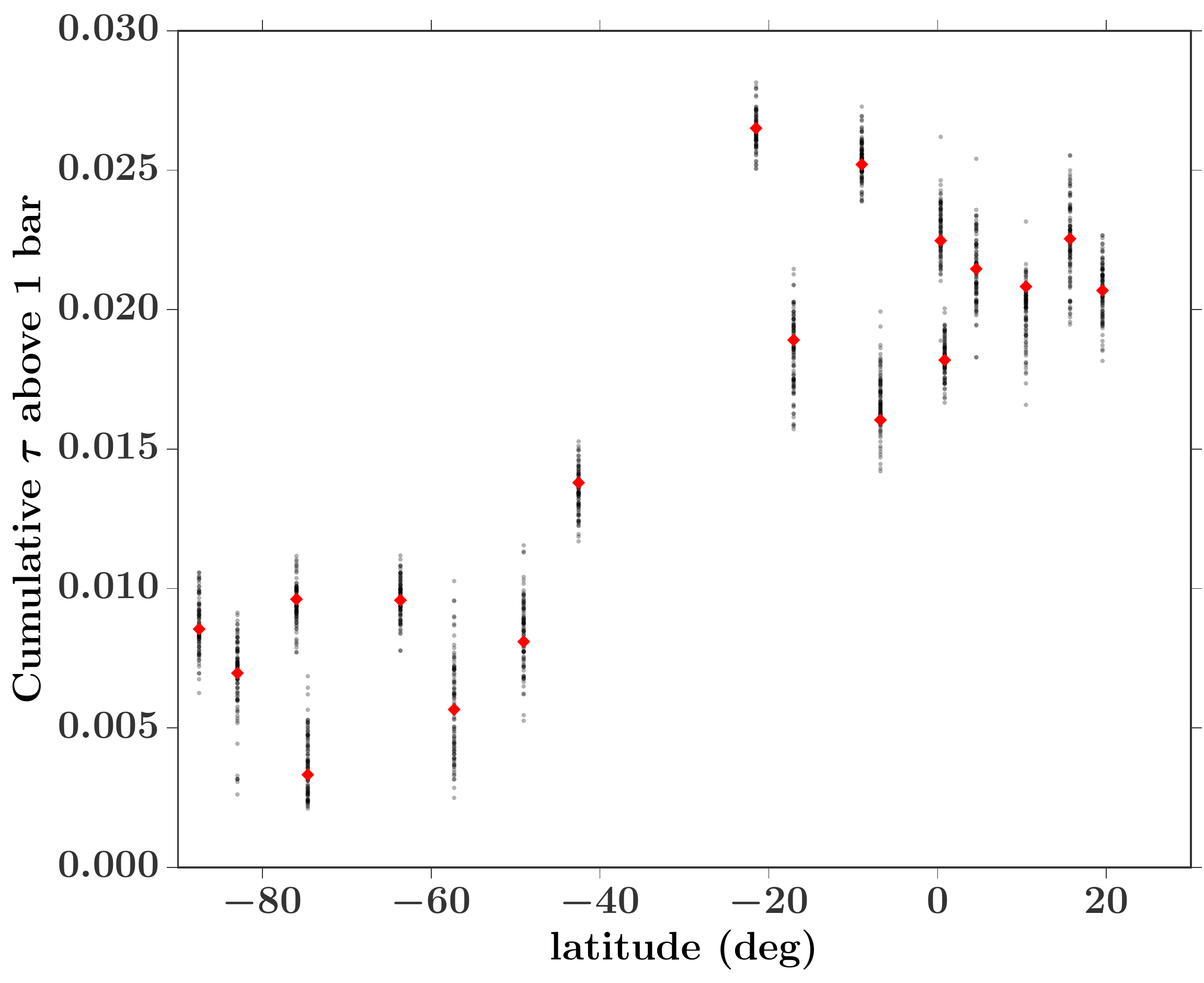} \includegraphics[width=0.5\textwidth]{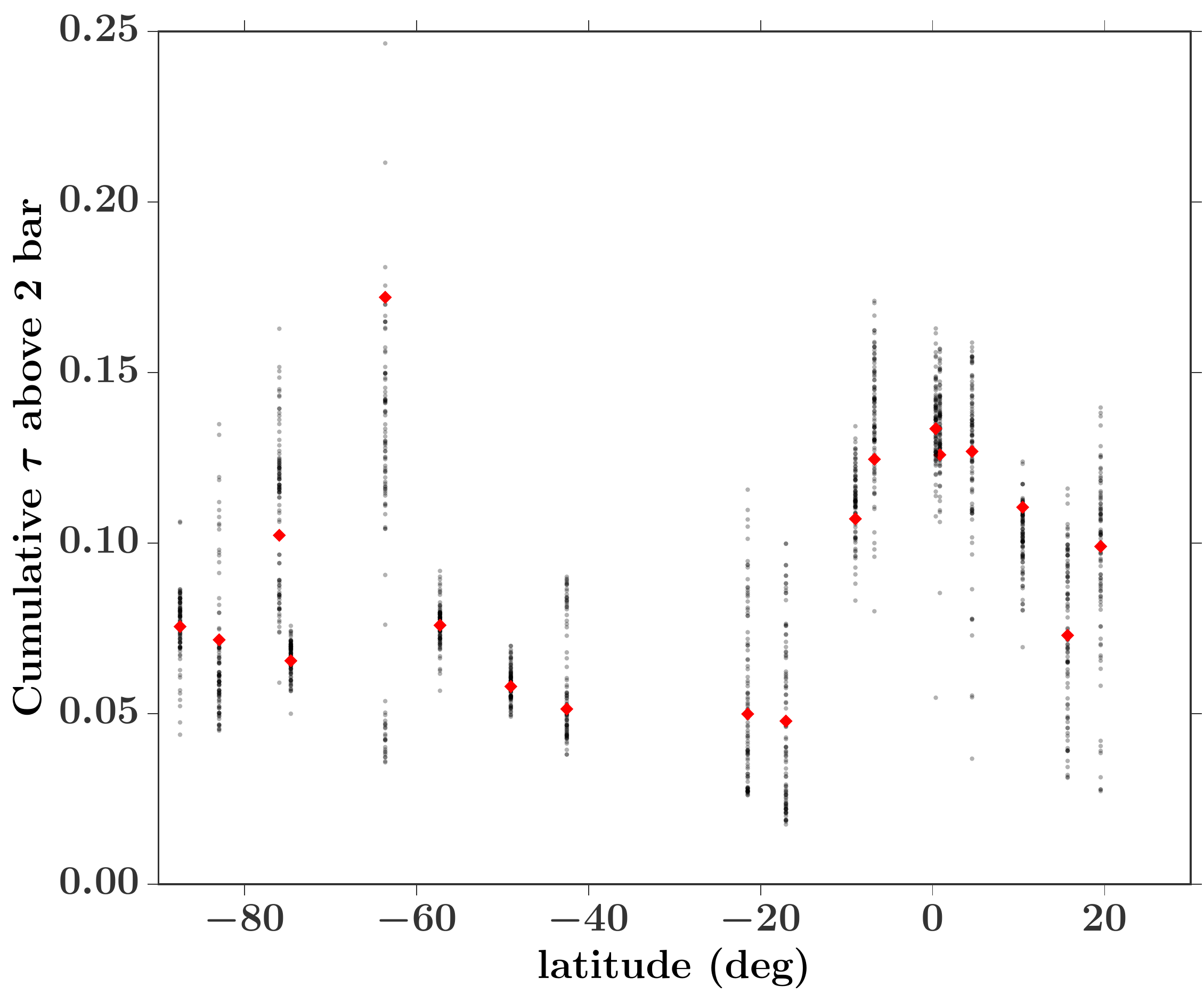}\\
\end{array}$
\end{center}
\caption[Cumulative optical depth vs. latitude, initial fits to cloud-free regions]{\label{fig:DR_tauCvlat} \small Cumulative optical depth of aerosols above four pressure levels, as a function of latitude, for the initial aerosol-only retrievals. Each vertical line of points represents a single location; the red diamonds are the cumulative optical depths corresponding to the best-fit solution, and the small black points are the cumulative optical depths  for 100 random solutions from the posterior probability distribution.    } 
\end{figure}

Using the nominal thermal profile, the {\it SN\_CH$_4$grid$_E$} methane profile, and the particle size distributions from Section \ref{sec:2L}, we retrieve the aerosol properties for each location A--R. Since the 18 cloud-free locations span a broad range of viewing geometries, we cannot utilize two stream for the radiative transfer; we instead use pyDISORT and adopt the best-fit values of $g$ and $\omega_0$ from the {\it 2L\_DISORT} retrieval (Section \ref{sec:disort}). The free parameters of these retrievals are the base pressure, optical depth, and fractional scale height of the two aerosol layers. The results are summarized in Table \ref{table:DR} and shown in Figs. \ref{fig:DRfits} -- \ref{fig:DR_tauCvlat}.

While the majority of the 18 retrievals result in well-behaved posterior probability distributions (e.g. Fig. \ref{fig:triangle_H}), three retrievals  (locations I, M, and O) have posterior probability distributions that are double-peaked in the parameters that describe aerosol layer $\alpha$ while otherwise meeting our criteria for convergence (see Fig. \ref{fig:triangle_O} for an example). We cannot rule out that with additional iterations, {\it emcee} may have converged on a single solution. The values quoted in Table \ref{table:DR} are, as always, the 16th, 50th, and 84th percentiles for the marginalized distributions, which, for double peaked distributions, do not clearly identify the highest probability values. For two of these locations (M, O), one of the two posterior probability distribution peaks corresponds to an aerosol layer based near the maximum allowed $P_{max,\alpha}$ value of 2.3 bar, such that layer $\alpha$ substantially overlaps with aerosol layer $\beta$ (Fig. \ref{fig:triangle_O}). Interpretation of a solution with such strongly overlapping aerosol layers is not straightforward and caution is recommended for these locations. Figs. \ref{fig:triangle_H} and \ref{fig:triangle_O} also highlight the strong correlations between parameters.

\begin{minipage}{1.1\textwidth} 
\setlength\LTleft{-0.5in}
\setlength\LTright{-1in plus 1 fill}
\renewcommand{\footnoterule}{}
\renewcommand{\thempfootnote}{\it \alph{mpfootnote}}
 \renewcommand{\thefootnote}{\it \alph{footnote}}
\begin{center}
\scriptsize
\setlength{\tabcolsep}{0.05in}
\begin{longtable}{+l | ^r ^l ^l | ^l  ^l ^l | ^l ^l ^l  |^l ^r}
\caption[Cloud-free regions, initial fits]{\small Results for initial, aerosol-only, fit to 18 cloud-free locations.\label{table:DR}} \tabularnewline
location	& lat ($^\circ$)	& $\mu_H$ & $\mu_K$		&\multicolumn{3}{l }{layer $\alpha$ (top)}	 	 &\multicolumn{3}{l }{ layer $\beta$ (bottom)	} 	     & $\log{\sigma_t}$	& 	$DIC$\\
	&  &	& &$P_{max,\alpha}$ (bar)	 &$\tau_\alpha$ &$h_{frac,\alpha}$		& $P_{max,\beta}$ (bar)	 &$\tau_\beta$ &$h_{frac,\beta}$		&  & 	\\

A & $20$& $0.66$& $0.54$&$0.7^{+0.2}_{-0.1}$ &   $0.021^{+0.001}_{-0.001}$ &   $2.1^{+0.4}_{-0.4}$ &   $2.3^{+0.2}_{-0.1}$ &   $0.58^{+0.1}_{-0.07}$ &   $0.09^{+0.04}_{-0.04}$ &   $-6.74^{+0.06}_{-0.07}$ &$-2878$ \\
B & $16$& $0.46$& $0.63$&$1.2^{+0.2}_{-0.1}$ &   $0.026^{+0.002}_{-0.002}$ &   $1.2^{+0.1}_{-0.1}$ &   $2.2^{+0.2}_{-0.1}$ &   $0.61^{+0.1}_{-0.07}$ &   $0.06^{+0.04}_{-0.03}$ &   $-6.94^{+0.08}_{-0.08}$ &$-2904$ \\
C & $10$& $0.26$& $0.52$&$0.9^{+0.2}_{-0.2}$ &   $0.020^{+0.002}_{-0.001}$ &   $0.9^{+0.2}_{-0.2}$ &   $3.4^{+1.0}_{-0.6}$ &   $4^{+30}_{-3}$ &   $0.15^{+0.02}_{-0.02}$ &   $-7.07^{+0.09}_{-0.08}$ &$-2994$ \\
D & $5$& $0.80$& $0.84$&$0.6^{+0.2}_{-0.1}$ &   $0.021^{+0.001}_{-0.001}$ &   $0.8^{+0.3}_{-0.3}$ &   $3.4^{+0.6}_{-0.6}$ &   $10^{+50}_{-10}$ &   $0.12^{+0.01}_{-0.02}$ &   $-6.90^{+0.07}_{-0.07}$ &$-2941$ \\
E & $1$& $0.88$& $0.82$&$0.8^{+0.1}_{-0.1}$ &   $0.0174^{+0.001}_{-0.0009}$ &   $1.6^{+0.2}_{-0.2}$ &   $3.7^{+0.9}_{-0.7}$ &   $12^{+50}_{-9}$ &   $0.14^{+0.01}_{-0.01}$ &   $-7.21^{+0.07}_{-0.08}$ &$-3042$ \\
F & $0$& $0.51$& $0.74$&$0.8^{+0.1}_{-0.1}$ &   $0.022^{+0.002}_{-0.002}$ &   $0.7^{+0.2}_{-0.2}$ &   $2.5^{+1.0}_{-0.2}$ &   $0.8^{+5.0}_{-0.2}$ &   $0.12^{+0.04}_{-0.04}$ &   $-6.96^{+0.08}_{-0.08}$ &$-2946$ \\
G & $-7$& $0.93$& $0.89$&$0.3^{+0.3}_{-0.1}$ &   $0.014^{+0.001}_{-0.001}$ &   $0.7^{+0.7}_{-0.5}$ &   $3.3^{+2.0}_{-0.5}$ &   $2.2^{+20.0}_{-0.9}$ &   $0.18^{+0.02}_{-0.03}$ &   $-6.62^{+0.06}_{-0.06}$ &$-2832$ \\
H & $-9$& $0.57$& $0.80$&$0.7^{+0.2}_{-0.2}$ &   $0.024^{+0.001}_{-0.001}$ &   $0.8^{+0.3}_{-0.4}$ &   $3.0^{+0.9}_{-0.4}$ &   $1.8^{+7.0}_{-0.9}$ &   $0.14^{+0.02}_{-0.02}$ &   $-6.77^{+0.07}_{-0.07}$ &$-2893$ \\
I & $-17$& $0.95$& $0.81$&$0.7^{+0.3}_{-0.2}$ &   $0.019^{+0.001}_{-0.001}$ &   $1.4^{+0.7}_{-1.0}$ &   $2.6^{+0.4}_{-0.1}$ &   $12^{+50}_{-8}$ &   $0.05^{+0.02}_{-0.02}$ &   $-6.44^{+0.06}_{-0.06}$ &$-2770$ \\
J & $-21$& $0.99$& $0.94$&$0.41^{+0.04}_{-0.03}$ &   $0.0265^{+0.0007}_{-0.0008}$ &   $0.10^{+0.1}_{-0.06}$ &   $2.4^{+0.2}_{-0.1}$ &   $20^{+60}_{-10}$ &   $0.03^{+0.01}_{-0.02}$ &   $-6.75^{+0.06}_{-0.06}$ &$-2882$ \\
K & $-43$& $0.91$& $0.73$&$1.86^{+0.08}_{-0.1}$ &   $0.038^{+0.005}_{-0.006}$ &   $0.70^{+0.09}_{-0.07}$ &   $2.5^{+0.2}_{-0.2}$ &   $12^{+40}_{-8}$ &   $0.04^{+0.02}_{-0.01}$ &   $-6.64^{+0.06}_{-0.06}$ &$-2831$ \\
L & $-49$& $0.79$& $0.60$&$2.04^{+0.09}_{-0.1}$ &   $0.065^{+0.01}_{-0.008}$ &   $0.37^{+0.04}_{-0.03}$ &   $2.6^{+0.2}_{-0.1}$ &   $10^{+50}_{-8}$ &   $0.023^{+0.01}_{-0.009}$ &   $-6.47^{+0.06}_{-0.06}$ &$-2741$ \\
M & $-57$& $0.87$& $0.85$&$2.1^{+0.4}_{-0.2}$ &   $0.08^{+0.1}_{-0.02}$ &   $0.28^{+0.06}_{-0.05}$ &   $2.7^{+0.7}_{-0.2}$ &   $2^{+20}_{-2}$ &   $0.06^{+0.06}_{-0.03}$ &   $-6.15^{+0.05}_{-0.06}$ &$-2605$ \\
N & $-64$& $0.53$& $0.34$&$1.78^{+0.08}_{-0.06}$ &   $0.024^{+0.005}_{-0.003}$ &   $0.7^{+0.1}_{-0.1}$ &   $2.3^{+0.2}_{-0.1}$ &   $11^{+40}_{-7}$ &   $0.04^{+0.01}_{-0.02}$ &   $-6.63^{+0.06}_{-0.07}$ &$-2840$ \\
O & $-75$& $0.68$& $0.67$&$2.13^{+0.3}_{-0.09}$ &   $0.10^{+0.1}_{-0.02}$ &   $0.24^{+0.04}_{-0.03}$ &   $2.6^{+3.0}_{-0.2}$ &   $2^{+20}_{-2}$ &   $0.03^{+0.1}_{-0.02}$ &   $-6.37^{+0.05}_{-0.05}$ &$-2706$ \\
P & $-76$& $0.36$& $0.30$&$1.79^{+0.1}_{-0.07}$ &   $0.025^{+0.004}_{-0.003}$ &   $0.69^{+0.1}_{-0.09}$ &   $2.5^{+0.2}_{-0.2}$ &   $3^{+10}_{-2}$ &   $0.07^{+0.02}_{-0.01}$ &   $-6.84^{+0.08}_{-0.07}$ &$-2921$ \\
Q & $-83$& $0.52$& $0.56$&$1.88^{+0.06}_{-0.08}$ &   $0.039^{+0.009}_{-0.007}$ &   $0.40^{+0.09}_{-0.06}$ &   $2.3^{+0.2}_{-0.1}$ &   $5^{+30}_{-4}$ &   $0.03^{+0.02}_{-0.01}$ &   $-6.42^{+0.05}_{-0.05}$ &$-2740$ \\
R & $-87$& $0.48$& $0.47$&$1.91^{+0.05}_{-0.07}$ &   $0.036^{+0.006}_{-0.004}$ &   $0.50^{+0.08}_{-0.06}$ &   $2.5^{+0.2}_{-0.2}$ &   $4^{+8}_{-2}$ &   $0.05^{+0.02}_{-0.01}$ &   $-6.57^{+0.06}_{-0.05}$ &$-2812$ \\
\hline

\end{longtable}
 \end{center}
\end{minipage}

The plots of optical depth per bar (Fig. \ref{fig:DRtaubar}) and cumulative optical depth (Fig. \ref{fig:DRtauC}) illustrate the range of likely aerosol distributions for each of the 18 locations.  In these plots, we also show the best-fit solution from the  {\it 2L\_DISORT} retrieval for reference. It is clear from these plots that a single spectrum provides a much weaker constraint on the aerosol distribution, relative to the more extensive limb-darkening dataset considered in the  {\it 2L\_DISORT} retrieval -- in particular the fractional scale height of the upper aerosol layer is poorly constrained at many locations. However, we do observe a number of interesting trends in the single-spectrum retrievals: at every location, the deep aerosol layer ($\beta$) dominates in terms of total optical depth, and is generally observed to be physically compact ($h_{frac,\beta} <0.2$) and optically thick. The exceptions are locations A and B in the north, and perhaps location F at the equator, which have best-fit values of the optical depth  $\tau_\beta <1$. The base pressure of this bottom cloud layer is always below 2.2 bar; since the cloud is generally found to be optically thick, solutions with higher base pressures (and correspondingly larger total optical depths) are allowed in many locations. 

The upper aerosol layer ($\alpha$) exhibits a more pronounced latitudinal variability than the deeper cloud. At location A, at 20$^\circ$N, the aerosols increase in concentration with altitude ($h_{frac,\alpha}\sim2$), indicating the presence of aerosols above the tropopause. Locations B--I, between 16$^\circ$N and 17$^\circ$S have aerosol scale heights similar to the gas scale height ($h_{frac,\alpha}\sim1$) and the aerosol distribution is consistent with the {\it 2L\_DISORT}  best-fit solution. Location J is an outlier, with a preference for a vertically compact aerosol layer near 0.4 bar, rather than a more extended haze. This location may be influenced by reflectivity from a nearby bright cloud -- see Fig. \ref{fig:DRcubes}. Further south (locations K--R, 43--87$^\circ$S), the optical depth of aerosols in the stratosphere and upper tropopause drops below the reference  {\it 2L\_DISORT}  solution (Fig.  \ref{fig:DRtauC}), and plots of $\tau$/bar  (Fig. \ref{fig:DRtaubar}) show no gap between aerosol layer $\alpha$ and the deep cloud. The details of the retrieved solution may depend somewhat on viewing geometry: locations M and O, for example, are at relatively low emission angles ($\mu_H > 0.6$), and have more compact solutions for the upper aerosol layer than neighboring locations N and P, which are nearer the limb. 

Figure \ref{fig:DR_tauCvlat} presents another visualization of the latitudinal trend in the aerosol structure. In this set of plots, we show the cumulative optical depth of aerosols at 0.1, 0.5, 1, and 2 bar, as a function of latitude. Above a pressure of 0.1 bar, we observe that the cumulative optical depth in aerosols south of 20$^\circ$S is very low. At low latitudes we observe a large scatter in the retrieved optical depths; however, much higher cumulative optical depths are permitted. Above 0.5 and 1 bar, there is a clear difference between cumulative optical depths observed for locations north and south of  20$^\circ$S. By a pressure  depth of 2 bar, the cumulative optical depth in aerosols is similar near the equator and mid- and high southern latitudes. 

\subsection{Local methane profile}\label{sec:depletion}

Next, we revisit Neptune's methane profile, to search for latitudinal variations. Using Hubble STIS spectroscopy, \cite{kark11a} found evidence of a shallow depletion of  CH$_4$  below $\sim1.2$ bar at some latitudes. To investigate whether our data exhibit signatures of similar behavior, we added to the parameterization of the previous section a methane depletion parameter, $P_{dep}$, as defined in Appendix \ref{app:composition} and Table \ref{table:model}, which controls the CH$_4$ profile below the methane condensation pressure of $\sim 1.7$ bar. The results of this set of retrievals were difficult to interpret:  retrieved values of the methane depletion parameter ranged from 2--3 bar to more than 50 bar -- well below our optically thick cloud $\beta$. There was no observable latitudinal trend in the retrieved parameters. Furthermore, we found very different depletion depths for spectra from similar latitudes, and high values of depletion near the equator, where it was not observed by \cite{kark11a}. We concluded that this model parameterization was too poorly constrained by our data to yield useful information.

Recalling the results from Section \ref{sec:CH4tests}, we hypothesize that we are less sensitive to the relatively deep CH$_4$ depletion than we are to the methane relative humidity near the tropopause.  To allow for more freedom in the CH$_4$ profile, we perform a set of retrievals in which both  $RH_{ac}$ (which controls the CH$_4$ mole fraction near the tropopause)  and $P_{dep}$ (which controls the CH$_4$ mole fraction below $\sim 1.7$ bar)  are free parameters. For this test, we fix the base pressure, optical depth, and fractional scale height of layer $\beta$ to be $P_{max,\beta}=3.3$ bar, $\tau_\beta=5$, and $h_{frac,\beta}=0.1$. In other words, we assume that the properties of the bottom cloud are spatially constant, and that any observed variations in the 1.6 $\mu$m spectral peak are due to spatial variations in the methane column above that cloud rather than due to variations in aerosol layer $\beta$. We also fix $h_{frac, \alpha}$ to be $0.8$; while there is no physical motivation for forcing the scale height of the top aerosol layer to be constant, doing so makes it easier to interpret the remaining free parameters in our model. As in the previous set of retrievals, $\omega_0$ and $g$ for both layers are fixed to the {\it 2L\_DISORT} best-fit values. The results of these retrievals are summarized in Table \ref{table:CH4fit}.

\begin{minipage}{1.1\textwidth}
\setlength\LTleft{-0.1in}
\setlength\LTright{-1in plus 1 fill}
\renewcommand{\footnoterule}{}
\renewcommand{\thempfootnote}{\it \alph{mpfootnote}}
 \renewcommand{\thefootnote}{\it \alph{footnote}}
\begin{center}
\scriptsize
\setlength{\tabcolsep}{0.05in}
\begin{longtable}{+l | ^r ^l ^l | ^l ^l | ^l ^l ^l ^l ^r }
\caption[Cloud-free regions, methane fits]{\small Results for variable-methane fits to 18 cloud-free locations. These retrievals include limited free parameters for aerosols and two free parameters describing the methane profile. \label{table:CH4fit}} \tabularnewline
location	& lat ($^\circ$)	& $\mu_H$ & $\mu_K$		&\multicolumn{2}{l }{layer $\alpha$ (top)}	    &  $RH_{ac}$& $P_{dep}$ (bar) & $\log{\sigma_t}$	& 	$DIC$ & $\Delta DIC$\footnote{Change in the $DIC$ relative to the initial aerosol-only fit for that location}\\
	        &      &	           &                            &$P_{max,\alpha}$ (bar)	                                     &$\tau_\alpha$   &  &  & & 	\\
A & $20$& $0.66$& $0.54$&$0.56^{+0.07}_{-0.07}$ &   $0.0208^{+0.001}_{-0.0009}$ &   $0.01^{+0.02}_{-0.01}$ &   $1.76^{+0.01}_{-0.04}$ &   $-6.71^{+0.06}_{-0.07}$ &$-2857$ & $21$ \\
B & $16$& $0.46$& $0.63$&$1.2^{+0.1}_{-0.1}$ &   $0.0219^{+0.001}_{-0.0009}$ &   $0.05^{+0.03}_{-0.02}$ &   $1.98^{+0.04}_{-0.04}$ &   $-6.95^{+0.07}_{-0.08}$ &$-2923$ & $-19$\\
C & $10$& $0.26$& $0.52$&$1.36^{+0.07}_{-0.09}$ &   $0.0190^{+0.001}_{-0.0009}$ &   $0.04^{+0.02}_{-0.01}$ &   $2.02^{+0.03}_{-0.04}$ &   $-7.17^{+0.08}_{-0.08}$ &$-3029$ & $-35$\\
D & $5$& $0.80$& $0.84$&$0.9^{+0.1}_{-0.1}$ &   $0.0181^{+0.0009}_{-0.0008}$ &   $0.02^{+0.02}_{-0.01}$ &   $2.25^{+0.03}_{-0.02}$ &   $-7.06^{+0.08}_{-0.08}$ &$-2997$ &$-57$\\
E & $1$& $0.88$& $0.82$&$0.73^{+0.09}_{-0.08}$ &   $0.0165^{+0.0008}_{-0.0007}$ &   $0.02^{+0.02}_{-0.01}$ &   $2.15^{+0.03}_{-0.03}$ &   $-7.32^{+0.08}_{-0.08}$ &$-3084$ &$-42$\\
F & $0$& $0.51$& $0.74$&$1.27^{+0.09}_{-0.1}$ &   $0.022^{+0.001}_{-0.001}$ &   $0.05^{+0.04}_{-0.02}$ &   $2.11^{+0.04}_{-0.03}$ &   $-7.05^{+0.07}_{-0.08}$ &$-2979$ & $-33$\\
G & $-7$& $0.93$& $0.89$&$0.4^{+0.2}_{-0.1}$ &   $0.015^{+0.002}_{-0.001}$ &   $0.05^{+0.09}_{-0.03}$ &   $2.01^{+0.05}_{-0.05}$ &   $-6.58^{+0.05}_{-0.06}$ &$-2817$ & $14$\\
H & $-9$& $0.57$& $0.80$&$1.3^{+0.1}_{-0.1}$ &   $0.0222^{+0.0007}_{-0.0007}$ &   $0.012^{+0.007}_{-0.004}$ &   $2.02^{+0.03}_{-0.04}$ &   $-7.04^{+0.08}_{-0.09}$ &$-2982$ & $-89$\\
I & $-17$& $0.95$& $0.81$&$1.62^{+0.04}_{-0.04}$ &   $0.0118^{+0.0006}_{-0.0007}$ &   $0.00004^{+6e-05}_{-3e-05}$ &   $2.16^{+0.04}_{-0.03}$ &   $-6.68^{+0.06}_{-0.06}$ &$-2863$&$-92$ \\
J & $-21$& $0.99$& $0.94$&$0.68^{+0.1}_{-0.06}$ &   $0.0237^{+0.0009}_{-0.002}$ &   $0.17^{+0.09}_{-0.1}$ &   $2.7^{+0.3}_{-0.3}$ &   $-6.72^{+0.06}_{-0.06}$ &$-2870$& $12$\\
K & $-43$& $0.91$& $0.73$&$1.77^{+0.05}_{-0.06}$ &   $0.029^{+0.002}_{-0.002}$ &   $0.41^{+0.1}_{-0.08}$ &   $3.9^{+0.9}_{-0.8}$ &   $-6.68^{+0.06}_{-0.06}$ &$-2846$ & $-16$\\
L & $-49$& $0.79$& $0.60$&$2.05^{+0.04}_{-0.04}$ &   $0.006^{+0.003}_{-0.002}$ &   $0.016^{+0.009}_{-0.006}$ &   $2.18^{+0.08}_{-0.06}$ &   $-6.48^{+0.06}_{-0.06}$ &$-2747$ &$-6$\\
M & $-57$& $0.87$& $0.85$&$2.08^{+0.04}_{-0.06}$ &   $0.006^{+0.004}_{-0.003}$ &   $0.03^{+0.03}_{-0.02}$ &   $2.38^{+0.08}_{-0.08}$ &   $-6.16^{+0.05}_{-0.05}$ &$-2605$ &$0$\\
N & $-64$& $0.53$& $0.34$&$1.88^{+0.05}_{-0.04}$ &   $0.021^{+0.002}_{-0.002}$ &   $0.39^{+0.09}_{-0.07}$ &   $4.0^{+0.9}_{-0.6}$ &   $-6.69^{+0.06}_{-0.06}$ &$-2862$ &$-22$ \\
O & $-75$& $0.68$& $0.67$&$2.30^{+0.05}_{-0.07}$ &   $0.0012^{+0.0009}_{-0.0003}$ &   $0.015^{+0.005}_{-0.004}$ &   $2.23^{+0.04}_{-0.05}$ &   $-6.42^{+0.05}_{-0.05}$ &$-2735$&$-28$ \\
P & $-76$& $0.36$& $0.30$&$1.94^{+0.1}_{-0.07}$ &   $0.023^{+0.003}_{-0.003}$ &   $0.3^{+0.1}_{-0.1}$ &   $3^{+1}_{-1}$ &   $-6.94^{+0.07}_{-0.07}$ &$-2934$ &$-13$\\
Q & $-83$& $0.52$& $0.56$&$2.21^{+0.05}_{-0.05}$ &   $0.0013^{+0.0007}_{-0.0003}$ &   $0.006^{+0.003}_{-0.002}$ &   $2.19^{+0.04}_{-0.05}$ &   $-6.54^{+0.06}_{-0.06}$ &$-2794$ &$-54$\\
R & $-87$& $0.48$& $0.47$&$2.18^{+0.05}_{-0.03}$ &   $0.006^{+0.002}_{-0.002}$ &   $0.013^{+0.01}_{-0.006}$ &   $2.05^{+0.04}_{-0.04}$ &   $-6.68^{+0.06}_{-0.06}$ &$-2850$ &$-38$\\

\end{longtable}
 \end{center}
\end{minipage}

\begin{figure}[h!]
\begin{center}$
\begin{array}{cccc}
\includegraphics[width=0.5\textwidth]{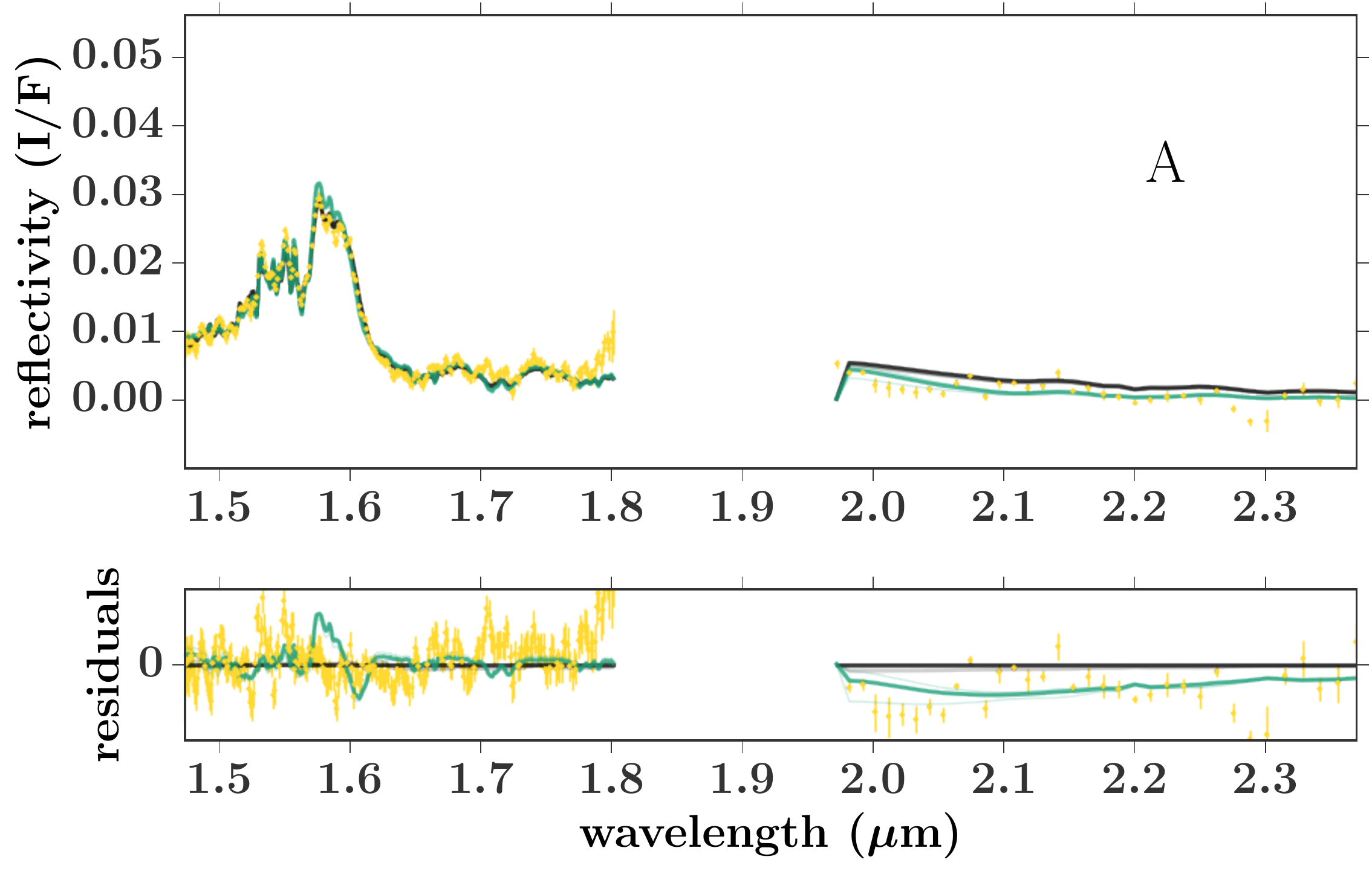} \includegraphics[width=0.5\textwidth]{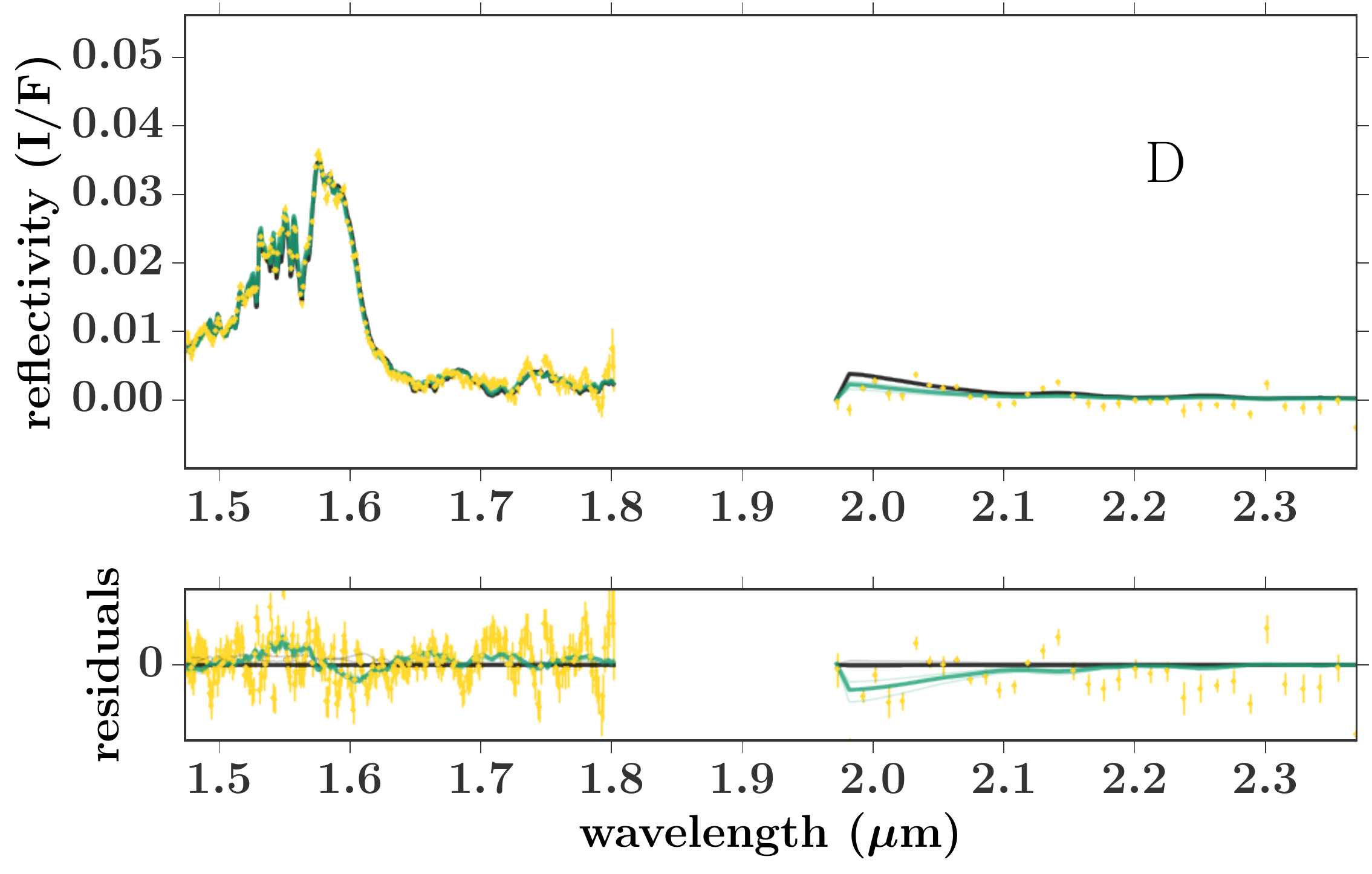}\\
\includegraphics[width=0.5\textwidth]{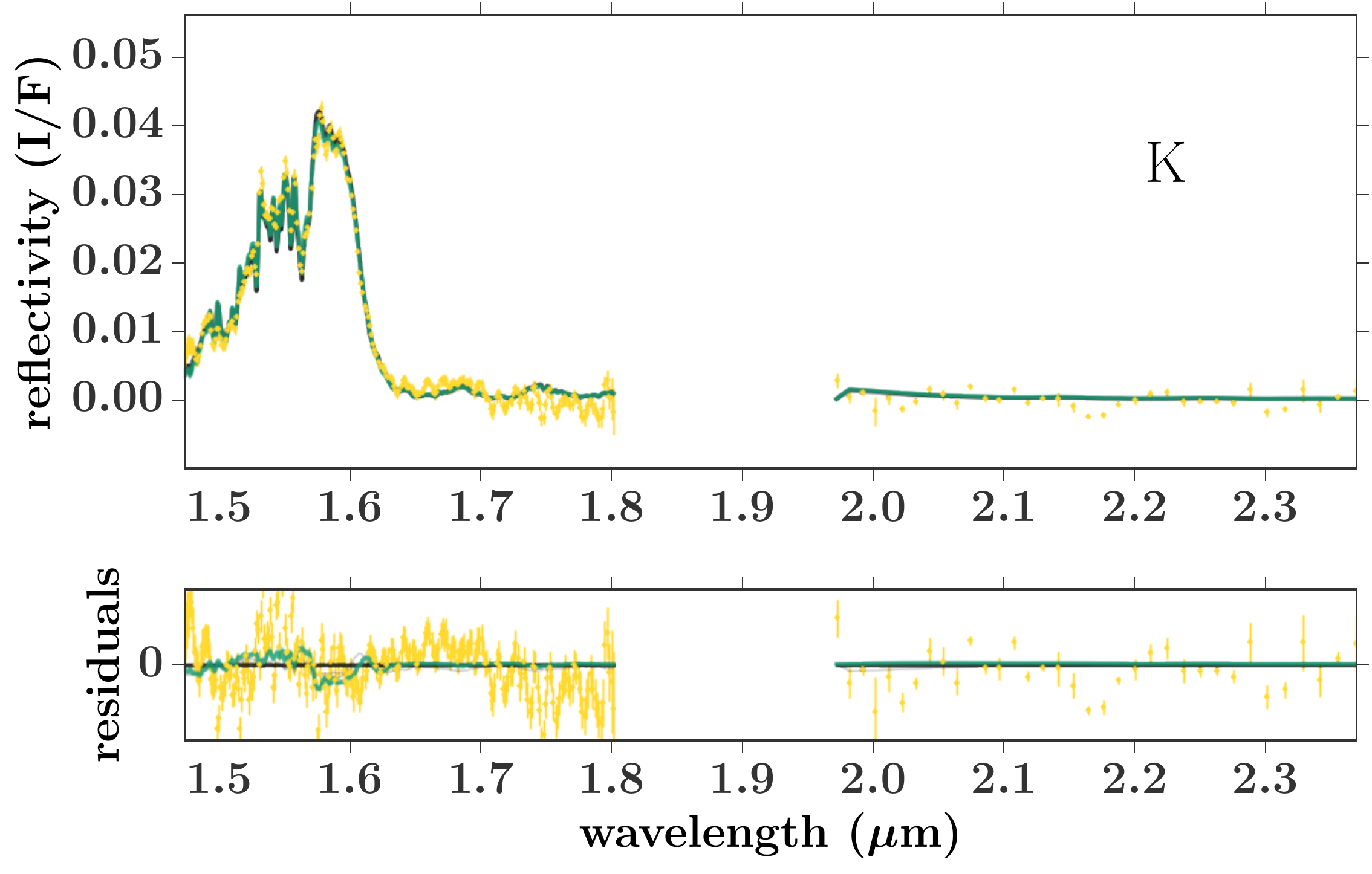}\includegraphics[width=0.5\textwidth]{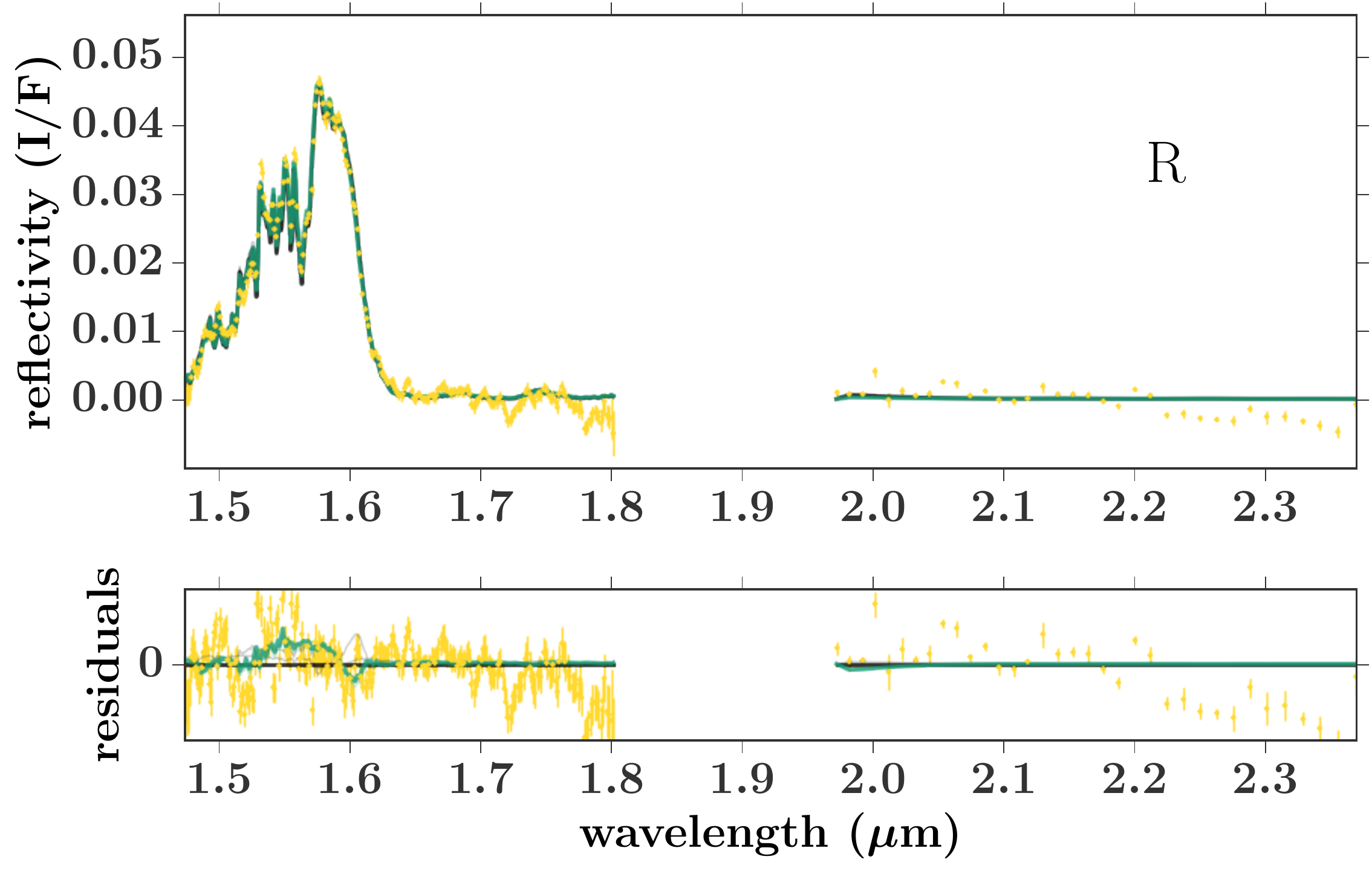}\\

\end{array}$
\end{center}
\caption[Model fits to dark regions, CH$_4$ fixed and fit]{\label{fig:DR_CH4fits} \small Comparison of two fits to Neptune's dark regions to the data (yellow), for four of the cloud-free locations. The initial aerosol-only fit, described in Section \ref{sec:defaulthaze} is shown in black, and the best-fit model for that case is used as the comparison point for the residual plots. The second fit (green) represents the variable-methane case, in which two parameters describing the methane profile are allowed to vary, as described in Section \ref{sec:depletion}. For location A, the variable-methane retrieval is of worse quality than the aerosol-only fit; in the other three cases, the variable-methane fit is preferred.   } 
\end{figure}

Of our 18 locations, we find that three locations (A, G, and J) are less well modeled by this parameterization relative to the original aerosol-only parameterization. Location A, demonstrates the greatest decrease in fit quality ($\Delta DIC = 21$); the retrieved models for this location are in particularly poor agreement with the data near 1.6 $\mu$m. The likely cause of the decrease in fit quality is that location A is not well-matched by the assumption that cloud $\beta$ is optically thick. For location J, the initial aerosol-only retrieval finds layer $\alpha$ to be vertically compact; the decreased fit quality in the latter retrieval may be due to forcing $h_{frac, \alpha} = 0.8$. We do not identify an obvious reason for the decrease in fit quality at location G. Of the remaining 15 locations, 13 are significantly better matched by the retrievals that allow the methane to vary, whereas the $DIC$ at two locations are comparable ($|DIC|\leq10$) for the aerosol-only and variable-methane parameterizations. For the locations demonstrating an improvement in fit quality, this improvement is greatest near 1.5--1.6 $\mu$m (Fig. \ref{fig:DR_CH4fits}). 

\begin{figure}[h!]
\begin{center}$
\begin{array}{cc}
\includegraphics[width=0.45\textwidth]{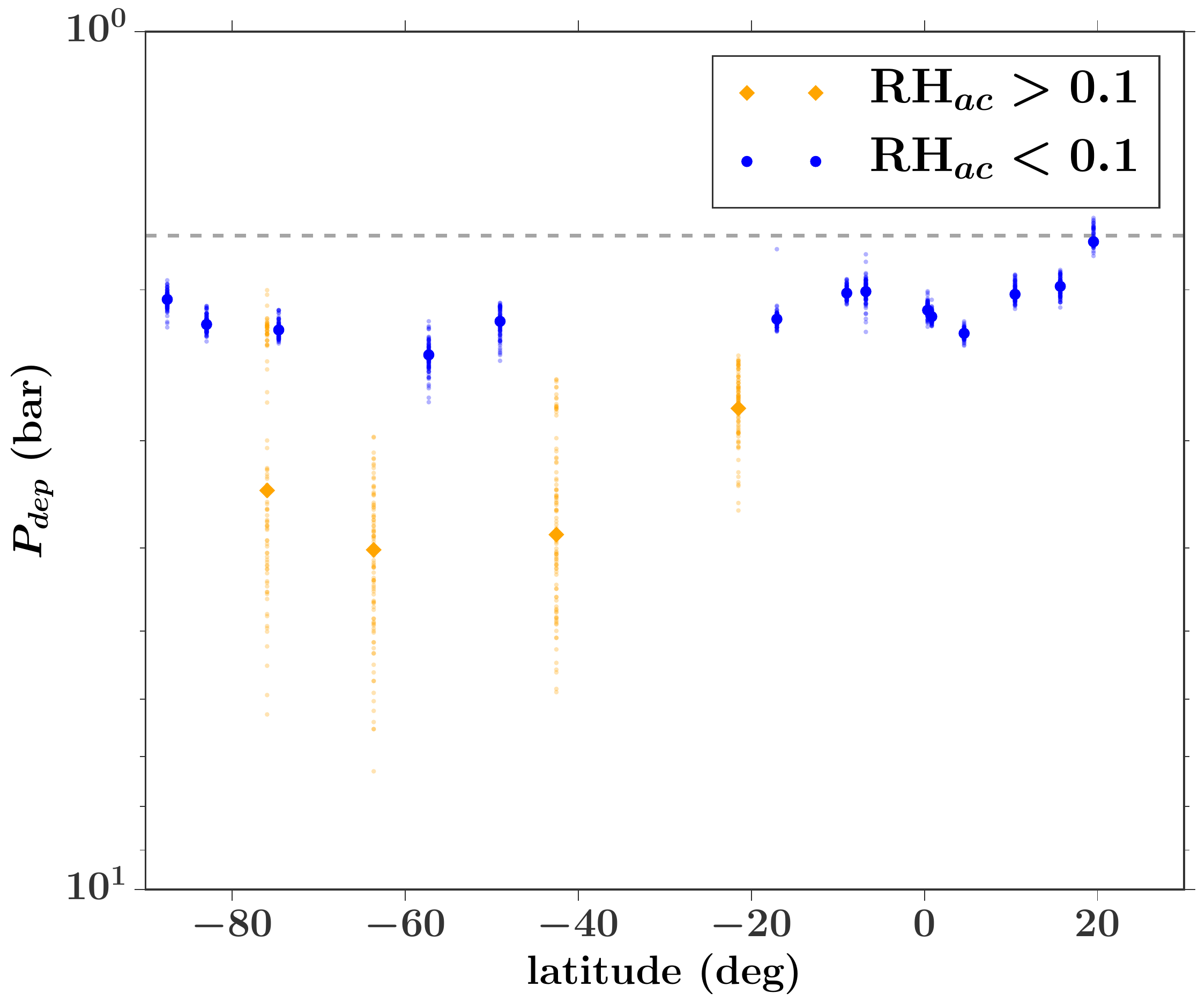}\includegraphics[width=0.45\textwidth]{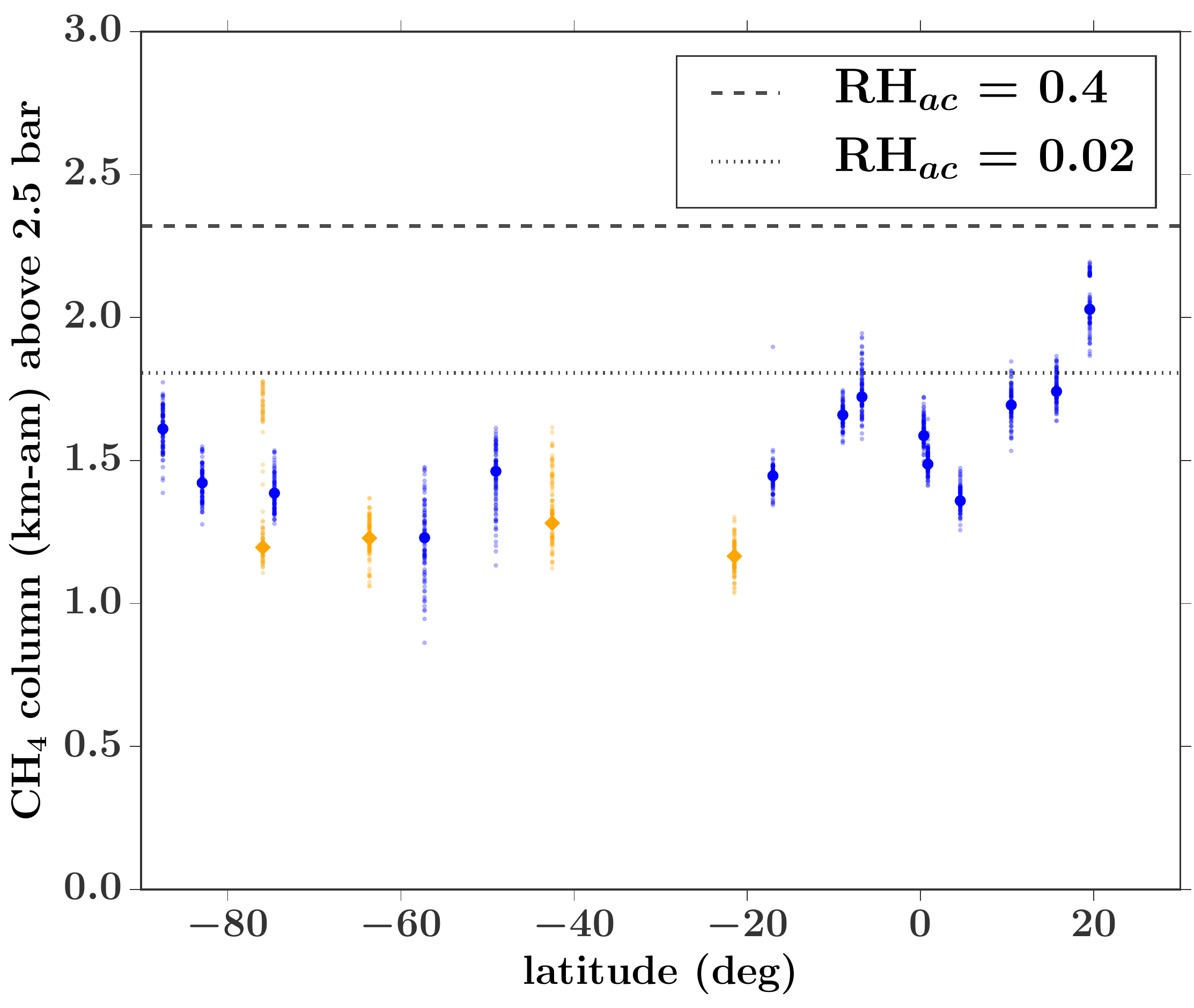}
\end{array}$
\end{center}
\caption[Methane depletion vs. latitude]{\label{fig:CH4dep} \small  Methane as a function of latitude. Left: retrieved depletion depth ($P_{dep}$) as a function of latitude.  The grey dashed line indicates the CH$_4$ condensation pressure. The large symbols indicate the best-fit value of $P_{dep}$ for each location; the smaller symbols represent 100 random draws from the posterior probability distribution.  In most cases (blue circles) the increased CH$_4$ depletion depth below condensation is coupled with a decrease in CH$_4$ relative humidity ($RH_{ac}$) above condensation. However, in four cases (orange diamonds), the decrease in CH$_4$ is achieved with a higher $RH_{ac}$ but a deeper CH$_4$ depletion-- see Section \ref{sec:depletion} for a discussion. Right: CH$_4$ column above 2.5 bar. Colors and symbols are defined as in the left-hand panel. The two horizontal lines indicate reference values of the CH$_4$ column, for the {\it SN\_CH$_4$grid$_E$} (used in the aerosol-only retrievals, dashed line) and {\it SN\_CH$_4$RHfree$_A$} ($RH_{ac}$=0.02, dotted line) methane profiles. The reference methane profiles have the same values of {\it mCH$_{4,t}$}	 and {\it mCH$_{4,s}$} that are used in the variable-methane retrievals, and do not include methane depletion below the condensation pressure. The {\it SN\_CH$_4$grid$_E$} and {\it SN\_CH$_4$RHfree$_A$}  models have methane relative humidities of $RH_{ac}$=0.4 and  $RH_{ac}$=0.02, respectively.    }
\end{figure}

At every location in our analysis, the variable-methane retrieval favors a tropospheric CH$_4$ mole fraction below our nominal {\it SN\_CH$_4$grid$_E$} methane profile (see Supplementary Fig. \ref{fig:CH4fit}). This is achieved through a combination of subsaturation near the tropopause and depletion below the CH$_4$ condensation pressure. In the majority of cases, we find a best-fit value of $RH_{ac}$ below 10\% coupled with CH$_4$ depletion to a depth of $P_{dep} \approx $2.0--2.5 bar. For four locations, an alternative solution is observed: the relative humidity remains higher (best-fit values of 20--40\%) but the retrieved depletion depth is higher as well. For three of these four cases (locations J, K, and P), the posterior probability distributions are actually double peaked, with one local minimum at the higher relative humidity and higher depletion pressure and a second at lower relative humidity and lower depletion pressure.  As before when we had double-peaked posterior probability distributions, we note that the MCMC chains may have converged with additional steps.  In Fig. \ref{fig:CH4dep} (left) we plot depletion pressure as a function of latitude. While latitudinal variations in the depletion depth are not ruled out, differences between the two families of solutions (low $RH_{ac}$ and $P_{dep}$ vs. higher $RH_{ac}$ and $P_{dep}$)  dominate the scatter in $P_{dep}$. Latitudinal trends are more compelling when both relative humidity and depletion below the condensation level are considered: in the righthand panel of Fig. \ref{fig:CH4dep}, we show the CH$_4$ abundance above 2.5 bar for the retrieved CH$_4$ profiles. This plot shows that both families of solutions represent similar CH$_4$ columns above 2.5 bar; at pressures greater than 2.5 bar, the deep cloud generally becomes optically thick, and we are not sensitive to variations in the CH$_4$ abundance. We also indicate in this plot the CH$_4$ columns for the nominal  {\it SN\_CH$_4$grid$_E$} methane profile ($RH_{ac}$=0.4) and for the {\it SN\_CH$_4$RHfree$_A$} retrieval from Section \ref{sec:CH4tests}, in which  $RH_{ac}$  (but not $P_{dep}$) was allowed to vary ($RH_{ac}\approx 0.02$). The retrieved CH$_4$ column at many locations near the equator is roughly consistent with the {\it SN\_CH$_4$RHfree$_A$} CH$_4$ column. While not conclusive, this plot may suggest a local minimum in the upper tropospheric CH$_4$ abundance just north of the equator, and a decreased abundance at high southern latitudes relative to low latitudes.

\begin{figure}[h!]
\begin{center}$
\includegraphics[width=0.9\textwidth]{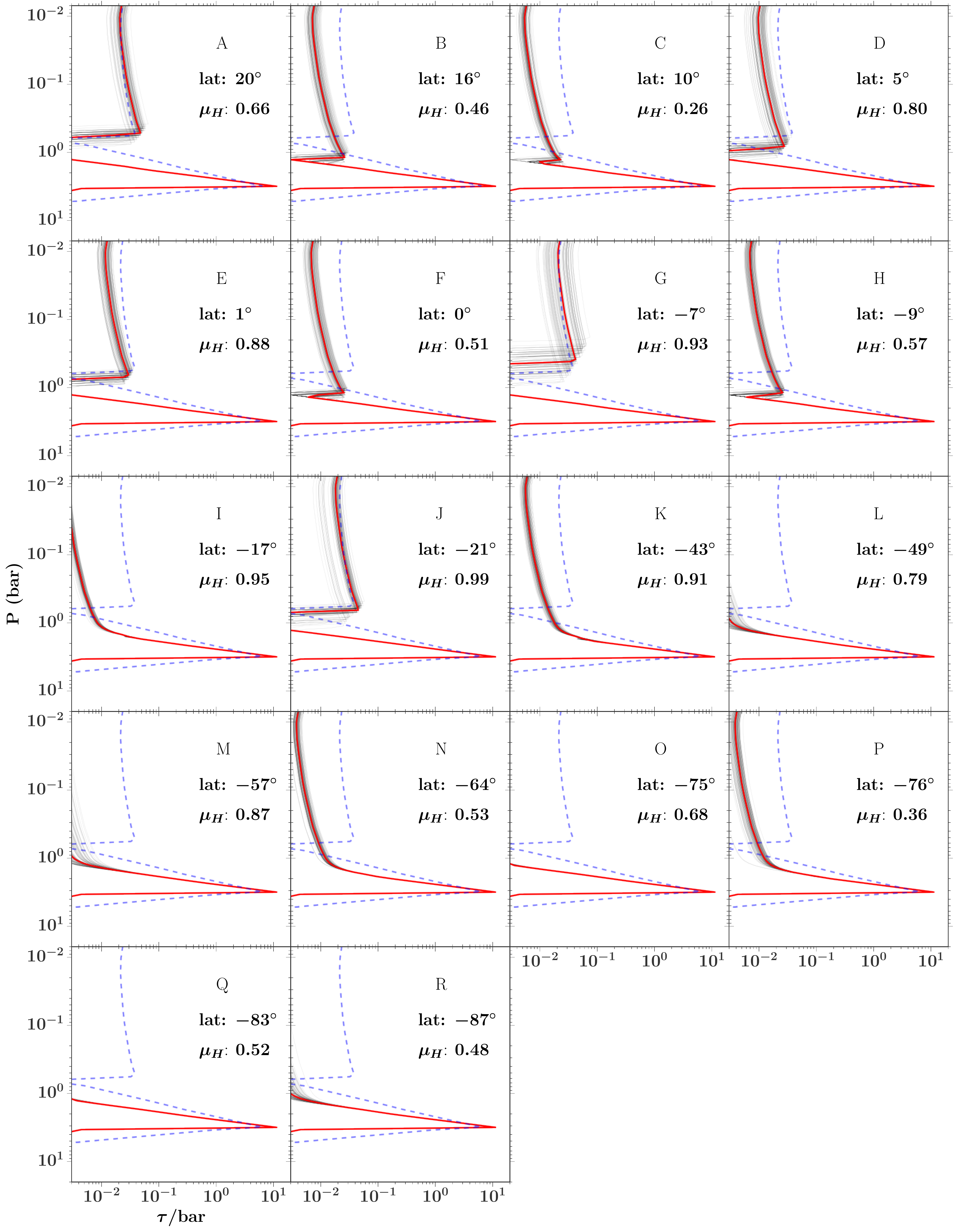}
$
\end{center}
\caption[Optical depth/bar, fits including methane depletion]{\label{fig:taubarCH4} \small  Same as Fig. \ref{fig:DRtaubar} except for the variable-methane retrieval (Section \ref{sec:depletion}). }
\end{figure}

Finally, we observe that the variable-methane parameterization does not qualitatively influence the trend in the upper aerosol layer $\alpha$ observed in the previous set of retrievals: this is evident by comparing Fig. \ref{fig:taubarCH4} to Fig. \ref{fig:DRtaubar}.  However, the optical depth of the upper aerosol layer is, in general, lower in the variable-methane retrieval than in the aerosol-only retrieval, with best-fit values of  $\tau_\alpha < 0.01$ at a number of locations in the south for the variable-methane case.

\subsection{Thermal variations}\label{sec:DR_TP}
In Section \ref{sec:TPtests} we found that perturbations to the stratospheric temperature did not cause appreciable changes to the posterior probability distribution of the aerosol model parameters. We also found that the 1.5 K adjustment to the tropopause temperature from its nominal value of 54.9 K to the Orton case B value (56.4 K) had a negligible effect on the retrieved aerosol structure. We now revisit the question of possible effects due to variations in the tropopause temperature, for three specific locations at which the local temperature is expected, from \cite{orton07}, to exhibit large deviations from the nominal value. Two of the locations (P, R) are at high southern latitudes and should have elevated temperatures relative to the nominal value.The third location (M) is at mid southern latitudes and should have a low tropopause temperature, relative to the nominal value. We repeat the initial (aerosol-only, no methane depletion) fits for these three locations, and present the retrieved parameters for these revised thermal profiles in Table \ref{table:DR_TP}.  None of the three locations exhibit an improvement in the {\it DIC} for retrievals with the revised thermal profile. The posterior probability distributions from the nominal retrievals and the temperature-perturbed retrievals are effectively the same for all three locations. 

\begin{minipage}{1.1\textwidth}
\setlength\LTleft{-0.5in}
\setlength\LTright{-1in plus 1 fill}
\renewcommand{\footnoterule}{}
\renewcommand{\thempfootnote}{\it \alph{mpfootnote}}
 \renewcommand{\thefootnote}{\it \alph{footnote}}
\begin{center}
\scriptsize
\setlength{\tabcolsep}{0.05in}
\begin{longtable}{+l | ^r ^l |  ^l ^l ^l |  ^l ^l ^l |^l ^l ^l}
\caption[Table of results, haze retrieval with temperature change]{\small Results for initial retrievals to the haze parameters (CH$_4$ fixed), including variations in the tropopause temperature as observed by \cite{orton07}.  \label{table:DR_TP}} \tabularnewline
location	& lat	& $\Delta T_{trop} (K)$	&\multicolumn{3}{l }{layer $\alpha$ (top)}	 	&\multicolumn{3}{l }{ layer $\beta$ (bottom)	} 	  & $\log{\sigma_t}$	& 	$DIC$& $\Delta DIC$\footnote{relative to initial model fit}\\
	&  &	 &$P_{max,\alpha}$ (bar)	 &$\tau_\alpha$ &$h_{frac,\alpha}$		& $P_{max,\beta}$ (bar)	 &$\tau_\beta$ &$h_{frac,\beta}$  & & &\\

M & $-57$& $-3.8$&$2.1^{+0.3}_{-0.2}$ &   $0.09^{+0.08}_{-0.03}$ &   $0.26^{+0.03}_{-0.03}$ &   $2.8^{+0.7}_{-0.3}$ &   $1^{+8}_{-1}$ &   $0.06^{+0.09}_{-0.04}$ &   $-6.15^{+0.06}_{-0.06}$ &$-2606$ & $0$ \\
P & $-76$& $+4.3$&$1.78^{+0.07}_{-0.06}$ &   $0.025^{+0.003}_{-0.002}$ &   $0.9^{+0.1}_{-0.1}$ &   $2.4^{+0.2}_{-0.1}$ &   $2^{+4}_{-1}$ &   $0.06^{+0.02}_{-0.01}$ &   $-6.80^{+0.07}_{-0.08}$ &$-2894$ &$28$\\
R & $-87$& $+7.5$&$1.90^{+0.06}_{-0.07}$ &   $0.034^{+0.006}_{-0.004}$ &   $0.62^{+0.1}_{-0.09}$ &   $2.5^{+0.2}_{-0.1}$ &   $4^{+9}_{-3}$ &   $0.05^{+0.02}_{-0.02}$ &   $-6.55^{+0.07}_{-0.07}$ &$-2796$ &$16$ \\

\hline

\end{longtable}
 \end{center}
\end{minipage}

\section{Summary of findings}\label{sec:summ}
We present a comprehensive analysis of OSIRIS Hbb and Kbb integral field spectrograph observations of the planet Neptune from 26 July 2009, focusing on regions free of discrete NIR-bright clouds. Our analysis involves a series of atmospheric retrievals using an atmospheric model and radiative transfer code, coupled to an MCMC algorithm that provides an estimate of the posterior probability distribution of the model parameters. The Deviance Information Criterion ({\it DIC}) is used to compare different atmospheric models for the same set of data. Taking advantage of the high spatial resolution of these data, we first construct a set of six high S/N spectra, representing the atmosphere in a latitude range of 2--12$^\circ$N at each of six values of the cosine of the emission angle. Our atmospheric retrievals using this high-quality limb darkening dataset reveal the following: 
\begin{itemize}
\item Neptune's cloud opacity at these wavelengths and latitudes is dominated by a compact cloud layer with a base near 3 bar. Using pyDISORT for the radiative transfer and assuming a Henyey-Greenstein phase function, we find that the base of this layer is located at $P_{max,\beta} =3.3^{+0.4}_{-0.3}$ bar, and the layer is optically thick ($\tau_\beta= 5^{+4}_{-2}$ at 1.6 $\mu$m for the   {\it 2L\_DISORT} retrieval). Using pyDISORT, we observe this cloud to be composed of low albedo ($\omega_{0,\beta}  =0.45^{+0.01}_{-0.01}$), moderately forward scattering ($g_\beta = 0.50^{+0.02}_{-0.02}$) particles, compared to the Mie theory prediction of $\omega_0=1.0$ and $g=0.76$ at 1.6 $\mu$m for $r_p = 1.0$ $\mu$m. 

\item A second aerosol layer, at lower pressures but also with a pressure base in the troposphere, is required in our models to match the data. The  {\it 2L\_DISORT} model retrieves  $P_{max,\alpha} =0.59^{+0.04}_{-0.03}$ bar, $\tau_\alpha =0.019^{+0.002}_{-0.001}$, and $h_{frac} = 0.85^{+0.07}_{-0.06}$ for the upper aerosol layer, indicating nearly uniform mixing of the aerosols with the gas, up into the stratosphere. For the wavelength-dependent cross section of upper aerosol particles, a particle size distribution of $r_p = 0.1$ $\mu$m is preferred over one dominated by larger ($r_p = 0.5$ or 1.0 $\mu$m) particles.  Using pyDISORT and a Henyey-Greenstein phase function, we retrieve $\omega_{0,\alpha}=0.91^{+0.06}_{-0.05}$ and $g_\alpha = 0.24^{+0.02}_{-0.03}$. 
\end{itemize}

Focusing on the highest S/N spectrum from this set of six, we investigate Neptune's methane and thermal profiles. We find: 

\begin{itemize}
\item Our retrievals indicate a strong preference for a methane relative humidity of $RH_{ac} =$40\% over 100\% relative humidity. When $RH_{ac}$ is allowed to be a free parameter, the retrieved relative humidity is less than 2\% ($0.016^{+0.01}_{-0.008}$ for the  {\it SN\_CH$_4$RHfree$_A$} retrieval). Our retrievals indicate a possible preference for a stratospheric CH$_4$ mixing ratio at the low end of the range found by previous authors ($3.5 \times 10^{-4}$ or less); however, based upon our {\it RHfree} retrievals, we interpret this result as a preference for models with a decreased CH$_4$ abundance near the tropopause, rather than a sensitivity to the CH$_4$ abundance in the upper stratosphere. We are not able to constrain the value of the tropospheric CH$_4$ abundance, $mCH_{4,t}$, with our data. We caution against over-interpreting the meaning of the  values of the methane parameters retrieved by this study, in light of the very simple parameterization used for the methane profile (Appendix \ref{app:composition}).
\item Overall, our retrieved aerosol structure is qualitatively insensitive to the assumed methane profile, being comprised in all cases of a vertically compact bottom cloud with a base pressure 3--4 bar, and a vertically extended, optically thin upper aerosol layer with a base pressure of 0.4--0.7 bar. The parameters most affected by the assumed methane profile are the base pressures of both aerosol layers and the scale height of the upper aerosol layer. The base pressure of the deeper aerosol layer depends on the tropospheric methane mixing ratio, with the highest base pressures corresponding to the lowest tropospheric methane mixing ratios. This result is not surprising, since a lower methane mixing ratio implies that greater distances must be traveled in order to encounter the same total methane column. The base pressure of the upper aerosol layer is slightly but systematically influenced by the value of the stratospheric methane mole fraction, with a higher values of {\it mCH$_{4,s}$} resulting in a higher value of $P_{max,\alpha}$. The compactness of the upper aerosol layer, as parameterized by  $h_{frac,\alpha}$, depends most strongly on the relative humidity of CH$_4$ near the tropopause, with higher values of $RH_{ac}$ resulting in a more vertically extended upper haze layer.  
\item For the same methane profile, a model with a higher stratospheric temperature results in an improved fit, with no change to the retrieved aerosol parameters. An increase to the tropopause temperature or decrease to the stratospheric temperature result in worse fits. As with CH$_4$, we consider only a simple parameterization to the thermal profile and only a few different test cases. 

\end{itemize}

We follow this analysis of the strip from 2--12$^\circ$N with a study of 18 cloud-free locations across Neptune, spanning latitudes from 20$^\circ$N to 87$^\circ$S and viewing geometries from $\mu=0.26$ to $\mu=0.99$. We use these data to evaluate the information content of a single spectrum at a single viewing geometry relative to the more complete and higher S/N limb-darkening dataset, and to explore latitude variations in atmospheric properties. 
We find:
\begin{itemize}
\item While we do not attempt to fit the optical properties of the aerosols using single spectra, retrievals of the other aerosol parameters ($P_{max}$, $\tau$, and $h_{frac}$) for locations C and D from within the limb-darkening study region (latitudes of 10$^\circ$N and 5$^\circ$N, respectively) are remarkably consistent with -- though not strictly within 1$\sigma$ of -- the values found by the full {\it 2L\_DISORT} retrieval. The posterior probability distributions for the layer $\alpha$ base pressure and optical scale height have greater uncertainties in the fits to individual locations than observed for the limb darkening dataset retrievals. This is likely due to the fact that most of the information about the upper aerosol layer is derived from low S/N portions of the spectrum.
\item The properties of the deep aerosol layer $\beta$ are qualitatively consistent across all locations and with the  full {\it 2L\_DISORT} retrieval. At  latitudes of 10$^\circ$N--20$^\circ$S, retrieved values of $P_{max,\beta}$ are within $1\sigma$ of the $3.3^{+0.4}_{-0.3}$ bar found by {\it 2L\_DISORT}; outside of this latitude range, the cloud base is typically at a slightly lower pressure, with best-fit values of 2.2--2.7 bar. As in the {\it 2L\_DISORT} case, we observe this cloud to be optically thick, with the exceptions of locations A and B at 20 and 16$^\circ$N, respectively, and possibly location F at the equator, for which this bottom cloud may be optically thin. At all locations, the retrieval favors a highly compact layer $\beta$, with $h_{frac,\beta}$ values similar to or less than the value retrieved in the full {\it 2L\_DISORT} case.
\item The structure of aerosol layer $\alpha$ varies with latitude: at low latitudes (20$^\circ$N -- 20$^\circ$S), we observe a vertically extended haze with a base in the troposphere above layer $\beta$, with  $P_{max, \alpha}$ typically 0.3--1.4 bar. A gap is observed between the top of layer $\beta$ and the base of layer $\alpha$, and the upper layer typically extends into the stratosphere. At mid- and high-southern latitudes, the base of aerosol layer $\alpha$ is deeper (from 1.7 to $>2$ bar), such that this aerosol layer overlaps in most cases with the top of layer $\beta$. We retrieve much lower aerosol optical depths in the upper troposphere for mid- and high-southern latitudes relative to low latitudes.
\item Fits in which we fix all aerosol parameters except for $P_{max, \alpha}$ and $\tau_\alpha$, but allow two parameters describing the CH$_4$ profile -- $RH_{ac}$ and $P_{dep}$ -- to vary, result in improved fits to the spectra from most locations. The predominant solution has $RH_{ac} <10\%$ and  $P_{dep} \approx$ 2.0--2.5 bar. At four locations, our retrievals favor a solution with a higher methane relative humidity but a deeper methane depletion. These results indicate that the CH$_4$ abundance in the upper troposphere is lower than our nominal CH$_4$ profile at all locations, with tentative evidence of meridional variations in the methane column above 2.5 bar. Our results may also suggest that our simple parameterization does not fully characterize the true shape of Neptune's CH$_4$ profile and variations within. 
\item  Thermal variations near the tropopause, reaching 7.5K near the south pole \citep{orton07} do not appear to have a dominant effect on the retrieved aerosol properties. 
\end{itemize}

\section{Discussion}\label{sec:disc}
\subsection{Comparison with previous work}

To our knowledge, this work represents the most thorough NIR characterization of Neptune's cloud-free regions. Resolving the aerosol structure and composition in these regions is important for characterizing Neptune's large-scale meridional circulation patterns and determining the processes responsible for aerosol production. These efforts also serve as context for characterizing localized storm activity that occurs across the planet. Comparison of this analysis with previous studies is complicated by differences in the information content of the data (due to different wavelength coverage and spatial extent) and differences in the modeling assumptions. Furthermore, temporal variations in Neptune's aerosol structure and atmospheric composition are likely:  {\it Voyager}, along with complementary Hubble and ground-based observations since, have shown that Neptune's visible and near-infrared (NIR) appearance exhibit both rapid, localized change \citep{limaye91} and decadal-scale brightness trends \citep{lockwood06, hammel07a, kark11b}. As a result, discrepancies between studies separated by several years may reflect real changes in atmospheric properties over time. 

Despite these complications, we note a number of key similarities and differences between our results and previous efforts. In contrast to \cite{kark11a}, who fit their 2003 HST-STIS  (300--1000 nm wavelength)  observations with a semi-infinite haze below 1.4 bar, our analysis strongly favors a compact and generally optically thick cloud layer at $\sim3$ bar. This result is consistent with earlier work \citep[e.g.,][] {hammel89, baines90, baines95a, baines95b, roe01b}, and more recently, with an analysis of Gemini NIFS  (1.477--1.803 $\mu$m wavelength) observations by \cite{irwin11, irwin14}. As noted in the literature \citep[e.g., by ][]{srom01c, roe01b}, an optically thick 3-bar cloud must be dark in the NIR to match observations: we retrieve a NIR single scattering albedo of $0.45^{+0.01}_{-0.01}$ for this layer assuming a Henyey-Greenstein phase function and using pyDISORT for the radiative transfer. This is intermediate between the values of 0.1 -- 0.2 found by \cite{srom01c, roe01b}  and the value of $\sim0.75$ assumed by \cite{irwin11, irwin14}. We note that the retrieval of $\omega_{0,\beta}$ is correlated with the asymmetry parameter of the scattering phase function (Fig. \ref{fig:disort_triangle}), which may partly explain the discrepancy: for example,  \cite{irwin11}  fix the asymmetry parameter $g$ of their Henyey-Greenstein phase function to be 0.7, whereas \cite{srom01c} assume an isotropic phase function. We retrieve a value of $0.50^{+0.02}_{-0.02}$ for the Henyey-Greenstein asymmetry parameter.  

Previous aerosol models of Neptune's NIR-dark regions typically include a second, optically thin tropospheric aerosol layer above the  $\sim3$ bar cloud deck, consistent with our findings \citep{pryor92,baines94,srom01b,gibbard02,irwin11,irwin14}. One exception to this is \cite{kark11a}; these authors find the troposphere to be clear above 1.4 bar. However, their visible-wavelength observations are most sensitive to altitudes below methane condensation. In the \cite{kark11b} analysis of HST/WFPC2 images from 1994 to 2008, the author allows for a possible upper tropospheric haze between the CH$_4$ condensation level and the tropopause, which is 10 times thinner than the main tropospheric aerosol layer -- similar to the model proposed by \cite{baines94}. Our best-fit model for the limb-darkening dataset from 2--12$^\circ$N likewise suggests the presence of a vertically extended upper tropospheric aerosol layer, although we find a somewhat higher bottom pressure of $P_{max,\alpha}=0.59^{+0.04}_{-0.03}$ (for the {\it 2L\_DISORT} retrieval), roughly a scale height above the CH$_4$ condensation level. Other models \citep[e.g.,][]{srom01b,gibbard02,irwin11} have presupposed that the upper tropospheric aerosol layer is vertically compact; our retrievals do not favor this possibility.  The reanalysis of the \cite{irwin11} data by \cite{irwin14} shows that a vertically extended upper haze is equally allowed by their data. Our analysis suggests two likely contributors to the variations in solutions across previous studies: first, the results of our methane profile tests show that the retrieved base pressure and scale height of the upper aerosol layer depend strongly on the adopted CH$_4$ profile. Secondly, many studies consider either a global average of Neptune's aerosol structure, or analyze only one specific latitude band. Our analysis demonstrates that this upper aerosol layer is spatially variable. Whereas \cite{irwin11,irwin14} use a single scattering albedo of 0.45 and asymmetry parameter of 0.7 for the upper aerosol layer, we retrieve values of $\omega_{0,\alpha}=0.91^{+0.06}_{-0.05}$ and $g_\alpha=0.24^{+0.02}_{-0.03}$. These numbers are similar to what one would expect from Mie theory for small particles ($r_p \approx 0.1 \mu$m). The single scattering albedo that we retrieve for the upper haze is also more consistent with what is observed for the particles in Neptune's discrete clouds \citep[e.g.,][]{irwin11}. 

Despite some discrepancies between the \cite{kark11a,kark11b} vertical aerosol profile and our results, these previous efforts provide a valuable point of comparison for our analysis of latitudinal variations in Neptune's discrete-cloud free atmosphere. For one, the  \cite{kark11a,kark11b}  analyses represent the most comprehensive previous investigations of Neptune's dark regions. Secondly, these studies, which utilize 300--1000 nm HST data, provide complementary constraints on Neptune's upper troposphere and stratosphere to our Hbb and Kbb NIR data. Whereas our observations are most sensitive to the aerosol structure between mbar pressures and the top of the  $\sim3$ bar cloud deck, the peak sensitivity of the \cite{kark11a} data is below the CH$_4$ condensation level. Likewise, \cite{kark11a} are most sensitive to Neptune's CH$_4$ distribution at deeper pressures.

\cite{kark11a} identify several prominent features in Neptune's reflectivities: the first is an increase in reflectivity from low latitudes to high southern latitudes, with the transition occurring around $-30^\circ$ latitude. A similar trend is clearly evident in our own dark-region spectra (Fig. \ref{fig:DRfits}). \cite{kark11a} explain this feature as the result of latitude variations in the methane mixing ratio around the 2-bar layer.  Our analysis shows a signature, at low significance, of a decreased CH$_4$ column above the 2.5-bar pressure level at mid- and high-southern latitudes, in qualitative agreement with the \cite{kark11a} findings. However, our results suggest that the dominant cause of the variation from low latitudes to mid- and high-southern latitudes is variation in Neptune's upper tropospheric haze: we find that the optical depth of this haze is higher at low latitudes ($-20$ to $+20^\circ$) than at more southern latitudes. The second feature observed by \cite{kark11a} is a narrow dark belt south of Neptune's equator, which these authors attribute to a dip in the optical depth of Neptune's tropospheric haze at this latitude. While our sensitivity to variations in the opacity of the deep tropospheric cloud deck is limited, we note that the three dark locations in our data for which we retrieve a bottom cloud opacity $\tau_\beta<1$ are at 20$^\circ$S, $16^\circ$S, and the equator -- these latitudes roughly correspond to the minima in tropospheric haze optical depth/bar found by \cite{kark11a}. Further, we tentatively retrieve a local maximum in the CH$_4$ column near 10$^\circ$S, which might correspond to the dark band identified by \cite{kark11a}. In \cite{kark11b}, the author finds evidence for either an increase in methane relative humidity or decrease in upper tropospheric aerosols in a band from 5--$20^\circ$N;  this may correspond to the peak in CH$_4$ column above 2.5 bar that we find at 10--20$^\circ$N. Finally, \cite{kark11a} observe a dark belt near 60$^\circ$S in Neptune's visible continuum, which they ascribe to a decrease in the single scattering albedo of tropospheric aerosols at long wavelengths. We do not identify this feature in our data. 

Taken together, our results and the \cite{kark11a,kark11b} analyses highlight the complementary nature of data from the optical and NIR wavelength regimes, and also the need to consider both the CH$_4$ and aerosol vertical profiles in order to understand meridional variations in Neptune's feature-free atmosphere. For CH$_4$, both studies are consistent in finding latitude variations in the methane abundance around the 2-bar layer, with the \cite{kark11a} data demonstrating this result at higher significance. Further, the \cite{kark11a} data are critical for showing that the depletion is shallow, reaching a maximum depletion depth of 3.3 bar at high southern latitudes, whereas our data are most robust at higher altitudes, showing that the CH$_4$ abundance near the tropopause, as parameterized by relative humidity, is lower than expected at all latitudes.  For the aerosols, resolving discrepancies between the \cite{kark11a,kark11b} studies and our own work, particularly regarding the compactness of the deep aerosol layer, is more challenging. Since the two wavelength regimes are optimally sensitive to different particle sizes and atmospheric depths, it is perhaps unsurprising that these discrepancies exist. Simultaneous observations and coincident analysis of data from both wavelength regimes would allow us to better elucidate Neptune's true tropospheric aerosol structure, and to rule out differences caused by temporal variability.

\subsection{Comparison with Uranus}
Studies of this nature, which involve extracting spectra far from discrete bright clouds across a range of viewing angles and latitudes, are far easier for the relatively cloud-free Uranus. Increasingly complex models of the aerosol structure -- some with 5 aerosol layers or more -- have been used to describe the available Uranus data. As discussed by \cite{dekleer15}, these models tend towards the same vertical profile of aerosols regardless of the parameterization. Furthermore, a two-layer model is sufficient to match the Uranus Hbb (and Kbb) data considered by \cite{dekleer15}. Therefore, we focus on comparisons with a two-cloud parameterization of the Uranus aerosol structure. 

Uranus is observed to have a deep, compact cloud in the 2--3 bar range \citep{kark09, irwin12b, tice13, dekleer15}, topped by a vertically diffuse haze with a base near 1 bar, increasing in concentration with altitude ($h_{frac} >1$). Both aerosol layers on Uranus decrease in optical depth from equator to pole, with the lower cloud exhibiting a sharp peak in optical depth near the equator and the upper haze trending more gradually \citep[e.g.,][]{irwin07, srom07a, kark09}. The latitudinal trends in aerosol structure are mirrored by an increase in methane depletion from no depletion near the equator to $P_{dep} >10 $ bar near the poles \citep[e.g.,][]{kark09,srom11u}. Considering the marked differences in (discrete) cloud activity between these two planets, the similarities between the background aerosol structures -- in particular, in terms of base pressures and aerosol scale heights -- are notable. The total optical depth of Neptune's upper aerosol layer is greater than observed on Uranus, consistent with Neptune's higher abundance of CH$_4$ and other hydrocarbons in the upper atmosphere. Like for Uranus, we observe a dramatic decrease in the optical depth of Neptune's upper haze from the equator towards the south pole. However, we do not observe strong signatures of decreased opacity in Neptune's bottom cloud like are observed for Uranus \citep{srom11u, dekleer15}. While CH$_4$ depletion towards the poles may be present for Neptune, we do not find evidence for the  relatively deep tropospheric methane depletion observed for Uranus. Further, the ``proportionally descended gas'' model of CH$_4$ depletion \citep{srom11u}, which matches the Uranus data well, may not be appropriate for Neptune: our data suggests that Neptune's atmosphere is relatively dry at altitudes above the CH$_4$ condensation pressure, and models including both variations in relative humidity (above the CH$_4$ condensation level) and CH$_4$ depletion (below the condensation level) are better at reproducing the observations than models in which only depletion is considered.

\subsection{Implications for global circulation pattern}

A primary goal of studies of Neptune's aerosol structure and composition is to understand the global circulation patterns. Enhanced cloud activity at midlatitudes, patterns in temperature, ortho/para H$_2$ ratios, and mid-infrared and radio brightness temperature measurements have led to a picture in which air rises above southern and northern midlatitudes, and dry air sinks over Neptune's equator and poles \citep{bezard91, conrath91, depater14}. \cite{depater14} suggest that this circulation extends all the way from the deep troposphere up into the stratosphere.  This circulation pattern contrasts dramatically with the circulation proposed for Uranus, in which air is expected to rise at the equator and sink at the poles, and may be structured in three vertically stacked layers rather than extending across many scale heights as a single circulation cell \citep{srom14}. 

Our retrieved latitude trends in Neptune's background aerosol structure and methane profile are hard to reconcile with the current model of its circulation. An upper tropospheric/stratospheric haze at low latitudes, which is either confined to deeper pressures or not present at all at mid-and high southern latitudes, does not seem to be a natural outcome of a two-cell-per-hemisphere circulation pattern that extends from the deep troposphere to upper stratosphere. It is also surprising, in the context of this presumed circulation pattern, that the deep aerosol layer on Neptune appears to be so consistent in its properties as a function of latitude, with the exception of a decrease in its optical depth at the highest observed northern latitudes. While we do tentatively observe a meridional trend in Neptune's 1--3 bar methane abundance, our data suggest that southern midlatitudes are as dry as Neptune's south pole. Other observations have hinted that this relatively simple circulation model does not adequately describe Neptune's circulation: for example, 1-cm maps of Neptune show an enhancement in the radio brightness temperature at mid-southern latitudes, indicative of relatively dry air in a region where moist, rising air is predicted \citep{depater14}. Additionally, clouds are often observed near the south pole and equator -- regions of apparent subsidence \citep[e.g.,][]{luszcz10a}.  This is similar to the situation for Uranus, in which models have difficulty reproducing all of the detected aerosol layers \citep{srom14}. In summary, a coherent dynamical picture for either Neptune or Uranus remains elusive; our observations reinforce this point. The vertical and latitudinal distribution of discrete, NIR-bright storms provides additional constraints on circulation models; in a followup paper, we will analyze these features in our OSIRIS data, and combine our findings for both cloud-free and cloudy regions to place improved constraints on Neptune's dynamics.

\section*{Acknowledgements}
The data presented were obtained at the W.M. Keck Observatory, which is operated as a scientific partnership among the California Institute of Technology, the University of California and the National Aeronautics and Space Administration. The Observatory was made possible by the generous financial support of the W.M. Keck Foundation. The authors wish to recognize and acknowledge the very significant cultural role and reverence that the summit of Mauna Kea has always had within the indigenous Hawaiian community. We are most fortunate to conduct observations from this mountain. This work was supported by NSF Grant AST-0908575 to UC Berkeley; SHLC was also supported by NASA Headquarters under the NASA Earth and Space Science Fellowship program - Grant NNX10AT17H; and a Kalbfleisch Postdoctoral Fellowship at the American Museum of Natural History.  SHLC would like to thank A. Veicht and D. Zurek for computing assistance and support. The authors would like to thank L. Fletcher for providing his temperature profiles, and two anonymous reviewers for their insightful and valuable comments, which improved this paper.

\footnotesize

\bibliographystyle{apalike}
\bibliography{osirisbib}

\newpage
\appendix
\section{Atmospheric retrieval code}\label{app:model}
In this appendix, we provide additional details of our atmospheric retrieval code, introduced in Section \ref{sec:modeling}.

\subsection{Inputs to the forward model}\label{app:forward}
As described in Section \ref{sec:forward}, we solve the radiative transfer equation for a  model atmosphere using either a two-stream approximation or a Python implementation of the discrete ordinate method for radiative transfer (pyDISORT).  The fundamental input parameters to the forward model include a thermal structure (temperature-pressure profile); the atmospheric composition as a function of depth; gas opacities as a function of temperature and pressure; and a description of the aerosols.  

\subsubsection{Thermal profile}\label{app:TP}
A range of thermal profiles have been derived for Neptune; many of these are summarized in Figs. 5 and 6 of \cite{luszcz13a}. For this analysis, we assume the \cite{lindal92} value of 71.5 K for the temperature at 1 bar. Below 1 bar, we extrapolate adiabatically to higher pressures assuming a dry adiabat with a deep methane abundance of 4\%; H$_2$O and H$_2$S enrichments equivalent to 50 times solar O and S; and a solar enrichment in N as NH$_3$ \citep{romani89,depater91,deboer96}. We find that adjustments to the deep abundance of CH$_4$ (Appendix \ref{app:composition}) have a negligible effect on the adiabatic thermal profile at the pressures considered here. Above the 1-bar level, we adopt the thermal profile described by \cite{fletcher14}: the tropospheric profile is consistent with the model of \cite{moses05}. This default thermal profile is shown in Fig. \ref{fig:TP_CH4}. 

Latitudinal temperature variations near the tropopause have been observed by  \cite{conrath98, orton07,fletcher14}. These authors find a clear temperature minimum around 40--60$^\circ$S and maxima around the equator and at the south pole. In order to investigate the possible influence of thermal variations of this type, the atmospheric model accepts offsets to the tropopause temperature, which are applied as a sinusoidal perturbation to the thermal profile, such that the profile matches the nominal profile at 0.01 and 1.0 bar, but is higher or lower at the tropopause by the specified {\it tropopause temperature offset} (Fig. \ref{fig:TP_CH4}). Such variations are considered in Sections \ref{sec:TPtests} and \ref{sec:DR_TP}.

In Neptune's stratosphere, disk-averaged thermal profiles differ from one another by as much as $\sim 20$K \citep[see][]{luszcz13a}, and stratospheric temperatures also may vary spatially.  For example, \cite{bezard91} measured a broad local maximum in stratospheric emission at low latitudes. This was not seen in the 2005 ground-based data of \cite{hammel07a}; rather, these authors detected evidence for a 4--5 K temperature enhancement near the south pole.  \cite{orton07} found evidence of a localized stratospheric warm region near 70$^\circ$ S, and \cite{hammel06} noted short-term variations in the disk-integrated ethane emission which could indicate discrete features. In contrast, \cite{fletcher14} found that, with the exception of the south pole, the stratosphere is latitudinally uniform, and temporal variability, if present, is limited to $<\pm$5 K.   

In this analysis, we adopt the mean stratospheric temperature structure from fits to Keck/LWS (2003) data \citep{fletcher14}. The model will accept an offset to the stratospheric temperatures, which is specified in terms of two values: a {\it stratosphere temperature offset} (K), and a {\it stratosphere offset pressure} (bar, set to $10^{-3}$ in this analysis). The full temperature offset is applied at pressures lower than the offset pressure. To ensure a smooth output thermal profile, smaller offsets, linear with log(pressure) are applied at pressures between the offset pressure and a pressure of 100 mbar.  See Fig. \ref{fig:TP_CH4} for examples. The effects of varying the stratospheric temperature on our fits are briefly investigated in Section \ref{sec:TPtests}.

\subsubsection{Composition}\label{app:composition}
Neptune's upper atmosphere is composed primarily of H$_2$ and He.  We maintain a He/H$_2$ ratio of 0.15/0.847 and an N$_2$/H$_2$ ratio of 0.003/0.847 by number throughout our model atmosphere. These values are consistent with \cite{conrath93} reanalysis of {\it Voyager} occultation measurements \citep{conrath91}, which placed constraints on the atmospheric mean molecular weight, and the \cite{burgdorf03} analysis of spectra from the Infrared Space Observatory Long-Wavelength Spectrometer (ISO-LWS). We assume `intermediate' hydrogen in the troposphere, in which the {\it ortho} and {\it para} states of hydrogen are in equilibrium at the local temperature, but the specific heat is near that of `normal' hydrogen. This situation is described in \cite{massie82}.  Fast vertical mixing from the interior {\it could} bring the {\it ortho/para} ratio closer to the 3:1 ratio expected for normal hydrogen, but previous results indicate that that the {\it ortho/para} ratio is in fact close to that of equilibrium hydrogen \citep{orton86,baines95a,burgdorf03}. We do not consider the effects of varying the {\it ortho/para} ratio. However, previous studies have found evidence for spatial variations in the {\it ortho/para} ratio:  \cite{conrath98}, in their analysis of {\it Voyager 2}/IRIS data, found sub equilibrium para ratios from the equator down to 50$^\circ$S, and higher para fractions relative to equilibrium from 50-90$^\circ$S and in the northern hemisphere. \cite{fletcher14}  performed a reanalysis of the {\it Voyager} data using updated absorption spectra; they found a more symmetric para-H$_2$ distribution, with super-equilibrium para ratios from 20$^\circ$S to 20$^\circ$N and near the poles; and sub-equilibrium para ratios at midlatitudes.

Methane absorption has a dominant effect on Neptune's NIR spectrum. A number of estimates have been made of Neptune's CH$_4$ abundance (mole fraction) at various atmospheric depths: based on {\it Voyager} occultation measurements, \cite{lindal92} determine a methane mole fraction of $0.02 \pm 0.02$ below the methane condensation level at the occultation ingress latitude of $\sim60^\circ$N. The \cite{baines95a}  analysis of hydrogen quadrupole and methane lines yields a disk-averaged tropospheric CH$_4$ mole fraction of $0.022^{+0.005}_{-0.006}$, in good agreement with the \cite{lindal92} result. A CH$_4$ mole fraction of 0.022 has since been adopted by many studies of Neptune's upper atmosphere \citep[e.g.][]{roe01a,srom01,gibbard03,luszcz10, irwin11}; this value implies an enrichment factor of $\sim50$ over the protosolar C/H ratio \citep{asplund09}. More recently, \cite{kark11a} find a significantly higher best-fit value of $0.04 \pm 0.01$ for the CH$_4$ mole fraction in the troposphere at low latitudes; they also find that at high southern latitudes, the methane abundance is depressed between 1.2 and 3.3 bar. A similar shallow high-latitude depletion has been observed on Uranus \citep{kark09,srom11,srom14,dekleer15}.

At the cooler temperatures of the upper troposphere, methane will condense out of the atmosphere; thermochemical models predict that the cloud base will be at the 1--2 bar level. This is consistent with the \cite{lindal92} interpretation of the {\it Voyager} occultation data, which placed the cloud base at 1.9 bar for their nominal model. The relative humidity above the methane condensation layer in the \cite{lindal92} nominal model is 20\%; \cite{kark11a} found that humidities of 40-100\% are required for good fits to their data. 

The stratospheric CH$_4$ abundance should be set by the saturation value at the temperature minimum of the tropopause-- $\sim2\times10^{-4}$ for our thermochemical equilibrium calculations. However, observations indicate the stratospheric CH$_4$ mole fraction is much higher than this: \cite{orton87,orton90,orton92,orton05} found values of 0.75--1.5$\times 10^{-3}$ from analyses of Neptune's thermal spectrum; \cite{yelle93} found a similar range (0.6--5.0 $\times 10^{-3}$ in the lower stratosphere). \cite{baines94} favored a somewhat lower mole fraction of 3.5$\times 10^{-4}$, whereas the more recent studies of \cite{fletcher10} and \cite{lellouch10} derived abundances of $(9 \pm 3) \times 10^{-4}$ and $(1.5 \pm 0.2) \times 10^{-3}$, respectively, for the middle of the stratosphere. 

Like in the troposphere, the stratospheric methane abundance may vary with latitude: \cite{conrath98,orton07,fletcher14} showed that the tropopause temperature varies, with warmer temperatures near the south pole; this could allow methane to leak into the stratosphere at high southern latitudes, and then be redistributed across the planet \citep{orton07}. However, \cite{greathouse11} suggest that stratospheric CH$_4$ abundances may actually peak near the equator; consistent with the distribution found by \cite{kark11a} for the troposphere.

We parameterize the vertical methane profile in the following way: we define the methane mole fraction below the condensation level, {\it mCH$_4$$_{t}$}. At the methane condensation pressure, the methane mole fraction drops, and we adopt a constant relative humidity $RH_{ac}$. We allow the methane abundance to increase again above the tropopause, holding the relative humidity constant, until the stratospheric methane mole fraction {\it mCH$_4$$_{s}$}, is reached.  Our initial CH$_4$ profile assumes  {\it mCH$_4$$_{t}$}=0.04,  $RH_{ac}$ = 1.0, and {\it mCH$_4$$_{s}$}=0.00035. These choices are reevaluated in Section \ref{sec:CH4tests}. 

The model also includes an additional parameter, $P_{dep}$, the pressure depth down to which the methane abundance is smoothly depleted. $P_{dep}$ is defined using the proportionally descended gas model described in \cite{srom11, srom14} and \cite{dekleer15}, with a shape parameter $vx=2$. We note that this depletion model was developed for Uranus and may not accurately represent the physical situation on Neptune. Methane depletion is considered in Section \ref{sec:depletion}. Examples of the parameterized CH$_4$ profile and CH$_4$ depletion are shown in Fig. \ref{fig:TP_CH4}.

\subsubsection{Gas opacity}\label{app:opacity}
The gas opacity in the NIR is dominated by H$_2$ collision-induced absorption (CIA) and CH$_4$ opacity. For CIA,  we use the coefficients for hydrogen, helium  and methane from \cite{borysow85,borysow88,borysow91,borysow92,borysow93}. For CH$_4$ we use tabulated k-coefficients based on the improved line list and recommendations for outer planet NIR spectra described in \cite{srom12}: shortwards of 1.80 $\mu$m we adopt the line shape between that of \cite{hartmann02} and \cite{debergh12} as described by these authors. Between 1.80 and  2.08 $\mu$m, we use the  \cite{kark10} k-coefficients, and at longer wavelengths we adopt  the \cite{srom12} line list with the \cite{hartmann02} line shape. All tabulated coefficients have a 10 cm$^{-1}$ resolution and 5 cm$^{-1}$ spacing.

\subsubsection{Aerosols}\label{app:aerosols}
\begin{figure}[h!]
\begin{center}$
\begin{array}{lcr}
\includegraphics[width=0.28\textwidth]{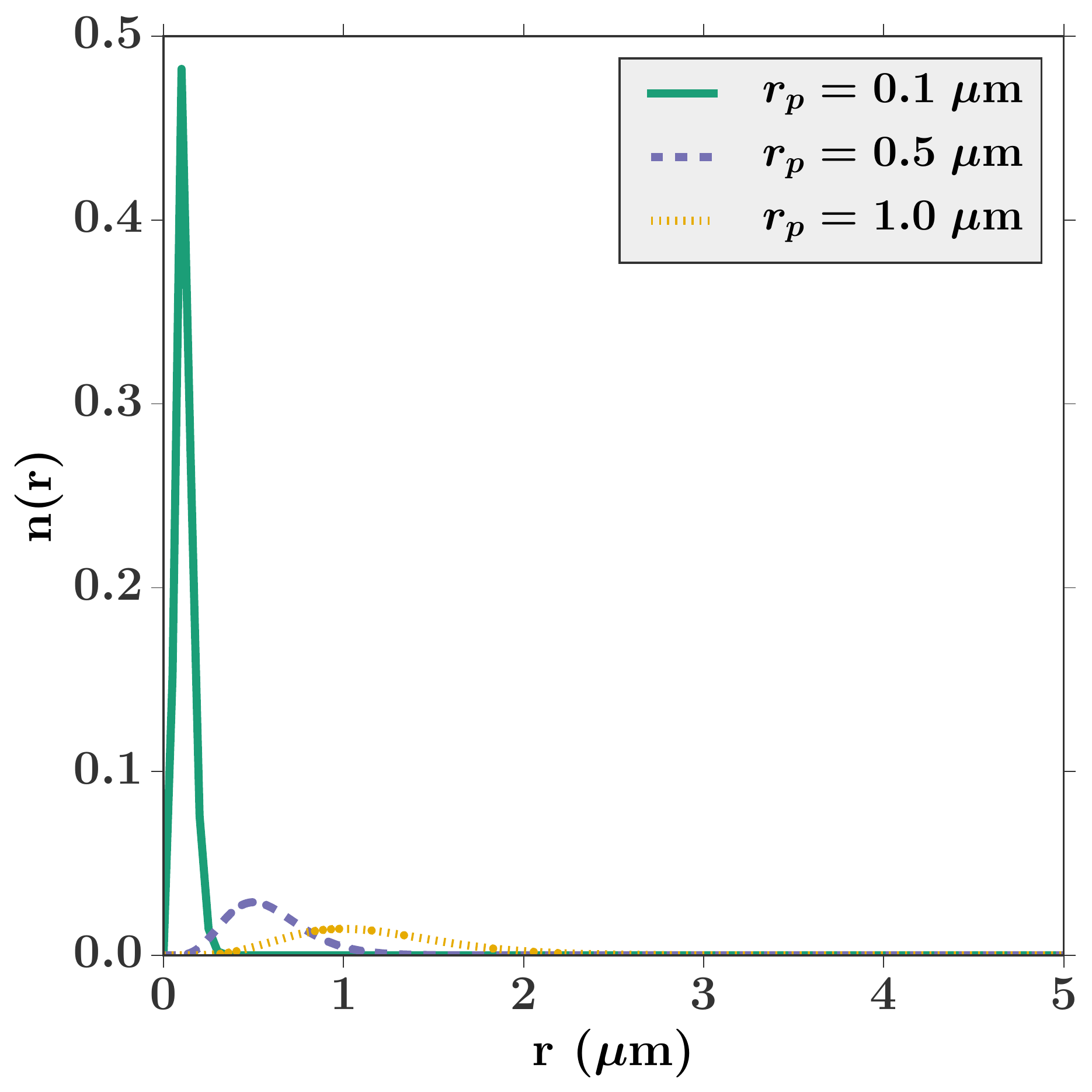} \includegraphics[width=0.28\textwidth]{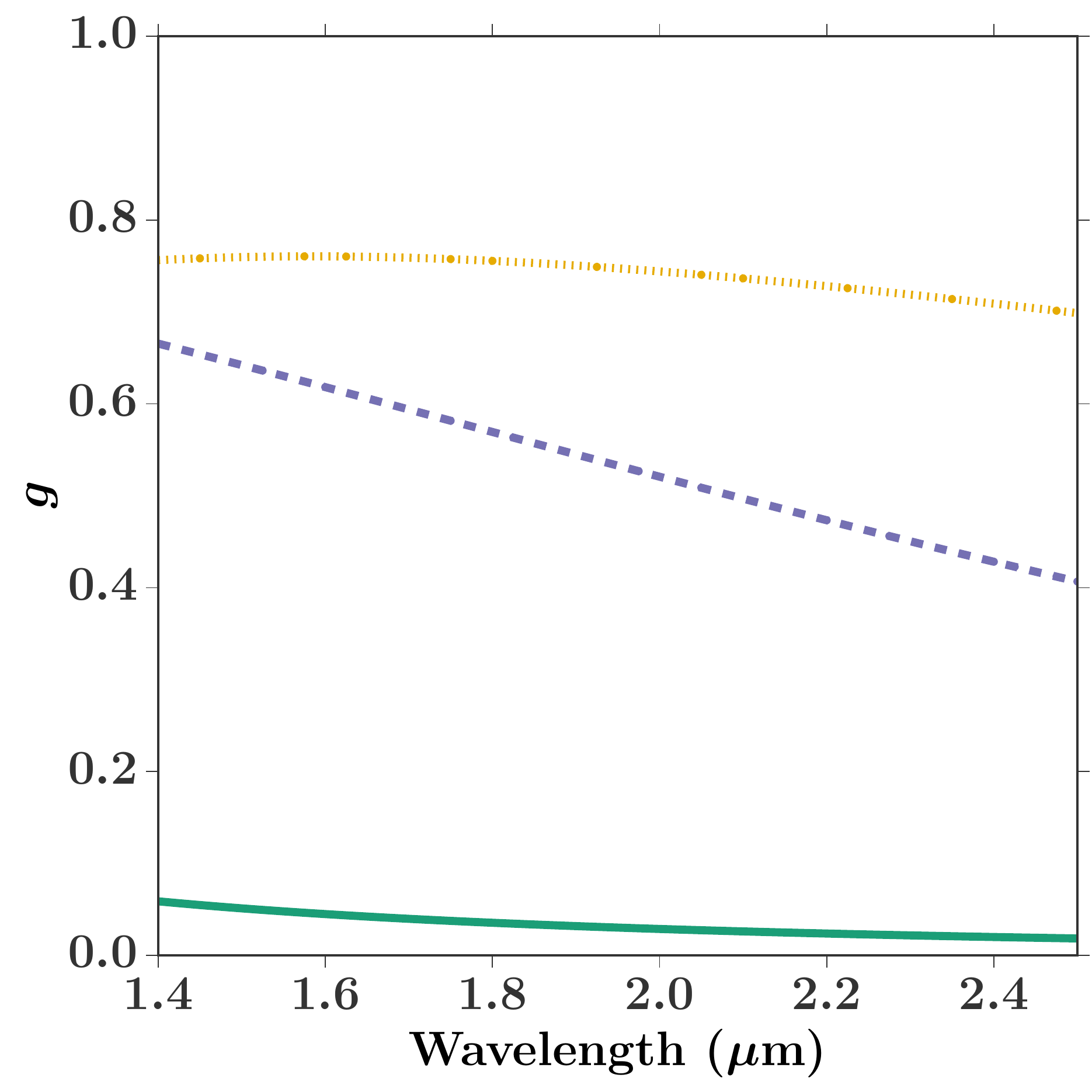} \includegraphics[width=0.28\textwidth]{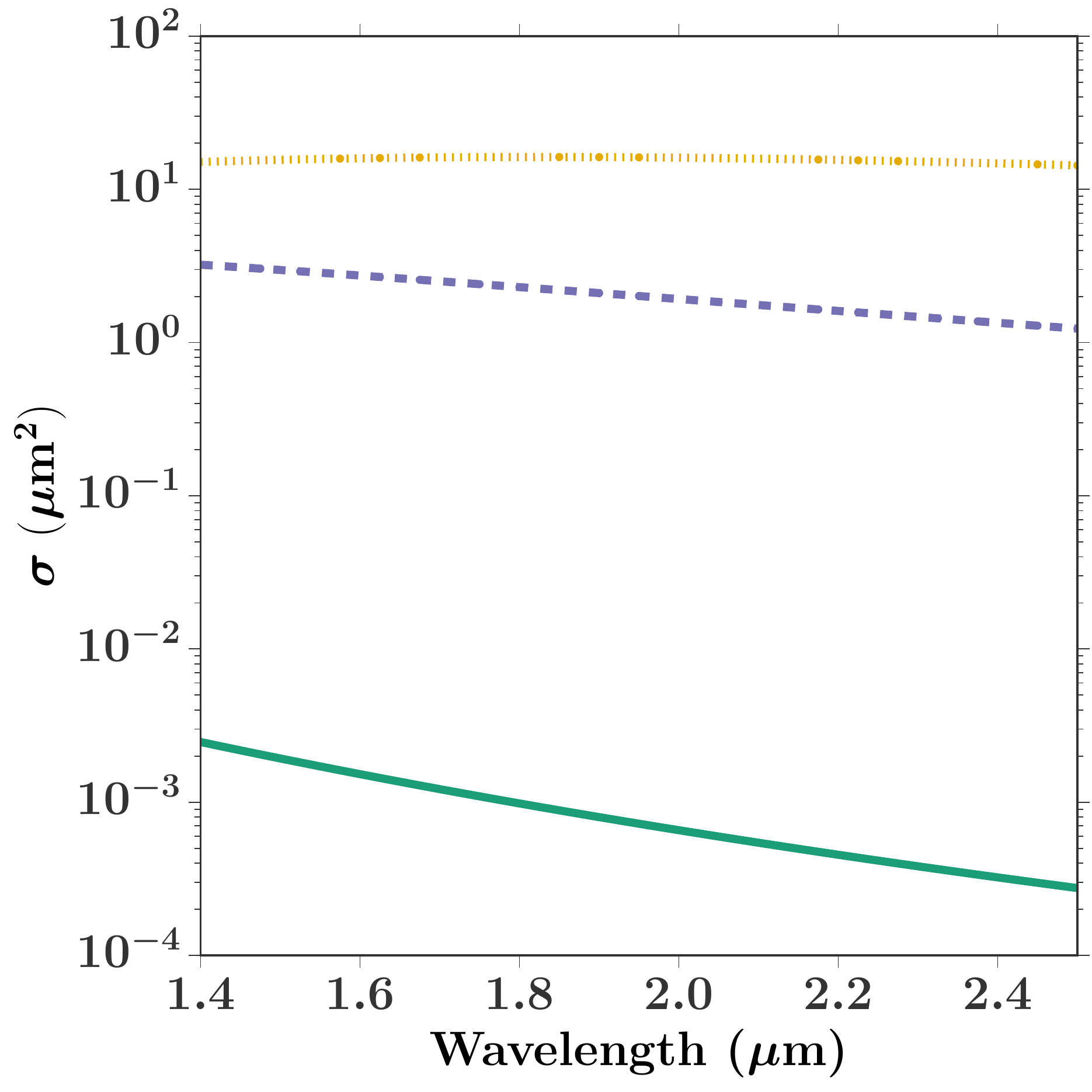} 
\end{array}$
\end{center}
\caption[Properties of particle ensembles]{\label{fig:particles} \small Properties of the particle ensembles considered in this analysis. Left: particle size distributions given by Eq. \ref{eqn:nr} for the specified peak particle size, $r_p$. Center: asymmetry factor as a function of wavelength for these particle size distributions, as determined by Mie theory. Right: mean extinction cross section as a function of wavelength for the three distributions, as determined by Mie theory.}
\end{figure}

A wide range of haze and cloud structures have been used in relation to Neptune. In our model, we aim to parameterize the aerosols in a simple, straightforward way that is nonetheless capable of representing realistic aerosol structures and reproducing previous models. Our model accepts any number of aerosol layers, which can be placed at any depth within the model atmosphere and can overlap one another. The parameters that define each aerosol layer are summarized in Table \ref{table:model}. They are: 

\begin{itemize}
\item{\it P$_{max}$}: Maximum (bottom) pressure of the aerosol layer. 
\item{\it P$_{min}$}: Minimum (top) pressure of the aerosol layer. Unless indicated, {\it P$_{min}$} is set to the top of the model atmosphere at $10^{-5}$ bar.
\item{\it h$_{frac}$}:  Fractional scale height of the aerosol layer. The number density of haze particles therefore varies across a layer of thickness $z$ (cm) as:
\begin{eqnarray}
N=N_0 *\exp{\left(\frac{z}{H*h_{frac}}\right)}
\end{eqnarray}
where $N_0$ is the number density at the bottom of the atmospheric layer being considered,  and $H$ is the pressure scale height of the gas. A very thin aerosol layer (``cloud") might have $h_{frac}<0.1$, while a ``haze" might have  $h_{frac}\sim 1$. 
\item{$\tau$}: Total optical depth for the haze layer at 1.6 $\mu$m.

\item{$\sigma$}: mean extinction cross section of the aerosols. For this analysis, we consider ensembles of particles with distributions of the form:
\begin{eqnarray} \label{eqn:nr}
n(r) \propto r^6 exp{\left(-6* \frac{r}{r_{p}}\right)} 
\end{eqnarray}
where $n(r)$ is the number density of particles of radius $r$ and $r_p$ is the peak in the particle distribution \citep{hansen70}\footnote{This particle size distribution can be written in terms of an equivalent gamma distribution. For $r_p=1.0$, the equivalent gamma distribution has a mean of $7/6$ and a variance of $7/36$; for $r_p=0.1$, the equivalent gamma distribution has a mean of $0.7/6$ and variance of $0.07/36$.}. A range of values have been adopted for the characteristic size of Neptune's aerosols in the past; these values generally range from 0.1 to 2.5 $\mu$m \citep[e.g.][]{pryor92,baines94, moses95,irwin14}. We test three characteristic particle sizes in this study: $r_p = 0.1, 0.5$ and 1.0 $\mu$m (Fig. \ref{fig:particles}). The mean extinction cross section is calculated from the ensemble of particles using Mie theory; since the compositions of the condensates in Neptune's atmosphere are not well known, and the refractive indices of the proposed condensates are also not well known, we use a `typical' hydrocarbon refractive index of $1.4+0i$ for all particles. 

\item{\it g}: the asymmetry factor, which describes the shape of the phase function. Values of -1 (strongly backscattering) to 1 (strongly forward scattering) are accepted. In this study, $g$ is assumed to be wavelength-independent, and is generally a free parameter in the retrievals (or fixed according to the results of earlier retrievals). Figure \ref{fig:particles} illustrates the wavelength-dependent value of $g$ derived by Mie theory for our three test particle size distributions. In cases where the two stream approximation is used, {\it g} fixes the fraction of forward and backscattered light according to the quadrature method as described in \cite{meador80}. For the pyDISORT algorithm, we interpret {\it g} as the asymmetry factor defining the Henyey-Greenstein phase function; the first four moments of the Legendre polynomial expansion are used in the radiative transfer calculation. 
\item{$\omega_{0}$}: single scattering albedo. Values between 0 (all extinction due to absorption) and 1 (all extinction due to scattering) are permitted. Mie theory finds $\omega_{0}=1.0$ for the particle ensembles considered here; however, this value is inconsistent with limb darkening data (see \cite{irwin11} and Section \ref{sec:limb}). Therefore, we allow $\omega_{0}$ to be a free parameter (constant with wavelength) in the retrievals.  
\end{itemize}

Once the properties of each aerosol layer are specified, the total optical depth and the weighted mean asymmetry parameter and asymmetry factor are calculated for each depth in the atmospheric model. 

\subsection{Retrieval}\label{app:retrieval}
To estimate model parameters and uncertainties, we pair our forward model with an affine-invariant Markov chain Monte Carlo (MCMC) ensemble sampler called {\it emcee} \citep{foreman13}.  {\it Emcee} generates a sampling approximation to the posterior probability function $p(\boldsymbol{x}|\boldsymbol{d})$, where $\boldsymbol{x}$ is the vector of model parameters and $\boldsymbol{d}$ are the data, by constructing an ensemble of chain(s) sampled from the desired probability distribution by random walk. As noted in Section \ref{sec:retrieval}, this algorithm possesses several advantages over the simpler and more common Metropolis-Hastings MCMC sampling. The method involves stepping an ensemble of K ``walkers" according to a proposal distribution that is based on the current position of the other K-1 walkers; a more complete description of the algorithm can be found in \cite{foreman13} and references within; here we explain how {\it emcee} is employed in our retrievals. A similar implementation is described in \cite{dekleer15}.

\subsubsection{Implementation of the MCMC algorithm}

Implementation of {\it emcee} requires specification of the posterior probability function, which is, up to a constant:
\begin{eqnarray}
p(\boldsymbol{x}|\boldsymbol{d}) \propto p(\boldsymbol{x}) p(\boldsymbol{d}|\boldsymbol{x})
\end{eqnarray}
where $p(\boldsymbol{x})$ is the prior probability distribution function and $p(\boldsymbol{d}|\boldsymbol{x})$ is the probability of drawing the observed values given the model parameters, or {\it likelihood function}. Any preexisting knowledge about the model parameters is encoded in  $p(\boldsymbol{x})$; for the parameters in this analysis, we use either uniform or log-uniform priors, within an acceptable parameter range. For example, the prior on $\omega_0$ is uniform between 0 and 1; the prior on $\ln{\tau}$ is defined to be uniform between $-7$ and 5. 

We assume a gaussian likelihood function:

\begin{eqnarray}
p(\boldsymbol{d}|\boldsymbol{x})= \prod_i {\left(\frac{1}{2\pi\sigma_i^2}\right)^{1/2} \exp{\left[-\frac{\left(d_i-d_{\boldsymbol{x},i}\right)^2}{2\sigma_i^2}\right]}}
\end{eqnarray}

where $d_i$ is the reflectivity of the $i$th data point, $d_{\boldsymbol{x},i}$ is the value returned by the model for point $i$ given parameters $\boldsymbol{x}$, and $\sigma_i$ is the uncertainty for point $i$. In terms of log-likelihood (which is the quantity accepted by {\it emcee}), this becomes:
\begin{eqnarray}
\ln{p(\boldsymbol{d}|\boldsymbol{x})}= -\frac{1}{2} \sum_i{\left[\frac{\left(d_i-d_{\boldsymbol{x},i}\right)^2}{\sigma_i^2} + \ln\left(2\pi\sigma_i^2\right)\right]}
\end{eqnarray}

As noted in Section \ref{sec:retrieval}, we define the uncertainty as:
\begin{eqnarray}
\sigma^2=\sigma_n^2+\sigma_p^2+\sigma_t^2
\end{eqnarray}
where $\sigma_n$ is the random noise component,  $\sigma_p$ is the relative photometry uncertainty, and  $\sigma_t$ is a model ``tolerance", representing any unknown uncertainties that prevent the model from matching the data. The primary effect of including $\sigma_t$ is to cause the parameter uncertainties to be more realistic. 

Each atmospheric retrieval proceeds in the following way: we select a dataset of interest $\boldsymbol{d}$. We then set up the model atmosphere as described in Section \ref{sec:forward}, and identify the set of parameters $\boldsymbol{x}$ that may vary in the model, specifying whether each parameter will be considered in linear or log space. We also specify an allowed range and initial guess value for each parameter. The starting position of each of $>100$ walkers (30--75 walkers per free parameter) in the {\it emcee} ensemble is initialized to have a small random offset from that initial guess. We generally run {\it emcee} for 500--750 steps, and then check whether the ``burn-in" phase -- in which the walkers lose memory of the initial conditions and begin to reasonably sample the posterior probability distribution -- is complete, 2) the chains have run for at least 10 autocorrelation times after burn-in, and 3) acceptance fraction is reasonably high (0.2--0.5) for all walkers. If these conditions are not yet met- for example, if the walkers still appear to be in the transient ``burn-in" phase, we continue to run the chains. In some cases, a low acceptance fraction is observed for one or more walkers, which have gotten stuck in a local maximum in the posterior probability. In these cases, we either continue to run {\it emcee} until all walkers appear to have found the global maximum, or we initialize a new ensemble of walkers and begin a new run.

\begin{figure}[t]
\begin{center}$
\includegraphics[width=1.0\textwidth]{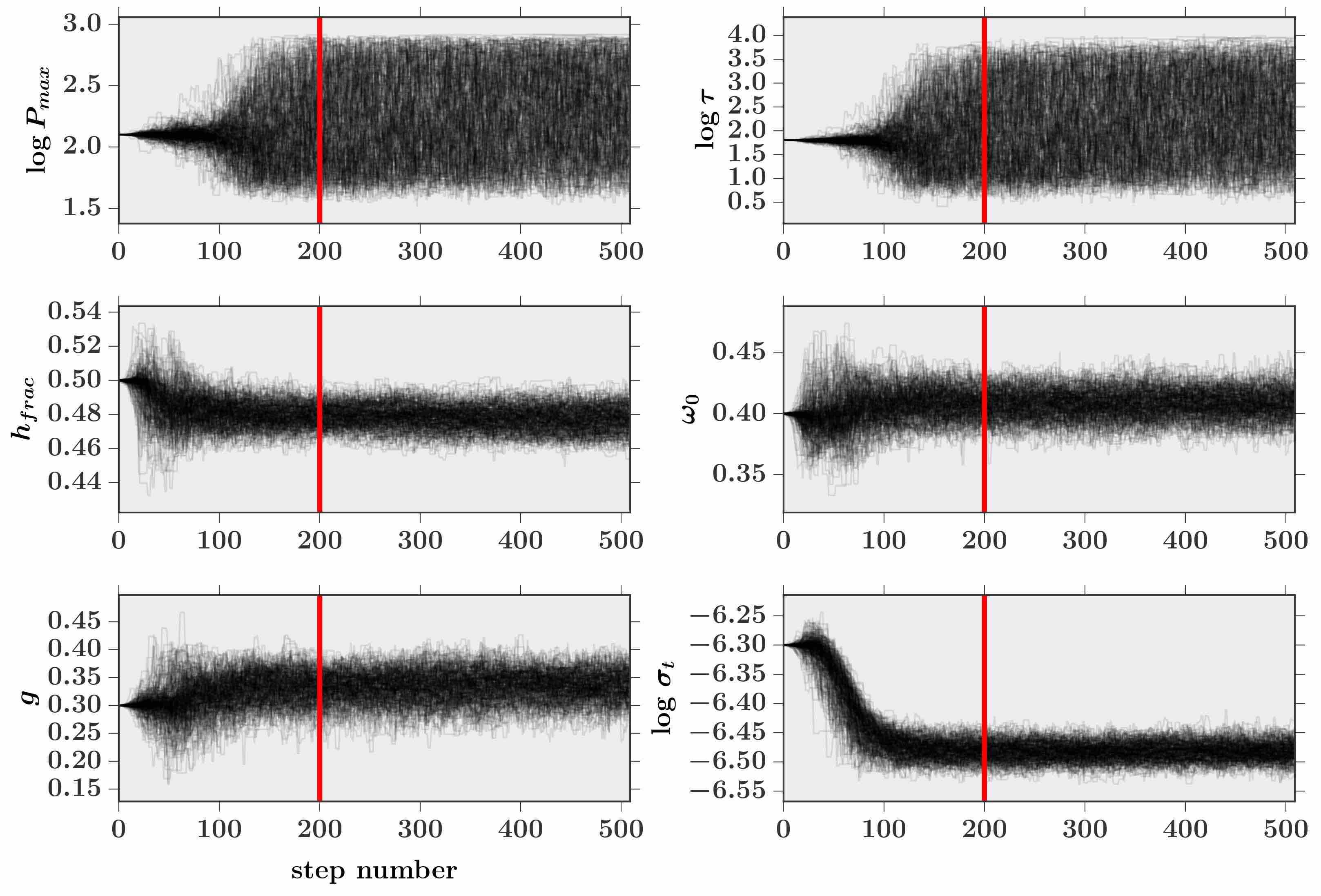}
$
\end{center}
\caption[Burnin]{\label{fig:burnin} \small Demonstration of the MCMC retrieval method for parameter estimation. Plots show the values taken on by each of 256 (in this case) walkers at each step of an example simulation ({\it 1L\_nom}, see Section \ref{sec:1L}.) The red line indicates the step at which we begin to use the points to generate the posterior probability distribution of the parameters. Points before this line (the `burn-in' phase) are discarded. }
\end{figure}

Once a run is complete, we produce a final set of samples by discarding the initial steps (burn-in) and combining the remaining samples from all walkers (see Fig. \ref{fig:burnin} for an example). Our results are represented by plots of the one- and two- dimensional projections of the posterior probability distributions (e.g. Fig. \ref{fig:1L_triangle}), and by the 16th, 50th and 84th percentiles from the marginalized parameter distributions.

\subsubsection{Model comparison}
While MCMC algorithms are a powerful tool for estimating model parameters and uncertainties, they do not provide a direct metric for evaluating the efficacy of a model, or for comparing different models (different sets of assumptions and free parameters). Large values of $\sigma_t$ imply the differences between the data and model are not well captured by the known uncertainties ($\sigma_n$ and $\sigma_p$), and therefore serve as one indicator that a model is not a good match to the data. A more robust way to evaluate the relative success of a model fit is to calculate the Deviance Information Criterion ($DIC$) \citep{spiegelhalter02}. The $DIC$ includes a term for goodness-of-fit with a penalty term for model complexity. The $DIC$ is based on the deviance, $D(\boldsymbol{x})$, defined as the difference in log-likelihoods between the model fit and a perfect fit to the data. The deviance is given by:
\begin{eqnarray}
D(\boldsymbol{x})= -2\ln{\left\{p(\boldsymbol{d}|\boldsymbol{x})\right\}} +2\ln{\left\{f(\boldsymbol{d})\right\}}
\end{eqnarray}
where $f(\boldsymbol{d})$ is a function of the data alone, and therefore cancels when considering the difference in deviance between models for the same dataset. The log-likelihood $\ln{p(\boldsymbol{d}|\boldsymbol{x})}$ is defined in Eq. 5.  $DIC$ is then defined as 
\begin{eqnarray}
DIC= \overline{D(\boldsymbol{x}})+p_D
\end{eqnarray}

$\overline{D(\boldsymbol{x})}$ is the mean deviance, which is easily calculated from the log-likelihoods found by {\it emcee};  $p_D$ is an effective number of parameters.  \cite{spiegelhalter02} defined $p_D$ in the following way:

\begin{eqnarray}
p_D=\overline{D(\boldsymbol{x}})-D(\overline{\boldsymbol{x}})
\end{eqnarray}
where $\overline{\boldsymbol{x}}$ is the posterior mean of the parameters (the 50th percentiles from the marginalized parameter distributions). We find that, for cases where the posterior probability distribution is not multivariate normal, this formulation of $p_D$ does not produce an accurate estimate of the degrees of freedom of a model (and is often negative). Therefore, we adopt the alternative formulation of $p_D$ following \cite{gelman04}:

\begin{eqnarray}
p_D=\frac{1}{2}{\mathrm{Var}}(D(\boldsymbol{x}))
\end{eqnarray}
which we find to be substantially more robust.

The goodness of fit, as estimated by $ \overline{D(\boldsymbol{x}})$, will improve for better models, but also with the addition of more parameters; $p_D$ compensates for this by favoring models with fewer parameters. Therefore, a smaller value of the $DIC$ (which may be a positive or negative number) indicates a better model.  We consider a difference in the $DIC$ ($\Delta DIC$) of 10 or more to indicate a preference for the model with the lower $DIC$. 

We note that the $DIC$ for a single model cannot be interpreted in terms of the success or failure of that model in reproducing the data; the value of the $DIC$ is in comparing two different models. Furthermore, the models being compared must be for the same dataset: values of the $DIC$ cannot be compared for models of different datasets (in this case, different spectra or sets of spectra) since $f(\boldsymbol{d})$ in the equation for  $D(\boldsymbol{x})$ varies between datasets, and is not known.

\section{Comparison and analysis with pyDISORT}\label{app:disort}

 Ideally, the errors introduced by our radiative transfer method will be less than, or at worst comparable to, other sources of uncertainty in the retrievals, including errors in the gas opacity coefficients, uncertainties and approximations in the model atmosphere properties, and data uncertainties. Using this same software and comparable data for Uranus, \cite{dekleer15} found that the differences between two stream and the more accurate pyDISORT algorithm varied with wavelength, from $<5$\% to $\sim$10\%. Here we perform a similar comparison of the two stream and pyDISORT models using the best-fit {\it 2L\_hazeA} model parameters.

Since two stream and pyDISORT treat the asymmetry factor $g$ differently, we do not expect $g_\alpha$ and $g_\beta$ to be the same for the two algorithms; therefore, we perform a simple, unweighted $\chi^2$ fit to calculate $g_{\alpha,D}$ and $g_{\beta,D}$: the pyDISORT asymmetry factors that provide the best match between the $\mu=0.8$ models from the two algorithms. We find that  $g_{\alpha,D}= 0.5$ and $g_{\beta,D}=-0.3$ (vs. $g_{\alpha}=0.4$ and $g_{\beta}=-0.6$ for two stream). The absolute difference between the two models, $|\delta m|\equiv | I/F_{DISORT}- I/F_{twostream} |$, is shown in Fig. \ref{fig:error}. It can be seen in this figure that the errors introduced by the two stream approximation are wavelength dependent, and comparable in magnitude to the other sources of known error in our retrievals, $\sigma_n$ and $\sigma_p$. In Hbb, where the model differences are greatest, $|\overline{\delta m}|=1.7e-4$, compared to $\overline{\sigma_n}=1.5e-4$, for the  $\mu=0.8$ spectrum. 
 
 \begin{figure}[h]
\begin{center}$
\begin{array}{cccccc}
\includegraphics[width=0.45\textwidth]{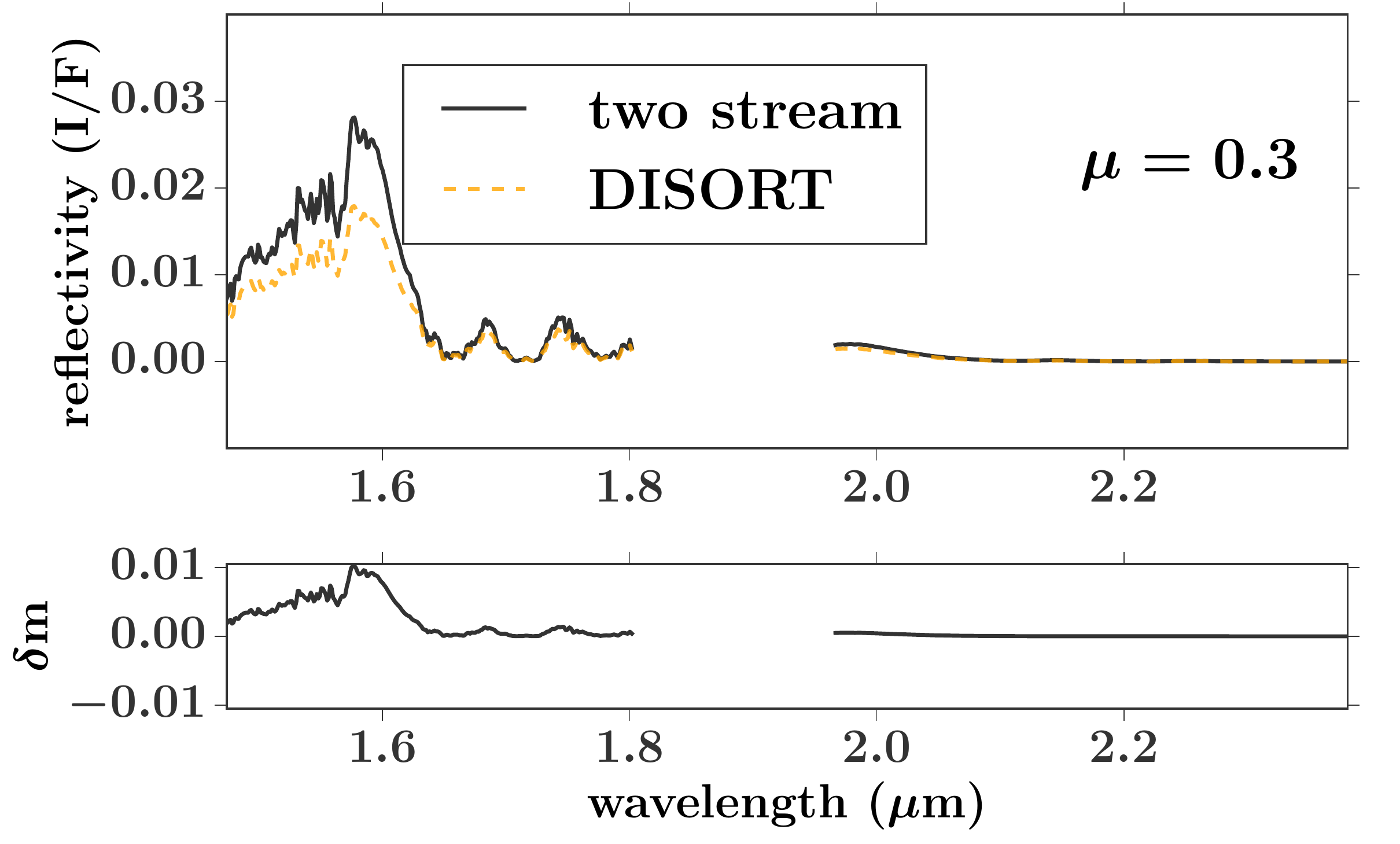} \includegraphics[width=0.45\textwidth]{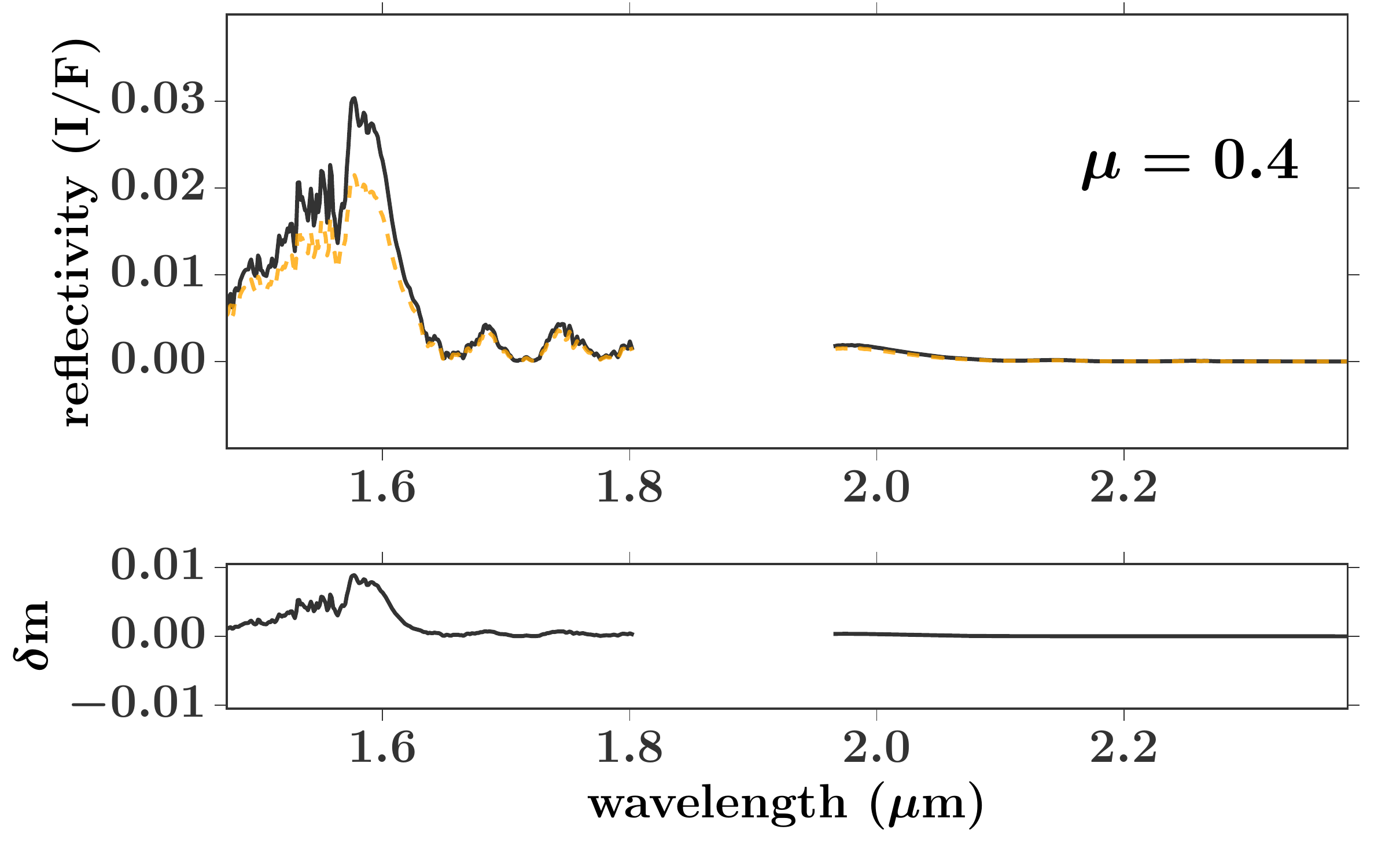}\\
\includegraphics[width=0.45\textwidth]{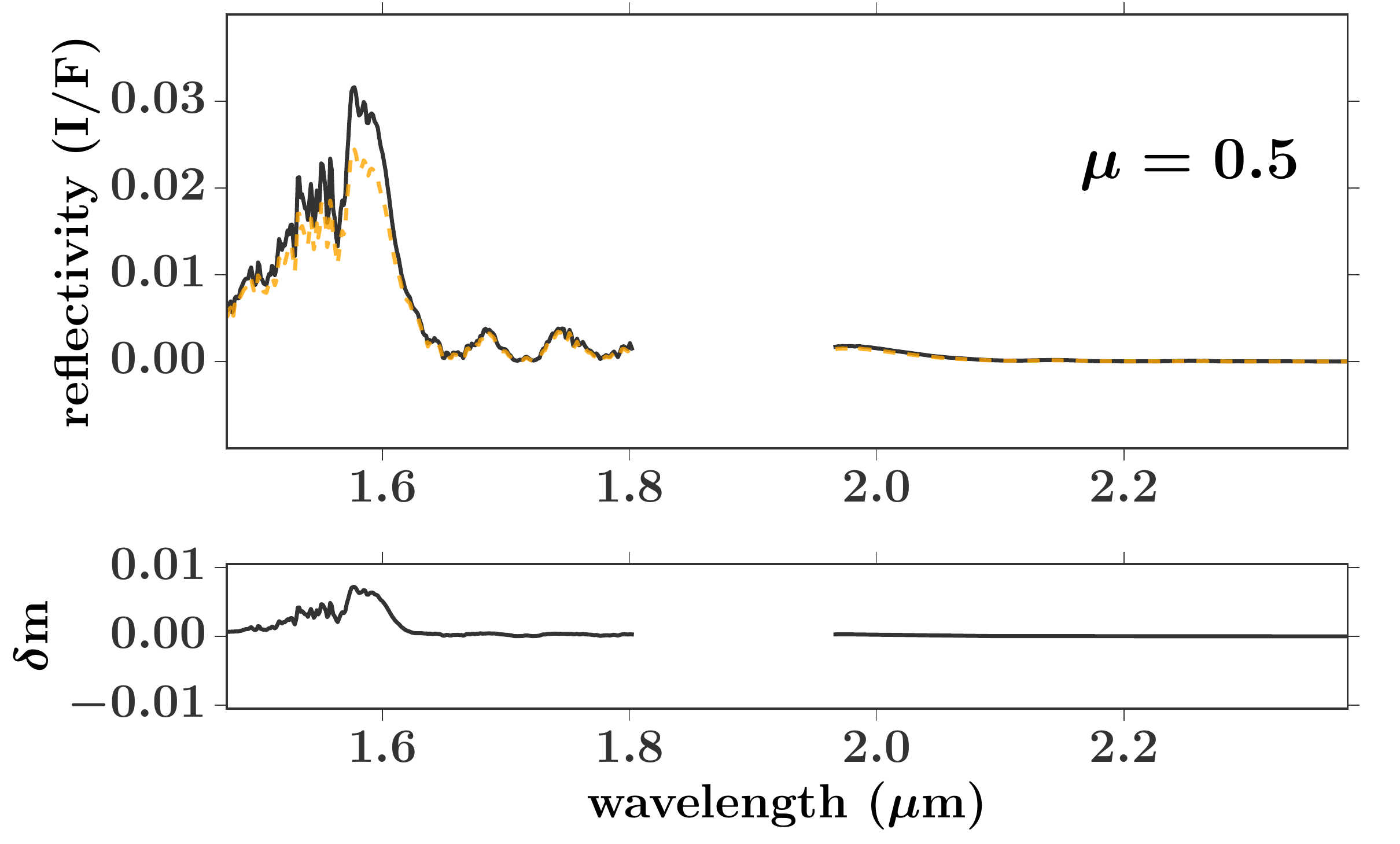}\includegraphics[width=0.45\textwidth]{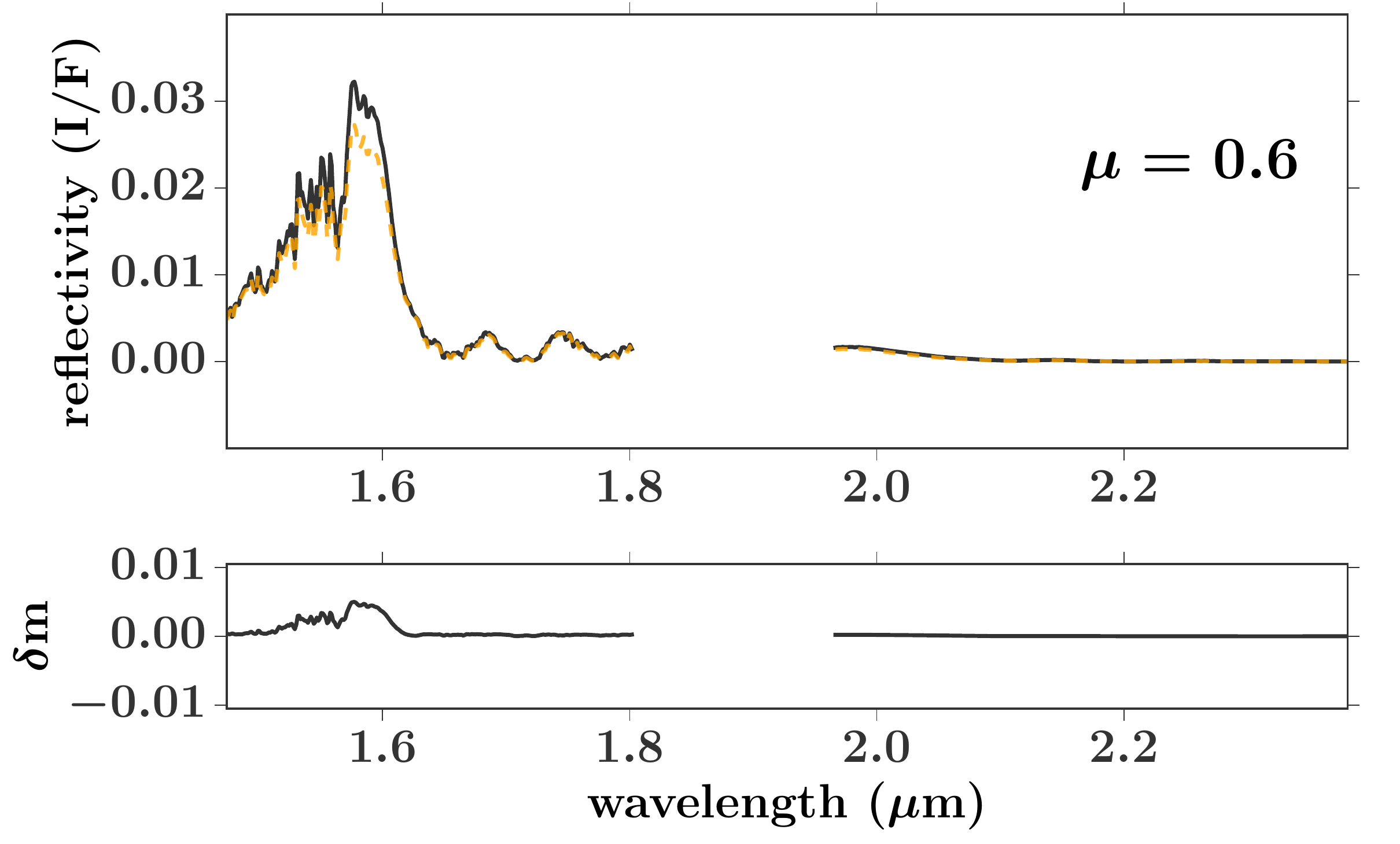}\\
\includegraphics[width=0.45\textwidth]{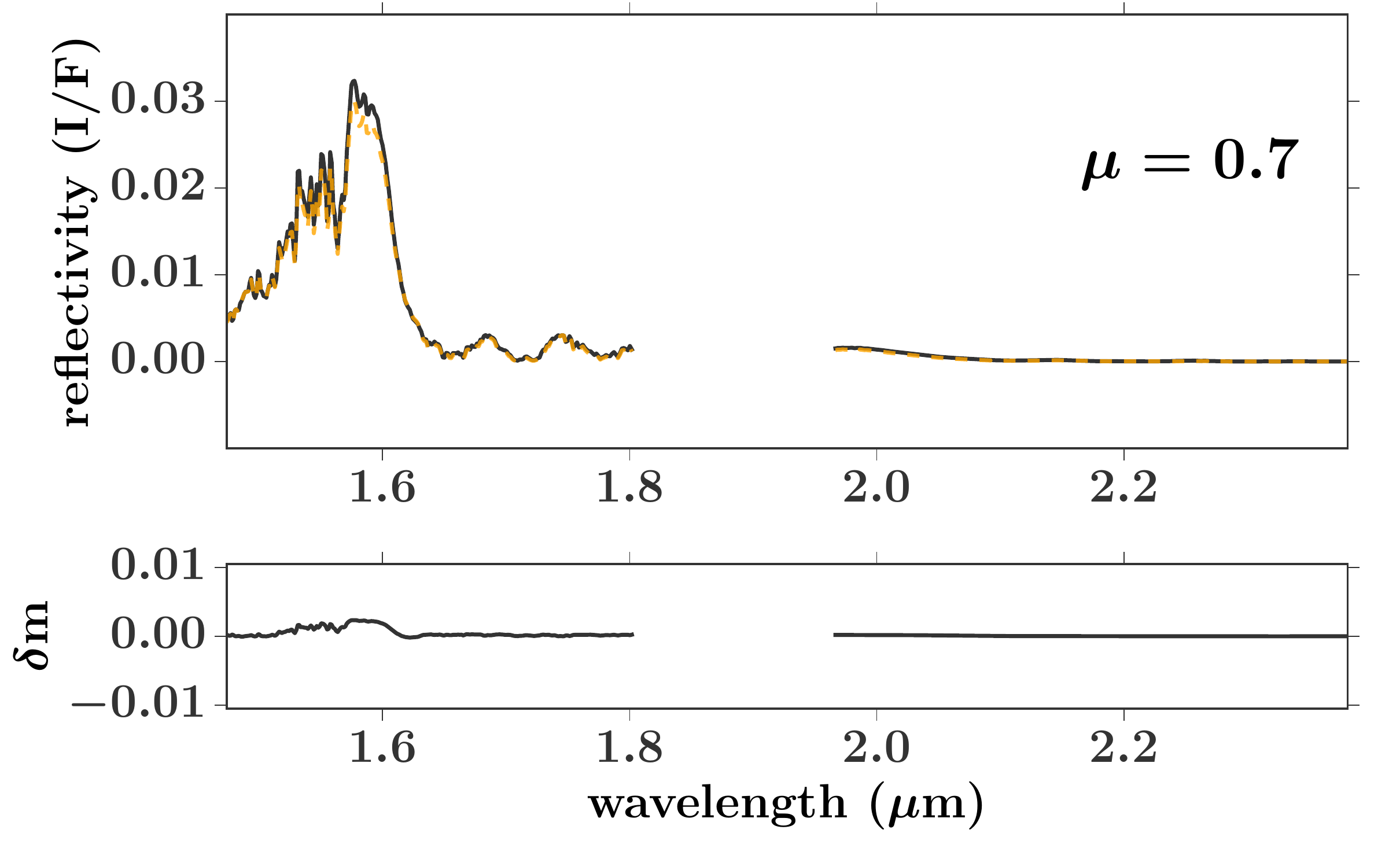}\includegraphics[width=0.45\textwidth]{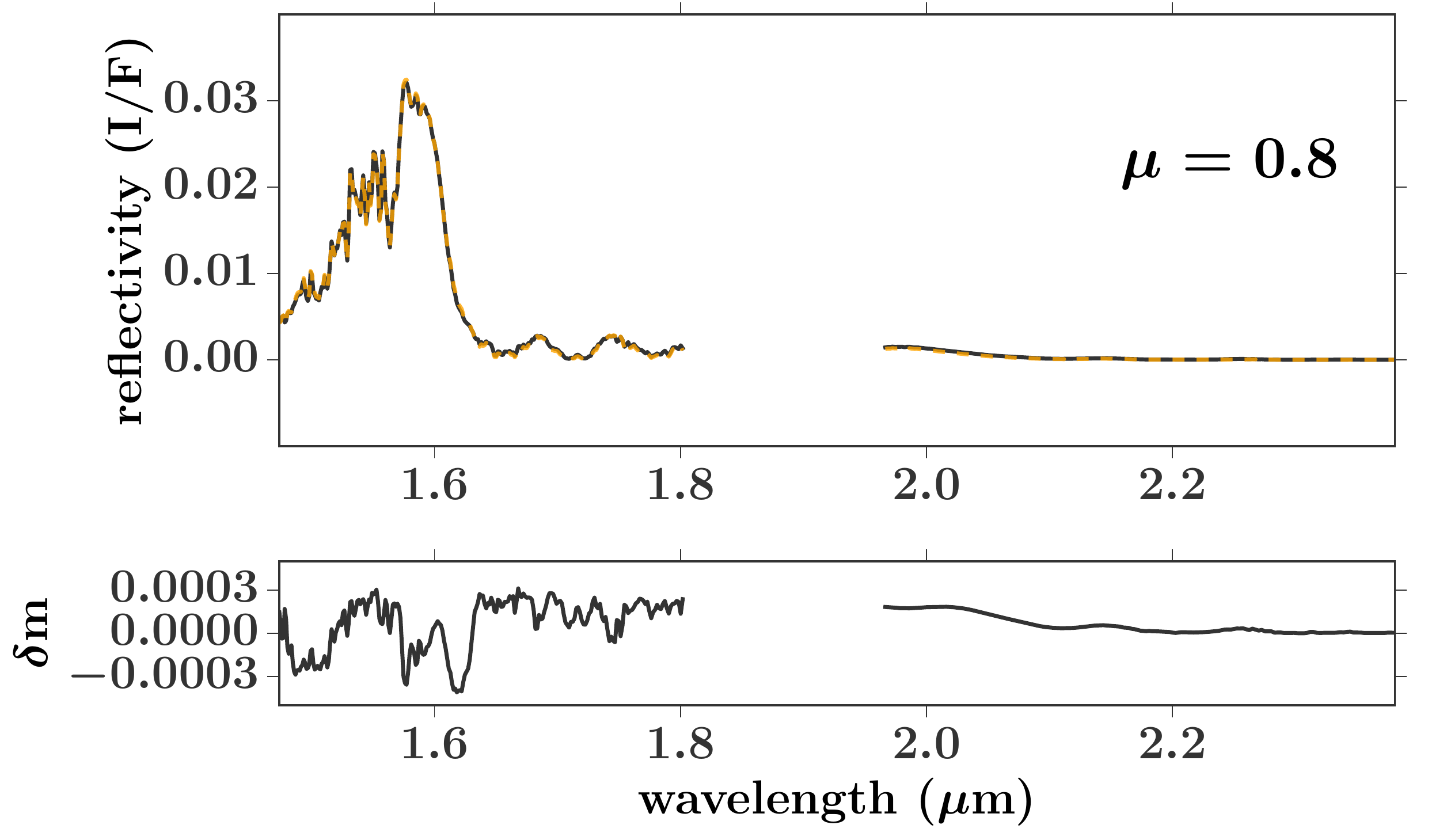}\\
\end{array}$
\end{center}
\caption[Two stream vs. pyDISORT for the {\it 2L\_hazeA}  model]{\label{fig:disort1} \small Comparison of best-fit  {\it 2L\_hazeA} two stream model to the equivalent pyDISORT model. The asymmetry parameters have been adjusted for pyDISORT to best match the $\mu=0.8$ two stream model, as discussed in Appendix \ref{app:disort}. $\delta m$ is defined as $I/F_{disort}- I/F_{twostream}$. Note that the vertical scale of the $\delta m$ plot changes in the last panel, to better highlight the small differences between the $\mu=0.8$ models.  } 
\end{figure}

\begin{figure}
\begin{center}$
\begin{array}{cccccc}
\includegraphics[width=0.45\textwidth]{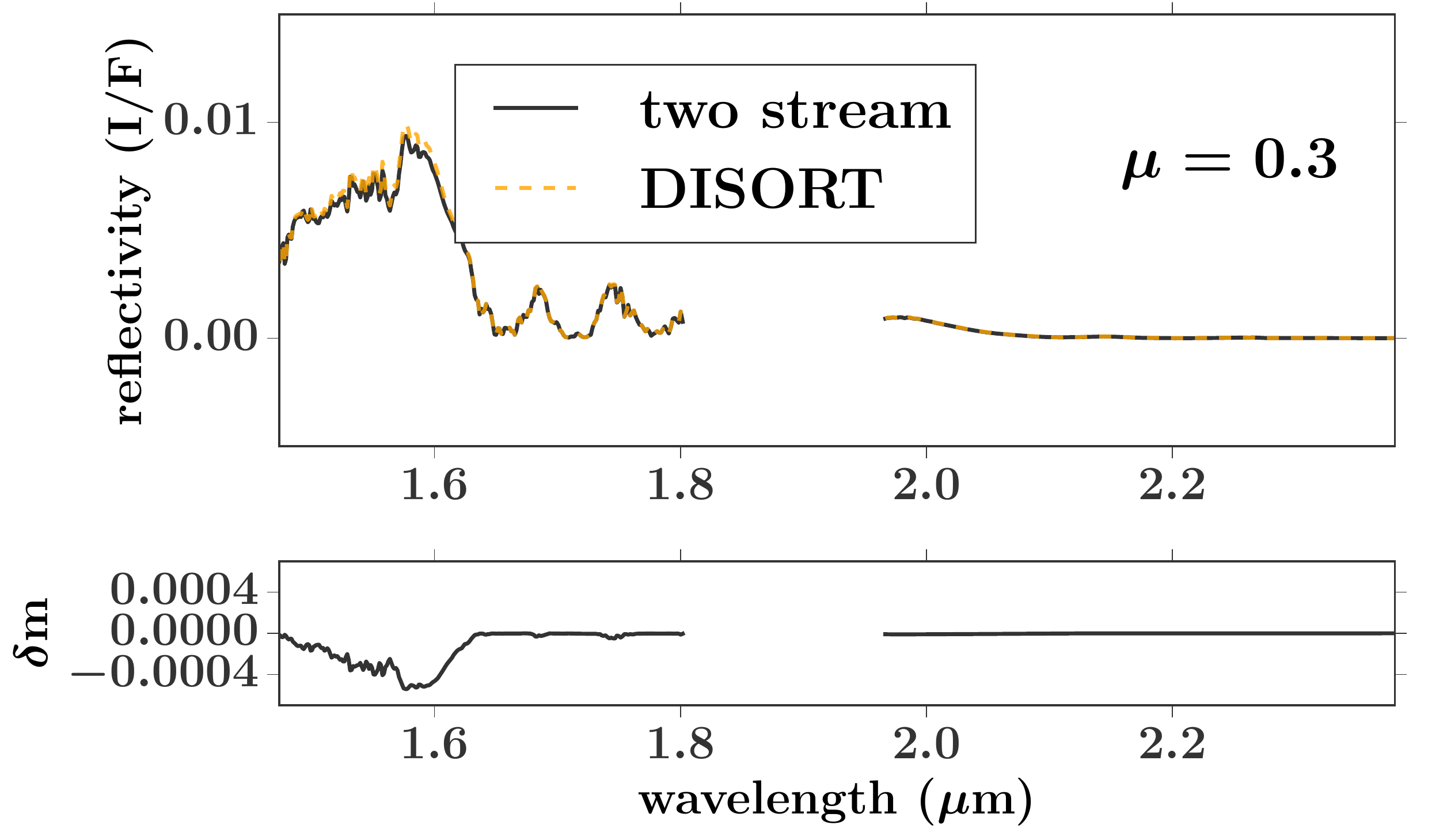} \includegraphics[width=0.45\textwidth]{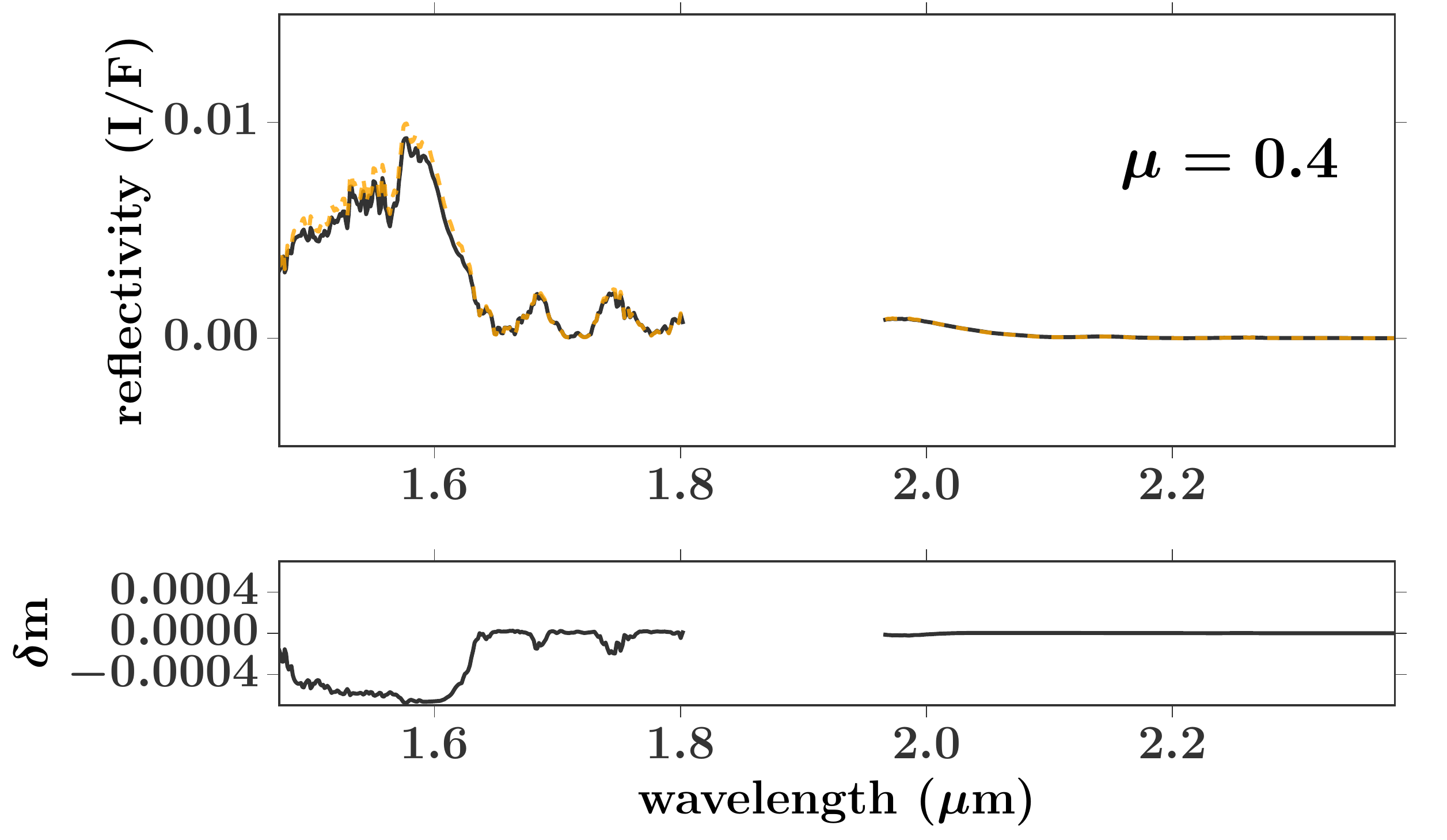}\\
\includegraphics[width=0.45\textwidth]{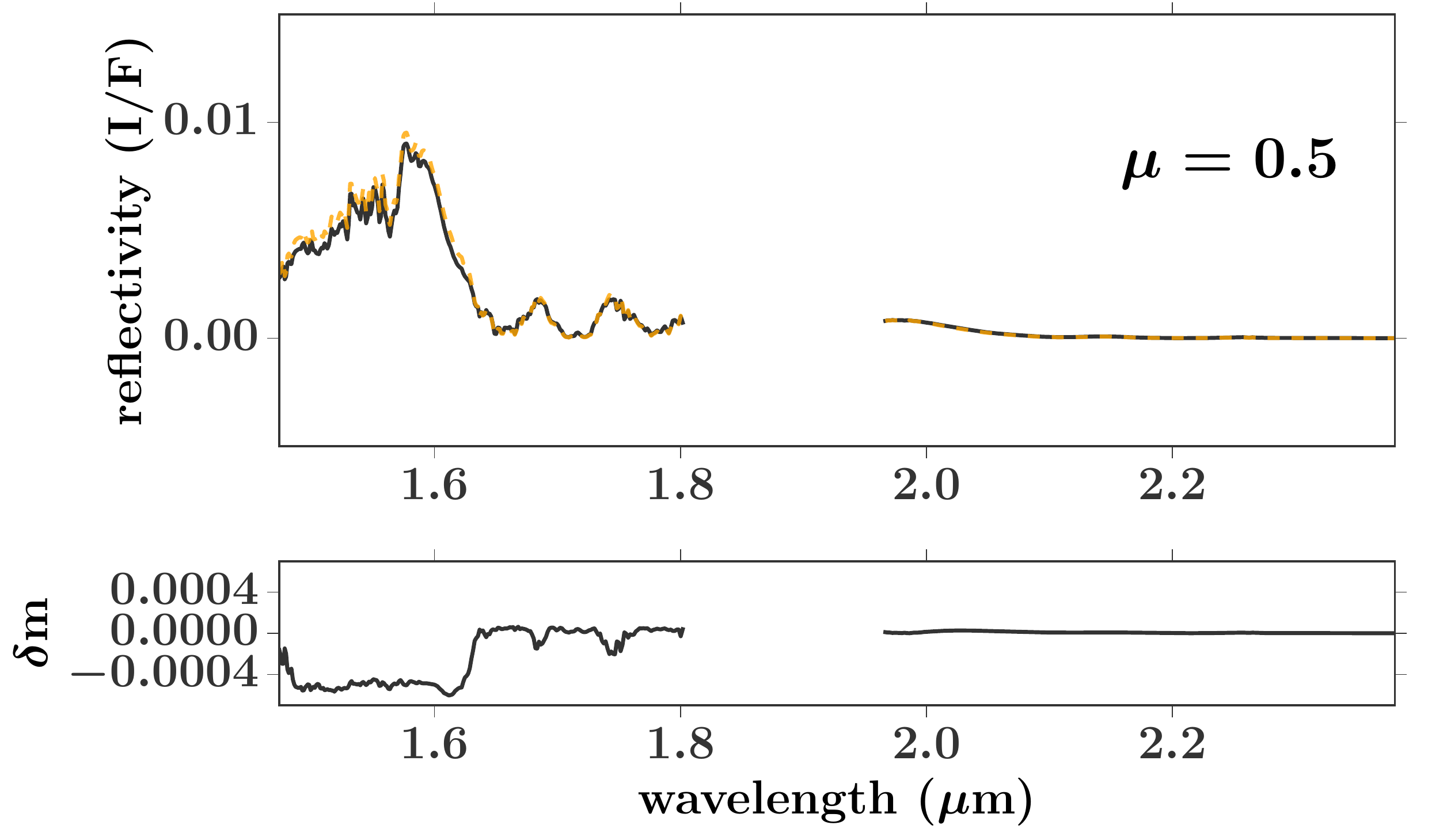}\includegraphics[width=0.45\textwidth]{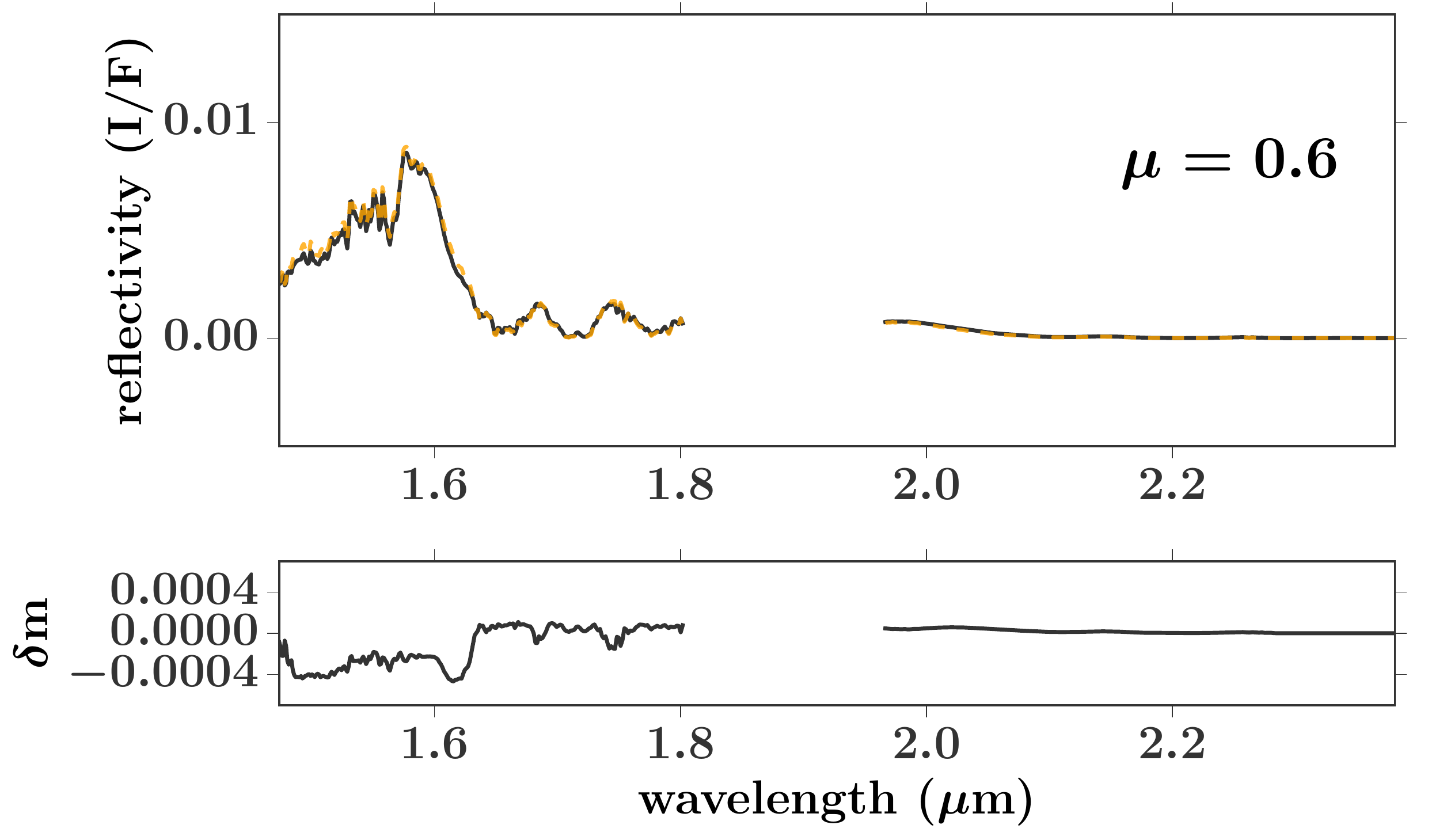}\\
\includegraphics[width=0.45\textwidth]{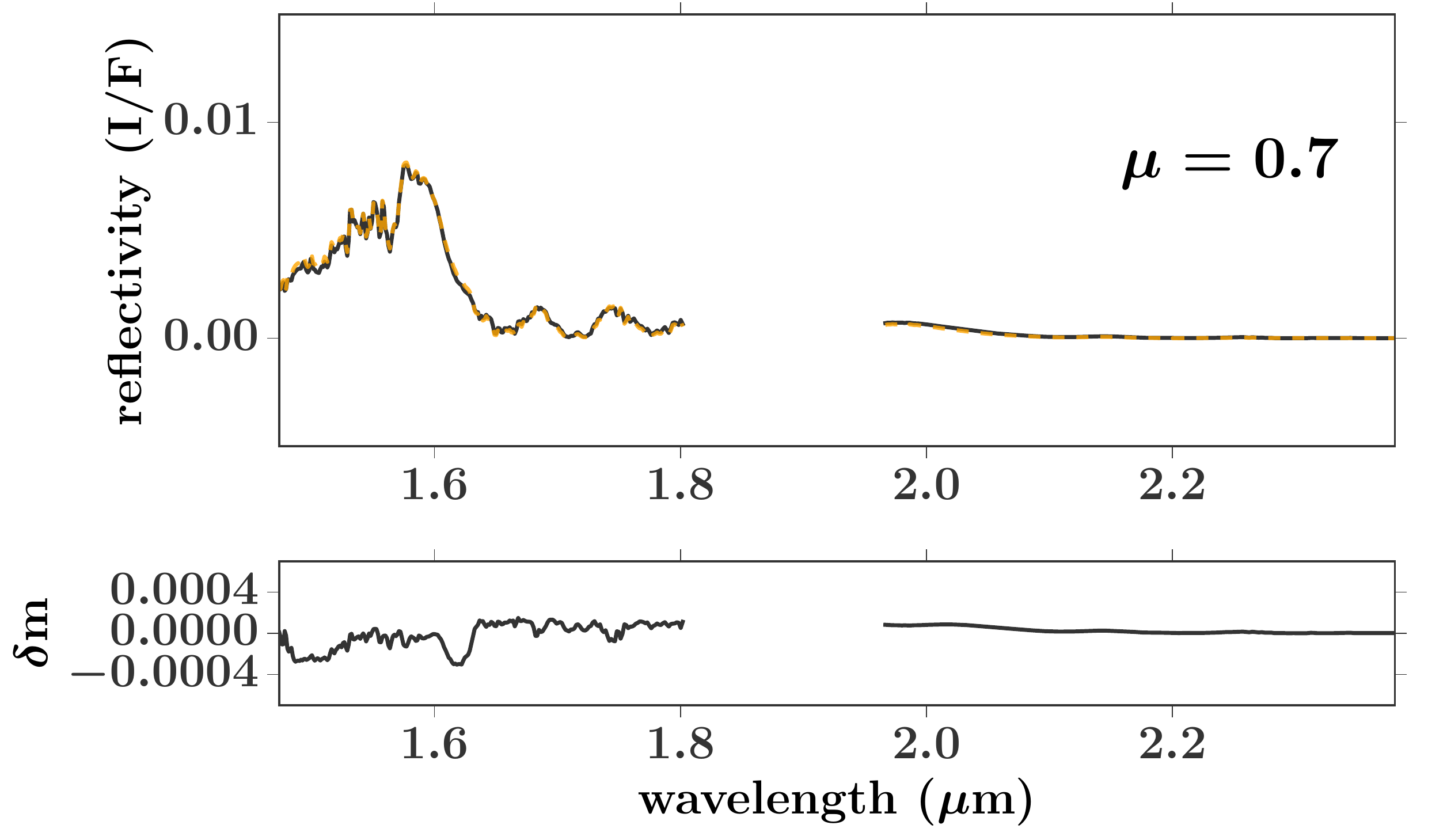}\includegraphics[width=0.45\textwidth]{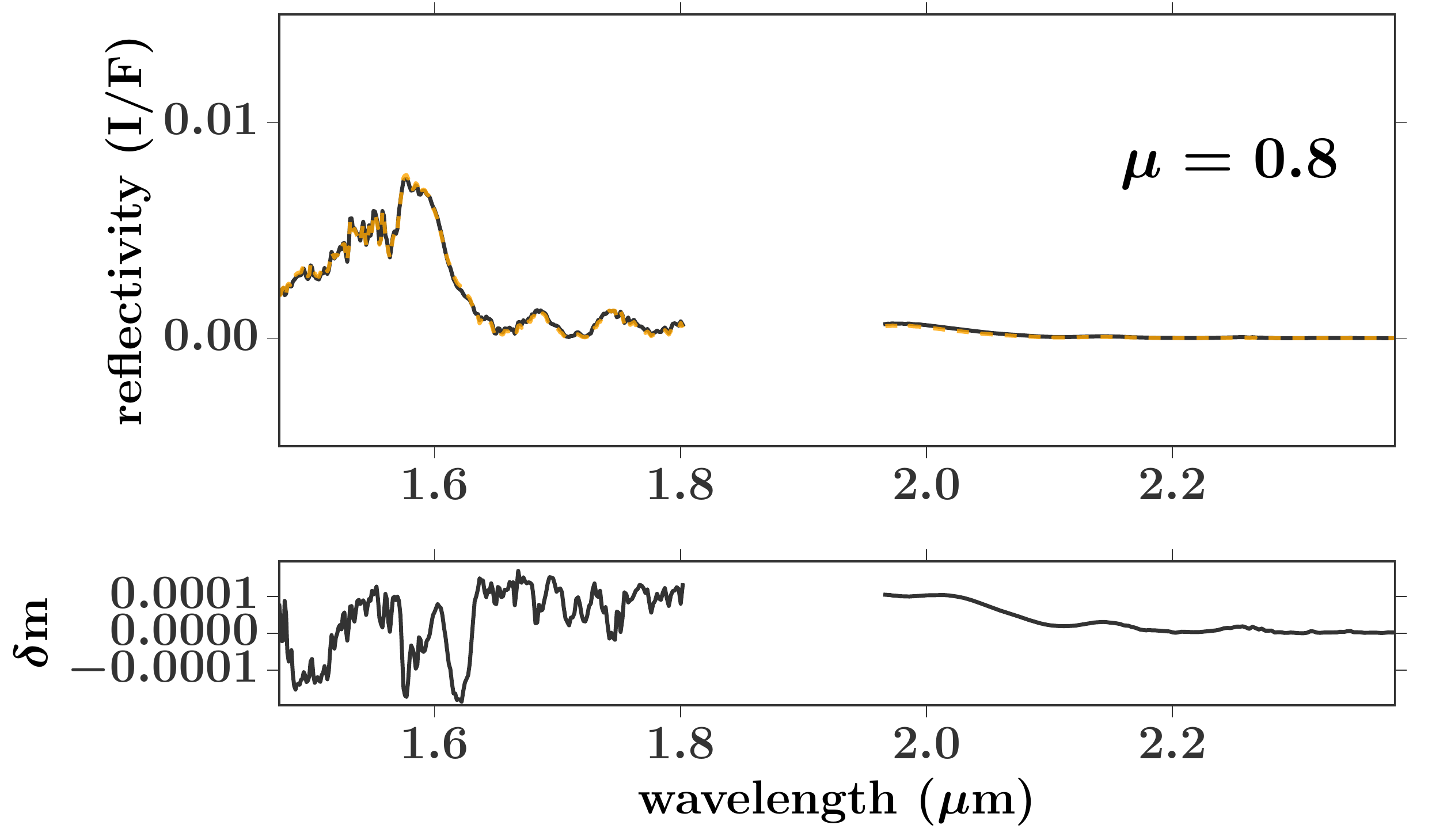}\\
\end{array}$
\end{center}
\caption[Two stream vs. pyDISORT for the {\it 2L\_hazeA}  model with $g\sim0.7$]{\label{fig:disort2}\small  Same as Fig. \ref{fig:disort1}, except $g$ has been set to 0.7 for both layers in the two stream model. The adjusted $g$ values for pyDISORT are $g_{\alpha,D}= 0.7$ and $g_{\beta,D}=0.6$. The scale on $\delta m$ is smaller than in the previous figure. } 
\end{figure}

More generally, with the exception of the largest emission angles, for which we were not able to match the pyDISORT and two stream models, we find that the two stream solution {\it at a single viewing geometry} is a good approximation to the more accurate pyDISORT model, allowing for different values of $g$ for the two algorithms. As emission angle varies, so do the values of  $g_{\alpha,D}$ and $g_{\beta,D}$ that generate the best agreement between the two models. In other words, a single two stream phase function does not match the pyDISORT Henyey-Greenstein phase function for all viewing geometries. This is perhaps not surprising, since the two algorithms treat the phase function differently. To illustrate the divergence of the two stream and pyDISORT solutions, we plot (Fig. \ref{fig:disort1})  the {\it 2L\_hazeA} two stream and pyDISORT models for all six $\mu$ values in the binned dataset, with $g_{\alpha}=0.4$ and $g_{\beta}=-0.6$ for two stream and $g_{\alpha,D}= 0.5$ and $g_{\beta,D}=-0.3$ for pyDISORT. We observe that the match between the two models diverges towards high emission angles, such that by $\mu=0.3$ the fractional difference between models is of order unity. This divergence appears to be much more extreme for the parameter values in this particular ({\it 2L\_hazeA}) model atmosphere than for models containing scatterers with more forward scattering phase functions: when we modify the two stream models such that $g=0.7$ in both aerosol layers, we find a good match to the $\mu=0.8$ model spectra for $g_{\alpha,D}= 0.7$ and $g_{\beta,D}=0.6$. Furthermore,  the resultant two stream and pyDISORT models are in very good agreement at all values of $\mu$ (Fig. \ref{fig:disort2}).

In light of these results, we conclude that while two stream is sufficient for modeling a single spectrum, particularly for locations far from the limb, a more accurate treatment of the radiative transfer is warranted when multiple viewing geometries are considered simultaneously. We emphasize that the two stream asymmetry factor $g$ does not apply for more than a single viewing geometry or to other radiative transfer algorithms, such as pyDISORT. 

\newpage
\section{Supplementary figures}\label{app:figs}

\begin{figure}[htb!]
\begin{center}$
\begin{array}{l r}
\includegraphics[width=0.49\textwidth]{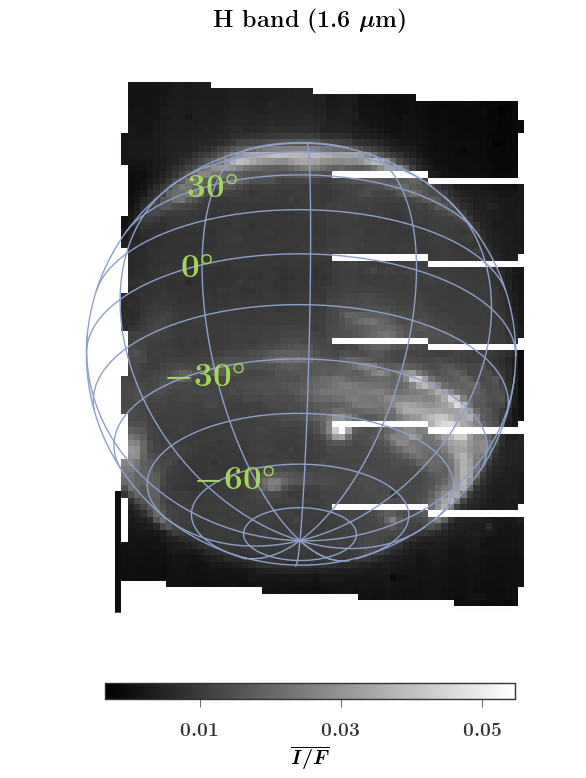}
\includegraphics[width=0.49\textwidth]{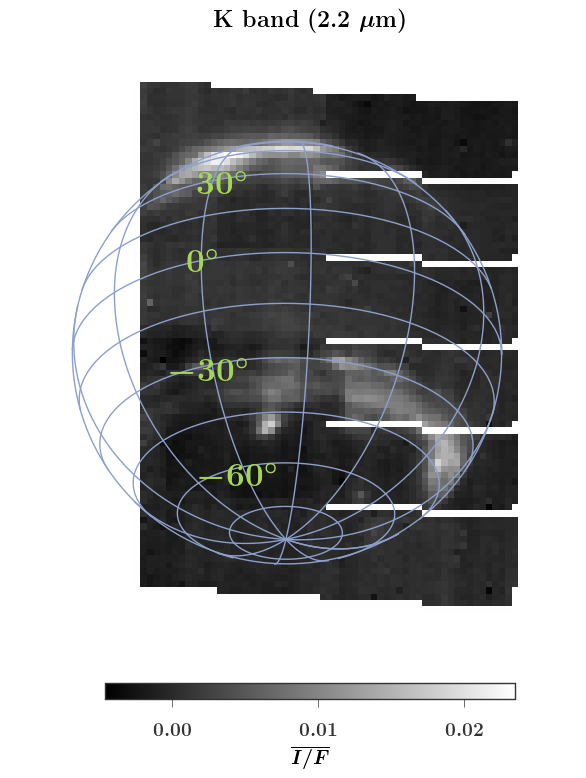}
\end{array}
$
\end{center}
\caption[26 July 2009 data]{\label{fig:cubes}\small Mosaicked and calibrated 26 July 2009 data cubes, averaged over wavelength, as in Fig. \ref{fig:cubes} except without labels. Color bars indicate average reflectance, in units of $I/F$. Pixels which are flagged at all wavelengths are shown in white.  }
\end{figure}
\begin{figure}
\begin{center}$
\begin{array}{ccc}
\includegraphics[width=0.43\textwidth]{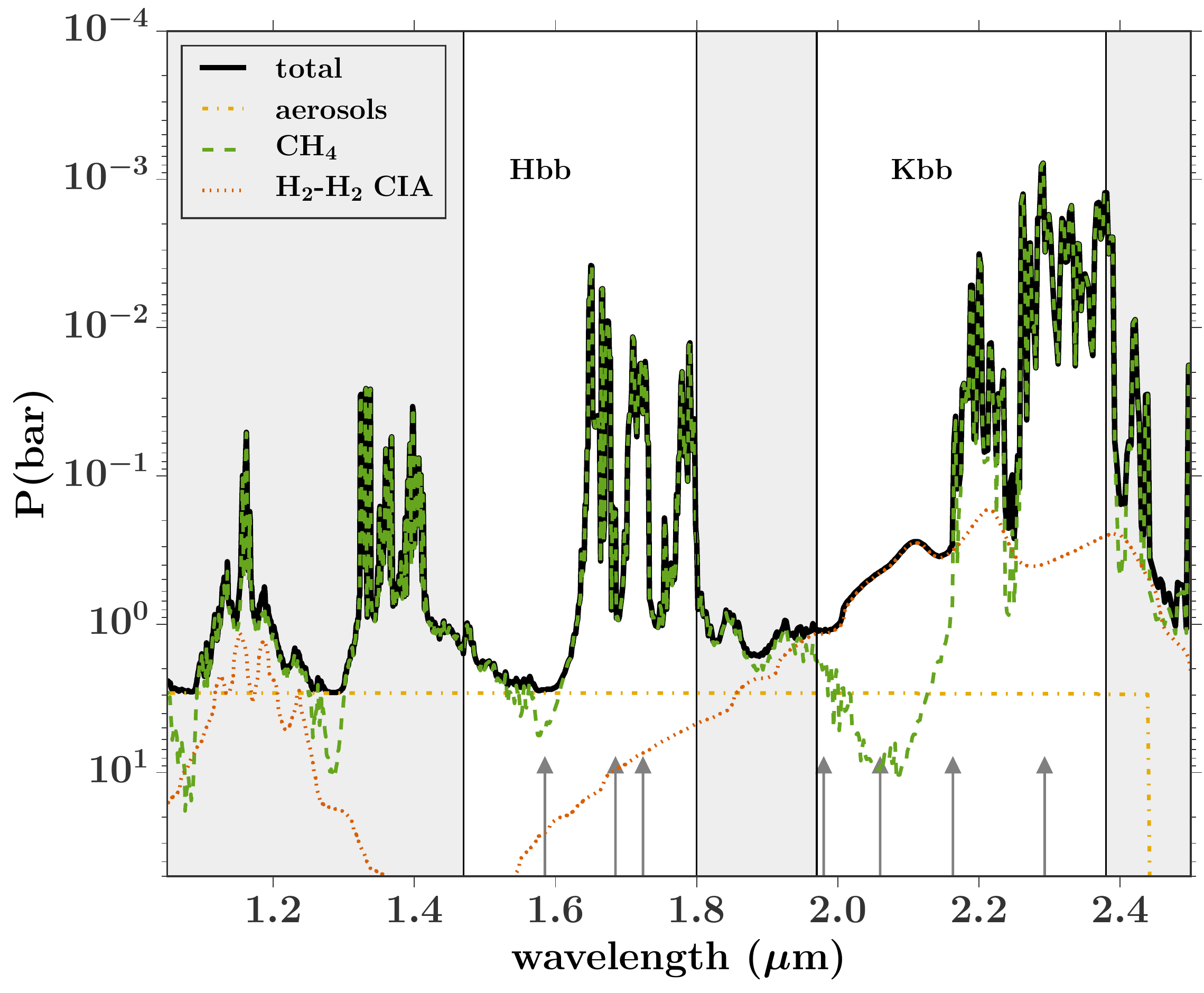} \includegraphics[width=0.28\textwidth]{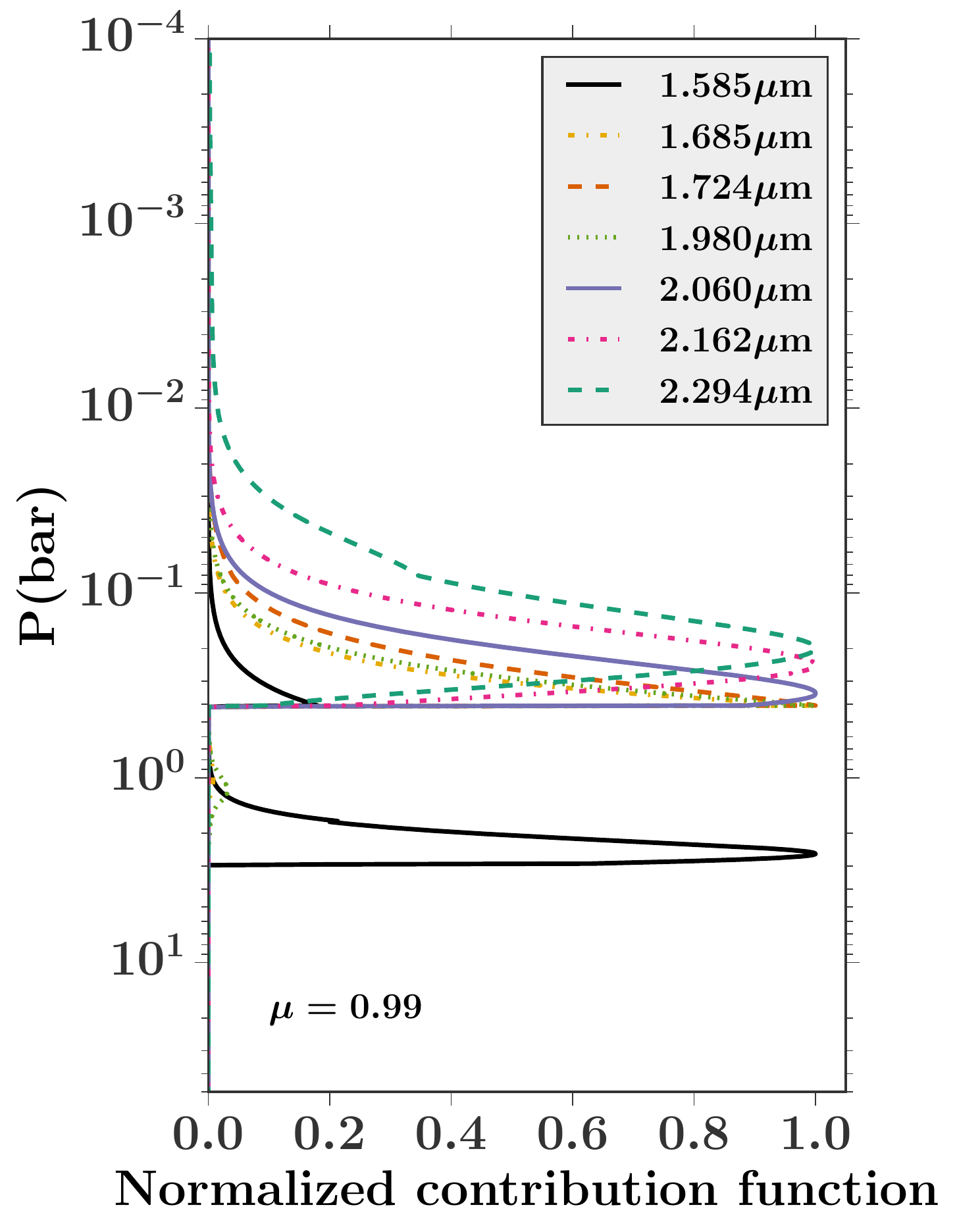}\includegraphics[width=0.28\textwidth]{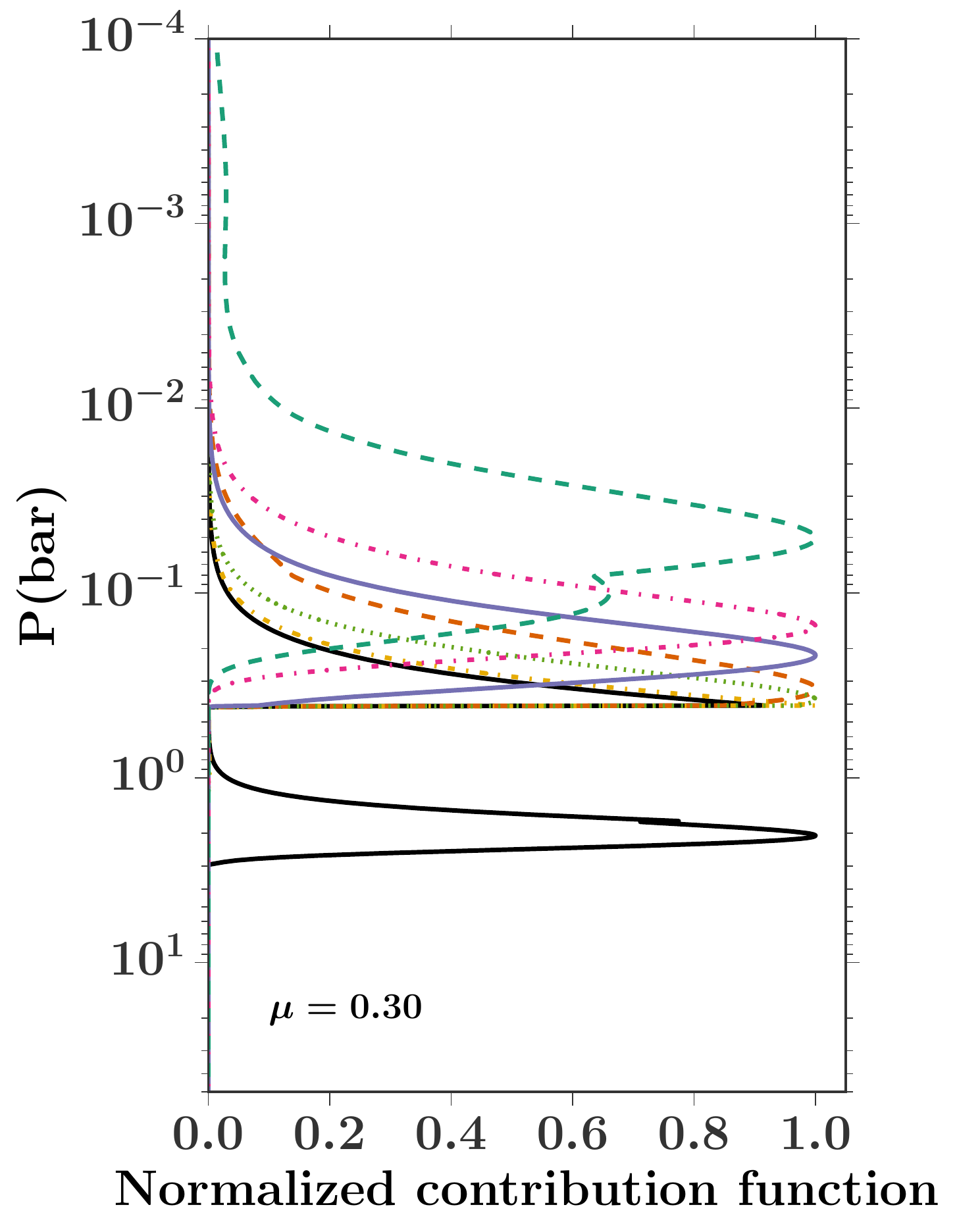}\\

\end{array}$
\end{center}
\caption[Transmission and contribution functions - 2L]{\label{fig:2L_contrib}\small Same as Fig. \ref{fig:1L_contrib}, but for the  {\it SN\_CH$_4$grid$_E$} atmospheric model, described in Section \ref{sec:mu8}. The primary difference in this model from the previous figure is a change in the aerosol distribution: the total aerosol opacity is near 1 at 1.6 $\mu$m, and since the aerosol cross section decreases as a function of wavelength (Fig. \ref{fig:particles}) it falls below 1 at the longest wavelengths, causing the observed falloff of the aerosol curve from the transmission plot.  Note that the contribution functions are very different than in Fig. \ref{fig:1L_contrib}, but there is a similar sensitivity of K-band wavelengths to the highest altitudes, and dependence of the contribution functions on viewing geometry. }
\end{figure}

\begin{figure}[htb!]
\begin{center}$
\includegraphics[width=0.9\textwidth]{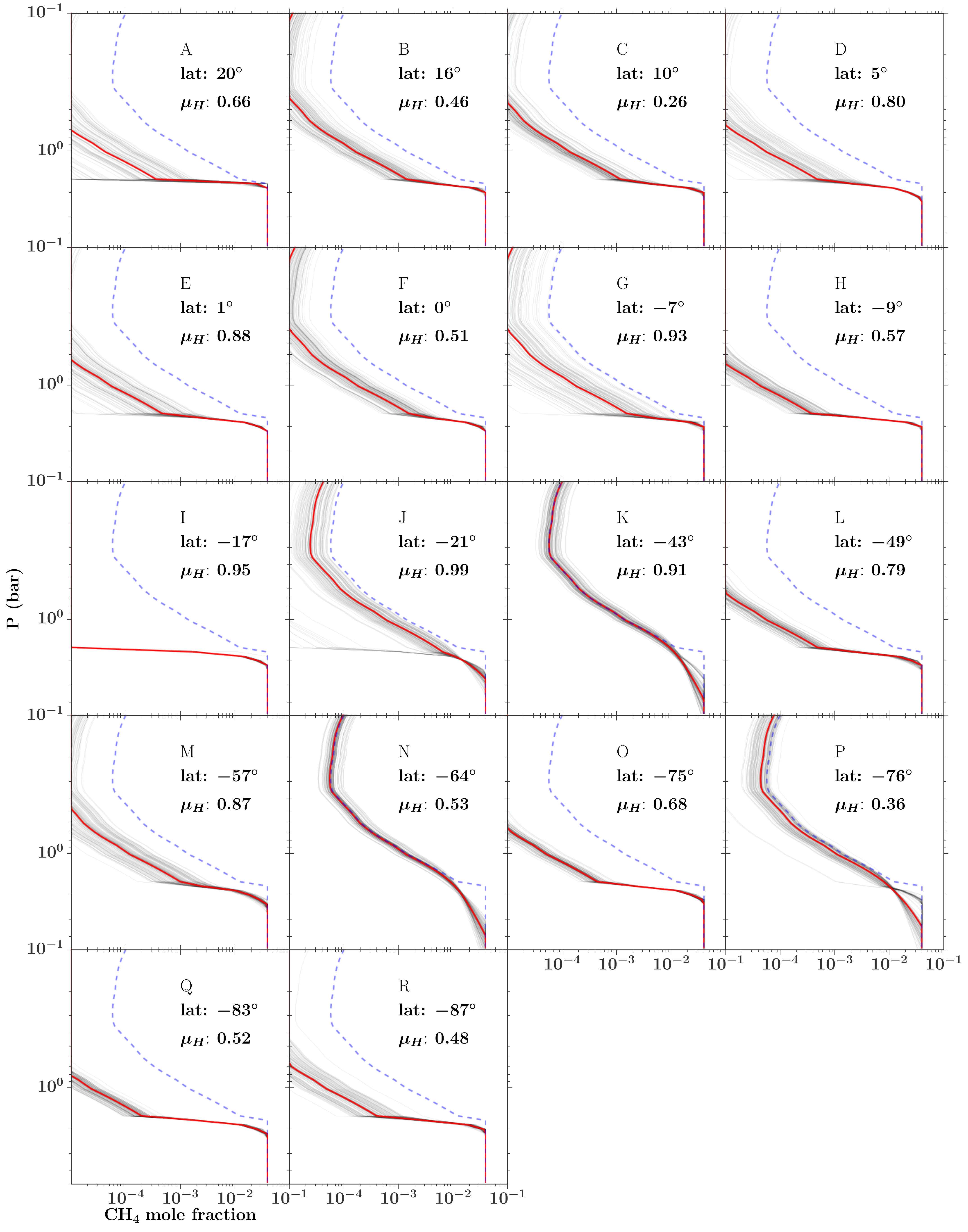}
$
\end{center}
\caption[Methane profile, fits including methane depletion]{\label{fig:CH4fit} \small  Retrieved CH$_4$ profiles, for the case described in Section \ref{sec:depletion}. Color scheme matches the optical depth plots: a thick red line indicates the best-fit model and thin black lines show 100 random solutions form the posterior probability distribution. In this figure, the reference profile, indicated by a blue dashed line, corresponds to the  {\it SN\_CH$_4$grid$_E$} methane profile: this is the profile adopted in the initial aerosol-only retrievals to the 18 locations considered in Section \ref{sec:DR}.    }
\end{figure}

\end{document}